\documentclass[10pt, aps, prx, superscriptaddress, showpacs, nofootinbib, longbibliography, floatfix, reprint]{revtex4-2}

\usepackage[utf8]{inputenc}     
\usepackage[english]{babel}     

\usepackage{amsmath, amssymb, amsthm, mathtools}    
\usepackage{array}
\usepackage{mathrsfs}                               
\usepackage{bbm}                                    
\usepackage{bm}                                     
\usepackage{latexsym}                               
\usepackage{xfrac}                                  
\usepackage{stmaryrd}                               
\usepackage[normalem]{ulem}
\usepackage{thmtools}
\usepackage{thm-restate}

\usepackage{physics}                                
\usepackage{siunitx}                                

\usepackage{booktabs}                               
\usepackage{longtable}                              
\usepackage{multirow}                               
\usepackage{makecell}

\newcolumntype{R}[1]{>{\raggedleft\arraybackslash}p{#1}}
\newcolumntype{C}[1]{>{\centering\arraybackslash}p{#1}}
\newcolumntype{L}[1]{>{\raggedright\arraybackslash}p{#1}}

\usepackage{graphicx}                               
\usepackage{epstopdf}                               
\usepackage{wrapfig}

\usepackage[caption=false]{subfig}
\newcommand{\phantomsubfloat}[1]{
    {%
        \captionsetup[subfigure]{labelformat=empty}
        \subfloat[][]{#1}
    }%
}

\usepackage{enumitem}                               

\usepackage{url}                                    
\usepackage{hyperref}                               
\usepackage{xcolor}

\definecolor{darkblue}{HTML}{043F5C}
\definecolor{lightblue}{HTML}{429EBD}
\definecolor{paleblue}{HTML}{E6F2F6}

\definecolor{darkestorange}{HTML}{934905}
\definecolor{darkerorange}{HTML}{A95406}
\definecolor{darkorange}{HTML}{D06D08}
\definecolor{lightorange}{HTML}{F27F0C}
\definecolor{paleorange}{HTML}{FAEADB}
\hypersetup{
    colorlinks=true,
    linktocpage=true,
    linkcolor=darkerorange,
    citecolor=darkblue,
    urlcolor=darkestorange
}

\usepackage{cleveref}
\crefname{equation}{Eq.}{Eqs.}
\crefname{section}{Sec.}{Secs.}
\crefname{subsection}{Sec.}{Secs.}
\crefname{appendix}{App.}{Apps.}
\crefname{figure}{Figure}{Figures}
\crefname{table}{Table}{Tables}
\crefname{result}{Result}{Results}
\crefname{algorithm}{Algorithm}{Algorithms}

\Crefname{equation}{Equation}{Equations}
\Crefname{appendix}{Appendix}{Appendices}

\crefname{definition}{Definition}{Definitions}
\crefname{proposition}{Proposition}{Propositions}
\crefname{theorem}{Theorem}{Theorems}
\crefname{lemma}{Lemma}{Lemmas}
\crefname{corollary}{Corollary}{Corollaries}
\crefname{remark}{Remark}{Remarks}
\crefname{fact}{Fact}{Facts}
\crefname{example}{Example}{Examples}
\crefname{construction}{Construction}{Constructions}

\crefname{enumi}{}{}
\Crefname{enumi}{}{}

\usepackage{tikz}
\usetikzlibrary{quantikz2}
\usetikzlibrary{tikzmark}

\usepackage{comment}                                
\usepackage{verbatim}                               
\usepackage{cancel}                                 
\usepackage{soul}                                   
\usepackage{indentfirst}                            
\allowdisplaybreaks                                 

\usepackage{tikz}
\usetikzlibrary{quantikz2}
\usetikzlibrary{tikzmark}

\declaretheorem[name=Definition,style=definition]{definition}
\declaretheorem[name=Remark,style=definition]{remark}
\declaretheorem[name=Example,style=definition]{example}
\declaretheorem[name=Fact,style=definition]{fact}

\declaretheorem[name=Proposition,style=definition]{proposition}
\declaretheorem[name=Theorem,style=definition]{theorem}
\declaretheorem[name=Lemma,style=definition]{lemma}
\declaretheorem[name=Corollary,style=definition]{corollary}

\declaretheorem[name=Construction,style=definition]{construction}

\DeclarePairedDelimiter\ceil{\lceil}{\rceil}
\DeclarePairedDelimiter\floor{\lfloor}{\rfloor}

\DeclareMathOperator*{\im}{im}

\DeclareMathOperator*{\diag}{diag}
\DeclareMathOperator*{\supp}{supp}

\DeclareMathOperator*{\rad}{rad}

\DeclareMathOperator*{\spn}{span}

\DeclareMathOperator*{\ord}{ord}

\DeclareMathOperator*{\id}{id}

\DeclareMathOperator*{\rs}{rs}
\DeclareMathOperator*{\cs}{cs}
\newcommand{\trans}[1]{#1^{\mathsf{T}}}

\DeclarePairedDelimiter\db{\llbracket}{\rrbracket}

\newcommand{\Stab}{\mathrm{Stab}}
\newcommand{\Isom}{\mathrm{Isom}}
\newcommand{\PCl}{\mathrm{PCl}} 
\newcommand{\Aut}{\mathrm{Aut}}

\definecolor{denim}{RGB}{59,110,167}

\definecolor{coral}{RGB}{254,125,106}

\newcommand{\bx}{\mathbf{x}}

\newcommand{\cL}{\mathcal{L}}
\newcommand{\ra}{\rangle}
\newcommand{\la}{\langle}
\newcommand{\rmx}{\mathrm{x}}
\newcommand{\rmz}{\mathrm{z}}

\newcommand{\CX}{\mathrm{CX}}
\newcommand{\SWAP}{\mathrm{SWAP}}

\newcommand{\CZ}{\mathrm{CZ}}

\newcommand{\F}{\mathbb{F}}

\newcommand{\GL}{\mathrm{GL}}

\newcommand{\GO}{\mathrm{GO}}
\newcommand{\GU}{\mathrm{GU}}
\newcommand{\Sp}{\mathrm{Sp}}
\newcommand{\SO}{\mathrm{SO}}

\newcommand{\PU}{\mathrm{PU}}

\newcommand{\nocontentsline}[3]{}
\newcommand{\tocless}[2]{\vspace{4ex}\bgroup\let\addcontentsline=\nocontentsline#1{#2}\egroup}

\makeatletter
\renewenvironment{acknowledgments}{%
  \begingroup
  \let\orig@addcontentsline\addcontentsline
  \renewcommand{\addcontentsline}[3]{}%
  \acknowledgments@sw{%
    \expandafter\section\expandafter*\expandafter{\acknowledgmentsname}%
  }{%
    \par
  }%
}{%
  \par
  \endgroup
}
\@booleantrue\acknowledgments@sw
\makeatother

\usetikzlibrary{positioning,arrows.meta,calc}
\usepackage{amssymb}
 
\definecolor{defFill}{HTML}{DBEAFE}\definecolor{defBorder}{HTML}{2563EB}\definecolor{defText}{HTML}{1E3A5F}
\definecolor{propFill}{HTML}{FFE4E6}\definecolor{propBorder}{HTML}{E11D48}\definecolor{propText}{HTML}{881337}
\definecolor{factFill}{HTML}{DCFCE7}\definecolor{factBorder}{HTML}{16A34A}\definecolor{factText}{HTML}{14532D}
\definecolor{remFill}{HTML}{FEF3C7}\definecolor{remBorder}{HTML}{D97706}\definecolor{remText}{HTML}{78350F}
\definecolor{exFill}{HTML}{F3E8FF}\definecolor{exBorder}{HTML}{9333EA}\definecolor{exText}{HTML}{581C87}
 
\tikzset{
  base/.style={rounded corners=3pt, align=center, inner sep=4pt, text width=3.2cm},
  def/.style ={base, draw=defBorder,  fill=defFill,  text=defText,  thick, font=\small},
  prop/.style={base, draw=propBorder, fill=propFill, text=propText, thick, font=\small},
  sat/.style={rounded corners=9pt, align=center, inner sep=3pt, text width=2.9cm, thin, font=\scriptsize},
  fact/.style={sat, draw=factBorder, fill=factFill, text=factText},
  rem/.style ={sat, draw=remBorder,  fill=remFill,  text=remText},
  ex/.style  ={sat, draw=exBorder,   fill=exFill,   text=exText},
  dep/.style={-{Stealth}, semithick},
  anchorline/.style={dash pattern=on 3pt off 2pt, gray!80},
  uses/.style={dotted, thick, -{Stealth}, gray},
}

\usetikzlibrary{shapes.geometric, arrows.meta, positioning, calc}
\usetikzlibrary{backgrounds}
\begin{document}

\setlength{\lineskiplimit}{0pt}
\setlength{\lineskip}{0pt}

\title{Achieving the limits of automorphism gates}

\author{Jin~Ming~Koh} 
\thanks{These authors contributed equally}
\email{jkoh@fas.harvard.edu}
\affiliation{Department of Physics, Harvard University, Cambridge, Massachusetts 02138, USA}
\affiliation{Harvard--MIT Center for Ultracold Atoms, Cambridge, Massachusetts 02138, USA}
\affiliation{The NSF AI Institute for Artificial Intelligence and Fundamental Interactions}
\author{Shayan~Majidy}
\thanks{These authors contributed equally}
\email{smajidy@fas.harvard.edu}
\affiliation{Department of Physics, Harvard University, Cambridge, Massachusetts 02138, USA}
\affiliation{Harvard--MIT Center for Ultracold Atoms, Cambridge, Massachusetts 02138, USA}
\author{Aranya~Chakraborty}
\affiliation{Joint Center for Quantum Information and Computer Science, University of Maryland, College Park, Maryland 20742, USA}
\author{Anqi~Gong}
\affiliation{Institute for Theoretical Physics, ETH Z\"urich, Z\"urich 8093, Switzerland}
\author{Shi~Jie~Samuel~Tan}
\affiliation{Joint Center for Quantum Information and Computer Science, University of Maryland, College Park, Maryland 20742, USA}
\author{Norman~Y.~Yao}
\affiliation{Department of Physics, Harvard University, Cambridge, Massachusetts 02138, USA}
\affiliation{Harvard--MIT Center for Ultracold Atoms, Cambridge, Massachusetts 02138, USA}

\begin{abstract}
    Universal fault-tolerant quantum computing combines versatile but expensive operations with specialized but cheap ones. Its efficiency depends on how much computation can be pushed onto the cheap operations and on the size of the code needed to do so. Automorphism gates provide such cheap operations using only physical single-qubit Clifford gates and qubit permutations. Yet no general theory characterizes their maximum logical power or the minimum code size needed to attain it. We develop such a theory. For stabilizer codes encoding $k\geq3$ logical qubits, we show that the largest logical group attainable by automorphisms is generated by all addressable $S$ and $\mathrm{CX}$ gates, and we construct codes attaining it. While this group contains exponentially fewer gates than the full Clifford group, adding one suitable non-Clifford gate yields universality. We further classify the largest logical groups attainable using qubit permutations, physical single-qubit Cliffords, or both across general stabilizer and CSS codes, and derive refined bounds for self-dual CSS subclasses. Achieving the maximum-size logical group through automorphisms requires $n=\Theta(2^k)$ physical qubits. By contrast, all addressable diagonal Clifford gates, generated by $S$ and $\mathrm{CZ}$, require only $n=\Theta(k^2)$ physical qubits when implemented using physical single-qubit Cliffords alone. Both bounds are tight. This polynomial qubit cost extends beyond Cliffords to all addressable diagonal gates at any fixed level of the Clifford hierarchy, using physical single-qubit diagonal gates. Thus, for full addressability, the sharpest physical-qubit cost divide lies between diagonal and $\mathrm{CX}$-type gates, not between Clifford and non-Clifford gates.
\end{abstract}

\maketitle

\tocless\section{Introduction
\label{sec:intro}}

\begin{figure*}
    \centering
    \includegraphics[width=\linewidth]{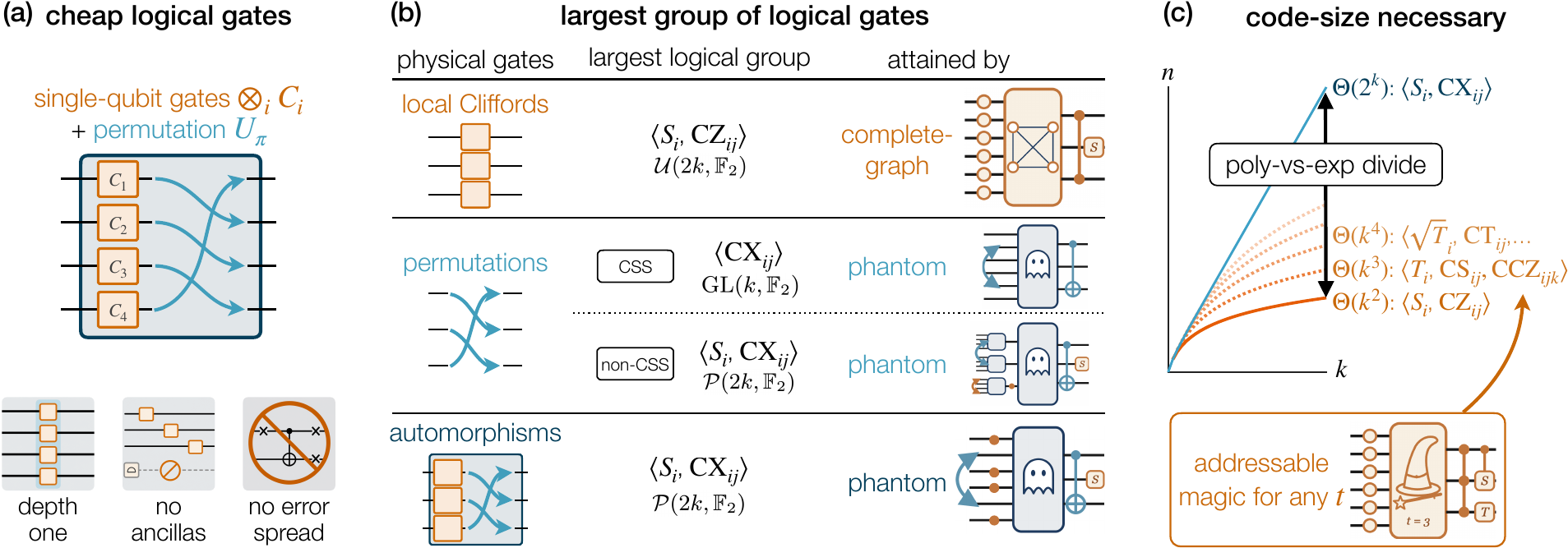}
    \phantomsubfloat{\label{fig:project_snapshot/auto_circuit}}
    \phantomsubfloat{\label{fig:project_snapshot/groups_table}}
    \phantomsubfloat{\label{fig:project_snapshot/costs_order}}
    \vspace{-20pt}
    \caption{\textbf{Largest logical groups from physical single-qubit gates and qubit permutations, and their physical-qubit costs.}
    \textbf{(a)}
    Logical gates implemented using physical single-qubit gates and permutations are cheap: they are depth one, require no ancillas, and do not spread errors when the permutations are performed by rearranging qubits or relabelling them in software.
    \textbf{(b)}
    For single-qubit Cliffords, permutations, and their combination (automorphisms), we show the largest logical group achievable, a generating gate set, and a code construction attaining it; all maxima are tight.
    \textbf{(c)}
    The minimum physical-qubit costs for implementing all addressable logical $S$ and CZ gates using only physical single-qubit Cliffords is $n=k(k+1)/2$. For $k\geq3$, all addressable logical $\mathrm{CX}$ gates via automorphisms, as well as the largest automorphism logical group, require $n=2^k-1$. Beyond Cliffords, complete-hypergraph codes realize all addressable level-$t$ diagonal logical gates using only physical diagonal single-qubit gates with $\Theta(k^t)$ physical qubits, for any fixed $t$.
    }
    \label{fig:project_snapshot}
\end{figure*}

In fault-tolerant quantum computing, there is a trade-off between how versatile an approach is for computation and the resources required to implement it~\cite{majidy2024building,gottesman2010introduction}. At one extreme are universal approaches that perform arbitrary computations but are expensive. Their cost comes from ancilla qubits, the many physical operations involved, and the error rates of those operations. At the other extreme are approaches that reduce these costs but are limited in capability. Universal approaches are indispensable, but using them when cheap operations suffice wastes resources. Modern architectures combine the two, and their efficiency depends on how much computation can be relegated to the cheap operations~\cite{yoder2025tour, webster2026pinnacle, cain2026shor, tripier2026fault}.

Among these cheap operations are logical gates implemented using only physical single-qubit operations and qubit permutations. Such gates require only shallow circuits, no ancillary qubits or measurements, and do not spread errors between physical qubits when permutations are implemented by rearrangement or relabelling~\cite{bluvstein2026fault, dasu2026computing, rosenfeld2025magic}. They also use the lowest-error physical operations available: single-qubit gates typically have the highest fidelities across platforms, while permutations can achieve comparable fidelities on atomic platforms. A particularly important class of cheap operations consists of \emph{automorphism gates}, which use single-qubit Clifford gates and qubit permutations~\cite{calderbank1998quantum, grassl2013leveraging, bravyi2024high}. The Clifford restriction makes codes supporting these logical gates easier to construct analytically or discover numerically than those supporting gates involving arbitrary single-qubit rotations. Automorphism gates have therefore become the principal class of cheap gates studied in recent fault-tolerant architecture proposals~\cite{yoder2025tour, webster2026pinnacle, cain2026shor, tripier2026fault}.

On the other hand, at the versatile but expensive extreme are approaches based on \textit{code surgery}~\cite{litinski2019game, bravyi2005universal, beverland2021cost, he2025extractors, tan2025single, hong2026quantum, bilokur2024thermodynamic, campbell2017roads}. Code surgery entails preparing a resource state and performing rounds of joint measurements between it and one or more error-correcting codes. These procedures require ancillary qubits, multi-qubit measurements, and entangling interactions that spread errors, but can implement universal computation. Because using surgery for an entire computation is expensive, architectures aim to push as much workload as possible onto automorphism gates.

The central question is how far this division of labour can be taken: how much of the logical computation can be performed using automorphism gates, and at what cost to the code parameters? Recent work has produced constructions realizing particular automorphism gates~\cite{xu2025fast,
malcolm2026computing, quintavalle2023partitioning, berthusen2025automorphism, koh2026entangling, sayginel2025fault, mac2026exhaustive}, no-go results for certain logical operations~\cite{guyot2026addressability, tansuwannont2025clifford, dasu2025classification, chakraborty2026nogo}, and bounds on achievable gate sets and code parameters for particular code families~\cite{holmes2026quantum, morris2026constraints}. However, there remains no complete theory that, across all stabilizer codes, determines the largest logical gate group realizable by automorphisms and the minimum number of physical qubits needed to attain it. The former reports on the maximum number of distinct logical operations realizable by automorphism gates, so measures their computational power, and the latter is their cost. Only such a theory can distinguish fundamental limits from limitations of current code constructions.

We develop such a theory. We first determine the largest logical gate group realizable by automorphisms across all stabilizer codes. A key notion is \emph{addressability}: an addressable gate can act on any chosen subset of logical qubits while leaving all others unchanged. We prove that this group is generated by all addressable $S$ and $\mathrm{CX}$ gates, and we construct codes attaining this maximum for any number $k$ of logical qubits. This group contains exponentially fewer gates in $k$ than the full Clifford group. Nevertheless, similar to the full Clifford group, adding a \textit{single} suitable non-Clifford gate yields universality.

We next ask how this boundary changes when we restrict either the physical operations implementing the logical gate or the code structure. Because platforms differ in the convenience and fidelities of physical single-qubit gates and qubit permutations, we consider three operation sets: permutations alone, single-qubit Cliffords alone, and both. The literature also often focuses on the broad class of Calderbank--Shor--Steane (CSS) codes~\cite{calderbank1996good}, which are important in fault-tolerant architectures. We therefore study general stabilizer codes~\cite{gottesman1997stabilizer}, CSS codes, and specialized CSS subclasses. For each combination of physical operations and code class, we determine the largest attainable logical gate group and construct codes that attain it.

We then ask how large a code must be to attain these largest logical gate groups. We show that using automorphisms to realize all addressable $S$ and $\mathrm{CX}$ gates---or even all addressable $\mathrm{CX}$ gates alone---requires $n=\Theta(2^k)$ physical qubits. A more scalable alternative is to restrict the gate set to all addressable $S$ and $\mathrm{CZ}$ gates, which are diagonal in the computational basis. We show that realizing this group of addressable diagonal Clifford gates using physical single-qubit Cliffords alone requires only $n=\Theta(k^2)$ physical qubits; we provide code constructions attaining both bounds. These bounds reveal a sharp separation in qubit cost between all addressable $\mathrm{CX}$ gates and all addressable diagonal Clifford gates.

This cost separation raises a broader question: does it extend beyond the Clifford group? We answer in the affirmative. We further study addressable diagonal non-Clifford gates, including $T$, CS, and CCZ. Codes supporting such logical gates are typically thought to be larger than those supporting only Clifford gates. Yet we prove that all such gates can be realized with only polynomially many physical qubits. This also shows that logical gate-group size does not determine qubit cost: the non-Clifford groups can be much larger than the full $\CX$ group, yet require fewer qubits.

Together, these results reshape the design objective for quantum error-correcting codes. First, they complete a long line of work on the fundamental limits of automorphism gates~\cite{koh2026entangling, sayginel2025fault, berthusen2025automorphism, mac2026exhaustive, guyot2026addressability, tansuwannont2025clifford, dasu2025classification, holmes2026quantum, chakraborty2026nogo, morris2026constraints}. The focus can now shift from determining what automorphisms can realize in principle to identifying useful subsets compatible with practical hardware and code constraints. Second, many logical gates whose attainability was previously unresolved are not merely expensive but impossible through automorphisms at any code size. Third, the sharpest code-size divide is not between Clifford and non-Clifford gates, but between diagonal and $\mathrm{CX}$-type gates. Finally, there is a trade-off between automorphism power and encoding rate $k/n$. When physical qubits are scarce, high-rate codes are desirable, and their native gate sets should therefore be chosen selectively for the target application. When sufficiently many physical qubits are available, however, lower-rate codes may instead be used to obtain richer native gate sets, an option that may become increasingly relevant as architectures scale through modular interconnects~\cite{li2024high, sinclair2025fault}.

\tocless\section{Overview
\label{sec:overview}}

We now summarize our main results in \cref{sub:overview/summary_of_results} and their implications in \cref{sub:overview/implications}. We develop the underlying ideas and intuition in the subsequent sections; formal statements and proofs appear in the appendices.

\tocless\subsection{Summary of main results
\label{sub:overview/summary_of_results}}

Our first two results determine the largest logical groups realizable by automorphisms, and the latter two their physical-qubit costs. \Cref{fig:project_snapshot} summarizes all four.

\par\bigskip
\begin{enumerate}[
    label=\textbf{1.},
    leftmargin=*,
    labelsep=1em,
    topsep=0pt
]
\item \textbf{The maximum power of automorphisms.}
\end{enumerate}
\par\medskip

We fix the number of logical qubits $k$, and ask for the largest group of logical gates realizable by automorphisms on stabilizer codes, with no restriction on code size or distance. For $k\geq 3$, we show that this is the group generated by all addressable logical $S$ and $\mathrm{CX}$ gates.

The bound is tight, as explicit CSS codes attain it for every $k\geq 3$, and allowing general stabilizer codes does not produce a larger group. This maximizing group is unique up to a change of logical basis, which has a conjugation effect on the logical group $G \rightarrow CG C^{\dagger}$, where $G$ is the logical group and $C$ is a logical Clifford.

Here and throughout, Clifford gates that differ only by Paulis or a global phase are considered as the same element. The maximum-size logical groups at $k=1$ and $k=2$ are exceptional: for $k=1$, all single-qubit Clifford gates are attainable, giving a maximum of six elements; for $k=2$, some Clifford gates become unattainable and the maximum is $72$ elements.

\par\bigskip
\begin{enumerate}[
    label=\textbf{2.},
    leftmargin=*,
    labelsep=1em,
    topsep=0pt
]
\item \textbf{How restrictions on physical operations and code structure change the maximum.}
\end{enumerate}
\par\medskip

In addition to identifying the maximum logical power of automorphism gates, we explore a number of finer-grained restrictions. In particular, we consider qubit permutations alone, physical single-qubit Cliffords alone, and automorphisms combining the two, across general stabilizer codes, CSS codes, and two \emph{self-dual} CSS subclasses whose code structure is restricted by symmetries. Our classifications for single-qubit Cliffords on stabilizer and CSS codes, and automorphisms on CSS codes, apply to \emph{indecomposable} codes---those that cannot be split into independent codes on disjoint sets of physical qubits.

For CSS codes that are not self-dual, the largest logical gate groups take a simple form: all addressable $S$ and $\mathrm{CZ}$ gates for single-qubit Cliffords, all addressable $\mathrm{CX}$ gates for permutations, and both when the two are combined.

Dropping the CSS restriction leaves the maximum unchanged for single-qubit Cliffords, and for automorphisms for $k\geq3$. But, it allows permutations alone to attain the automorphism maximum of $\langle S_i, \mathrm{CX}_{ij}\rangle$.

\begin{figure}
    \centering
    \includegraphics[width=\linewidth]{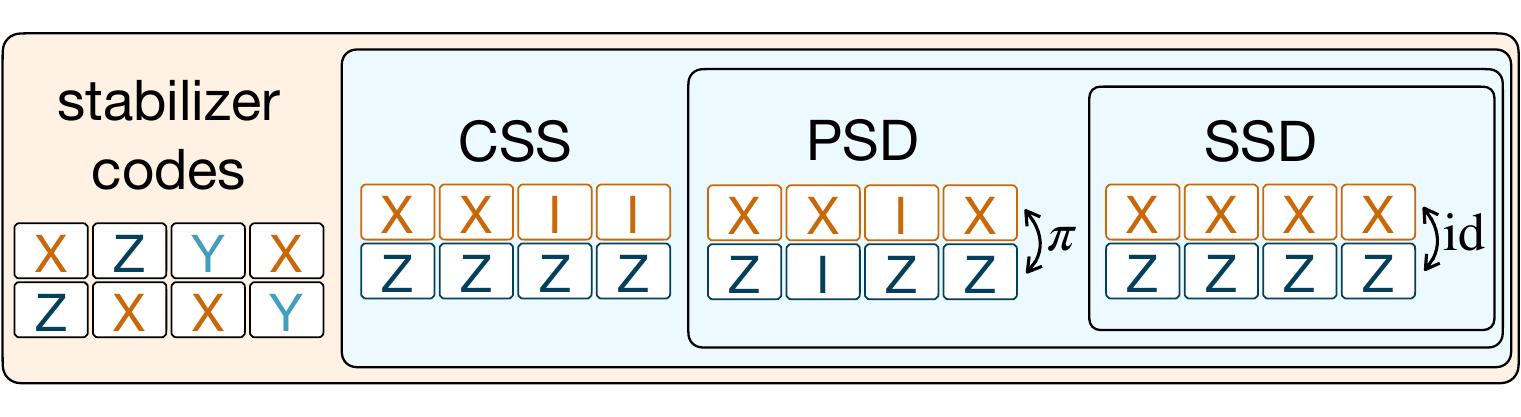}
    \caption{\textbf{Classes of codes.}
    CSS codes admit separate $X$- and $Z$-type stabilizers. We distinguish permutationally self-dual (PSD) codes, whose $X$- and $Z$-stabilizer spaces coincide up to a qubit permutation, and strictly self-dual (SSD) codes, for which they coincide directly.}
    \label{fig:nested_groups}
\end{figure}

Within CSS codes, imposing self-duality restricts the available gates. We distinguish \emph{permutationally self-dual} (PSD) codes, whose $X$- and $Z$-stabilizer spaces coincide up to a qubit permutation, and \emph{strictly self-dual} (SSD) codes, for which they coincide directly (\Cref{fig:nested_groups}). \Cref{tab:summary_of_bounds} summarizes the resulting bounds for pairs of physical operations and code class. All bounds are attained by code constructions at every $k$ except for permutations and automorphisms on the self-dual subclasses.

\par\medskip
\begin{enumerate}[
    label=\textbf{3.},
    leftmargin=*,
    labelsep=1em,
    topsep=0pt
]
\item \textbf{How many physical qubits are required to attain these maxima.}
\end{enumerate}
\par\medskip

Our work also determines the explicit number of physical qubits (code size) necessary to support each maximum-size logical gate group. Any stabilizer code supporting all addressable logical $S$ and $\mathrm{CZ}$ gates using single-qubit Cliffords must have $n\geq k+\binom{k}{2}=k(k+1)/2$. By contrast, for $k\geq3$, any stabilizer code supporting all addressable logical $\mathrm{CX}$ gates through automorphisms must have $n\geq 2^k-1$; the same bound holds when these gates are implemented by permutations alone. Since the maximum automorphism group identified above contains this full set of logical $\mathrm{CX}$ gates, it inherits the same exponential lower bound.

We meet both bounds with explicit constructions. These lower bounds hold at every distance and are tight at $Z$-distance one. The codes supporting these logical gate groups, even at distance one, are genuine encodings rather than collections of bare physical qubits: single-qubit physical gates induce entangling logical gates, an effect that is impossible without encoding. Both gate groups remain attainable at any distance by concatenating the distance-one codes with suitable inner codes.

\par\medskip
\begin{enumerate}[
    label=\textbf{4.},
    leftmargin=*,
    labelsep=1em,
    topsep=0pt
]
\item \textbf{The polynomial-versus-exponential cost divide persists beyond the Clifford group.}
\end{enumerate}
\par\medskip

As stated, realizing all addressable logical diagonal Clifford gates requires quadratically many physical qubits, whereas realizing all addressable logical $\mathrm{CX}$ gates requires exponentially many. We investigate whether this polynomial scaling persists for diagonal magic gates higher in the Clifford hierarchy. At the third level, for example, we find that the qubit cost of supporting all addressable logical $T$, $\mathrm{CS}$, and $\mathrm{CCZ}$ gates is only $\Theta(k^3)$.

More generally, fix any Clifford hierarchy level $t\leq k$ and consider all addressable diagonal gates at that level. Any stabilizer code realizing this gate group with physical diagonal single-qubit gates must have $ n\geq\sum_{s=1}^{t}\binom{k}{s}=\Theta(k^t)$.
This bound is attained by a code construction with $Z$-distance one. As in the preceding result, these distance-one codes are genuine encodings, and concatenation gives arbitrary-distance codes.

\tocless\subsection{Implications
\label{sub:overview/implications}}

\begin{table}[!t]
    \centering
    \setlength{\tabcolsep}{3.5pt}
    \renewcommand{\arraystretch}{1.25}
    \newcommand{\stack}[2]{\makecell{#1\\[2pt]\footnotesize #2}}
    \begin{tabular}{llccc}
        \toprule
        \multicolumn{2}{c}{\textbf{Code class}} & \multicolumn{3}{c}{\textbf{Physical gate class}} \\
        \cmidrule(l){3-5}
        & & Transversal
        & Permutation
        & Automorphism
        \\
        \midrule
        \multicolumn{2}{l}{\textbf{Stabilizer}}
            & \stack{$\mathcal{U}(2k, \mathbb{F}_2)$}{\cref{thm:stab_codes_trans_logical_groups}$^{\dag}$}
            & \stack{$\mathcal{P}(2k,\mathbb{F}_2)$}{\cref{thm:stab_codes_perm_logical_groups}}
            & \stack{$\mathcal{P}(2k,\mathbb{F}_2)$}{\cref{thm:stab_codes_auto_logical_max_order_k_geq_3}$^{\ddag}$} \\
        \midrule[\heavyrulewidth]
        \multirow{8}{*}{\rotatebox[origin=c]{90}{\textbf{CSS}}}
          & General
            & \stack{$\mathcal{U}(2k, \mathbb{F}_2)$}{\cref{thm:css_codes_trans_logical_groups_non_ssd_indecomp}}
            & \stack{$\GL(k, \mathbb{F}_2)$}{\cref{thm:css_codes_perm_logical_group}}
            & \stack{$\mathcal{P}(2k,\mathbb{F}_2)$}{\cref{thm:css_codes_auto_logical_groups}} \\
        \cmidrule(l){2-5}
        & PSD
            & \stack{$\mathcal{U}(2\lfloor k/2\rfloor, \mathbb{F}_2)^2$}{\cref{thm:css_codes_trans_logical_groups_non_ssd_indecomp}}
            & \stack{
                $\begin{gathered}
                \mathcal{F}^{\Sp}_m(k, \mathbb{F}_2) \\[-2pt]
                \text{or }
                \mathcal{F}^{O}_m(k, \mathbb{F}_2)
                \end{gathered}$}
                {\cref{thm:css_codes_perm_logical_group_psd}}
            & \stack{
                $\begin{gathered}
                \mathrm{Trans}_\mathcal{L}(\mathcal{C}) \rtimes \\[-2pt]
                \mathrm{Perm}_\mathcal{L}(\mathcal{C}).2
                \end{gathered}$}
                {\cref{thm:css_codes_auto_logical_groups}} \\
        \cmidrule(l){2-5}
        & SSD
        & \stack{$S_3$}{\cref{thm:css_codes_trans_logical_groups_ssd_indecomp}}
        & \stack{
            $\begin{gathered}
            \Sp(k, \mathbb{F}_2) \\[-2pt]
            \text{or } 
            O(k, \mathbb{F}_2)
            \end{gathered}$}
            {\cref{thm:css_codes_perm_logical_group_psd}$^\ast$}
        & \stack{
            $\begin{gathered}
            \mathrm{Trans}_\mathcal{L}(\mathcal{C}) \rtimes \\[-2pt]
            \mathrm{Perm}_\mathcal{L}(\mathcal{C})
            \end{gathered}$}
            {\cref{thm:css_codes_auto_logical_groups}} \\
        \bottomrule
    \end{tabular}
    \caption{\textbf{Largest logical groups by code and physical gate class.}
    $\mathrm{Aut}_\mathcal{L}(\mathcal{C})$, $\mathrm{Trans}_\mathcal{L}(\mathcal{C})$, and $\mathrm{Perm}_\mathcal{L}(\mathcal{C})$ denote the automorphism, transversal, and permutation logical groups of code $\mathcal{C}$ in logical basis $\mathcal{L}$; the theorem establishing each entry is listed below it. Transversal gates are implemented by tensor products of physical single-qubit Cliffords, and automorphism gates by single-qubit Cliffords and qubit permutations. Rows: general stabilizer, general CSS, permutationally self-dual (PSD) CSS, and strictly self-dual (SSD) CSS codes. Exceptions at $k = 1$ and $k \le 2$ are marked by $^\dag$ and $^\ddag$, respectively; $^\ast$ marks non-attainability at $k \ge 9$. For brevity, the code $\mathcal{C}$ is assumed to be indecomposable, and the CSS code subclasses in this table are understood as mutually exclusive: generic codes are non-PSD, while PSD codes are non-SSD. Group notation is defined in the main text and \cref{app:preliminaries}.
    }
    \label{tab:summary_of_bounds}
\end{table}

Our results inform the design of fault-tolerant architectures and direct research priorities.

First, this work completes a line of research on the logical power of automorphism gates by establishing an exact extremal theory. Earlier constructions and no-go results~\cite{guyot2026addressability, tansuwannont2025clifford, dasu2025classification, holmes2026quantum, chakraborty2026nogo, morris2026constraints} left open whether the largest known automorphism gate groups reflected merely limitations of code design or fundamental ones. Our results remove this ambiguity across stabilizer codes and shift the focus toward determining which intermediate logical gate sets are attainable and at what cost.

Second, our work suggests that automorphism gate sets should be selected for utility, not necessarily maximized, when using high-rate codes. Indeed, we show that supporting all addressable $S$ and $\mathrm{CX}$ gates through automorphisms forces the code rate to vanish quickly as the number of logical qubits $k$ per codeblock grows. While rich gate sets may be worthwhile for codes encoding a moderate number of logical qubits~\cite{koh2026entangling}, high-rate codes must instead target the automorphism gates that replace the most expensive procedures for a given workload and hardware platform. A narrowly scoped automorphism gate set, when carefully chosen, can dramatically reduce the overhead for universal fault-tolerant architectures~\cite{yoder2025tour, xu2025fast}. The relevant objective for high-rate code design is therefore not gate-set size but architectural benefit relative to the underlying code overhead. However, lower-rate codes with richer gate sets may become an increasingly viable route as architectures scale~\cite{li2024high, sinclair2025fault}. 

Third, our classifications yield several no-go results for logical gates implemented using automorphisms, physical single-qubit Cliffords alone, or qubit permutations alone. Code size and structure place further restrictions on the achievable logical groups. Notable consequences are:\footnote{The no-gos for CSS codes naturally assume a CSS logical basis is used---that is, the physical representatives of $X$- and $Z$-logical operators comprise $X$- and $Z$-physical Paulis.}
\begin{itemize}[before=\vspace{6pt},leftmargin=*,labelsep=0.75em,itemsep=0pt]
    \item No addressable logical $H$, $HS$, and $SH$ gates on $k > 1$ indecomposable CSS codes from automorphisms.
    \item PSD, and SSD, code structure is required for logical $H^{\otimes k}$, $(HS)^{\otimes k}$, $(SH)^{\otimes k}$ gates on CSS codes from automorphisms, and single-qubit Cliffords, respectively.
    \item No logical $\mathrm{CX}$ and $\mathrm{SWAP}$ gates on CSS codes from single-qubit Cliffords; and no logical $S$ and $\mathrm{CZ}$ gates from qubit permutations.
    \item High-rate $n=\Theta(k)$ stabilizer codes are far from attaining the largest logical groups. For example, their logical groups from single-qubit Cliffords are smaller by a factor of ${\sim}2^{k^2/2}$ than the largest possible.
\end{itemize}
These obstructions constrain the design space of fault-tolerant platforms. For example, the lack of Hadamard-type logical gates through automorphisms implies that architectures must use another mechanism to perform addressable basis-change gates, or divide their logical qubits among smaller blocks. 

\begin{figure}
    \centering
    \includegraphics[width=\linewidth]{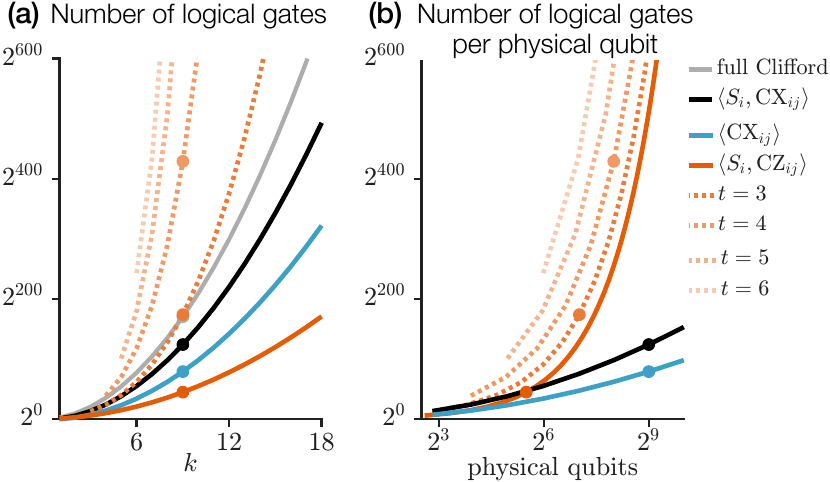}
    \caption{\textbf{Number of logical gates in each logical group.}
    Solid lines (other than grey) denote tight logical groups for automorphism, transversal, and permutation gates; dotted lines denote additional diagonal logical groups achievable by code constructions at level $t$ of the Clifford hierarchy. We plot the number of logical gates (i.e.~order of the logical groups) against the number of logical qubits $k$ in \textbf{(a)} and the minimum required number of physical qubits in \textbf{(b)}. For scale, markers indicate $k=9$ on each curve when the corresponding point lies within the displayed range.}
    \label{fig:cost_and_order}
\end{figure}

Lastly, Cliffordness or the number of logical gates alone is a poor guide to qubit cost. The Clifford-versus-non-Clifford distinction remains central to quantum computational power but does not reliably predict physical-qubit cost. At any fixed level of the Clifford hierarchy, the full set of addressable diagonal gates requires only polynomially many physical qubits, while the full set of addressable $\mathrm{CX}$ gates requires exponentially many. These non-Clifford groups are also much larger (\cref{fig:cost_and_order}). The overhead is thus sharply governed by the structure of the logical action, not by whether the gates are Clifford or how many are realized.

\tocless\section{CSS code results
\label{sec:css_codes}}

We begin with CSS codes, whose stabilizer generators and logical Pauli representatives can be chosen to comprise purely $X$ or $Z$ physical Pauli operators. This structure underlies the practical advantages of CSS codes, such as the availability of transversal interblock $\mathrm{CX}$ gates, construction from classical codes, and simpler syndrome extraction, decoding and state preparation~\cite{steane1997active, delfosse2022toward, tan26syndrome, berthusen2025adaptive, manes2025distance, tan2025effective, majidy2025scalable, roberts2026cored, liu2026achieving, tan2026generalized, hong2025single, wang2026multi}. The same structure makes their automorphism gates particularly transparent---namely, every automorphism gate of an indecomposable CSS code factorizes into a transversal layer and a permutation-containing part, \emph{each} a valid logical gate. This reduces their classification to the two constituent gate classes. 

In summary, we determine the largest logical groups realizable by transversal single-qubit Clifford gates, qubit permutations, and their combination (i.e.~automorphisms). For generic CSS codes, the three maxima are generated, respectively, by all addressable logical $S$ and $\mathrm{CZ}$ gates, all addressable logical $\mathrm{CX}$ gates, and all addressable logical $S$ and $\mathrm{CX}$ gates. We refine this classification across the non-PSD, PSD-but-not-SSD, and SSD classes of CSS codes shown in \cref{fig:nested_groups}, with the resulting bounds summarized in \cref{tab:summary_of_bounds}.

We first establish the factorization property (\cref{sec:css_codes/method}), then classify the transversal and permutation logical groups (\cref{sec:css_codes/trans,sec:css_codes/perm}) and combine them for automorphisms (\cref{sub:css_auto}), before extending the analysis beyond within-block Clifford gates (\cref{sec:css_codes/extensions}). Henceforth, for brevity, we denote on $k$ logical qubits the set of all addressable $S$, $\mathrm{CX}$, and $\mathrm{CZ}$ logical gates by $\mathbf{S}_k$, $\mathbf{CX}_k$, and $\mathbf{CZ}_k$, respectively.

\tocless\subsection{Automorphism structure
\label{sec:css_codes/method}}

We work in the binary symplectic representation. We label an $n$-qubit Pauli by $(x\,{\mid}\,z) \in \mathbb{F}_2^n \oplus \mathbb{F}_2^n$, where the vectors $x$ and $z$ record which qubits carry an $X$ and a $Z$ component; we call these the $X$ and $Z$ sectors. For example, on one qubit, $Y = iXZ$ has label $(1\,{\mid}\,1)$. Up to phases, a Clifford acts linearly on these labels, so it is a symplectic matrix of four $n\times n$ sector-to-sector blocks,
\begin{equation}
    \begin{pmatrix} A & C \\ B & D \end{pmatrix}.
    \label{eq:sec_to_sec}
\end{equation}
This matrix representation identifies Cliffords up to Paulis, matching the projective convention mentioned in \cref{sec:overview}.

We classify Cliffords by which blocks of \cref{eq:sec_to_sec} are invertible (\cref{def:preserving_exchange_multi_qubit_cliffords}). We call a Clifford \emph{preserving-type} when both diagonal blocks are invertible and \emph{exchange-type} when both off-diagonal blocks are invertible. This classification is dichotomous for single-qubit Cliffords: $I$, $S$, and $HSH$ are preserving-type, while $H$, $HS$, and $SH$ are exchange-type.

This classification enables a simple all-or-nothing rule for exchange-type physical Cliffords (\cref{lem:css_codes_auto_all_or_nothing_exchange_cliffords}). We call a stabilizer code \emph{decomposable} if it comprises two independent codes on disjoint qubit subsets, and \emph{indecomposable} otherwise. Write an automorphism of an indecomposable CSS code as $\overline{U}=U_\pi\bigotimes_{i=1}^n C_i$, where $U_\pi$ is the qubit-permutation unitary and each $C_i$ a single-qubit Clifford, and let $E$ be the subset of qubits whose $C_i$ is exchange-type. Applying $\overline{U}$ and its inverse forces the projection of each $X$- and $Z$-type stabilizer generator onto $E$ to remain a stabilizer of the same type, splitting the code across the cut between $E$ and its complement. Indecomposability forbids the split, so $E$ is empty or all $n$ qubits: the $C_i$ are exchange-type on every qubit or on none.

By this rule, every automorphism of an indecomposable CSS code factorizes into a permutation-containing part and a transversal Clifford layer, \emph{each itself} a valid logical gate (\cref{lem:css_codes_auto_physical_splitting_property}). In the case where the transversal Clifford gates are all preserving-type, the two logical gates are simply $U_\pi$ and $\bigotimes_i C_i$. Otherwise, in the all-exchange case, inserting $H^{\otimes n}H^{\otimes n}=I$ and regrouping gives the logical gates $U_\pi H^{\otimes n}$ and $\bigotimes_i H_i C_i$. This latter case is possible only for PSD codes: $H^{\otimes n}$ exchanges the $X$- and $Z$-stabilizer groups and $U_\pi$ relabels qubits, so $U_\pi H^{\otimes n}$ preserves the code only if the two groups coincide up to relabelling. In both cases, the transversal layer is preserving-type: immediately so in the all-preserving case, and in the all-exchange case because each $H_i C_i$ composes two single-qubit exchange-type gates to yield a preserving-type one.

If the code is not PSD, only the all-preserving case occurs, so every automorphism factors into a physical qubit permutation and a transversal Clifford layer. On a code $\mathcal{C}$ with logical basis $\mathcal{L}$, the logical actions of permutation and transversal gates form the groups $\mathrm{Perm}_\mathcal{L}(\mathcal{C})$ and $\mathrm{Trans}_\mathcal{L}(\mathcal{C})$. Their product is \emph{semidirect} for two reasons. First, the two groups meet only in the identity, so the factorization is unique. Second, conjugating a transversal layer by a permutation only relabels its single-qubit Cliffords, so permutations normalize transversal layers. Hence the automorphism logical group is (\cref{thm:css_codes_auto_logical_groups})
\begin{equation}
    \mathrm{Aut}_\mathcal{L}(\mathcal{C}) = \mathrm{Trans}_\mathcal{L}(\mathcal{C}) \rtimes \mathrm{Perm}_\mathcal{L}(\mathcal{C}).
    \label{eq:css_main_factorization}
\end{equation}
This reduces the analysis of automorphism logical groups to determining the transversal and permutation ones.

An SSD code yields the same factorization, \cref{eq:css_main_factorization}, even though automorphisms with all exchange-type Cliffords can now occur. Here $H^{\otimes n}$ is itself a valid logical gate, so removing it from $U_\pi H^{\otimes n}$ and $\bigotimes_i H_i C_i$ leaves valid permutation and transversal logical gates. Hence, the logical action of every automorphism still factorizes into an element of $\mathrm{Trans}_\mathcal{L}(\mathcal{C})$ and of $\mathrm{Perm}_\mathcal{L}(\mathcal{C})$.

A PSD-but-not-SSD code requires a minor amendment to \cref{eq:css_main_factorization}. Here $H^{\otimes n}$ is not a valid logical gate, so it cannot be removed from $U_\pi H^{\otimes n}$, leaving an exchange-type logical action outside the semidirect product. The correct accounting is an index-two extension of the semidirect product: replace $\mathrm{Perm}_\mathcal{L}(\mathcal{C})$ with $\mathrm{Perm}_\mathcal{L}(\mathcal{C}).2$.

\tocless\subsection{Transversal gates
\label{sec:css_codes/trans}}

We now classify the transversal logical group $\mathrm{Trans}_\mathcal{L}(\mathcal{C})$. On an indecomposable non-SSD code, every transversal gate is preserving-type on each physical qubit. This is because, by the all-or-nothing rule, the only alternative is an all-exchange layer, which swaps the $X$- and $Z$-stabilizer groups and is possible only on SSD codes. In a CSS logical basis this fixes the diagonal blocks to $I$, while symplecticity demands that the off-diagonal blocks be symmetric. Thus the logical action has the form
\begin{equation}
    \begin{pmatrix}
        I & C \\
        B & I
    \end{pmatrix},
    \qquad
    B=B^\top,\quad C=C^\top.
    \label{eq:css_trans_term}
\end{equation}
That is, a transversal gate acts trivially on logical Paulis within a sector but can couple sectors. Moreover, since the product of two transversal gates is itself transversal, the allowed $B$ and $C$ blocks form spaces that must mutually annihilate---$BC=CB=0$ for every pair.

The largest possible group occurs when one off-diagonal block vanishes. This is the unipotent radical of the Siegel parabolic,
\begin{equation}
    \mathcal{U}(2k,\mathbb{F}_2) = \left\{
    \begin{pmatrix}
        I & 0 \\
        B & I
    \end{pmatrix}: B=B^\top \right\}.
    \label{eq:css_trans_unipotent_radical_standard_form}
\end{equation}
In fact, up to Clifford conjugation dependent on the logical basis, every logical action lies in the unipotent radical (\cref{thm:css_codes_trans_logical_groups_non_ssd_indecomp}),
\begin{equation}
     \mathrm{Trans}_{\mathcal{L}}(\mathcal{C}) 
     \leq 
     \mathcal{U}(2k,\mathbb{F}_2)
     \cong
     \langle \mathbf{S}_k, \mathbf{CZ}_k \rangle.
     \label{eq:css_trans_logical_group_unipotent_radical}
\end{equation}
As written, the unipotent radical is generated by all addressable logical $S$ and $\mathrm{CZ}$ gates, and has order
\begin{equation}
    |\mathcal{U}(2k,\mathbb{F}_2)| = 2^{k(k+1)/2}.
    \label{eq:trans_order}
\end{equation}
This bound is tight, attained by a code construction in \cref{sec:constructions} for every $k$.

Transversal gates cannot, however, realize CX or SWAP logical actions in any CSS logical basis (\cref{corr:css_codes_trans_no_cx_swap}): these gates act nontrivially on logical Paulis within a sector, which \cref{eq:css_trans_term} forbids. These logical gates, which generate the linear subgroup of the Siegel parabolic, must therefore come from qubit permutations.

PSD-but-not-SSD codes cannot attain the maximum of \cref{eq:css_trans_logical_group_unipotent_radical}. Here the exchange-type self-duality $U_\pi H^{\otimes n}$ swaps the two logical Pauli sectors, and conjugating a transversal gate by it yields another transversal gate---so the $B$- and $C$-spaces are images of each other, hence of equal dimension. The annihilation condition then bounds their common dimension to $\ell \le\lfloor k/2\rfloor(\lfloor k/2\rfloor+1)/2$. Although the spaces are tied, each gate picks its $B$ and $C$ independently, so the group has up to $2\ell$ free binary parameters (\cref{thm:css_codes_trans_logical_groups_non_ssd_indecomp}). Its order is at most $2^{\lfloor k/2\rfloor(\lfloor k/2\rfloor+1)}$, versus the non-PSD maximum $2^{k(k+1)/2}$. At maximal $\ell$, a suitable CSS logical basis splits the maximum-size transversal logical group across two disjoint subsets $a$ and $b$ of $\lfloor k/2\rfloor$ logical qubits each:
\begin{equation}
    \langle\mathbf S_{a},\mathbf{CZ}_{a}\rangle
    \times
    H_{b}\langle\mathbf S_{b},\mathbf{CZ}_{b}\rangle H_{b},
\end{equation}
where $H_{b}\coloneqq\bigotimes_{i\in b}H_i$, and $\mathbf S_{a}$ and $\mathbf{CZ}_{a}$ denote the groups generated by all addressable $S$ and CZ gates on the qubits in $a$, respectively. This bound is also tight: a CSS code construction attains it---see \cref{sec:constructions}.

SSD codes face the strongest limit, and one different in form from \cref{eq:css_trans_logical_group_unipotent_radical}. A transversal gate could, a priori, apply different single-qubit Cliffords on different qubits. On an indecomposable SSD code, however, differing Cliffords would split the code into independent pieces, contradicting indecomposability (\cref{lem:trans_exchange_uniformity_indecomp_css_codes}). So every qubit must carry the \emph{same} Clifford, up to Paulis. All six Clifford choices are in fact admissible on any indecomposable SSD code, and the transversal logical group is exactly the six-element $\Sp(2,\mathbb{F}_2)\cong S_3$, independent of $k$ (\cref{thm:css_codes_trans_logical_groups_ssd_indecomp}). In a CSS logical basis, this group is generated by $S^{\otimes k}$ and $H^{\otimes k}$, up to extraneous $S$, $\mathrm{CZ}$ and $\mathrm{CX}$ logical actions fixed by the logical basis. These extraneous actions are absent when each $X$- and $Z$-logical representative has odd overlap only with itself.

For $k\geq 2$, strict self-duality shrinks the maximum transversal group order from $\abs{\mathcal{U}(2k,\mathbb{F}_2)} \geq 8$ to $\abs{S_3}=6$. Moreover, $S_3$ lies in no unipotent radical, as it contains both preserving- and exchange-type logical actions. At $k=1$, $S_3$ is the full projective Clifford group, so strict self-duality instead enlarges the transversal group.

\tocless\subsection{Permutation gates
\label{sec:css_codes/perm}}

We next classify the permutation logical group $\mathrm{Perm}_\mathcal{L}(\mathcal{C})$. The logical $X$ and $Z$ sectors are complementary \emph{Lagrangian subspaces}: each is a maximal subspace on which the logical symplectic form vanishes, and the two intersect trivially. On a CSS code, these sectors have pure $X$- and $Z$-type physical representatives. Because qubit permutations preserve physical Pauli types, they preserve both logical Lagrangians and map each sector onto itself. The logical action then has the form
\begin{equation}
    \begin{pmatrix}
        A & 0 \\
        0 & A^{-\top}
    \end{pmatrix},
    \qquad
    A\in\GL(k,\mathbb{F}_2),
    \label{eq:css_perm_term}
\end{equation}
where symplecticity fixes the $Z$-sector action $A^{-\top}$. That is, a permutation gate can act nontrivially on logical Paulis within a sector but cannot couple sectors---the mirror image of a transversal gate. Up to Clifford conjugation dependent on the logical basis, \cref{eq:css_perm_term} implies
\begin{equation}
    \mathrm{Perm}_\mathcal{L}(\mathcal{C}) 
    \leq 
    \GL(k,\mathbb{F}_2)
    \cong
    \langle \mathbf{CX}_k \rangle.
\end{equation}
As written, the linear group is generated by all addressable logical CX gates and has order
\begin{equation}
    \abs{\GL(k,\mathbb F_2)}
    =
    2^{k(k-1)/2} \prod_{i=1}^k (2^i - 1),
    \label{eq:perm_order}
\end{equation}
which exceeds $\abs{\mathcal{U}(2k,\mathbb F_2)}$ by a factor $\Theta(2^{k(k-1)/2})$. This bound is tight, attained by \emph{phantom codes}~\cite{koh2026entangling} where qubit permutations alone realize every logical CX.

Permutation gates cannot, however, realize the unipotent radical part of the parabolic in any CSS logical basis. A logical $S$ or CZ gate has nonzero off-diagonal blocks, which \cref{eq:css_perm_term} forbids. The unipotent radical must instead come from transversal gates.

For an SSD code, $A$ no longer ranges over all of $\GL(k,\mathbb F_2)$ because it must also preserve a \emph{logical overlap form} $\beta$. Letting $C \subseteq \mathbb{F}_2^n$ be the stabilizer space of the code, we consider $\beta(u,v)\coloneqq u\cdot v \pmod 2$ for any two logical-$X$ operators with supports $u,v \in C^\perp/C$. On SSD codes, logical-$X$ operators have even overlaps with $X$-stabilizers: they have even overlap with the $Z$-stabilizers by commutation and the $X$- and $Z$-stabilizer spaces are identical. So $\beta$ is invariant to stabilizer deformation of the logical representatives and is a well-defined logical form. Moreover, qubit permutations preserve $\beta$.

Since $A$ preserves $\beta$, it is an isometry of $\beta$. The form $\beta$ is nondegenerate---no nontrivial logical operator has even overlap with every logical operator (\cref{prop:strictly_self_dual_css_codes_properties})---so Albert's classification~\cite{Albert1938SymmetricAlternate} applies: over $\mathbb F_2$, a nondegenerate symmetric bilinear form is classified by whether it is alternating. Here $\beta$ is alternating exactly when every logical vector is even-weight, equivalently when $X^{\otimes n}$ is a stabilizer (\cref{prop:strictly_self_dual_css_codes_properties}); moreover $\beta$ is congruent to the standard symplectic form when alternating and the standard inner product when not. Therefore (\cref{thm:css_codes_perm_logical_group_psd}),
\begin{equation}
    \mathrm{Perm}_\mathcal{L}(\mathcal{C})
    \leq
    \begin{cases}
        \Sp(k,\mathbb{F}_2),
        & \vb{1}\in C,
        \\
        O(k,\mathbb{F}_2),
        & \vb{1}\notin C.
    \end{cases}
    \label{eq:css_perm_logical_groups_ssd}
\end{equation}

Lastly, for PSD-but-not-SSD codes, $\beta$ is not generally a well-defined form on the entire logical $X$ space. Letting $C_\mathrm{x}, C_\mathrm{z} \subseteq \mathbb{F}_2^n$ be the $X$- and $Z$-stabilizer spaces of the code, the appropriate replacement is the \emph{balanced} logical space $(C_\mathrm{x}^\perp \cap C_\mathrm{z}^\perp) / (C_\mathrm{x} \cap C_\mathrm{z})$: essentially the common support subspace of both logical sectors. Unlike for SSD codes, $\beta$ may be degenerate. Writing its radical dimension as $2m$, preserving the form reveals two nested subspaces of the logical $X$ space, of dimensions $m$ and $k-m$, that permutation gates preserve. Then, in a suitable basis, the logical action is block-upper-triangular. The middle quotient has dimension $k-2m$, carries a nondegenerate overlap form, and so supports either a symplectic or an orthogonal action. Concretely (\cref{thm:css_codes_perm_logical_group_psd}),
\begin{equation}
    \mathrm{Perm}_\mathcal{L}(\mathcal{C})
    \leq
    \begin{cases}
        \mathcal{F}_m^{\Sp}(k,\mathbb{F}_2), 
        & \vb{1}\in C_\mathrm{x}+C_\mathrm{z},
        \\
        \mathcal{F}_m^{O}(k,\mathbb{F}_2),   
        & \vb{1}\notin C_\mathrm{x}+C_\mathrm{z},
    \end{cases}
    \label{eq:css_perm_logical_groups_psd}
\end{equation}
where $\mathcal{F}_m^{\Sp}(k,\mathbb{F}_2)$ and $\mathcal{F}_m^{O}(k,\mathbb{F}_2)$ are the so-called flagged isometry subgroups (see \cref{def:flagged_isometry_groups}). When the form is nondegenerate, $m = 0$ and \cref{eq:css_perm_logical_groups_ssd} is recovered.

A consequence is that no $k \geq 3$ PSD code can be a phantom code, as the permutation logical groups become proper subgroups of $\GL(k,\mathbb{F}_2)$. This settles an open question of Ref.~\cite[Footnote~9]{koh2026entangling}. The bound in \cref{eq:css_perm_logical_groups_ssd} is also not generally tight, as self-duality imposes a parity constraint that full symplectic and orthogonal groups violate in large enough dimension. No PSD code realizes the full $\Sp(k,\mathbb{F}_2)$ for $k\ge8$, and none the full $O(k,\mathbb{F}_2)$ for $k\ge9$ (\cref{thm:css_codes_perm_logical_group_psd}). Tightness in the intermediate range $6\le k\le8$ remains open.

\tocless\subsection{Automorphism gates
\label{sub:css_auto}}

We now return to automorphism gates. By \cref{sec:css_codes/method}, every automorphism logical action of a non-PSD code is the unique product of one element from each of the transversal and permutation logical groups. Explicitly, in a CSS logical basis,
\begin{equation}
    \begin{pmatrix}
        I & C \\
        B & I
    \end{pmatrix}
    \begin{pmatrix}
        A & 0 \\
        0 & A^{-\top}
    \end{pmatrix}
    =
    \begin{pmatrix}
        A & CA^{-\top} \\
        BA & A^{-\top}
    \end{pmatrix}.
    \label{eq:css_auto_general_product}
\end{equation}
That is, an automorphism gate can act nontrivially on logical Paulis within a sector and can couple sectors, in the pattern prescribed above. Combining the two logical groups gives the Siegel parabolic,
\begin{equation}
    \mathcal{P}(2k,\mathbb{F}_2)
    =
    \mathcal{U}(2k,\mathbb{F}_2)
    \rtimes
    \GL(k,\mathbb{F}_2),
    \label{eq:css_auto_semidirect_product}
\end{equation}
which is precisely its Levi decomposition into the unipotent radical and linear subgroups. Up to Clifford conjugation dependent on the logical basis, \cref{eq:css_main_factorization} then gives (\cref{thm:css_codes_auto_logical_groups})
\begin{equation}
    \mathrm{Aut}_\mathcal{L}(\mathcal{C}) 
    \leq 
    \mathcal{P}(2k,\mathbb{F}_2)
    \cong
    \langle \mathbf{S}_k, \mathbf{CX}_k \rangle.
\end{equation}
As written, the Siegel parabolic is generated by all addressable logical $S$ and CX gates, and its order is the product of \cref{eq:trans_order,eq:perm_order}. This bound is also tight: code constructions in \cref{sec:constructions} based on phantom codes attain the full parabolic at any $k$ and distance.

Self-duality replaces the parabolic with smaller groups, obtained by substituting the constrained transversal and permutation logical groups (\cref{sec:css_codes/trans,sec:css_codes/perm}) into the semidirect product. For an SSD code, the transversal logical group is fixed to $S_3$, so up to Clifford conjugation, $\mathrm{Aut}_\mathcal{L}(\mathcal{C}) = S_3 \rtimes \mathrm{Perm}_\mathcal{L}(\mathcal{C})$. For a PSD-but-not-SSD code, the index-two extension of \cref{eq:css_main_factorization} gives $\mathrm{Aut}_\mathcal{L}(\mathcal{C}) = \mathrm{Trans}_\mathcal{L}(\mathcal{C}) \rtimes \mathrm{Perm}_\mathcal{L}(\mathcal{C}).2$. Their orders fall short of the parabolic's by factors of $\smash{\Omega(2^{k^2})}$ and $\smash{\Omega(2^{\ceil{k/2}^2})}$, respectively, for $k\ge2$.

However, self-duality allows automorphisms to induce exchange-type logical actions (\cref{corr:css_codes_auto_logical_hadamards_only_on_self_dual_codes}). These include actions with block-antidiagonal form
\begin{equation}
    \begin{pmatrix}
        0 & \Gamma^{-\top} \\
        \Gamma & 0
    \end{pmatrix},
\end{equation}
which swaps the logical $X$ and $Z$ sectors. Exchange-type logical gates are the extra feature self-duality confers even as it shrinks the largest attainable automorphism logical group.

These exchange-type logical actions are necessarily global. By the all-or-nothing rule (\cref{lem:css_codes_auto_all_or_nothing_exchange_cliffords}), an automorphism gate on an indecomposable CSS code has logical action either preserving- or exchange-type on its $k$ logical qubits, in any CSS logical basis. It therefore cannot realize exchange-type $H$, $SH$, or $HS$, or any tensor product of them, on a proper subset of the logical qubits (\cref{corr:css_indecomp_no_add_h_sh_hs}). In particular, self-dual codes can realize $H^{\otimes k}$, $(SH)^{\otimes k}$, and $(HS)^{\otimes k}$---transversally on SSD codes, or through an automorphism on PSD codes---but never their addressable versions.

\tocless\subsection{Extensions
\label{sec:css_codes/extensions}}

Finally, we consider two natural extensions of our analysis. First, our results thus far concern gates within a single codeblock. Allowing transversal \emph{interblock} CX gates between CSS codeblocks does not much change the picture. On $N$ codeblocks of a non-PSD code, each encoding $k$ logical qubits, the same logical-group ceilings simply scale to the combined logical space: $\GL(Nk,\mathbb{F}_2)$ for permutations and $\mathcal{P}(2Nk,\mathbb{F}_2)$ for automorphisms, with both bounds attainable (\cref{thm:css_codes_tcxs_logical_group_promotion}). The full Clifford group on all $Nk$ logical qubits is attainable only for SSD codes with $k=1$.

Interblock transversal CX gates can nevertheless improve addressability. In particular, additional codeblocks can serve as workspace to turn otherwise global exchange-type logical actions, such as $H^{\otimes k}$, into actions on specific logical qubits (\cref{rem:css_codes_tcxs_exchange_addressability}). Thus interblock coupling can make specific gates more useful without overcoming the underlying logical-group limits.

Second, the factorization property of automorphism gates into valid transversal and permutation parts extends beyond Cliffords. On any CSS code, a physical operation comprising a qubit permutation and arbitrary diagonal single-qubit gates is a logical gate if and only if the permutation and diagonal layer are separately valid logical gates (\cref{lem:css_codes_auto_splitting_property_beyond_clifford}). That is, a permutation cannot make an otherwise invalid diagonal transversal layer logical. Accordingly, the logical action is a product of that arising from the permutation, characterized by $\mathrm{Perm}_\mathcal{L}(\mathcal{C})$, and that arising from the diagonal layer.

\tocless\section{Stabilizer code results
\label{sec:stab_codes}}

We now ask whether removing the CSS restriction on codes enlarges the logical groups realizable by transversal single-qubit Clifford gates, qubit permutations, and their combination (i.e.~automorphisms). We find that it changes which gate classes can attain the overall ceiling but not the ceiling itself. Specifically, non-CSS code structure raises the permutation maximum from $\GL(k,\mathbb{F}_2)$ to $\mathcal{P}(2k,\mathbb{F}_2)$: permutations alone can do what CSS codes require automorphisms to do. However, the maximum is unchanged for transversal single-qubit Cliffords, and for automorphisms for $k \geq 3$. At $k=2$, the maximum-size automorphism logical group on indecomposable non-CSS codes is $O^+(4,\mathbb{F}_2) \cong \langle H_1, S_1, \mathrm{SWAP}_{12} \rangle$ of order $72$, exceeding the CSS maximum of order $48$.

Unlike CSS codes, an automorphism gate of a general stabilizer code need not factor into separately valid transversal and permutation logical gates. A simple $\db{4,1,2}$ code given in \cref{ex:stab_code_auto_semidirect_product_counterexample} illustrates this. We therefore analyze the three gate classes directly, proving logical group upper bounds and matching them with code constructions. We first bound the permutation logical group using a preserved logical Lagrangian (\cref{sec:stab_codes/perm}), then prove that automorphisms do not exceed the permutation maximum (\cref{sec:stab_codes/auto}), and finally classify the transversal logical groups (\cref{sec:stab_codes/trans}), including the exceptional possibilities.

\tocless\subsection{Permutation gates
\label{sec:stab_codes/perm}}

Qubit permutations preserve physical Pauli type. A stabilizer code can be placed in standard form (\cref{def:stab_code_standard_form}), wherein the physical representatives of $Z$-logical operators are $Z$-type strings and span a logical Lagrangian. Qubit permutations preserve this Lagrangian, so, up to Clifford conjugation dependent on the choice of logical basis,
\begin{equation}
    \mathrm{Perm}_\mathcal{L}(\mathcal{C}) 
    \leq 
    \mathcal{P}(2k,\mathbb{F}_2)
    \cong
    \langle \mathbf{S}_k, \mathbf{CX}_k \rangle.
    \label{eq:stab_perm_bound_siegel_parabolic}
\end{equation}
This bound is tight---we give a stabilizer code construction in \cref{sec:constructions} that realizes it. In contrast, permutation gates on CSS codes preserve \emph{two} complementary logical Lagrangians instead of one, restricting them to $\GL(k,\mathbb{F}_2)$.

What enables this expansion from $\GL(k, \mathbb{F}_2)$ to $\mathcal{P}(2k,\mathbb{F}_2)$ is that, on a stabilizer code, even in standard form, logical $X$ representatives may be mixed-type Pauli strings, so a qubit permutation may produce logical $Z$ components. This enables, for example, a logical $S$ action. The smallest code on which a qubit relabelling acts as a logical phase gate is a $\db{4,1,2}$ code, generated by cyclic shifts of the $IXYZ$ stabilizer generator (\cref{cons:non_css_code_permutation_logical_s}). In the logical basis $\overline{X} = IIZX$ and $\overline{Z} = ZZZZ$, the qubit cycle $(4321)$ implements a logical $S$ gate. This is, in fact, the unique $n \le 4$ code with this property, up to qubit indexing and local-Clifford code deformation (i.e.~$\mathrm{LC\Pi}$ code equivalence~\cite{cross2025small}).

In fact, permutations can realize even exchange-type logical gates on stabilizer codes. For example, a $\db{8,1,3}$ code implements a logical $H$ by a qubit permutation (\cref{cons:non_css_code_permutation_logical_h}), which no CSS code can do. Recall that on CSS codes, qubit permutations induce pure preserving-type logical actions in any CSS logical basis, because the $X$ and $Z$ logical sectors have pure $X$- and $Z$-type physical representatives and they both are preserved (\cref{sec:css_codes/perm}). Going to other logical bases Clifford-conjugates the preserving-type action, which does not reach the form of addressable nor $H^{\otimes k}$ logical actions. On non-CSS codes, however, pure $X$- and $Z$-type physical representatives need not belong to complementary logical sectors, as mixed-type stabilizers can cast them into the same logical class. Indeed, in the $\db{8,1,3}$ code, they correspond to the $Y$ logical class, which is fixed by the $H$ logical action. This does not contradict \cref{eq:stab_perm_bound_siegel_parabolic} which holds up to conjugacy and is modulo logical Pauli actions.

\tocless\subsection{Automorphism gates
\label{sec:stab_codes/auto}}

We now ask whether automorphisms raise the logical group ceiling set by permutations. For \(k \geq 3\), they do not (\cref{thm:stab_codes_auto_logical_max_order_k_geq_3}):
\begin{equation}
    \abs{\mathrm{Aut}_\mathcal{L}(\mathcal{C})} 
    \leq 
    \abs{\mathcal{P}(2k, \mathbb{F}_2)},
\end{equation}
and the maximum is attained uniquely by $\mathrm{Aut}_\mathcal{L}(\mathcal{C}) = \mathcal{P}(2k, \mathbb{F}_2)$ up to Clifford conjugation dependent on the choice of logical basis. Our proof strategy makes use of known classifications of finite groups at various points. The Siegel parabolic is so large inside \(\Sp(2k,\mathbb{F}_2)\) that only a few subgroup families can exceed it, and the structure of automorphism gates rules them out.

We bound the automorphism logical group $\Gamma \coloneqq \mathrm{Aut}_\mathcal{L}(\mathcal{C}) \leq \Sp(2k,\mathbb{F}_2)$ by induction on \(k\) from \(k=3\). We first assume $\Gamma$ reducible (\cref{lem:stab_codes_auto_reducible_logical_group_order_bound}): it preserves a nonzero proper subspace $V$ of the logical symplectic space, and the radical $R \coloneqq V \cap V^\perp$ gives three subcases. If $R$ is a Lagrangian, $\Gamma$ lies in a Siegel parabolic. Otherwise $\Gamma$ decomposes---acting on $R$ and residually on the quotient $R^\perp/R$ for nontrivial $R$, or on the summands of \(V \oplus V^\perp\) for trivial $R$---and each piece is bounded by a symplectic subgroup on a smaller space or by the automorphism logical group of a code with fewer logical qubits. The induction hypothesis then applies, and direct comparisons show that the order of Siegel parabolic cannot be exceeded.

We next assume that \(\Gamma\) is irreducible and larger than the Siegel parabolic. The full symplectic group \(\Sp(2k,\mathbb{F}_2)\) is impossible, due to a prime-order obstruction described below, and indeed also recently shown in Ref.~\cite{chakraborty2026nogo}. Thus \(\Gamma\) lies in a proper maximal subgroup \(M\) of \(\Sp(2k,\mathbb{F}_2)\), and \(M\) is also larger than the Siegel parabolic. This leaves few possibilities. The key observation is \(\abs{M}^3 \geq \abs{\mathcal{P}(2k,\mathbb{F}_2)}^3>\abs{\Sp(2k,\mathbb{F}_2)}\), so $M$ is a so-called \emph{large} maximal symplectic subgroup. The recent Yin--Lan--Liu--Chen refinement of the Aschbacher classification of finite groups~\cite{yin2025large} then leaves a short list of candidates (\cref{prop:aschbacher_symplectic_maximal_subgroups_at_least_size_of_siegel_parabolic}): the orthogonal groups \(O^\pm(2k,\mathbb{F}_2)\) and an exceptional Chevalley group \(G_2(2)\) when $k = 3$. The rest of the proof excludes these.

Both exclusions start with a prime-order obstruction. Let \(p\geq 5\) be prime, and consider an order-\(p\) automorphism logical action $\Sigma$, generating the cyclic group $\langle \Sigma \rangle$. After a Clifford deformation to the code dependent on the physical implementation of $\Sigma$, the logical subgroup $\langle \Sigma \rangle$ can be implemented purely by qubit permutations (\cref{lem:stab_codes_auto_perm_reduction_cyclic_group_logical_action})---this is a stronger version of Ref.~\cite[Lem.~6]{chakraborty2026nogo}. Two main consequences arise. First, automorphism gates on a stabilizer code encoding $k$ logical qubits cannot be of order $p$ for any prime divisor $p \ge 5$ of $\abs{\Sp(2k, \mathbb{F}_2)}$ but not $\abs{\GL(k, \mathbb{F}_2)}$ (\cref{lem:stab_codes_auto_prime_factor_nogo_1}). At least one such prime exists for every $k \geq 2$, so $\mathrm{Aut}_\mathcal{L}(\mathcal{C})$ must be a proper subgroup of $\Sp(2k, \mathbb{F}_2)$. Second, as qubit permutations do not change physical Pauli types, the gate $\Sigma$ on the original code must preserve some logical Lagrangian (\cref{lem:stab_codes_auto_coprime_six_preserves_lagrangian}). 

This preservation of a logical Lagrangian results in a characteristic-polynomial constraint. In a logical basis adapted to the preserved Lagrangian, the logical action is block-triangular with diagonal blocks \(A\) and \(A^{-\top}\). Consequently, the irreducible factors of its characteristic polynomial occur in reciprocal pairs, and self-reciprocal factors must have even multiplicity (\cref{lem:stab_codes_auto_prime_order_reciprocal_factor_constraint}). In particular, when \(\ord_p(2)\) is even, every irreducible factor of the cyclotomic polynomial $\Phi_p(x) \coloneqq (x^p-1)/(x-1)$ must occur with even multiplicity (\cref{corr:stab_codes_auto_prime_order_even_factor_constraint}).

The orthogonal groups \(O^\pm(2k,\mathbb{F}_2)\) are then ruled out with \(p=5\). Since \(\ord_5(2)=4\) is even, every irreducible factor of \(\Phi_5\) must occur with even multiplicity in the characteristic polynomial of an order-five automorphism logical action. But the derived subgroups \(\Omega^\pm(2k,\mathbb{F}_2)\coloneqq [O^\pm(2k,\mathbb{F}_2),O^\pm(2k,\mathbb{F}_2)]\) each contains an order-five element whose characteristic polynomial has a single irreducible factor of \(\Phi_5\). So, an automorphism logical group cannot contain \(\Omega^\pm(2k,\mathbb{F}_2)\) (\cref{lem:stab_codes_auto_logical_groups_cannot_contain_orthogonal_cores}). Moreover, by the recent Yin--Lan--Liu--Chen classification of finite groups~\cite{yin2025large}, no irreducible subgroup of \(O^\pm(2k,\mathbb{F}_2)\) larger than the Siegel parabolic but not containing \(\Omega^\pm(2k,\mathbb{F}_2)\) exists (\cref{lem:group_theory_irreducible_orthogonal_subgroups_larger_than_siegel_parabolic_contain_orthogonal_cores}). We conclude that no orthogonal candidate can exceed the parabolic.

It remains to exclude the exceptional \(G_2(2)\) possibility at \(k=3\). Because $G_2(2)$ is just barely larger than $\mathcal{P}(6, \mathbb{F}_2)$, if \(\Gamma\leq G_2(2)\) is larger than the Siegel parabolic, then \(\Gamma = G_2(2)\) exactly. Unfortunately, while \(G_2(2)\) contains an order-seven element, \(\ord_7(2)=3\) is odd, so the self-reciprocal obstruction above does not apply. We overcome this by further observing that the dihedral group \(D_{14} < G_2(2)\). For automorphism logical actions, the inversion element in \(D_{14}\) restores an even-multiplicity constraint on the irreducible factors of \(\Phi_7\) in the characteristic polynomial of the order-seven logical action (\cref{lem:stab_codes_auto_square_characteristic_polynomial_inverted_actions}). But order-seven elements of \(\Sp(6,\mathbb{F}_2)\) have characteristic polynomial exactly \(\Phi_7(x)\), whose two cubic factors each appear once. This violates the constraint, thus \(\Gamma\neq G_2(2)\) (\cref{corr:stab_codes_auto_logical_group_k_eq_3_cannot_contain_g22}). All candidates for $M$ are now ruled out, so the induction closes.

\tocless\subsection{Transversal gates
\label{sec:stab_codes/trans}}

Lastly, on an indecomposable stabilizer code, what a transversal gate can do is determined by the order of its physical Cliffords. Order-one and -two Cliffords fix Pauli axes: $I$ fixes all three, $S$ fixes $Z$, $H$ fixes $Y$, and $HSH$ fixes $X$. The order-three $HS$ and $SH$ fix none. 

When every Clifford fixes an axis, the $n$ fixed axes span a physical Lagrangian, and the transversal gate fixes it \emph{pointwise}: it returns every element (i.e.~physical Pauli string) in the Lagrangian to itself. In contrast, a permutation fixes a Lagrangian only as a subspace and elements may get shuffled within. Without loss of generality, we may associate the pointwise-fixed Lagrangian to the $Z$ logical sector. Then, since Cliffords preserve Pauli commutation relations and every $Z$-logical is frozen, each $X$ logical is free only to pick up $Z$ factors. These actions are generated by logical $S$ and $\mathrm{CZ}$ gates, so $\mathrm{Trans}_\mathcal{L}(\mathcal{C}) \leq \mathcal{U}(2k,\mathbb{F}_2)$ up to Clifford conjugation.

It remains to consider the order-three physical Cliffords $HS$ and $SH$. Here, another all-or-nothing rule holds: order-three gates must sit on every qubit or none (\cref{lem:stab_codes_trans_all_or_nothing_physical_c3_cliffords})\footnote{The all-or-nothing rule for transversal gates on indecomposable CSS codes (\cref{lem:css_codes_auto_all_or_nothing_exchange_cliffords}) works the same way but isolates a larger subset of physical gates: the exchange-type $\{H, HS, SH\}$ must sit on every qubit or none. A general stabilizer code does not have distinguished $X$- and $Z$-type sectors, so the notion of preserving- and exchange-type Cliffords is not productive here.}. Otherwise, they would split the code in two, which indecomposability forbids. Moreover, any two tensor products of order-three physical Cliffords must either coincide or be inverses; otherwise, their product would be trivial on some physical qubits but nontrivial on others, violating the all-or-nothing rule. Hence the order-three transversal gates generate at most one logical \(C_3\). On each physical qubit, a three-cycle in the Clifford group is self-centralizing---the subgroup of Cliffords commuting with it is itself---so any transversal gate commuting with this \(C_3\) adds nothing to the logical group. A genuinely new gate must instead conjugate the cycle to its inverse, giving $S_3$. These are precisely the two automorphisms of \(C_3\). 

Putting these together, up to Clifford conjugation dependent on the choice of logical basis (\cref{thm:stab_codes_trans_logical_groups}),
\begin{equation}
    \mathrm{Trans}_\mathcal{L}(\mathcal{C}) \leq \mathcal{U}(2k,\mathbb{F}_2)
    \,\, \text{or} \,\,
    \mathrm{Trans}_\mathcal{L}(\mathcal{C}) \cong C_3
    \,\, \text{or} \,\,
    S_3.
\end{equation}
The \(\mathcal{U}(2k,\mathbb{F}_2)\) bound is tight and is attained by CSS codes (see \cref{sec:constructions}). The $S_3$ logical group is likewise attained by SSD CSS codes. The bare $C_3$, however, occurs only on non-CSS codes. An order-three physical Clifford is exchange-type, so a CSS code that hosts such a transversal gate must be SSD (\cref{sec:css_codes/trans})---but such a code also carries a $H^{\otimes n}$ gate, lifting the logical group to $S_3$. In contrast, nothing forces the extra Hadamard on stabilizer codes, and $C_3$ can be attained---for example, in the $\db{5,1,3}$ perfect code~\cite{gottesman1997stabilizer}.

\tocless\subsection{Universality
\label{sec:stab_codes/universality}}

Automorphism gates by themselves cannot support universal fault-tolerant quantum computation. Given the full Clifford group, the addition of \emph{any} non-Clifford gate promotes the gate set to be universal~\cite{jozsa2013classical}. While the maximum-size automorphism logical group achievable on stabilizer codes falls exponentially short in size compared to the full Clifford group, this has only a minor effect on the ingredient needed for universality: the addition of a \emph{single} non-Clifford gate is nonetheless sufficient.

In particular, for the maximum-size automorphism logical group $\langle\mathbf S_k,\mathbf{CX}_k\rangle$ attainable on CSS and stabilizer codes, with logical Paulis restored, the addition of a single-qubit logical gate $V$ confers universality for $k\geq2$ provided only that it is non-Clifford and does not preserve the Pauli-$Z$ axis up to signs (\cref{thm:universality_s_cx}). For $k = 1$, and for the maximum-size transversal logical group $\langle\mathbf S_k,\mathbf{CZ}_k\rangle$ with $V$ available addressably, an additional obstruction appears: $T V T^\dag$ must not be Clifford (\cref{thm:universality_s_cz}). These conditions are not difficult to satisfy. For example, the $HT$ or $TH$ gate satisfies all criteria (but not the plain $T$ gate).

\tocless\section{Physical-qubit costs
\label{sec:cost}}

Having determined the largest logical groups attainable by transversal, permutation, and automorphism gates, we now examine how many physical qubits are required to realize them. For $k\geq3$, realizing all addressable logical $\mathrm{CX}$ gates through automorphisms requires $n \ge 2^k-1$ physical qubits, and hence so does attaining the maximum-size automorphism logical group. By contrast, realizing all addressable logical $S$ and $\mathrm{CZ}$ gates through transversal Cliffords requires only $n \ge k+\binom{k}{2}=k(k+1)/2$. These lower bounds hold at every distance and are exactly attained by distance-one codes.

The polynomial scaling, in fact, persists for diagonal gates at every fixed level $t$ of the Clifford hierarchy: realizing the complete addressable diagonal logical gate group through transversal physical single-qubit diagonal gates requires $n\geq\sum_{s=1}^{t}\binom{k}{s}=\Theta(k^t)$, with equality attainable at distance one.
The resulting sharp cost divide is therefore between $\mathrm{CX}$ and diagonal gate groups, rather than between Clifford and non-Clifford gates.
We first establish the exponential lower bound for the complete $\mathrm{CX}$ gate group (\cref{sec:costs/linear}), then determine the exact cost of diagonal Clifford gate groups and extend the analysis to arbitrary levels of the Clifford hierarchy (\cref{sec:costs/diagonal}).

\tocless\subsection{Exponential cost of $\mathrm{CX}$ gates
\label{sec:costs/linear}}

We begin with the logical group $\GL(k,\mathbb{F}_2)\cong\langle\mathbf{CX}_k\rangle$. Attaining this group on a stabilizer code with automorphisms requires
\begin{equation}
    n \ge 2^k - 1,
    \label{eq:stab_cost_general_linear}
\end{equation}
for all distances $d$ and $k$, with a single $\db{2,2,1}$ exception (\cref{thm:stab_codes_auto_saturating_general_linear_n_scaling}). Recently, Ref.~\cite{morris2026constraints} proved this bound for $d > 1$ and $k \neq 4$. We close $k = 4$ by exhaustive computational verification and $d = 1$ by a counting argument.

For the latter, we note that the $\langle \mathbf{CX}_k \rangle$ logical group preserves the $X$- and $Z$-logical Lagrangians. Some logical Pauli then has a weight-one physical representative, and since automorphisms preserve weight, so does every logical class in its orbit. Suppose that this Pauli is $Z$-type, say $Z_1$. The $\GL(k,\mathbb{F}_2)$ logical actions send $Z_1$ to all $2^k - 1$ nontrivial products of $Z_i$, so each must have weight-one representatives\footnote{This observation is similar to \cite[Lem.~1]{koh2026entangling} but for automorphisms on stabilizer codes here.}. These logical classes commute, and each physical qubit can host at most one of them---different single-qubit Paulis anticommute, while the same Pauli represents the same logical class---giving \cref{eq:stab_cost_general_linear}. The $X$-type case is identical, and the mixed-type case follows from counting the \(3n\) weight-one physical Paulis against the $\ge (2^k-1)(2^{k-1}-1)$ classes in the orbit, giving also \cref{eq:stab_cost_general_linear}. This counting weakens to the trivial $n \geq k$ bound for $k \le 2$. Under permutations instead of automorphisms, however, the $\db{2,2,1}$ code cannot attain $\GL(2, \mathbb{F}_2)$ because $\abs{S_2} < \abs{\GL(2, \mathbb{F}_2)}$, so \cref{eq:stab_cost_general_linear} holds for all $k$ (\cref{thm:stab_codes_perm_saturating_general_linear_n_scaling}).

The bound in \cref{eq:stab_cost_general_linear} is tight. Distance-one CSS phantom codes achieve the $\langle \mathbf{CX}_k \rangle$ logical group through qubit permutations with \(n=2^k-1\), and concatenation reaches arbitrary distances---see \cref{sec:constructions}. These \(d=1\) codes are not bare qubits: unentangled qubits cannot realize entangling logical \(\mathrm{CX}\) gates by permutations.

\Cref{eq:stab_cost_general_linear} carries over to stabilizer codes realizing the Siegel parabolic through automorphisms or permutations because $\GL(k,\mathbb{F}_2) < \mathcal{P}(2k, \mathbb{F}_2) \cong \langle \mathbf{S}_k, \mathbf{CX}_k \rangle$ (\cref{thm:stab_codes_auto_saturating_siegel_parabolic_n_scaling,thm:stab_codes_perm_saturating_siegel_parabolic_n_scaling}); the $\db{2,2,1}$ exception does not occur here. Distance-one CSS phantom codes can realize $\mathcal{P}(2k, \mathbb{F}_2)$ through automorphisms with $n = 2^k - 1$, and concatenated variants reach arbitrary distances. Further concatenation with a $\db{4,1,2}$ gadget produces non-CSS codes hosting $\mathcal{P}(2k, \mathbb{F}_2)$ through permutations with $n = \Theta(2^k)$---see \cref{sec:constructions}.

\tocless\subsection{Polynomial cost of diagonal gates
\label{sec:costs/diagonal}}

\begin{figure*}
    \centering
    \includegraphics[width=\linewidth]{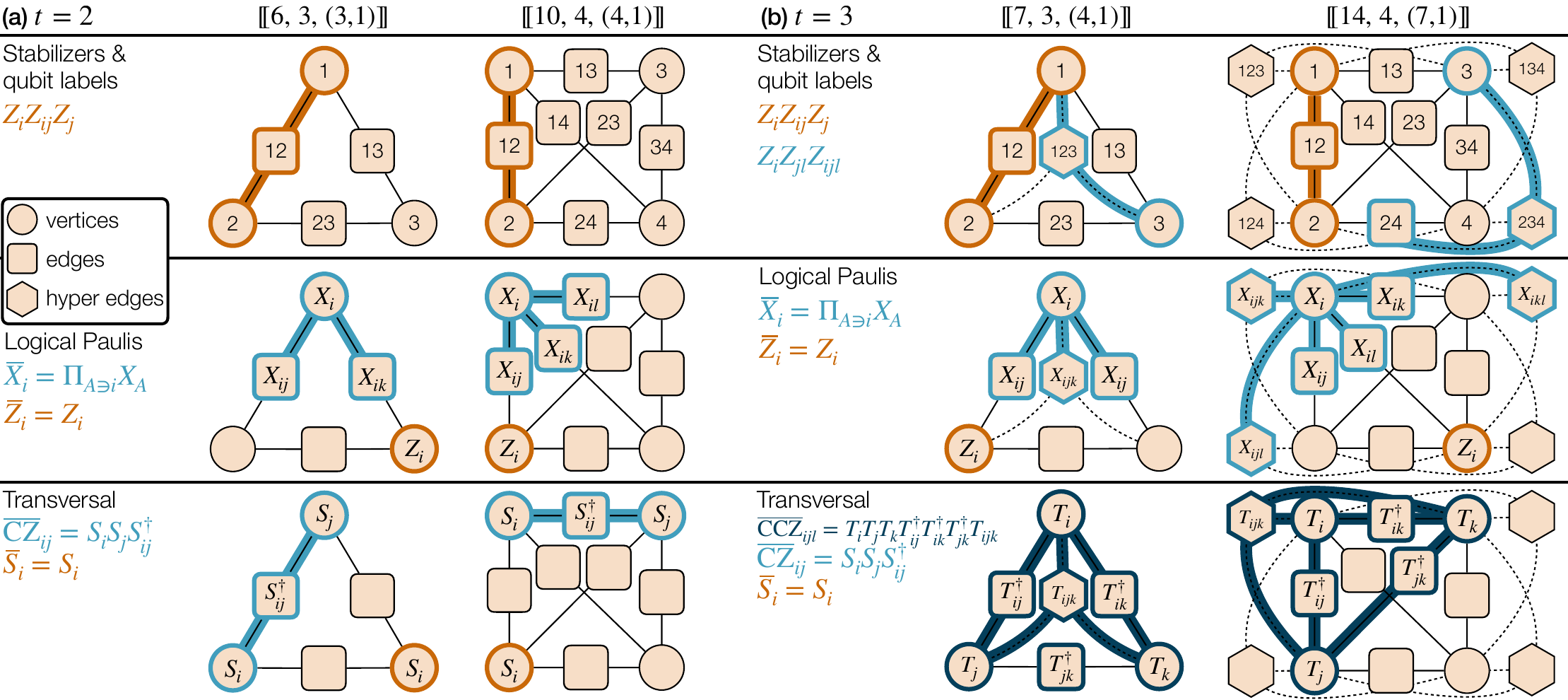}
    \phantomsubfloat{\label{fig:complete_hypergraph_codes/graph}}
    \phantomsubfloat{\label{fig:complete_hypergraph_codes/hypergraph}}
    \vspace{-20pt}
    \caption{\textbf{Complete-hypergraph CSS codes.}
    For any $t$, these codes realize every diagonal logical gate in the $t^\text{th}$ level of the Clifford hierarchy transversally, through tensor products of single-qubit rotations. The code is defined on the rank-$t$ complete hypergraph of $k$ vertices---with one hyperedge per nonempty subset of $\le t$ vertices---placing one qubit per hyperedge. Here qubits are labelled by the vertices in the hyperedges omitting set notation (e.g.~$12$ for $\{1,2\}$). Rows: stabilizers, logical Pauli representatives, and transversal gates (examples highlighted).
    \textbf{(a)} $t=2$ at $k=4,5$, giving all addressable logical $S$ and $\mathrm{CZ}$ gates and products thereof.
    \textbf{(b)} $t=3$ at $k=3,4$, giving all addressable logical $T$, $\mathrm{CS}$, and $\mathrm{CCZ}$ gates and products thereof.
    }
    \label{fig:complete_hypergraph_codes}
\end{figure*}

The $\mathcal{U}(2k,\mathbb{F}_2)$ logical group is much cheaper than $\GL(k,\mathbb{F}_2)$. Since $\mathcal{U}(2k, \mathbb{F}_2)$ is an elementary abelian 2-group---every non-identity logical action is order-two and commutative---the transversal gates realizing this group can be taken to comprise physical Cliffords of order at  most two  (\cref{thm:stab_codes_trans_saturating_n_scaling}). Moreover, the product of two transversal gates is itself a transversal gate, but two distinct order-two Cliffords multiply to an order-three one. Thus, on each qubit, every gate in the group must act by either $I$ or a single order-two Clifford---two choices per qubit. Then $n$ qubits realize at most $2^n$ distinct logical actions, and covering all $\abs{\mathcal{U}(2k, \mathbb{F}_2)} = 2^{k(k+1)/2}$ of them requires
\begin{equation}
    n \ge \frac{k(k+1)}{2} = \Theta(k^2).
    \label{eq:stab_cost_unipotent_radical}
\end{equation}
The bound is tight at $d = 1$, attained by a complete-graph CSS code (see \cref{sec:constructions}).
Conversely, small block lengths on indecomposable stabilizer codes constrain the size of their transversal logical groups (\cref{lem:stab_codes_trans_logical_group_bound_by_n}): when $n \ge 3$ falls short of \cref{eq:stab_cost_unipotent_radical}, $\smash{\abs{\mathrm{Trans}_\mathcal{L}(\mathcal{C})} \leq 2^n}$. Thus a high-rate code family with $n=\Theta(k)$ can support transversal logical groups of order $2^{\order{k}}$, smaller than the $\mathcal{U}(2k,\mathbb{F}_2)$ maximum by a factor of $2^{k^2/2+\order{k}}$.

The polynomial qubit cost of the transversal logical group persists beyond Cliffords. Fix any level $t$ of the Clifford hierarchy, and let $\mathbf{D}_k^{(t)}$ be the group of all diagonal level-\(t\) gates on \(k\) logical qubits (modulo $Z$ Paulis): level three, for example, contains all addressable \(T\), \(\mathrm{CS}\), \(\mathrm{CCZ}\), and all products thereof. The group has one natural generator for each nonempty subset of at most \(t\) qubits: single-qubit phases \(\smash{Z^{(t)} \coloneqq Z^{1/2^{t-1}}}\), controlled-phases on pairs, controlled-controlled-phase on triples, and higher-weight analogues. Hence the group has \(\sum_{s=1}^{t} \binom{k}{s} = \Theta(k^t)\) independent generators in total. For \(t=2\), this recovers the \(k(k+1)/2\) count of \(\mathcal{U}(2k,\mathbb{F}_2)=\langle \mathbf{S}_k,\mathbf{CZ}_k\rangle\).

We consider implementing the \(\mathbf{D}_k^{(t)}\) logical group transversally through tensor products of continuous-angle single-qubit diagonal rotations. These physical gates thereby lie in the \(n\)-torus. Continuous freedom, however, cannot genuinely help generate a finite logical group: every connected family of physical gates induces a constant logical action. The logical group is therefore associated with the component group of a closed subgroup of the $n$-torus, which is generated by at most \(n\) elements. Thus necessarily (\cref{thm:stab_codes_trans_any_clifford_hierarchy_level_n_scaling})\footnote{For \(t \ge 3\), this bound assumes the physical single-qubit gates are diagonal. The \(t=2\) bound of \cref{thm:stab_codes_trans_saturating_n_scaling} considers arbitrary physical single-qubit Cliffords and has no such assumption.}
\begin{equation}
    n\geq
    \sum_{s=1}^{t}\binom{k}{s}
    =
    \Theta(k^t).
    \label{eq:stab_cost_transversal_level_t}
\end{equation}
Equality is attained by complete-hypergraph CSS codes (see \cref{sec:constructions}). The complete group of addressable diagonal logical gates is therefore polynomially cheap at every fixed level of the Clifford hierarchy.

\tocless\section{Code constructions
\label{sec:constructions}}

We now construct explicit stabilizer-code families that attain the maximum-size logical gate groups and qubit-cost limits established above. At $Z$-distance one, these families exactly saturate the polynomial and exponential qubit bounds of \cref{sec:cost}; suitable concatenations preserve the corresponding maximum-size logical groups and reach arbitrary distance. The qubit-cost divide between diagonal and $\mathrm{CX}$ logical gate groups is expounded by two code construction approaches. First, complete-hypergraph CSS codes realize the full addressable diagonal logical group $\mathbf{D}_k^{(t)}$ at any fixed Clifford hierarchy level $t$ through transversal physical single-qubit rotations, using exactly $n=\sum_{s=1}^{t}\binom{k}{s}$ physical qubits. At $t=2$, they attain $\langle\mathbf{S}_k,\mathbf{CZ}_k\rangle$ with $n=k(k+1)/2$; at $t=3$, they realize all addressable logical $T$, $\mathrm{CS}$, and $\mathrm{CCZ}$ magic gates with cubic block length. Second, distance-one CSS phantom codes realize all addressable logical $\mathrm{CX}$ gates by qubit permutations, and all logical $S$ gates transversally, using the optimal $n=2^k-1$ physical qubits.

For both, concatenating with a single-logical-qubit code supporting transversal $S$ increases the distance while preserving their Clifford logical groups. Such inner codes can be \emph{good}, in that their distance grows linearly with code size~\cite{haah2018towers,jain2025transversal}. Lastly, using the non-CSS $\db{4,1,2}$ gadget, we can further implement the required $S$ actions with physical qubit permutations, allowing qubit permutations alone to attain $\langle \mathbf{S}_k, \mathbf{CX}_k \rangle$ on stabilizer codes.

We begin with the complete-hypergraph constructions (\cref{sec:constructions/hypergraph}), build the phantom codes and their concatenated extensions (\cref{sec:constructions/phantom}), and conclude with the non-CSS, small-$k$, and self-dual refinements (\cref{sec:constructions/exceptions}).

\tocless\subsection{Complete-hypergraph codes
\label{sec:constructions/hypergraph}}

The $\mathcal{U}(2k, \mathbb{F}_2) \cong \langle \mathbf{S}_k, \mathbf{CZ}_k \rangle$ transversal logical group is realized by a family of CSS codes built on the complete graph of $k$ vertices (\cref{cons:complete_hypergraph_css_t_eq_2}). We place one qubit on each vertex $i$ and each edge $\{i,j\}$, so $n=k(k+1)/2$; and for each edge, we impose the stabilizer generator $Z_{\{i\}}Z_{\{j\}}Z_{\{i,j\}}$, thereby fixing the edge qubit to the parity of its endpoints (see \cref{fig:complete_hypergraph_codes/graph}). This way, the $k$ vertex qubits host weight-one $Z$-logical representatives of the $k$ logical qubits and the edge qubits store parities of all logical qubit pairs. Then, as diagonal actions can be written as phase polynomials, all diagonal Clifford logical gates can be implemented transversally: a physical $S_{\{i\}}$ implements a logical $S_i$, while $\smash{S_{\{i\}} S_{\{j\}} S_{\{i,j\}}^\dag}$ implements a logical $\mathrm{CZ}_{ij}$. This code family therefore realizes $\mathcal{U}(2k, \mathbb{F}_2)$ transversally at the minimum possible block length of \cref{eq:stab_cost_unipotent_radical}. As-is, the code has distance $(d_\mathrm{x}, d_\mathrm{z}) = (k, 1)$, which can be increased by concatenating with a $k=1$ inner code hosting a transversal $S$ logical gate.

This construction extends to any level $t$ of the Clifford hierarchy by replacing the complete graph with a rank-$t$ complete hypergraph. We place one qubit on each nonempty subset of $\{1, \ldots, k\}$ of size at most $t$, and use weight-three $Z$-checks to fix the parity of each subset qubit to that of its elements (see \cref{fig:complete_hypergraph_codes/hypergraph}). Each element of the $\mathbf{D}_k^{(t)}$ logical group is then implemented by a tensor product of physical $\smash{Z^{(t)}}$ rotations and their powers (\cref{cons:complete_hypergraph_css_generalized}). For $t=3$, these are physical $T$ gates, giving all addressable $T$, $\mathrm{CS}$, $\mathrm{CCZ}$ logical gates and their products (\cref{cons:complete_hypergraph_css_t_eq_3}). The qubit counts of these complete-hypergraph codes meet exactly the lower bound of \cref{eq:stab_cost_transversal_level_t}.

\tocless\subsection{Concatenated phantom codes
\label{sec:constructions/phantom}}

\begin{figure}
    \centering
    \includegraphics[width=\linewidth]{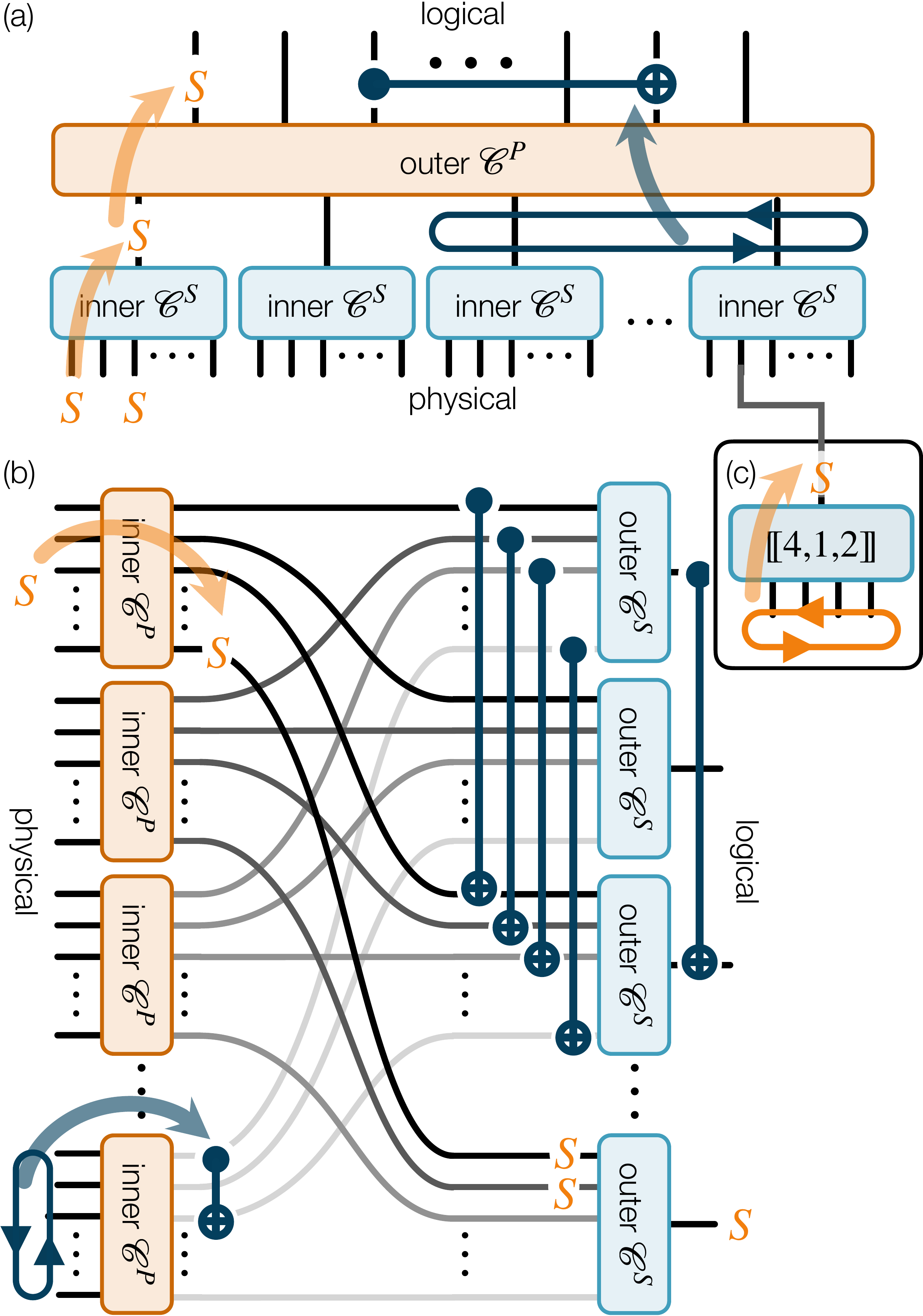}
    \phantomsubfloat{\label{fig:concatenated_phantom_codes/as_outer}}
    \phantomsubfloat{\label{fig:concatenated_phantom_codes/as_inner}}
    \phantomsubfloat{\label{fig:concatenated_phantom_codes/412_non_css}}
    \vspace{-20pt}
    \caption{\textbf{Concatenated phantom codes}.
    These codes realize every addressable logical CX by qubit permutations and logical $S$ transversally, at arbitrary distance.
    The constructions concatenate a distance-one phantom code $\mathcal{C}^\mathrm{p}$ and a code $\mathcal{C}^\mathrm{s}$ supporting a transversal $S$ logical gate.
    \textbf{(a)} With the phantom code as the outer code, its permutation gates shuffle whole inner $\mathcal{C}^\mathrm{s}$ codeblocks.
    \textbf{(b)} Alternatively, the phantom code can serve as the inner code. A transversal CX between outer $\mathcal{C}^\mathrm{s}$ codeblocks becomes within-block logical CX gates, which the inner phantom codes implement by qubit permutations.
    Either way, permutations alone realize all logical CX gates, and transversal rotations supply logical $S$ gates, giving the full $\mathcal{P}(2k, \mathbb{F}_2) \cong \langle \mathbf{S}_k, \mathbf{CX}_k \rangle$ parabolic (\cref{cons:concatenated_phantom_clifford}). Generalizations to any level of the Clifford hierarchy are natural (\cref{cons:concatenated_phantom_as_inner_css_generalized,cons:concatenated_phantom_as_outer_css_generalized}).
    }
    \label{fig:concatenated_phantom_codes}
\end{figure}

We now turn to the \(\mathcal{P}(2k,\mathbb{F}_2) \cong \langle \mathbf{S}_k, \mathbf{CX}_k \rangle\) automorphism logical group. This is attained by distance-one CSS phantom codes on which the logical-$Z$ operator of each logical qubit has a weight-one representative. As the code is phantom, qubit permutations realize all addressable logical CX gates~\cite{koh2026entangling}. Moreover, a physical $S$ on the support of each logical-$Z$ representative gives a corresponding logical $S$ gate. Thus the code supports $\mathcal{P}(2k,\mathbb{F}_2)$ through automorphisms. An example of such codes is the $\db{2^k - 1, k, (d_\mathrm{x} = 2^{k-1}, d_\mathrm{z} = 1)}$ simplex code family, which meets the qubit lower-bound of \cref{eq:stab_cost_general_linear}.

To reach arbitrary distance, the phantom code $\mathcal{C}^\mathrm{p}$ can be concatenated with a $k=1$ CSS code $\mathcal{C}^\mathrm{s}$ hosting a transversal logical $S$ gate, with the phantom as either the outer or inner code (\cref{fig:concatenated_phantom_codes}). As the outer code, its permutation gates shuffle whole inner $\mathcal{C}^\mathrm{s}$ codeblocks (\cref{fig:concatenated_phantom_codes/as_outer,cons:concatenated_phantom_as_outer_css_generalized}); as the inner code, it implements by permutations the within-block logical CX gates into which the transversal CX between outer $\mathcal{C}^\mathrm{s}$ codeblocks decomposes (\cref{fig:concatenated_phantom_codes/as_inner,cons:concatenated_phantom_as_inner_css_generalized}). Either way, the code retains the $\mathcal{P}(2k,\mathbb{F}_2)$ automorphism logical group with distance conferred by $\mathcal{C}^\mathrm{s}$; admissible $\mathcal{C}^\mathrm{s}$ include the two-dimensional colour codes and, more generally, SSD codes. We provide a list of $\mathcal{C}^\mathrm{p}$ and $\mathcal{C}^\mathrm{s}$ options, and further technical details, in \cref{cons:concatenated_phantom_concrete}.

In fact, these codes saturate all three CSS logical group ceilings simultaneously: $\mathcal{U}(2k,\mathbb{F}_2) \cong \langle \mathbf{S}_k, \mathbf{CZ}_k \rangle$ transversally, $\GL(k,\mathbb{F}_2) \cong \langle \mathbf{CX}_k \rangle$ by permutations, and $\mathcal{P}(2k,\mathbb{F}_2) \cong \langle \mathbf{S}_k, \mathbf{CX}_k \rangle$ by automorphisms. They also generalize beyond the Clifford setting. The weight-one logical-$Z$ representatives host every single-qubit logical $Z^{(t)}$ rotation via a physical $Z^{(t)}$, and composing with the logical $\mathrm{CX}$s present yields the full $\smash{\langle \mathbf{D}^{(t)}_k, \mathbf{CX}_k \rangle}$ by qubit permutations and transversal $Z^{(t)}$ layers (\cref{cons:concatenated_phantom_as_inner_css_generalized,cons:concatenated_phantom_as_outer_css_generalized}). For arbitrary distance, $\mathcal{C}^\mathrm{s}$ must then host a transversal logical $Z^{(t)}$.

\tocless\subsection{Other constructions
\label{sec:constructions/exceptions}}

Non-CSS codes can attain \(\mathcal{P}(2k,\mathbb{F}_2)\) by qubit permutations alone: take any code attaining \(\mathcal{P}(2k,\mathbb{F}_2)\) through permutations and transversal powers of $S$, and concatenate with the $\db{4,1,2}$ non-CSS code (\cref{cons:non_css_code_permutation_logical_s}), whose logical $S$ is a permutation. Every required operation then becomes a permutation (\cref{fig:concatenated_phantom_codes/412_non_css}). Permutations on non-CSS codes can reach even exchange-type logical actions: two LC$\Pi$-inequivalent $\db{8,1,2}$ codes and an $\db{8,1,3}$ code implement exactly a logical $H$ by permutations (\cref{cons:non_css_code_permutation_logical_h}), and these are the smallest stabilizer codes with this property. 

Lastly, we discuss the $k \le 2$ exceptions to the $\mathcal{P}(2k,\mathbb{F}_2)$ automorphism logical group bound. At $k = 1$, every SSD CSS code attains the full Clifford group transversally (see \cref{thm:css_codes_trans_logical_groups_ssd_indecomp}), the smallest with nontrivial distance being the $\db{7,1,3}$ Steane code. The smaller non-CSS $\db{4,1,2}$ code above and the $\db{5,1,3}$ perfect code~\cite{gottesman1997stabilizer,sayginel2025fault,chakraborty2026nogo} attain the same through automorphisms, and are the unique smallest stabilizer codes at their distances to do so, up to $\mathrm{LC\Pi}$ equivalence. At $k = 2$, indecomposable non-CSS codes can attain $O^+(4,\mathbb{F}_2) \cong \langle H_1, S_1, \mathrm{SWAP}_{12} \rangle$; the $\db{6,2,2}$ and $\db{9,2,3}$ codes of \cref{cons:non_css_code_auto_k_eq_2_h1_h2_cz} are the unique smallest stabilizer codes at their distances to do so.

The appendix also supplies codes for the CSS self-dual refinements of \cref{sec:css_codes}. Namely, PSD codes attain the transversal maximum $\mathcal{U}(2\floor{k/2}, \mathbb{F}_2)^2$ split on two halves of the logical qubits (\cref{cons:css_codes_maximum_trans_logical_group_psd}), and SSD codes attain the $\Sp(k,\mathbb{F}_2)$ and $O(k,\mathbb{F}_2)$ permutation branches at small $k$ (\cref{cons:css_codes_maximum_perm_logical_group_ssd}). Whether these permutation logical group branches are tight at intermediate $k$ remains open (\cref{sec:discussion_and_outlook}).

\tocless\section{Discussion and Outlook
\label{sec:discussion_and_outlook}}

Taken together, our results determine the extremal logical power of automorphism gates and reveal a qubit-cost divide that persists beyond the Clifford group. The resulting logical-group landscape separates into three regimes: polynomial qubit cost, exponential qubit cost, and impossibility. At any fixed level of the Clifford hierarchy, the full group of addressable diagonal gates---including magic gates such as $T$, $\mathrm{CS}$, $\mathrm{CCZ}$---requires only polynomially many qubits. By contrast, full addressability of logical $S$ and $\mathrm{CX}$ gates requires exponentially many qubits, while some gates cannot be realized by automorphisms at any block length. Thus, neither Cliffordness nor logical-group size alone predicts qubit cost.

These results motivate evaluating native gate sets by their architectural utility relative to their physical-qubit cost, rather than by automorphism-group size alone. Rich automorphism groups may remain valuable for small codeblocks, where their relative cost is manageable~\cite{koh2026entangling}. Attaining the maximum requires a lower encoding rate and may remain worthwhile when the physical-qubit budget permits small $k$ codes. When high rate is required, however, native gate sets should instead be tailored to the intended workload, prioritizing operations that would otherwise be especially costly to implement.

This selective perspective points toward hybrid compilation, with automorphism gates supplying some operations at acceptable cost in code parameters and other fault-tolerant gadgets supplying the rest at a cost in circuit depth. Although the maximum-size automorphism group is exponentially smaller than the full logical Clifford group, adding a single suitable non-Clifford gate already yields a universal logical gate set. The challenge is therefore not universality itself, but achieving it at the lowest overall cost. Code surgery is a natural complement to automorphisms because it can mediate operations unavailable or prohibitively costly through automorphisms alone~\cite{xu2025fast,yoder2025tour}. Surgery-only compilation can be expensive in circuit depth, whereas automorphism-only compilation can be expensive in physical qubits. Because neither mechanism is optimal alone, the open problem is to determine their best division of labour for a given workload, block size, distance, and hardware platform.

A natural next step is therefore to determine the physical-qubit cost of selective logical gate sets at a prescribed code distance. Our cost results concern complete, fully addressable gate sets, but not smaller tailored ones. Targets include $\mathrm{CX}$ gates between subsets of selected logical pairs and selected mixtures of diagonal and non-diagonal gates. A theory of these intermediate costs would show how much overhead can be saved by sacrificing particular gates or forms of addressability. Even for the complete gate sets, our lower bounds hold at every distance but are known to be exactly attained only at $Z$-distance one. Concatenation yields codes at any target distance while retaining the same logical gates, but may use more physical qubits than necessary. Determining the optimal trade-off among logical power, encoding rate, and distance therefore calls for distance-explicit lower bounds and matching code constructions.

A complementary direction is to determine how the attainable logical gates and their physical costs change as this physical gate model is relaxed. The first natural extension is to allow arbitrary physical single-qubit gates rather than only Cliffords. These retain the motivating advantages of the present model: they require no ancillas, measurements, or physical two-qubit gates and can be implemented in depth one without spreading errors. Our diagonal-gate results address part of this broader model, but general non-diagonal non-Clifford gates remain unaddressed. Such gates fall outside both the discrete symplectic framework and the phase-polynomial description used for diagonal gates, so our methods do not extend directly; general methods to construct codes with such gates are also lacking. A classification would distinguish limits imposed by the depth-one architecture itself from those arising from the Clifford restriction.

A second extension is to admit physical entangling gates. Beyond the transversal interblock $\mathrm{CX}$ gates between CSS codeblocks studied here, more general interblock transversal gates and within-block fold gates are largely uncharacterized. The former couple corresponding physical qubits across blocks, while the latter apply multi-qubit gates on partitions of qubits within a codeblock. These operations may enable logical gates or addressability excluded by our within-block model~\cite{benhemou2026automated}, but their capabilities must be weighed against connectivity requirements, circuit depth, coordination between codeblocks, and fault propagation across qubits. The broader goal is to map out the logical power each relaxation affords and the physical costs exacted in return.

Alongside this architectural programme, our methods leave three mathematical questions. First, can the automorphism logical-group bound for stabilizer codes be proved without invoking classification results of finite groups? A classification-free proof could reveal the structural origin of the bound more transparently, and may extend more readily to other code and gate models. Second, do SSD or PSD CSS codes exist whose permutation logical groups saturate the $\Sp(k,\mathbb{F}_2)$ branch for $k=6$ or the $O(k,\mathbb{F}_2)$ branch for $6\leq k\leq8$? We have examples at smaller $k$ and no-go results at larger $k$, but the two regimes do not yet meet. Settling these remaining cases would complete the finite-$k$ picture for CSS logical groups. Third, CSS structure makes the SSD and PSD classification natural; is there an analogous classification for general stabilizer codes, with correspondingly finer-grained logical-group bounds? Similar questions arise in broader error-correction frameworks~\cite{poulin2005stabilizer, brun2006correcting, hastings2021dynamically, majidy2023unification}. Resolving these open questions would strengthen the theory's mathematical foundations and help turn structural constraints into design principles for minimizing the total cost of fault-tolerant architectures.

\bigskip

\begin{acknowledgments}

We thank Andrew Cameron, Ningping Cao, Philip Crowley, Sarang Gopalakrishnan, Daniel Gottesman, Michael Gullans, Yifan Hong, Milan Kornja\v{c}a, Anirudh Krishna, Mikhail Lukin, Arthur Morris, Chris Pattison, Adam Wills, and Harry Zhou for useful discussions and helpful feedback on the initial draft, with special thanks to Sajant Anand.

We also thank Hasti Majidy for taking on extra care of Kaius so S.M. could focus on the final push to finish this paper. We wish S.M. and H.M. a happy fifth wedding anniversary.

\textbf{Funding statement.} We acknowledge support from
the IARPA and the Army Research Office, under the Entangled Logical Qubits program (Cooperative Agreement Number W911NF-23-2-0219);
the U.S. Department of Energy (DOE Quantum Systems Accelerator, contract number 7568717);
the National Science Foundation (QLCI grant number 2553619).
We acknowledge additional support from the A*STAR Graduate Academy (J.M.K.), Banting Postdoctoral Fellowship (S.M.), NUS Development Grant (S.J.S.T.), and a Simons Investigator award (N.Y.Y.).

This material was funded in part by the U.S. Department of Energy, Office of Science, Office of Advanced Scientific Computing Research. Sandia National Laboratories is a multimission laboratory managed and operated by National Technology and Engineering Solutions of Sandia, LLC, a wholly owned subsidiary of Honeywell International Inc., for the U.S. Department of Energy’s National Nuclear Security Administration under contract DE-NA0003525. This paper describes objective technical results and analysis. Any subjective views or opinions that might be expressed in the paper do not necessarily represent the views of the U.S. Department of Energy or the United States Government.

\textbf{Use of AI tools.} GPT 5.1--5.4 Thinking and 5.5 Pro (OpenAI) and Claude Opus 4.6--4.8 (Anthropic) were used to generate candidate proofs for a subset of results in \cref{sec:css_codes/trans,sec:css_codes/perm,sec:stab_codes/trans,sec:stab_codes/auto,sec:cost} from target statements and context formulated by J.~M.~K.~and S.~M.
All proofs involving AI use were verified, refined, and rewritten by the authors, and common concepts and structure were manually factored out for integration into the broader project. 
The authors also acknowledge the use of AI tools to improve the writing of the manuscript. 
The authors take full responsibility for all content.

\textbf{Note added.} In the final stages of writing this manuscript, the authors were made aware of a parallel work~\cite{albert2026beyond} also studying the logical actions of automorphism gates on CSS codes.

\end{acknowledgments}

\clearpage

\let\oldaddcontentsline\addcontentsline
\renewcommand{\addcontentsline}[3]{}
\bibliography{references}
\let\addcontentsline\oldaddcontentsline

\clearpage

\makeatletter

\def\set@footnotewidth@full{%
  \hsize\textwidth
  \linewidth\hsize
}

\newcommand{\onecolumnwithfootnotes}{%
  \onecolumngrid
  %
  \let\set@footnotewidth\set@footnotewidth@full
  %
  \let\compose@footnotes\compose@footnotes@one
  \expandafter\let
    \csname combine@insert@1\endcsname
    \combine@insert@@ne
}

\newcommand{\twocolumnwithfootnotes}{%
  %
  \AddToHookNext{shipout/after}{%
    \global\let\set@footnotewidth\set@footnotewidth@two
    \global\let\compose@footnotes\compose@footnotes@two
    \global\expandafter\let
      \csname combine@insert@1\endcsname
      \combine@insert@tw@
  }%
  \twocolumngrid
}

\makeatother

\appendix
\onecolumnwithfootnotes

\begin{center}
    \textbf{APPENDICES}
\end{center}

\tableofcontents

\clearpage

\section{General definitions and preliminaries}
\label{app:preliminaries}

This appendix collects the background material used throughout the appendices; \cref{tab:summary_of_notation} summarizes notation. Projective Clifford logical actions can be represented by binary symplectic matrices, so we start with symplectic linear algebra over $\mathbb{F}_2$ (\cref{app:linear_algebra}). Our theorems constrain which subgroups of the symplectic group can arise as logical groups, and several proofs proceed by comparing group orders, so we define the relevant subgroups and record their orders (\cref{app:group_theory}). Then, we fix stabilizer-code conventions and define the three gate classes and their logical groups (\cref{app:stabilizer_codes}). Lastly, a number of no-go arguments hinge on characteristic polynomials, so we collect the needed facts about polynomials over $\mathbb{F}_2$ (\cref{app:polynomials}).

\begin{table}[!h]
    \centering
    \renewcommand{\arraystretch}{1.2}
    \begin{tabular}{p{3cm} p{15cm}}
        \toprule
        \textbf{Notation} & \textbf{Meaning} \\
        \midrule
        $\db{n,k}$ 
            & Number of physical qubits $n$ and logical qubits $k$ of a stabilizer code; any distance.
        \\
        $\db{n,k,d}$ 
            & Number of physical qubits $n$, logical qubits $k$, distance $d$ of a stabilizer code. 
        \\
        $\db{n, k, (d_\mathrm{x}, d_\mathrm{z})}$ 
            & Number of physical qubits $n$, logical qubits $k$, $X$- and $Z$-distances $(d_\mathrm{x}, d_\mathrm{z})$ of a CSS code.
        \\
        $[m]$
            & $[m] \coloneqq \{1, \ldots, m\}$, the set of the first $m$ positive integers.
        \\
        $\Omega_{k}$
            & Gram matrix of the standard symplectic form on $\mathbb{F}_2^{2k}$; $\Omega_{k} = \begin{psmallmatrix} 0& I_k \\ I_k & 0\end{psmallmatrix}$.
        \\
        $\mathrm{Cl}_k$
            & Clifford group on $k$ qubits.
        \\
        $\mathrm{PCl}_k$ 
            & Projective Clifford group on $k$ qubits (i.e.~$\mathrm{Cl}_k$ modulo Paulis and global phases).
        \\
        $C_m, S_m, A_m$
            & Cyclic group $C_m$ of order $m$; symmetric group $S_m$ and alternating group $A_m$ on $m$ elements.
        \\
        $\GL(m, \mathbb{F}_2)$
            & General linear group of degree $m$ over $\mathbb{F}_2$.
        \\
        $\Sp(2m, \mathbb{F}_2)$
            & Symplectic group of degree $2m$ over $\mathbb{F}_2$; $\mathrm{PCl}_k \cong \mathrm{Sp}(2k, \mathbb{F}_2)$.
            \textit{See \cref{def:group_theory_symplectic_group}.}
        \\
        $\mathcal{P}(2m, \mathbb{F}_2)$
            & The Siegel parabolic subgroup of $\Sp(2m, \mathbb{F}_2)$.
            \textit{See \cref{def:group_theory_siegel_parabolic_subgroups}.}
        \\
        $\mathcal{U}(2m, \mathbb{F}_2)$
            & The unipotent radical of the Siegel parabolic subgroup $\mathcal{P}(2m, \mathbb{F}_2)$.
            \textit{See \cref{def:group_theory_siegel_parabolic_unipotent_radicals}.}
        \\
        $O^\pm(2m, \mathbb{F}_2)$
            & Binary orthogonal groups of degree $2m$ preserving quadratic forms $Q^\pm$.
            \textit{See \cref{def:group_theory_binary_orthogonal_groups}.}
        \\
        $\Omega^\pm(2k, \mathbb{F}_2)$
            & Binary orthogonal cores: the derived subgroups $[O^\pm(2k, \mathbb{F}_2), O^\pm(2k, \mathbb{F}_2)]$.
            \textit{See \cref{def:group_theory_binary_orthogonal_cores}.}
        \\
        $O(m, \mathbb{F}_2)$
            & Binary orthogonal group of degree $m$ preserving the bilinear form with Gram matrix $I_m$.
            \textit{See \cref{def:flagged_isometry_groups}.}
        \\
        $\mathcal{F}^{\Sp/O}_m(k, \mathbb{F}_2)$
            & Flagged isometry subgroups of $\GL(k, \mathbb{F}_2)$ preserving a pair of nested subspaces.
            \textit{See \cref{def:flagged_isometry_groups}.}
        \\
        $\rs(H), \cs(H)$ 
            & Row span and column span of a matrix $H$ over the field of $H$.
        \\
        $\mathbb{F}_q$ 
            & Finite (i.e.~Galois) field over a prime power $q = p^m$ (where $p$ is prime and $m \geq 1$).
        \\
        $\overline{\mathbb{F}}_q$ 
            & Algebraic closure of $\mathbb{F}_q$; $\overline{\mathbb{F}}_q = \bigcup_{m \geq 1} \mathbb{F}_{q^m}$.
        \\
        $[v]_Q$
            & Equivalence class of $v \in V$ in a quotient space $Q=V/U$; $[v]_Q\coloneqq v+U$.
        \\
        $X$, $Y$, $Z$, $H$, $S$
            & Single-qubit Pauli gates, Hadamard gate, phase gate.
        \\
        $\mathrm{CX}, \mathrm{CY}, \mathrm{CZ}, \mathrm{SWAP}$
            & Two-qubit controlled-Pauli and swap gates.
        \\
        $\mathbf{S}_k$
            & The complete set of $S$ gates on $k$ qubits; $\mathbf{S}_k = \{ S_i: i \in [k] \}$.
        \\
        $\mathbf{CX}_k$
            & The complete set of $\mathrm{CX}$ gates on $k$ qubits; $\mathbf{CX}_k = \{ \mathrm{CX}_{ij}: i, j \in [k], i \neq j \}$.
        \\
        $\mathbf{CZ}_k$
            & The complete set of $\mathrm{CZ}$ gates on $k$ qubits; $\mathbf{CZ}_k = \{ \mathrm{CZ}_{ij}: i, j \in [k], i < j \}$.
        \\
        $\mathbf{D}^{(t)}_k$
            & Projective group of all diagonal gates at the $t^\text{th}$ level of the Clifford hierarchy on $k$ qubits.
            \textit{See \cref{def:clifford_hierarchy_diagonal_gates}.}
        \\
        $\mathrm{Perm}_\mathcal{L}(\mathcal{C})$
            & Group of logical actions induced by permutation gates of a code $\mathcal{C}$ in a logical basis $\mathcal{L}$.
            \textit{See \cref{def:transversal_permutation_automorphism_logical_groups}.}
        \\
        $\mathrm{Trans}_\mathcal{L}(\mathcal{C})$
            & Group of logical actions induced by transversal gates of a code $\mathcal{C}$ in a logical basis $\mathcal{L}$.
            \textit{See \cref{def:transversal_permutation_automorphism_logical_groups}.}
        \\
        $\mathrm{Aut}_\mathcal{L}(\mathcal{C})$
            & Group of logical actions induced by automorphism gates of a code $\mathcal{C}$ in a logical basis $\mathcal{L}$.
            \textit{See \cref{def:transversal_permutation_automorphism_logical_groups}.}
        \\
        $A \cong B$
            & The matrices $A$ and $B$ are congruent.
        \\
        $G \cong K$
            & The groups $G$ and $K$ are isomorphic.
        \\
        $G \leq K$ ($G < K$)
            & $G$ is a subgroup (proper subgroup) of $K$.
        \\
        $G \lhd K$ 
            & $G$ is a normal subgroup of $K$ (i.e.~$r g r^{-1}\in G$, $\forall r\in K, g\in G$).
        \\
        $G \sim_R K$ 
            & $G$ is $R$-conjugate to $K$ (i.e.~$\exists \, r \in R: r G r^{-1} = K$).
        \\
        $G \preceq_R K$ 
            & $G$ is $R$-conjugate to a subgroup of $K$ (i.e.~$\exists \, r \in R: r G r^{-1} \leq K$).
        \\
        $[G:K]$
            & The index of the subgroup $K$ in $G$, i.e~the number of left cosets of $K$ in $G$; $[G:K] \coloneqq |G|/|K|$.
        \\
        \bottomrule
    \end{tabular}
    \caption{Summary of notation used throughout this work.}
    \label{tab:summary_of_notation}
\end{table}

Notation introduced and used only within a single proof or section is defined locally. There remain notational overloading and potential ambiguities which we make explicit:

\begin{itemize}
    \item Throughout our work, the number of physical qubits $n$ and logical qubits $k$ are assumed to be nonzero ($n \geq k \ge 1$). While $k = 0$ stabilizer codes are well-defined, corresponding to stabilizer states, the notion of logical gates and groups is usually no longer meaningful.
    \item We use $H$ to denote both the Hadamard gate and a stabilizer generator matrix, following standard conventions. Context makes the intended meaning clear. We only use $H$ for subgroups in \cref{app:stab_codes/prelims} to be consistent with classification results from Refs.~\cite{aschbacher1984maximal,yin2025large}.
    \item $S_m$ can be the $S$ gate on the $m^\text{th}$ qubit or the symmetric group of $m$ elements. Context makes the intended meaning clear.
    \item For matrices, $A \cong B$ means that the two matrices are congruent (see \cref{def:linear_algebra_congruence_gram_matrices}); for groups, $G \cong K$ means that the groups are isomorphic. Context makes the intended meaning clear.
    \item $\mathrm{PCl}_k \cong \Sp(2k, \mathbb{F}_2)$ and, for brevity, we do not distinguish between them. Thus, for example, for a subgroup $G \leq \mathrm{PCl}_k$ and $K \leq \Sp(2k, \mathbb{F}_2)$, we may write that $G \leq K$. That is, conversion between projective Cliffords and their symplectic representations is implicit when not stated.
    \item $\comm{P_1}{P_2}$ for two Pauli operators denotes the commutator $P_1 P_2 - P_2 P_1$; $\comm{G}{K}$ for two subgroups $G$ and $K$ of a common group denotes the group generated by all group commutators between elements of $G$ and $K$, $\comm{G}{K}\coloneqq\langle ghg^{-1}h^{-1}:g\in G,\ h\in K\rangle$.
    \item We work with fixed coordinate systems (i.e.~bases) in most settings, often naturally supplied by the physical- or logical-qubit indexing of a code, and therefore usually use explicit matrix groups such as $\GL(m,\mathbb F_2)$ and $\Sp(2m,\mathbb F_2)$. In select places, however, it is clearer to refer directly to transformation groups of an abstract vector space. For a vector space $V$, we write $\GL(V)$ for the group of invertible linear transformations of $V$. If $V$ is equipped with a symplectic form (\cref{def:symplectic_form_space}), we write $\Sp(V)$ for the group of invertible linear transformations of $V$ that preserve this form.
    \item Zero-dimensional edge cases rarely occur in the setting of this paper; but to entirely avoid ambiguity, we declare zero-dimensional matrix groups to be the trivial group. That is, $\GL(0, \mathbb{F}_2) = \Sp(0, \mathbb{F}_2) = O^+(0, \mathbb{F}_2) = O(0, \mathbb{F}_2) = \{I\}$, with order one. See \cref{def:group_theory_symplectic_group,def:group_theory_binary_orthogonal_groups,def:flagged_isometry_groups} for the definitions of these groups.
\end{itemize}

\clearpage

\subsection{Binary symplectic and quadratic spaces}
\label{app:linear_algebra}

\subsubsection{Bilinear forms and binary symplectic spaces}
\label{app:linear_algebra/bilinear_symplectic_forms}

\begin{definition}
    [Bilinear form]
    \label{def:bilinear_form}
    Let $V$ be a finite-dimensional vector space over a field $\mathbb{F}$. A \emph{bilinear form} on $V$ is a map $\beta : V \times V \to \mathbb{F}$ that is linear in each argument. The form $\beta$ is
    \begin{enumerate}[noitemsep]
        \item \emph{symmetric} if $\beta(v,w) = \beta(w,v)$ for all $v, w \in V$;
        \item \emph{alternating} if $\beta(v,v) = 0$ for all $v \in V$;
        \item \emph{nondegenerate} if $\beta(v,w) = 0$ for all $w \in V$ implies $v = 0$.\footnote{Since $V$ is finite-dimensional, the same condition with the arguments swapped is equivalent.}
    \end{enumerate}
\end{definition}

\begin{definition}
    [Radical of a symmetric bilinear form]
    \label{def:radical_bilinear_form}
    Let $\beta$ be a symmetric bilinear form on a finite-dimensional vector space $V$ over $\F$. The \emph{radical} of $\beta$ is the subspace
    \begin{equation}
        \rad(\beta)
        \coloneqq
        \{v\in V: \beta(v,v')=0 \,\, \forall \,\, v'\in V\}
        \subseteq
        V.
    \end{equation}
    The form $\beta$ is nondegenerate iff $\rad(\beta)=\{0\}$.
    
    Moreover, $\beta$ induces a symmetric bilinear form $\overline{\beta}$ on the quotient space $V/\rad(\beta)$, defined by
    \begin{equation}
        \overline{\beta}
        \bigl(v+\rad(\beta),w+\rad(\beta)\bigr)
        \coloneqq
        \beta(v,w),
        \qquad
        v,w\in V.
    \end{equation}
    This is well-defined: if $v'=v+r$ and $w'=w+r'$ for $r,r'\in\rad(\beta)$, then $\beta(v',w')=\beta(v,w)+\beta(v,r')+\beta(r,w)+\beta(r,r')=\beta(v,w)$.
    The induced form $\overline{\beta}$ is nondegenerate on $V/\rad(\beta)$.
    Indeed, if a coset $v+\rad(\beta)$ lies in the radical of $\overline{\beta}$, then $\overline{\beta}(v+\rad(\beta),w+\rad(\beta))=0 \,\, \forall \,\, w\in V \Longrightarrow \beta(v,w)=0 \,\, \forall \,\, w\in V \Longrightarrow v\in \rad(\beta)$. Hence $v+\rad(\beta)=\rad(\beta)$ is the zero element of $V/\rad(\beta)$, showing that $\overline{\beta}$ is nondegenerate.
\end{definition}

\begin{definition}
    [Symplectic form and symplectic space]
    \label{def:symplectic_form_space}
    Let $V$ be a finite-dimensional vector space over a field $\mathbb{F}$. A \emph{symplectic form} on $V$ is an alternating nondegenerate bilinear form $\omega : V \times V \to \mathbb{F}$. The pair $(V, \omega)$ is a \emph{symplectic space} and is called \emph{binary} when $\mathbb{F} = \mathbb{F}_2$. 
    Every finite-dimensional symplectic space has even dimension, $\dim V=2k$. An ordered basis $(e_1,\ldots,e_k,f_1,\ldots,f_k)$ of $V$ is a \emph{symplectic basis} if $\omega(e_i, f_j) = \delta_{ij}$ and $\omega(e_i, e_j) = \omega(f_i, f_j) = 0$.
\end{definition}

\begin{definition}
    [Gram matrices, congruence, and isometry]
    \label{def:linear_algebra_congruence_gram_matrices}
    Let $\beta$ be a bilinear form on an $n$-dimensional vector space $V$ over a field $\mathbb{F}$ and fix an ordered basis $(v_1, \ldots, v_n)$ of $V$. The \emph{Gram matrix} of $\beta$ in this basis is the matrix $A \in \mathbb{F}^{n \times n}$ with entries
    \begin{equation}
        A_{ij} \coloneqq \beta(v_i, v_j).
    \end{equation}
    If $x, y \in \mathbb{F}^n$ are the coordinate rows of $a, b \in V$ in this basis, i.e.~$a=\sum_{i=1}^n x_i v_i$ and $b=\sum_{i=1}^n y_i v_i$, then $\beta(a, b)=x^\top A y$. Under a change of ordered basis $w_i=\sum_j P_{ij}v_j$, with $P\in\GL(n,\mathbb{F})$, the Gram matrix becomes $PAP^\top$. This motivates two notions of equivalence, one for matrices and one for spaces.
    \begin{enumerate}[noitemsep]
        \item Matrices $A, B \in \mathbb{F}^{n \times n}$ are \emph{congruent}, written $A \cong B$, if $B = PAP^\top$ for some $P \in \GL(n,\mathbb{F})$.
        \item Let $\beta'$ be a bilinear form on a vector space $V'$ over the same field $\mathbb{F}$. The spaces $(V,\beta)$ and $(V',\beta')$ are \emph{isometric} if there is a linear isomorphism $\varphi : V \to V'$ with $\beta'(\varphi(a),\varphi(b)) = \beta(a,b)$ for all $a,b \in V$; such a $\varphi$ is an \emph{isometry}.
    \end{enumerate}
    $(V,\beta)$ and $(V',\beta')$ are isometric exactly when $\dim V = \dim V'$ and their Gram matrices are congruent. The choice of ordered bases does not matter here, since changing basis replaces a Gram matrix by a congruent one.
\end{definition}

\begin{fact}
    [Gram matrix criteria]
    \label{fact:linear_algebra_matrix_criteria}
    Let $\beta$ be a bilinear form on an $n$-dimensional vector space over a field $\mathbb{F}$ with Gram matrix $A$ in some ordered basis (\cref{def:linear_algebra_congruence_gram_matrices}). Then $\beta$ is
    \begin{enumerate}[noitemsep]
        \item symmetric iff $A = A^\top$;
        \item nondegenerate iff $A \in \GL(n,\mathbb{F})$;
        \item alternating iff $A = -A^\top$ and $\diag(A) = \mathbf{0}$. The diagonal condition is automatic unless $\mathbb{F}$ has characteristic $2$. In particular, over $\mathbb{F}_2$, the criterion is $A = A^\top$ and $\diag(A) = \mathbf{0}$.
    \end{enumerate}
    Over $\mathbb{F}_2$ we have $-A^\top = A^\top$, so every alternating form is symmetric. In particular, every binary symplectic form is symmetric and nondegenerate.
\end{fact}

\begin{definition}
    [Standard binary symplectic space]
    \label{def:linear_algebra_standard_binary_symplectic_form}
    Let $\mathbb{F}_2^{2k}=\mathbb{F}_2^k\oplus\mathbb{F}_2^k$ be the $2k$-dimensional vector space over $\mathbb{F}_2$. For $(x,z),(x',z')\in \mathbb{F}_2^k\oplus\mathbb{F}_2^k$, define
    \begin{equation}
    \langle (x,z),(x',z')\rangle \coloneqq x\cdot z' + z\cdot x'.
    \end{equation}
    This is the \emph{standard symplectic form}, and $\mathbb{F}_2^{2k}$ equipped with this form is the \emph{standard binary symplectic space}. In the standard ordered basis of $\mathbb{F}_2^{2k}$, the Gram matrix of the standard symplectic form is
    \begin{equation}
    \Omega_k \coloneqq \mqty(0 & I_k \\ I_k & 0) \in\mathbb{F}_2^{2k\times 2k}.
    \end{equation}
    Since $\Omega_k=\Omega_k^\top$ and $\diag(\Omega_k)=\mathbf{0}$, the form is alternating by \cref{fact:linear_algebra_matrix_criteria}. Since $\Omega_k^2=I_{2k}$, it is nondegenerate. Hence the form is symplectic.
\end{definition}

\begin{fact}
    [Albert classification over $\mathbb{F}_2$]
    \label{fact:linear_algebra_albert_classification_f2}
    Let $\beta$ be a symmetric nondegenerate bilinear form on an $n$-dimensional vector space over $\mathbb{F}_2$, with Gram matrix $A$ in some ordered basis. Then $A$ is congruent to a standard matrix determined by whether $\beta$ is alternating~\cite{Albert1938SymmetricAlternate}:
    \begin{itemize}[noitemsep]
        \item if $\beta$ is alternating, then $n$ is even and $A \cong \Omega_{n/2}$;
        \item if $\beta$ is not alternating, then $A \cong I_n$.
    \end{itemize}
    For odd $n$ only the second case occurs; for positive even $n$ the two cases are mutually exclusive, since congruence preserves the alternating property and $\Omega_{n/2}$ is alternating while $I_n$ is not. Thus, in each positive dimension, alternation is a complete invariant of nondegenerate symmetric bilinear forms over $\mathbb{F}_2$ under congruence~\cite{Albert1938SymmetricAlternate}; see also Lebedev's restatement \cite[Sec.~1.5, Thm.~2.1]{lebedev2006}. Consequently, every $2k$-dimensional binary symplectic space is isometric to the standard binary symplectic space of dimension $2k$ in \cref{def:linear_algebra_standard_binary_symplectic_form}.
\end{fact}

\subsubsection{Binary quadratic forms and their classification}
\label{app:linear_algebra/quadratic_forms}

\begin{definition}
    [Binary quadratic form and quadratic space]
    \label{def:linear_algebra_binary_quadratic_forms_plus_minus}
    Let $V$ be a finite-dimensional vector space over $\mathbb{F}_2$. For a map $Q: V\to\mathbb{F}_2$, define its \emph{polar form} by
    \begin{equation}
        B_Q(v,w) \coloneqq Q(v+w) + Q(v) + Q(w). 
    \end{equation}
    The map $Q$ is a \emph{quadratic form} if $B_Q$ is bilinear. Given a bilinear form $B: V\times V\to\mathbb{F}_2$, we say that $Q$ \emph{polarizes} to $B$ if $B_Q=B$. The pair $(V,Q)$ is a \emph{quadratic space} and is \emph{nondegenerate} if $B_Q$ is nondegenerate. Two quadratic spaces $(V,Q)$ and $(V',Q')$ are \emph{isometric} if there is a linear isomorphism $\varphi : V \to V'$ with $Q'(\varphi(v)) = Q(v)$ for all $v \in V$; such a $\varphi$ is then also an isometry of the polar forms.
\end{definition}

\begin{fact}
    [Polar forms of quadratic forms]
    \label{fact:polar_of_quadratic}
    Let $Q : V \to \mathbb{F}_2$ be a quadratic form on a finite-dimensional vector space over $\mathbb{F}_2$, and let $B_Q$ be its polar form (\cref{def:linear_algebra_binary_quadratic_forms_plus_minus}). Bilinearity forces $B_Q(0,0) = 0$ and the defining formula gives $B_Q(0,0) = Q(0) + Q(0) + Q(0) = Q(0)$. Hence $Q(0) = 0$, and
    \begin{equation}
        B_Q(v,v) = Q(v+v) + Q(v) + Q(v) = Q(0) = 0
    \end{equation}
    is alternating for all $v \in V$. Consequently, if $(V,Q)$ is nondegenerate, then $(V,B_Q)$ is a binary symplectic space.
\end{fact}

\begin{definition}
    [Standard quadratic forms of plus and minus type]
    \label{def:standard_quad_form}
    On the standard binary symplectic space $\mathbb{F}_2^{2k}=\mathbb{F}_2^k\oplus\mathbb{F}_2^k$, define the forms
    \begin{equation}
        Q^+(x,z)
        \coloneqq
        \sum_{i=1}^k x_i z_i,
        \qquad
        Q^-(x,z)
        \coloneqq
        x_1z_1+x_1+z_1+\sum_{i=2}^k x_i z_i,
    \end{equation}
    for $x,z\in\mathbb{F}_2^k$, where $k \ge 1$; note that $Q^-$ singles out the first coordinate and so is undefined for $k = 0$. Both forms polarize to the standard symplectic form. We call $Q^+$ the standard quadratic form of \emph{plus type} and $Q^-$ the standard quadratic form of \emph{minus type}. For $k = 0$ we adopt the convention that the zero space, carrying the zero form, has plus type.
\end{definition}

\begin{definition}
    [Orthogonal direct sum]
    \label{def:linear_algebra_orthogonal_direct_sum}
    The \emph{orthogonal direct sum} of quadratic spaces $(V_1,Q_1)$ and $(V_2,Q_2)$ is
    \begin{equation}
        (V_1, Q_1) \perp (V_2, Q_2) \coloneqq (V_1 \oplus V_2,\, Q_1 + Q_2),
    \end{equation}
    where $(Q_1+Q_2)(v_1,v_2) \coloneqq Q_1(v_1)+Q_2(v_2)$. Its polar form restricts to $B_{Q_1}$ and $B_{Q_2}$ on the two summands, which are mutually orthogonal.
\end{definition}

\begin{fact}
    [Classification of nondegenerate binary quadratic spaces]
    \label{fact:classification_nondegen_bqs}
    Every nondegenerate binary quadratic space has even dimension: by \cref{fact:polar_of_quadratic} its polar form makes it a binary symplectic space, and by \cref{def:symplectic_form_space} such spaces have even dimension.
    For $k \geq 1$, a nondegenerate binary quadratic space $(V,Q)$ of dimension $2k$ is isometric to exactly one of the two standard quadratic spaces defined in \cref{def:standard_quad_form}~\cite{kleidman1990subgroup}:
    \begin{itemize}[noitemsep]
        \item $(\mathbb{F}_2^{2k},Q^+)$, in which case $(V,Q)$ has \emph{plus type}, written $\epsilon(V,Q)=+$;
        \item $(\mathbb{F}_2^{2k},Q^-)$, in which case $(V,Q)$ has \emph{minus type}, written $\epsilon(V,Q)=-$.
    \end{itemize}
    (For $k=0$, the zero space has plus type by convention, as in \cref{def:standard_quad_form}.)
    Thus, for each fixed dimension, the type is a complete invariant of nondegenerate binary quadratic spaces under isometry.
    The type is multiplicative under the orthogonal direct sum of \cref{def:linear_algebra_orthogonal_direct_sum}. If $(V_1,Q_1)$ and $(V_2,Q_2)$ are nondegenerate, then so is $(V_1,Q_1)\perp(V_2,Q_2)$, and
    \begin{equation}
        \epsilon\bigl((V_1,Q_1)\perp(V_2,Q_2)\bigr)
        =
        \epsilon(V_1,Q_1)\,\epsilon(V_2,Q_2),
    \end{equation}
    with the usual multiplication of signs.
\end{fact}

\subsubsection{Isotropic subspaces of symplectic spaces}
\label{app:linear_algebra/isotropic_subspaces}

\begin{definition}
    [Isotropic subspaces and symplectic complements]
    \label{def:linear_algebra_isotropic_vector_subspace}
    Let $V$ be a vector space with a bilinear form $\langle\cdot,\cdot\rangle$, and let $U\subseteq V$ be a subspace. The subspace $U$ is \emph{totally isotropic} (or just \emph{isotropic}\footnote{%
    Some authors call $U$ isotropic when it contains a nonzero vector $u$ with $\langle u, u \rangle = 0$, reserving \emph{totally isotropic} for the condition defined here. In settings where the bilinear form $\langle\cdot,\cdot\rangle$ is alternating, every vector $u$ satisfies $\langle u,u\rangle=0$, so the weaker notion is vacuous and the shorter name unambiguous.}) if $\langle u, u' \rangle = 0$, for all $u, u' \in U$. When $(V,\langle\cdot,\cdot\rangle)$ is symplectic, the \emph{symplectic complement} of $U$ is
    \begin{equation}
        U^\perp \coloneqq \{v \in V : \langle v, u \rangle = 0 \,\, \forall \,\, u \in U\}.
    \end{equation}
\end{definition}

\begin{definition}
    [Radical of a subspace]
    \label{def:linear_algebra_radical}
    Let $U$ be a subspace of a symplectic space $(V,\langle\cdot,\cdot\rangle)$. The \emph{radical} of $U$ is the subspace
    \begin{equation}
        \rad(U)\coloneqq U\cap U^\perp,
    \end{equation}
    and is isotropic. The form restricted to $U$ is nondegenerate iff $\rad(U)=\{0\}$, and $U$ is isotropic iff $\rad(U)=U$.
\end{definition}

\begin{fact}
    [Symplectic complements and reduction]
    \label{fact:linear_algebra_symplectic_complement_properties}
    Let $(V, \langle \cdot, \cdot \rangle)$ be a $2k$-dimensional symplectic space. Then, for every subspace $U \subseteq V$:
    \begin{itemize}[noitemsep]
        \item $\dim U + \dim U^\perp = 2k$ and $(U^\perp)^\perp = U$;
        \item $U$ is isotropic iff $U \subseteq U^\perp$;
        \item if $U$ is isotropic, then $\rad(U^\perp) = U^\perp \cap (U^\perp)^\perp = U$. Hence the form induces a nondegenerate symplectic form on $U^\perp/\rad(U^\perp)=U^\perp/U$, which has dimension $2k - 2\dim U$. The resulting symplectic space is called the \emph{symplectic reduction} of $V$ along $U$.
    \end{itemize}
\end{fact}

\begin{definition}
    [Lagrangian subspaces]
    \label{def:linear_algebra_lagrangian_vector_subspace}
     Let $(V, \langle \cdot, \cdot \rangle)$ be a $2k$-dimensional symplectic space. A subspace $L \subseteq V$ is \emph{Lagrangian} if it is a maximal isotropic subspace: $L$ is isotropic, and no isotropic subspace of $V$ properly contains it.
\end{definition}

\begin{fact}
    [Characterizations of Lagrangian subspaces]
    \label{fact:linear_algebra_lagrangian_vector_subspace_condition}
    For a subspace $L$ of a $2k$-dimensional symplectic space, the following are equivalent: (i) $L$ is Lagrangian; (ii) $L = L^\perp$; (iii) $L$ is isotropic and $\dim L = k$.
\end{fact}

\begin{definition}
    [Flags and isotropic flags]
    \label{def:linear_algebra_flags_and_isotropic_flags}
    Let $V$ be a vector space. A \emph{flag} $U_\bullet$ of \emph{length} $r \geq 1$ in $V$ is a strictly increasing sequence of subspaces
    \begin{equation}
        U_\bullet \coloneqq 
        \{0\} \subsetneq U_1 \subsetneq U_2 \subsetneq \cdots
        \subsetneq U_r \subseteq V.
    \end{equation}
     If $V$ carries a bilinear form, the flag is \emph{isotropic} if $U_r$ is; since isotropy passes to subspaces, every $U_i$ is then isotropic as well.
\end{definition}

\begin{fact}
    [Adapted symplectic bases]
    \label{fact:linear_algebra_adapted_symplectic_bases}
    Let $(V, \langle \cdot, \cdot \rangle)$ be a $2k$-dimensional binary symplectic space and $U \subseteq V$ a totally isotropic subspace of dimension $r$. Then there exists a symplectic basis $(e_1, \ldots, e_k, f_1, \ldots, f_k)$ of $V$, satisfying $\langle e_i, f_j \rangle = \delta_{ij}$ and $\langle e_i, e_j \rangle = \langle f_i, f_j \rangle = 0$, such that $U = \spn\{e_1, \ldots, e_r\}$; we call such a basis \emph{adapted} to $U$. Equivalently---since any symplectic basis is carried to any other by an element of $\Sp(2k, \mathbb{F}_2)$---the group acts transitively on totally isotropic subspaces of each fixed dimension. A basis can likewise be chosen adapted simultaneously to every subspace of an isotropic flag (\cref{def:linear_algebra_flags_and_isotropic_flags}); explicitly, $U_i=\spn\{e_1,\ldots,e_{\dim U_i}\}$ for each subspace $U_i$ in the flag.
\end{fact}

\clearpage

\subsection{Symplectic groups and their subgroups}
\label{app:group_theory}

\begin{definition}
    [$p$-groups and Sylow $p$-subgroups]
    \label{def:group_theory_p_groups}
    \label{def:group_theory_sylow_p_groups}
    Let $p$ be a prime. A finite group $G$ is a \emph{$p$-group} if $\abs{G} = p^a$ for some integer $a \geq 0$; equivalently, every element of $G$ has order a power of $p$. Writing $\abs{G} = p^a m$ with $\gcd(p, m) = 1$, a subgroup $P \leq G$ is a \emph{Sylow $p$-subgroup} of $G$ if $\abs{P} = p^a$, that is, a $p$-subgroup of maximum possible order in $G$.
    The Sylow theorems state that for every prime factor $p$ of $\abs{G}$, there exists a Sylow $p$-subgroup of $G$; moreover, all Sylow $p$-subgroups are conjugate to each other.
\end{definition}

\subsubsection{Binary symplectic groups and their parabolic subgroups}
\label{app:gt_subgroups}

\begin{definition}
    [Binary symplectic group]
    \label{def:group_theory_symplectic_group}
    Let $\mathbb{F}_2^{2k}$ be the standard binary symplectic space, equipped
    with the standard symplectic form $\langle \cdot,\cdot\rangle$. The
    \emph{binary symplectic group} is
    \begin{equation}\begin{split}
        \Sp(2k,\mathbb{F}_2)
        &\coloneqq \left\{
            M \in \GL(2k,\mathbb{F}_2)
            :
            \langle Mv, Mw\rangle = \langle v,w\rangle
            \,\, \forall \,\, v,w \in \mathbb{F}_2^{2k}
        \right\}
        \\
        &= \left\{
            M \in \GL(2k,\mathbb{F}_2)
            :
            M^\top \Omega_k M = \Omega_k
        \right\}
        \leq \GL(2k,\mathbb{F}_2).
    \end{split}\end{equation}
\end{definition}

\begin{definition}
    [Reducible subgroups of the binary symplectic group]
    \label{def:group_theory_binary_reducible_subgroups}
    Let $V = \mathbb{F}_2^{2k}$ be the standard binary symplectic space, and let $G \leq \Sp(2k,\mathbb{F}_2)$. We say that $G$ is \emph{reducible} if there exists a nonzero proper subspace $U \subsetneq V$ such that $g U = U$ for every $g \in G$. That is, $G$ preserves a nontrivial proper subspace of the ambient symplectic space. Otherwise, $G$ is called \emph{irreducible}.
    Every overgroup of an irreducible group is irreducible: suppose $G\le K\le \Sp(2k,\F_2)$ and $G$ is irreducible, then $K$ is irreducible as well.
\end{definition}

\begin{definition}
    [Parabolic subgroups of the binary symplectic group]
    \label{def:group_theory_parabolic_subgroups}
    Let $V=\mathbb{F}_2^{2k}$ be the standard binary symplectic space. A \emph{parabolic subgroup} of $\Sp(2k,\mathbb{F}_2)$ is the stabilizer of a totally isotropic flag (\cref{def:linear_algebra_flags_and_isotropic_flags}) of length  $r \geq 1$, $\{0\} \subsetneq U_1 \subsetneq U_2 \subsetneq \cdots \subsetneq U_r \subseteq V$,
    \begin{equation}
    \mathcal{P}_{U_\bullet} \coloneqq \left\{ M \in \Sp(2k,\mathbb{F}_2) : M U_i = U_i \,\, \forall \, i \in [r] \right\} \leq \Sp(2k,\mathbb{F}_2).
    \end{equation}
    If $r=1$, we call $\mathcal{P}_{U_\bullet}$ a \emph{maximal parabolic subgroup}. A maximal parabolic subgroup is determined up to conjugacy in $\Sp(2k,\mathbb{F}_2)$ by $m \coloneqq \dim U_1$ (\cref{fact:linear_algebra_adapted_symplectic_bases}), and we write $P_m(2k, \mathbb{F}_2)$ for any representative. For $m = k$, the stabilized subspace is a Lagrangian; these stabilizers form the Siegel parabolic subgroups, defined below.
\end{definition}

\begin{remark}
    [Reducibility and parabolic subgroups]
    \label{rem:invariant_subspaces_isotropic}
     If a subgroup of $\Sp(2k,\mathbb{F}_2)$ preserves a nonzero proper \emph{isotropic} subspace, it lies in a parabolic subgroup, namely the stabilizer of that subspace. However, reducibility (\cref{def:group_theory_binary_reducible_subgroups}) is weaker. The invariant subspace need not be isotropic, so a reducible subgroup need not lie in any parabolic subgroup.
\end{remark}

\begin{definition}
    [Siegel parabolic subgroups of the binary symplectic group]
    \label{def:group_theory_siegel_parabolic_subgroups}
    A \emph{Siegel parabolic subgroup} of $\Sp(2k, \mathbb{F}_2)$ is the stabilizer of a Lagrangian subspace (\cref{def:linear_algebra_lagrangian_vector_subspace}). 
    We give two standard representatives. 
    First, for a subspace $U$ of the standard binary symplectic space $V \coloneqq \mathbb F_2^{2k}$, we write its stabilizer as
    \begin{equation}
        \Stab_{\Sp(2k,\mathbb{F}_2)}(U)
        \coloneqq \left\{ M \in \Sp(2k,\mathbb{F}_2) : M U = U \right\}.
    \end{equation}
    Let $(e_1,\ldots,e_k,f_1,\ldots,f_k)$ denote the standard symplectic basis of $V$, and $\mathcal{X} = \mathbb{F}_2^k \oplus \{0\} = \spn\nolimits_{\mathbb{F}_2}\{e_1,\ldots,e_k\} \subseteq \mathbb{F}_2^{2k}$ and $\mathcal{Z} = \{0\} \oplus \mathbb{F}_2^k = \spn\nolimits_{\mathbb{F}_2}\{f_1,\ldots,f_k\} \subseteq \mathbb{F}_2^{2k}$ be the complementary Lagrangians supported on the first $k$ and last $k$ coordinates, respectively. Their stabilizers are:
    \begin{subequations}\begin{align}
        \mathcal{P}_\mathrm{x}(2k, \mathbb{F}_2)
        \coloneqq \Stab_{\Sp(2k,\mathbb{F}_2)}(\mathcal{X})
        &= \left\{
        \begin{pmatrix}
            A & B \\
            0 & A^{-\top}
        \end{pmatrix}
        :
        A \in \GL(k,\mathbb{F}_2),
        \;
        A^{-1}B = (A^{-1}B)^\top
        \right\}
        \notag
        \\
        &= \left\{
        \begin{pmatrix}
            A & AS \\
            0 & A^{-\top}
        \end{pmatrix}
        :
        A \in \GL(k,\mathbb{F}_2),
        \;
        S=S^\top
        \right\}
        \leq \Sp(2k, \mathbb{F}_2),
        \label{eq:group_theory_siegel_parabolic_subgroups_form_x}
        \\[1ex]
        \mathcal{P}_\mathrm{z}(2k, \mathbb{F}_2)
        \coloneqq \Stab_{\Sp(2k,\mathbb{F}_2)}(\mathcal{Z})
        &= \left\{
        \begin{pmatrix}
            A & 0 \\
            B & A^{-\top}
        \end{pmatrix}
        :
        A \in \GL(k,\mathbb{F}_2),
        \;
        A^\top B = (A^\top B)^\top
        \right\}
        \notag
        \\
        &= \left\{
        \begin{pmatrix}
            A & 0 \\
            SA & A^{-\top}
        \end{pmatrix}
        :
        A \in \GL(k,\mathbb{F}_2),
        \;
        S=S^\top
        \right\}
        \leq \Sp(2k, \mathbb{F}_2).
        \label{eq:group_theory_siegel_parabolic_subgroups_form_z}
    \end{align}\end{subequations}
\end{definition}

\begin{fact}
    [Conjugacy of the Siegel parabolic subgroups]
    \label{fact:group_theory_siegel_parabolic_conjugacy}
    Since $\Sp(2k, \mathbb{F}_2)$ acts transitively on Lagrangians (\cref{fact:linear_algebra_adapted_symplectic_bases}), any two Siegel parabolic subgroups are conjugate; that is, the Siegel parabolic subgroups form a single conjugacy class in $\Sp(2k, \mathbb{F}_2)$. For the standard representatives, $\mathcal{P}_\mathrm{x}(2k, \mathbb{F}_2) = \Omega_k \mathcal{P}_\mathrm{z}(2k, \mathbb{F}_2) \Omega_k^{-1}$. We use the shorthand $\mathcal{P}(2k, \mathbb{F}_2) \coloneqq \mathcal{P}_\mathrm{z}(2k, \mathbb{F}_2)$ since the choice of representative is immaterial.
\end{fact}

\begin{definition}
    [Unipotent radicals of the Siegel parabolic subgroups]
    \label{def:group_theory_siegel_parabolic_unipotent_radicals}
    The \emph{unipotent radical} $\mathcal{U}_\mu(2k, \mathbb{F}_2)$ of the Siegel parabolic subgroup $\mathcal{P}_\mu(2k, \mathbb{F}_2)$ is the subgroup
    obtained by setting $A = I_k$ in the block form of
    $\mathcal{P}_\mu(2k, \mathbb{F}_2)$, as written in \cref{eq:group_theory_siegel_parabolic_subgroups_form_x,eq:group_theory_siegel_parabolic_subgroups_form_z}, for $\mu \in \{\mathrm{x}, \mathrm{z}\}$. Equivalently, these are the kernels of the natural projections
    \begin{subequations}\begin{alignat}{2}
        \mathcal{P}_\mathrm{x}(2k,\mathbb{F}_2)
        &\rightarrow
        \GL(k,\mathbb{F}_2),
        \qquad&
        \begin{pmatrix}
            A & AS \\
            0 & A^{-\top}
        \end{pmatrix}
        &\mapsto A,
        \label{eq:group_theory_siegel_parabolic_projection_x}
        \\
        \mathcal{P}_\mathrm{z}(2k,\mathbb{F}_2)
        &\rightarrow
        \GL(k,\mathbb{F}_2),
        \qquad&
        \begin{pmatrix}
            A & 0 \\
            SA & A^{-\top}
        \end{pmatrix}
        &\mapsto A.
        \label{eq:group_theory_siegel_parabolic_projection_z}
    \end{alignat}\end{subequations}

    The unipotent radicals appear in the Levi decomposition $\mathcal{P}_\mu(2k, \mathbb{F}_2) \cong \mathcal{U}_\mu(2k, \mathbb{F}_2) \rtimes \GL(k, \mathbb{F}_2)$. 
    They are elementary abelian $2$-groups---in particular $\mathcal{U}_\mu(2k, \mathbb{F}_2) \cong \smash{C_2^{k(k+1)/2}}$ as abstract groups.
    Mirroring \cref{fact:group_theory_siegel_parabolic_conjugacy}, the two unipotent radicals are conjugate in $\Sp(2k, \mathbb{F}_2)$: $\mathcal{U}_\mathrm{x}(2k, \mathbb{F}_2) = \Omega_k \mathcal{U}_\mathrm{z}(2k, \mathbb{F}_2) \Omega_k^{-1}$. 
    We use the shorthand $\mathcal{U}(2k, \mathbb{F}_2) \coloneqq \mathcal{U}_\mathrm{z}(2k, \mathbb{F}_2)$ since the choice of representative is immaterial.
\end{definition}

\subsubsection{Binary orthogonal groups and their cores}
\label{app:gt_orthogonal}

\begin{definition}
    [Binary orthogonal groups]
    \label{def:group_theory_binary_orthogonal_groups}
    Let \(Q^\pm\) be the plus- and minus-type quadratic forms from
    \cref{def:standard_quad_form}. The corresponding \emph{binary orthogonal groups} are
    \begin{equation}
        O^\pm(2k,\mathbb{F}_2)
        \coloneqq
        \left\{
            M\in \GL(2k,\mathbb{F}_2) : Q^\pm(Mv)=Q^\pm(v) \,\, \forall \,\, v\in \mathbb{F}_2^{2k}
        \right\}.
    \end{equation}
    Since \(Q^\pm\) polarize to the standard symplectic form, every \(Q^\pm\)-preserving transformation also preserves its polar form (\cref{def:linear_algebra_binary_quadratic_forms_plus_minus}), so $O^\pm(2k,\mathbb{F}_2) \leq \Sp(2k,\mathbb{F}_2)$.
\end{definition}

\begin{definition}
    [Orthogonal cores of binary orthogonal groups]
    \label{def:group_theory_binary_orthogonal_cores}
    The groups \(\Omega^\pm(2k,\mathbb{F}_2)\coloneqq
    [O^\pm(2k,\mathbb{F}_2),O^\pm(2k,\mathbb{F}_2)]\) are the derived subgroups of the corresponding binary orthogonal groups; we call them the \emph{binary orthogonal cores} of the corresponding binary orthogonal groups.
    (Here, for elements \(g, h\) of a group \(G\), \([g,h]\coloneqq ghg^{-1}h^{-1}\) is the group commutator and \([G,G]\) is the subgroup generated by all such commutators, as notationally introduced in \cref{app:preliminaries}.)
    The notation $\Omega^\pm(2k,\mathbb{F}_2)$ should not be confused with the standard alternating matrix $\Omega_k$: the former are finite groups, while the latter is the $2k \times 2k$ Gram matrix of the symplectic form.
\end{definition}

\begin{fact}
    [Index of the binary orthogonal cores]
    \label{fact:group_theory_binary_orthogonal_core_index}
    For all $k \geq 1$, the binary orthogonal cores are normal subgroups of the corresponding binary orthogonal groups, $\Omega^\pm(2k,\mathbb{F}_2) \triangleleft O^\pm(2k,\mathbb{F}_2) \leq \Sp(2k,\mathbb{F}_2)$. With the single exception of the plus-type group in dimension four, they have index two, $[O^\pm(2k,\mathbb{F}_2):\Omega^\pm(2k,\mathbb{F}_2)]=2$; in the exceptional case, $[O^+(4,\mathbb F_2):\Omega^+(4,\mathbb F_2)]=4$.
\end{fact}

\begin{remark}
    [Conventions for the binary orthogonal cores in literature]
    \label{rem:group_theory_omega_convention}
    Two conventions for the notation $\Omega^\pm(2k,\mathbb{F}_2)$ occur in the finite-classical-group literature.
    One convention~\cite{aschbacher1984maximal,liebeck2010regular} defines $\Omega^\pm(2k,\mathbb{F}_2)$ as we do in \cref{def:group_theory_binary_orthogonal_cores}.
    Another convention defines $\Omega^\pm(2k,\mathbb{F}_2)$ via the kernel of the Dickson invariant~\cite[Prop.~2.5.7]{kleidman1990subgroup}; this convention is also used in, for example, Refs.~\cite{bray2013maximal,yin2025large}.
    The two conventions coincide except for $O^+(4,\mathbb{F}_2)$, where the Dickson-kernel subgroup has order $36$, while the derived subgroup has order $18$ (\cref{fact:group_theory_binary_orthogonal_core_index,fact:group_theory_relevant_group_orders}).
    Throughout this work, we use the derived-subgroup convention.
    In \cref{app:stab_codes/prelims/maximal_subgroups_aschbacher}, we use finite-group classification results from Ref.~\cite{yin2025large}; but because $k \geq 3$ there, the difference in definitions is immaterial.
\end{remark}

\subsubsection{Group orders and bounds}
\label{app:gt_orders}

\begin{fact}
    [Orders of relevant binary groups]
    \label{fact:group_theory_relevant_group_orders}
    Below, $\mathbb{I}_p$ denotes the indicator of a predicate $p$: it has value one when $p$ is true and zero otherwise. For all $k \geq 1$, we have the exact orders:
    \begin{subequations}\begin{align}
        \abs{\Sp(2k,\mathbb{F}_2)}
        &=
        2^{k^2}\prod_{i=1}^k (2^{2i}-1), 
        \label{eq:group_order_symplectic}
        \\
        \abs{\mathcal{P}(2k,\mathbb{F}_2)}
        &= 
        \abs{\mathcal{U}(2k,\mathbb{F}_2)} \cdot \abs{\GL(k,\mathbb{F}_2)}
        = 
        2^{k^2} \prod_{i=1}^k (2^i - 1),
        \label{eq:group_order_siegel_parabolic}
        \\
        \abs{\mathcal{U}(2k,\mathbb{F}_2)}
        &=
        2^{k(k+1)/2}, 
        \label{eq:group_order_unipotent_radical}
        \\
        \abs{\GL(k,\mathbb{F}_2)}
        &=
        2^{k(k-1)/2} \prod_{i=1}^k (2^i - 1),
        \label{eq:group_order_general_linear}
        \\
        \abs{\Omega^+(2k,\mathbb{F}_2)}
        &= 2^{k^2-k}(2^k-1)\prod_{i=1}^{k-1}(2^{2i}-1) \cdot \left(\frac{1}{2}\right)^{\mathbb{I}_{k=2}},
        \label{eq:group_order_orthogonal_core_plus}
        \\
        \abs{\Omega^-(2k,\mathbb{F}_2)}
        &= 2^{k^2-k}(2^k+1)\prod_{i=1}^{k-1}(2^{2i}-1),
        \label{eq:group_order_orthogonal_core_minus}
        \\
        \abs{O^+(2k,\mathbb{F}_2)}
        &= 2 \cdot \abs{\Omega^+(2k,\mathbb{F}_2)} \cdot 2^{\mathbb{I}_{k=2}} 
        = 2 \cdot 2^{k^2-k}(2^k-1)\prod_{i=1}^{k-1}(2^{2i}-1),
        \label{eq:group_order_orthogonal_plus}
        \\
        \abs{O^-(2k,\mathbb{F}_2)}
        &= 2 \cdot \abs{\Omega^-(2k,\mathbb{F}_2)}
        = 2 \cdot 2^{k^2-k}(2^k+1)\prod_{i=1}^{k-1}(2^{2i}-1).
        \label{eq:group_order_orthogonal_minus}
    \end{align}\end{subequations}
    In \cref{eq:group_order_unipotent_radical}, $\abs{\mathcal{U}(2k,\mathbb{F}_2)}$ equals the number of symmetric $k \times k$ matrices over $\mathbb{F}_2$, since each choice of symmetric matrix gives exactly one element of the group (\cref{def:group_theory_siegel_parabolic_unipotent_radicals}).
    The factorization $\abs{\mathcal{P}(2k,\mathbb{F}_2)} = \abs{\mathcal{U}(2k,\mathbb{F}_2)} \abs{\GL(k,\mathbb{F}_2)}$ follows from the Levi decomposition (\cref{def:group_theory_siegel_parabolic_unipotent_radicals}).
    Orders at small $k$ are listed in \cref{tab:group_theory_relevant_group_orders}. 
    Some useful bounds are:
    \begin{subequations}\begin{align}
        2^{k^2-2}
        <
        \abs{\GL(k,\mathbb{F}_2)}
        &\leq
        2^{k^2-1},
        \label{eq:group_order_general_linear_lower_and_upper_bound}
        \\
        \abs{\Sp(2k,\mathbb{F}_2)}
        &<
        2^{k^2}\prod_{i=1}^k 2^{2i}
        =
        2^{k^2 + k(k+1)}
        =
        2^{2k^2+k},
        \label{eq:group_order_symplectic_upper_bound}
        \\
        \abs{\mathcal{P}(2k, \mathbb{F}_2)} 
        &\geq 
        2^{k^2}\prod_{i=1}^k 2^{i-1} 
        = 
        2^{(3k^2-k)/2},
        \label{eq:group_order_siegel_parabolic_lower_bound}
        \\
        \abs{O^\pm(2k,\mathbb F_2)}
        &<
        2^{k(k-1)+1}
        \cdot
        2^{k+1}
        \cdot
        2^{k(k-1)}
        =
        2^{2k^2-k+2},
        \label{eq:group_order_orthogonal_upper_bound}
    \end{align}\end{subequations}
    for all $k \geq 1$. Furthermore,
    \begin{equation}
        \abs{\mathcal{P}(2k, \mathbb{F}_2)} 
        > 
        (2k + 2)!,
        \label{eq:group_order_siegel_parabolic_lower_bound_factorial}
    \end{equation}
    for all $k \geq 4$. The inequality holds at $k = 4$, since $\abs{\mathcal{P}(8, \mathbb{F}_2)} = 20\,643\,840 > 10! = 3\,628\,800$. By \cref{eq:group_order_siegel_parabolic},
    \begin{equation}\begin{split}
        \frac{|\mathcal P(2(k+1),\mathbb F_2)|}
             {|\mathcal P(2k,\mathbb F_2)|}
        &=
        2^{2k+1}(2^{k+1}-1),
        \qquad
        \frac{(2(k+1)+2)!}{(2k+2)!}
        = (2k+3)(2k+4),
    \end{split}\end{equation}
    and $2^{2k+1}(2^{k+1}-1) > (2k+3)(2k+4)$ for $k \geq 4$; hence the ratio $\abs{\mathcal{P}(2k, \mathbb{F}_2)} / (2k+2)!$ is strictly increasing, and the inequality persists for all $k \geq 4$.

    \begin{table}[h]
        \centering
        \footnotesize
        \begin{tabular}{p{0.5cm} R{4cm} R{4cm} R{4cm} R{4cm}}
            \toprule
            $k$
                & $\abs{\Sp(2k,\mathbb{F}_2)}$
                & $\abs{\mathcal{P}(2k,\mathbb{F}_2)}$
                & $\abs{\mathcal{U}(2k,\mathbb{F}_2)}$
                & $\abs{\GL(k,\mathbb{F}_2)}$
            \\
            \midrule
            1
                & 6
                & 2
                & 2
                & 1
            \\
            2
                & 720
                & 48
                & 8
                & 6
            \\
            3
                & 1\,451\,520
                & 10\,752
                & 64
                & 168
            \\
            4
                & 47\,377\,612\,800
                & 20\,643\,840
                & 1\,024
                & 20\,160
            \\
            5
                & 24\,815\,256\,521\,932\,800
                & 327\,659\,028\,480
                & 32\,768
                & 9\,999\,360
            \\
            6
                & 208\,114\,637\,736\,580\,743\,168\,000
                & 42\,275\,878\,490\,603\,520
                & 2\,097\,152
                & 20\,158\,709\,760
            \\
            \bottomrule
        \end{tabular}
        \\[6pt]
        \begin{tabular}{p{0.5cm} R{4cm} R{4cm} R{4cm} R{4cm}}
            \toprule
            $k$
                & $\abs{O^+(2k,\mathbb{F}_2)}$
                & $\abs{\Omega^+(2k,\mathbb{F}_2)}$
                & $\abs{O^-(2k,\mathbb{F}_2)}$
                & $\abs{\Omega^-(2k,\mathbb{F}_2)}$
            \\
            \midrule
            1
                & 2
                & 1
                & 6
                & 3
            \\
            2
                & 72
                & 18
                & 120
                & 60
            \\
            3
                & 40\,320
                & 20\,160
                & 51\,840
                & 25\,920
            \\
            4
                & 348\,364\,800
                & 174\,182\,400
                & 394\,813\,440
                & 197\,406\,720
            \\
            5
                & 46\,998\,591\,897\,600
                & 23\,499\,295\,948\,800
                & 50\,030\,759\,116\,800
                & 25\,015\,379\,558\,400
            \\
            6
                & 100\,055\,114\,296\,433\,049\,600
                & 50\,027\,557\,148\,216\,524\,800
                & 103\,231\,467\,131\,240\,448\,000
                & 51\,615\,733\,565\,620\,224\,000
            \\
            \bottomrule
        \end{tabular}
        \caption{Orders of relevant binary groups for $1 \le k \le 6$. Formulae are given in \cref{fact:group_theory_relevant_group_orders}.}
        \label{tab:group_theory_relevant_group_orders}
    \end{table}
\end{fact}

\clearpage

\subsection{Stabilizer formalism}
\label{app:stabilizer_codes}

\subsubsection{Symplectic representation of Cliffords}

\begin{definition}
    [Symplectic representation of Paulis and Cliffords]
    \label{def:stab_codes_symplectic_representation}
    A \emph{Pauli string} on $n$ qubits is an operator of the form $X^x Z^z \equiv X_1^{x_1} \cdots X_n^{x_n} Z_1^{z_1} \cdots Z_n^{z_n}$, where $x, z \in \mathbb{F}_2^n$; we consider Pauli strings up to overall phase. 
    Its \emph{symplectic representation} is the vector $(x \mid z) \in \mathbb{F}_2^{2n}$, the $2n$-component vector formed by concatenating $x$ and $z$, and we adopt this $(x \mid z)$ ordering throughout.
    For a Clifford operator $V$, consider its conjugation action on the single-qubit generators $X_j$ and $Z_j$ for each $j \in [n]$, modulo phase:
    \begin{equation}
        V X_j V^\dagger = X^{\alpha_j} Z^{\beta_j}, \qquad V Z_j V^\dagger = X^{\mu_j} Z^{\nu_j},
    \end{equation}
    where $\alpha_j, \beta_j, \mu_j, \nu_j \in \mathbb{F}_2^n$. 
    The \emph{symplectic representation} of $V$ is the matrix $S_V$ whose columns are the symplectic representations of these images:
    \begin{equation}
        \renewcommand{\arraystretch}{1.5}
        S_V = \left(
        \begin{array}{cccc|cccc}
            \alpha_1^\top & \alpha_2^\top & \cdots & \alpha_n^\top
            & \mu_1^\top & \mu_2^\top & \cdots & \mu_n^\top
            \\
            \hline
            \beta_1^\top & \beta_2^\top & \cdots & \beta_n^\top
            & \nu_1^\top & \nu_2^\top & \cdots & \nu_n^\top
        \end{array}
        \right),
    \end{equation}
    where $\alpha_j, \beta_j, \mu_j, \nu_j$ are regarded as row vectors, so their transposes appear as columns in $S_V$.
\end{definition}

\begin{fact}
    [Properties of the symplectic representation]
    \label{fact:stab_codes_symplectic_representation_properties}
    Two Pauli strings $X^x Z^z$ and $X^{x'} Z^{z'}$ commute iff their symplectic representations pair to zero under the standard symplectic form (\cref{def:linear_algebra_standard_binary_symplectic_form}), $\langle (x \mid z), (x' \mid z') \rangle = 0$.
    Let $V$ be a Clifford operator on $n$ qubits with symplectic representation $S_V$ (\cref{def:stab_codes_symplectic_representation}). 
    Then $S_V \in \Sp(2n,\mathbb{F}_2)$; that is, $S_V$ preserves the standard symplectic form and thereby the commutation relations of Pauli strings. 
    Moreover, ignoring overall phase, conjugation by $V$ acts on the symplectic representation of every Pauli string by left-multiplication:
    \begin{equation}
        V X^x Z^z V^\dagger = X^{x'} Z^{z'}
        \iff
        S_V\,(x \mid z)^\top = (x' \mid z')^\top.
    \end{equation}
\end{fact}

\begin{example}
    [Symplectic representation of $\mathrm{CX}$]
    \label{ex:stab_codes_symplectic_representation_cx}
    Consider the two-qubit $\mathrm{CX}_{12}$ gate with control qubit $1$ and target qubit $2$. 
    Its conjugation action on the Pauli generators is $\mathrm{CX}_{12}\, X_1\, \mathrm{CX}_{12}^\dagger = X_1 X_2$, and similarly $X_2 \mapsto X_2$, $Z_1 \mapsto Z_1$, and $Z_2 \mapsto Z_1 Z_2$. In symplectic representation, the corresponding matrix is
    \begin{equation}
        S_{\mathrm{CX}_{12}} =
        \left(
        \begin{array}{cc|cc}
        1 & 0 & 0 & 0 \\
        1 & 1 & 0 & 0 \\
        \hline
        0 & 0 & 1 & 1 \\
        0 & 0 & 0 & 1
    \end{array}
    \right),
    \end{equation}
    so that $(x_1,x_2 \mid z_1,z_2) \mapsto (x_1,\; x_1 + x_2 \mid z_1 + z_2,\; z_2)$ with arithmetic in $\mathbb{F}_2$.
\end{example}

\begin{remark}
    [Clifford realizations of the Siegel parabolic and orthogonal subgroups]
    \label{rem:group_theory_presentation_siegel_parabolic_as_cliffords}
    Under the symplectic representation of Clifford operators (\cref{def:stab_codes_symplectic_representation}), the Siegel parabolic subgroups and the plus-type binary orthogonal group correspond to the projective Clifford subgroups generated by all $\mathrm{CX}$ gates plus one extra single-qubit gate:
    \begin{equation}
        \mathcal{P}(2k,\mathbb{F}_2)
        = \mathcal{P}_\mathrm{z}(2k,\mathbb{F}_2) \cong \langle S_1, \mathbf{CX}_k \rangle,
        \qquad
        \mathcal{P}_\mathrm{x}(2k,\mathbb{F}_2) \cong \langle H_1 S_1 H_1, \mathbf{CX}_k \rangle,
        \qquad
        O^+(2k, \mathbb{F}_2) \cong \langle H_1, \mathbf{CX}_k \rangle.
    \end{equation}
    A single gate on the first qubit suffices because $\langle \mathbf{CX}_k \rangle$ contains all $\mathrm{SWAP}$ gates, and conjugating by $\mathrm{SWAP}$s moves the gate to any qubit. The two Siegel parabolic presentations are related by conjugation by the global Hadamard $H^{\otimes k}$, whose symplectic matrix is $\Omega_k$, realizing \cref{fact:group_theory_siegel_parabolic_conjugacy}.
\end{remark}

\subsubsection{Stabilizer codes and decomposability}

\begin{definition}
    [Stabilizer codes and binary stabilizer spaces]
    \label{def:stab_codes_stabilizer_code}
    A \emph{stabilizer group} on $n$ qubits is an abelian subgroup $\mathcal{S}$ of the $n$-qubit signed Pauli group with $-I \notin \mathcal{S}$. If $\mathcal{S}$ has $n - k$ independent generators, it defines an $\db{n,k}$ \emph{stabilizer code} $\mathcal{C}$, whose \emph{codespace} is the joint $+1$-eigenspace of $\mathcal{S}$. A \emph{stabilizer generator matrix} of $\mathcal{C}$ is a matrix $H \in \smash{\mathbb{F}_2^{(n-k) \times 2n}}$ whose rows are the symplectic representations (\cref{def:stab_codes_symplectic_representation}) of a choice of independent generators of $\mathcal{S}$. The \emph{binary stabilizer space} of $\mathcal{C}$ is the row span $\rs(H) \subseteq \mathbb{F}_2^{2n}$; it is independent of the choice of generators, and totally isotropic (\cref{def:linear_algebra_isotropic_vector_subspace}) because the elements of $\mathcal{S}$ pairwise commute (\cref{fact:stab_codes_symplectic_representation_properties}).
\end{definition}

\begin{definition}
    [Decomposability of stabilizer codes]
    \label{def:code_decomposability}
    An $\db{n,k}$ stabilizer code with stabilizer group $\mathcal{S}$ is \emph{decomposable} if there exists a nontrivial partition $A \sqcup B = [n]$ of the qubits such that 
    \begin{equation}
        \mathcal{S} 
        = 
        \left\{s_A \otimes s_B: 
            s_A \in \mathcal{S}_A, s_B \in \mathcal{S}_B\right\}
    \end{equation}
    for stabilizer groups $\mathcal{S}_A$ and $\mathcal{S}_B$ on the qubits in $A$ and $B$, respectively; the factors then define stabilizer codes on $A$ and $B$. Otherwise, the code is \emph{indecomposable}.
\end{definition}

\begin{proposition}
    [Projection criterion for decomposability of stabilizer codes]
    \label{prop:code_decomposability}
    Let $\mathcal{C}$ be an $\db{n,k}$ stabilizer code with stabilizer generator matrix $\smash{H \in \mathbb{F}_2^{(n-k) \times 2n}}$, and for $A \subseteq [n]$ let $H|_A$ denote the matrix obtained from $H$ by zeroing the $X$- and $Z$-columns of every qubit outside $A$---i.e.~the projection of $H$ onto coordinates $A$. Then $\mathcal{C}$ is decomposable iff there exists a nontrivial qubit partition $A \sqcup B = [n]$ such that $\rs(H|_A) \subseteq \rs(H)$ and $\rs(H|_B) \subseteq \rs(H)$.
\end{proposition}

\begin{proof}
    Let $S\coloneqq \rs(H)\subseteq \mathbb{F}_2^{2n}$ be the binary stabilizer space (\cref{def:stab_codes_stabilizer_code}), and let $\pi_A,\pi_B$ be the projections onto the coordinates supported on $A$ and $B$; thus $\rs(H|_A)=\pi_A(S)$ and $\rs(H|_B)=\pi_B(S)$. The map sending each stabilizer to its symplectic representation is a group isomorphism from $\mathcal{S}$ onto the additive group $S$: it is a surjective homomorphism by construction, and its kernel is trivial since $-I \notin \mathcal{S}$.
    
    First suppose the code is decomposable, so $\mathcal S=\{s_A\otimes s_B:s_A\in\mathcal S_A, s_B\in\mathcal S_B\}$ as in \cref{def:code_decomposability}. For any $s = s_A \otimes s_B \in \mathcal{S}$, the projection $\pi_A$ maps the symplectic representation of $s$ to that of $s_A \otimes I_B$, which is an element of $\mathcal{S}$ because $I_B \in \mathcal{S}_B$. Hence $\pi_A(S) \subseteq S$, and likewise $\pi_B(S)\subseteq S$.
    
    Conversely, suppose $\pi_A(S)\subseteq S$ and $\pi_B(S)\subseteq S$. Consider the stabilizers in $\mathcal{S}$ whose symplectic representations lie in $\pi_A(S)$. Each such stabilizer is supported on $A$, so it has the form $s_A \otimes I_B$ for a Pauli operator $s_A$ on the qubits in $A$. These operators $s_A$ form a group $\mathcal{S}_A$: it is abelian because $\mathcal{S}$ is, and it does not contain $-I_A$ because $\mathcal{S}$ does not contain $-I$. Hence $\mathcal{S}_A$ is a stabilizer group on the qubits in $A$; define $\mathcal{S}_B$ similarly.

    It remains to check $\mathcal{S}=\{s_A \otimes s_B: s_A \in \mathcal{S}_A, s_B \in \mathcal{S}_B\}$. Every such product lies in $\mathcal{S}$, since $s_A \otimes s_B = (s_A \otimes I_B)(I_A \otimes s_B)$ is a product of two stabilizers. In the other direction, take $s \in \mathcal{S}$ with symplectic representation $u$. Both $\pi_A(u)$ and $\pi_B(u)$ lie in $S$ by hypothesis, so they are the representations of stabilizers $s_A \otimes I_B$ and $I_A \otimes s_B$. The product $s_A \otimes s_B$ then has the same representation as $s$, namely $\pi_A(u) + \pi_B(u) = u$, and injectivity of the labelling map gives $s = s_A \otimes s_B$. Hence the code is decomposable.
\end{proof}

\subsubsection{Automorphism gates and logical groups}

\begin{definition}
    [Logical Pauli basis]
    \label{def:stab_codes_logical_pauli_basis}
    Let $\mathcal{C}$ be an $\db{n,k}$ stabilizer code with stabilizer group $\mathcal{S}$. A \emph{logical Pauli basis} for $\mathcal{C}$ is a set of Pauli operators $\{\overline X_i, \overline Z_i\}_{i=1}^k$, where $\overline{X}_i, \overline{Z}_i \in \{I,X,Y,Z\}^n$ up to sign, that commute with every element of $\mathcal{S}$ and satisfy the canonical Pauli commutation relations. 
    These operators induce Pauli logical actions and generate the logical Pauli group.
    Such a logical basis can be described by a pair of binary matrices $L_\mathrm{x}, L_\mathrm{z} \in \mathbb{F}_2^{k \times 2n}$, where the $i^\text{th}$ row of $L_\mathrm{x}, L_\mathrm{z}$ contains the symplectic representation of $\overline X_i, \overline{Z}_i$, respectively, and the canonical Pauli commutation relations require $L_\mathrm{x} \Omega_n L_\mathrm{x}^\top = L_\mathrm{z} \Omega_n L_\mathrm{z}^\top = 0$ and $L_\mathrm{x} \Omega_n L_\mathrm{z}^\top = I_k$.
\end{definition}

\begin{proposition}
    [Pauli phase correction for binary stabilizer automorphisms]
    \label{prop:pauli_phase_correction_binary_stabilizer_automorphisms}
    Let $\mathcal{C}$ be an $\db{n,k}$ stabilizer code with a \emph{signed} stabilizer group $\mathcal{S}$, and binary stabilizer space $C \subseteq \mathbb{F}_2^{2n}$. Let $W$ be a physical Clifford whose symplectic representation $\gamma \in \Sp(2n,\mathbb{F}_2)$ preserves $C$. Then there exists a physical Pauli operator $P$ such that $PW$ preserves the signed stabilizer group,
    \begin{equation}
        (PW) \mathcal{S} (PW)^\dagger = \mathcal{S}.
    \end{equation}
    The physical operation $PW$ has the same symplectic representation $\gamma$, and induces on the logical space $C^\perp/C$ the projective logical Clifford action determined by $\gamma$.
\end{proposition}

\begin{proof}
    This proposition is similar to Ref.~\cite[Alg.~3, Step 5]{rengaswamy2020logical} and Ref.~\cite[Sec.~IV]{sayginel2025fault}, but is presented here in a self-contained form.
    Since $\mathcal{S}$ contains no $-I$, for each $u \in C$ there is a unique signed stabilizer element $s_u \in \mathcal{S}$ whose symplectic representation, or \emph{binary label}, is $u \in \mathbb{F}_2^{2n}$. 
    Since $\gamma(C) = C$, the operators $W s_u W^\dagger$ and $s_{\gamma u}$ have the same binary label (\cref{fact:stab_codes_symplectic_representation_properties}). Moreover, both are Hermitian Pauli operators, so they can differ only by a sign. Hence there is a function $\epsilon: C \to \mathbb{F}_2$ such that $W s_u W^\dagger = (-1)^{\epsilon(u)} s_{\gamma u}$ for all $u \in C$.
    The function $\epsilon$ is linear. Indeed, the product $s_u s_v$ lies in $\mathcal{S}$ and has label $u+v$, so uniqueness gives $s_{u+v}=s_u s_v$ for all $u,v \in C$. Therefore
    \begin{equation}\begin{split}
        (-1)^{\epsilon(u+v)} s_{\gamma(u+v)}
        = W s_{u+v} W^\dagger
        = W s_u s_v W^\dagger
        = \left(W s_u W^\dagger\right)\left(W s_v W^\dagger\right)
        &= (-1)^{\epsilon(u)+\epsilon(v)} s_{\gamma u} s_{\gamma v} \\
        &= (-1)^{\epsilon(u)+\epsilon(v)} s_{\gamma(u+v)}.
    \end{split}\end{equation}
    Hence $\epsilon(u+v)=\epsilon(u)+\epsilon(v)$. Now, the map $v \mapsto \epsilon(\gamma^{-1} v)$ is a linear functional on $C$ (using $\gamma(C) = C$), and the physical symplectic pairing on $\mathbb{F}_2^{2n}$ is nondegenerate, so every linear functional on $C$ is realized by pairing with some vector. Choose $w \in \mathbb{F}_2^{2n}$ such that $\epsilon(\gamma^{-1} v) = \langle w, v \rangle$ for all $v \in C$; equivalently, $\epsilon(u) = \langle w,\gamma u\rangle$ for all $u \in C$. Let $P_w$ be a physical Pauli with binary label $w$. Conjugation by $P_w$ changes the sign of a Pauli with binary label $v$ by $(-1)^{\langle w,v\rangle}$ (\cref{fact:stab_codes_symplectic_representation_properties}). Hence
    \begin{equation}\begin{split}
        (P_wW)s_u(P_wW)^\dagger
        = P_w\left((-1)^{\epsilon(u)}s_{\gamma u}\right)P_w^\dagger
        = (-1)^{\epsilon(u)+\langle w,\gamma u\rangle}s_{\gamma u}
        = s_{\gamma u}.
    \end{split}\end{equation}
    Thus $P_wW$ preserves the signed stabilizer group exactly, since $\gamma$ restricts to a bijection of $C$. Since a physical Pauli has trivial symplectic representation, $P_wW$ has the same symplectic representation $\gamma$ as $W$. Moreover, since $\gamma$ preserves $C$, it also preserves $C^\perp$ and induces a symplectic transformation on the logical space $C^\perp/C$ determined by $\gamma$.
\end{proof}

\begin{definition}
    [Transversal, permutation, and automorphism gates]
    \label{def:transversal_permutation_automorphism_gate}
    Let $\mathcal{C}$ be an $\db{n,k}$ stabilizer code. Each gate below is a physical operation of the stated form that preserves the codespace of $\mathcal{C}$, and the logical gate it implements is the action it induces on the logical qubits.
    \begin{itemize}[noitemsep]
        \item A \emph{transversal gate} is a unitary of the form $\smash{\bigotimes_{i=1}^n V_i}$, where each $V_i$ acts on the $i^\text{th}$ physical qubit.
        \item A \emph{permutation gate} is a unitary $U_\pi$ corresponding to a permutation $\pi \in S_n$ of the physical qubits.
        \item An \emph{automorphism gate} is a unitary of the form $\smash{U_\pi \bigotimes_{i=1}^n V_i}$, where each $V_i$ is a single-qubit Clifford and $U_\pi$ is as above. Permutation gates ($V_i = I$) and transversal gates with single-qubit Cliffords ($\pi = \mathrm{id}$) are special cases.
    \end{itemize}
    For the logical groups below (\cref{def:transversal_permutation_automorphism_logical_groups}) and most of this work, we restrict transversal gates to single-qubit Clifford factors $V_i$.
    Note that, in transversal and automorphism gates, $V_i$ can be different on different physical qubits $i$.
\end{definition}

\begin{definition}
    [Transversal, permutation, and automorphism logical groups]
    \label{def:transversal_permutation_automorphism_logical_groups}
    Let $\mathcal{C}$ be an $\db{n,k}$ stabilizer code with a logical Pauli basis $\mathcal{L}$ (\cref{def:stab_codes_logical_pauli_basis}). 
    Because the inverse of an automorphism gate and the product of two automorphism gates are themselves automorphism gates, the automorphism gates of $\mathcal{C}$ form a group under composition; the same is true for transversal and permutation gates.
    Each gate induces a logical action in the basis $\mathcal{L}$.
    The \emph{automorphism logical group} $\mathrm{Aut}_\mathcal{L}(\mathcal{C})$ is the image of these logical actions in $\mathrm{PCl}_k \cong \Sp(2k, \mathbb{F}_2)$, i.e.~logical actions modulo logical Paulis and phases. 
    It is a group because closure under composition and the availability of inverses are inherited from the physical gate group.
    The \emph{permutation logical group} $\mathrm{Perm}_\mathcal{L}(\mathcal{C})$ and the \emph{transversal logical group} $\mathrm{Trans}_\mathcal{L}(\mathcal{C})$ are defined identically for permutation and transversal gates with single-qubit physical Cliffords, so that
    \begin{equation}
        \mathrm{Perm}_\mathcal{L}(\mathcal{C}),
        \mathrm{Trans}_\mathcal{L}(\mathcal{C})
        \leq
        \mathrm{Aut}_\mathcal{L}(\mathcal{C})
        \leq
        \Sp(2k, \mathbb{F}_2).
    \end{equation}
    By \cref{fact:stab_codes_logical_group_logical_basis_transformations}, a change of logical basis conjugates all three groups by a common element of $\mathrm{PCl}_k$, so their orders and abstract structures are independent of $\mathcal{L}$. 
    Note that transversal gates with non-Clifford single-qubit physical unitaries may induce logical actions outside the Clifford group---we exclude them here, and they reappear in \cref{app:constructions/css} with distinct notation.
\end{definition}

\begin{fact}
    [Logical group transformations under logical basis changes]
    \label{fact:stab_codes_logical_group_logical_basis_transformations}
    Let $\mathcal{C}$ be an $\db{n,k}$ stabilizer code, and fix a group of code-preserving physical Clifford operations on $\mathcal{C}$. For a logical Pauli basis $\mathcal{L}$, let $G_\mathcal{L}(\mathcal{C}) \leq \mathrm{PCl}_k$ be the induced logical group. Any two logical Pauli bases $\mathcal{L}$ and $\mathcal{L}'$ are related by some $R_{\mathcal L \mathcal L'} \in \mathrm{Cl}_k$. Letting $Q_{\mathcal L \mathcal L'} \in \mathrm{PCl}_k$ denote the image of $R_{\mathcal L \mathcal L'}$, we have
    \begin{equation}
        G_{\mathcal{L'}}(\mathcal{C}) = Q_{\mathcal L \mathcal L'}\, G_\mathcal{L}(\mathcal{C})\, Q_{\mathcal L \mathcal L'}^{-1}.
    \end{equation}
    In particular, the order and abstract group structure of a logical group do not depend on the logical basis.
\end{fact}

\clearpage

\subsection{Polynomials over \texorpdfstring{$\F_2$}{F2}}
\label{app:polynomials}

This subsection closely follows Lidl and Niederreiter~\cite[Chaps.~1--3]{Lidl_Niederreiter_1996}.

\subsubsection{Irreducible polynomials and Frobenius orbits}
\label{app:poly_frobenius}

\begin{definition}
    [Irreducible polynomials and unique factorization]
    \label{def:linear_algebra_irreducible_polynomials}
    A nonconstant polynomial $f(x)\in\F_2[x]$ is \emph{irreducible}\footnote{Over a general field, irreducible factors are usually taken to be monic to make the factorization unique up to the order of the factors. Over $\F_2$, every nonzero polynomial is monic, so we omit this qualifier throughout the text.} if it cannot be written as a product of two nonconstant polynomials in $\F_2[x]$. Every nonconstant polynomial $g(x)\in\F_2[x]$ factors uniquely, up to the order of the factors, as $g(x) = \prod_{i=1}^r f_i(x)^{m_i}$, where the $f_i(x)$ are distinct irreducible polynomials and $m_i\geq1$. For an irreducible polynomial $f(x)\in\F_2[x]$ and a nonzero polynomial $g(x)\in\F_2[x]$, we write $\nu_f(g)$ for the largest integer $m\geq0$ such that $f(x)^m$ divides $g(x)$. For nonconstant $g(x)$, this is the \emph{multiplicity} of $f(x)$ in the factorization above.
\end{definition}

\begin{definition}
    [Minimal and characteristic polynomials]
    \label{def:linear_algebra_minimal_and_characteristic_polynomials}
    Let $T$ be a linear operator on a finite-dimensional vector space $V$ over $\F_2$. A polynomial $f(x)=\sum_{j=0}^m a_jx^j\in\F_2[x]$ \emph{annihilates} $T$ if $f(T) = \sum_{j=0}^m a_jT^j = 0$.
    
    The \emph{minimal polynomial} $m_T(x)$ is the unique nonzero polynomial of least degree that annihilates $T$. It divides every polynomial that annihilates $T$~\cite{Lidl_Niederreiter_1996}. We use the same term for elements of $\overline{\F}_2$, the algebraic closure of $\F_2$. For $\alpha\in\overline{\F}_2$, the \emph{minimal polynomial} $m_\alpha(x)$ is the unique nonzero polynomial of least degree in $\F_2[x]$ such that $m_\alpha(\alpha)=0$. It is irreducible and divides every polynomial in $\F_2[x]$ that vanishes at $\alpha$~\cite{Lidl_Niederreiter_1996}. We describe the latter using Frobenius orbits in \cref{def:polynomials_frobenius}.

    The \emph{characteristic polynomial} of $T$ is
    \begin{equation}
        \chi_T(x) = \det(xI-T).
    \end{equation}
    It has degree $\dim V$, and $m_T(x)$ divides $\chi_T(x)$~\cite{Lidl_Niederreiter_1996}.
\end{definition}

\begin{definition}
    [Frobenius map and orbits]
    \label{def:polynomials_frobenius}
    The \emph{Frobenius map} on $\overline{\F}_2$ is $\operatorname{Fr}(\alpha)=\alpha^2$. Its fixed points are exactly the elements of $\F_2$.
    
    The \emph{Frobenius orbit} of $\alpha\in\overline{\F}_2$ is the set $\{\alpha,\alpha^2,\alpha^{2^2},\ldots\}$. Since $\alpha$ lies in a finite subfield of $\overline{\F}_2$, this orbit is finite. If it has length $d$, then the minimal polynomial of $\alpha$ from \cref{def:linear_algebra_minimal_and_characteristic_polynomials} is
    \begin{equation}
        m_\alpha(x) = \prod_{j=0}^{d-1}\left(x-\alpha^{2^j}\right).
    \end{equation} 
    Frobenius permutes the factors in this product, so it fixes every coefficient; thus the coefficients lie in the fixed field $\F_2$. Hence $\deg m_\alpha=d$~\cite{Lidl_Niederreiter_1996}. Primitive $m^\text{th}$ roots of unity exist in $\overline{\F}_2$ if and only if $m$ is odd~\cite{Lidl_Niederreiter_1996}.
\end{definition}

\subsubsection{Cyclotomic polynomials and finite-order operators}
\label{app:poly_cyclotomic}

\begin{definition}
    [Multiplicative and polynomial orders]
    \label{def:polynomials_multiplicative_order}
    \label{def:order_polynomials}
    Let $a,m\geq1$ be integers with $\gcd(a,m)=1$. The \emph{multiplicative order} $\ord_m(a)$ is the least positive integer $r$ such that $a^r\equiv1\pmod m$. For any $s\geq1$, we have $a^s\equiv1\pmod m$ if and only if $\ord_m(a)\mid s$. Thus, for an odd prime $p$,
    \begin{equation}
        p\mid 2^r-1
        \quad\Longleftrightarrow\quad
        \ord\nolimits_p(2)\mid r.
    \end{equation}
    We use the same term for elements of $\overline{\F}_2^\times$. The multiplicative order of $\alpha\in\overline{\F}_2^\times$ is the least positive integer $r$ such that $\alpha^r=1$. Likewise, for any $s\geq1$, $\alpha^s=1$ if and only if its multiplicative order divides $s$.

    Let $f(x)\in\F_2[x]$ satisfy $f(0)\neq0$. The \emph{order} $\ord(f)$ is the least positive integer $d$ such that $f(x)$ divides $x^d-1$. (Such a positive integer $d$ always exists.) If $f(x)$ is irreducible and $\alpha\in\overline{\F}_2^\times$ is a root of $f(x)$, then $f(x)=m_\alpha(x)$ by \cref{def:linear_algebra_minimal_and_characteristic_polynomials}, so its roots form one Frobenius orbit by \cref{def:polynomials_frobenius}. The order $\ord(f)$ equals the multiplicative order of $\alpha$~\cite{Lidl_Niederreiter_1996}, independently of the chosen root, since Frobenius is an automorphism of the finite field containing $\alpha$ and therefore preserves multiplicative order. It follows that
    \begin{equation}
        f(x)\mid x^r-1
        \quad\Longleftrightarrow\quad
        \ord(f)\mid r.
    \end{equation}
\end{definition}

\begin{definition}
    [Cyclotomic polynomials]
    \label{def:linear_algebra_cyclotomic_polynomials}
    For $m\geq1$, the $m^\text{th}$ \emph{cyclotomic polynomial} $\Phi_m(x)\in\F_2[x]$ is defined recursively by $\Phi_1(x)=x-1=x+1$ and
    \begin{equation}
        x^m-1
        =
        \prod_{d\mid m}\Phi_d(x).
    \end{equation}
    This determines $\Phi_m(x)$ from the cyclotomic polynomials for the proper divisors of $m$~\cite{Lidl_Niederreiter_1996}. If $m$ is odd, the roots of $\Phi_m(x)$ in $\overline{\F}_2$ are exactly the elements of multiplicative order $m$~\cite{Lidl_Niederreiter_1996}.

    For an odd prime $p$, the divisors of $p$ are $1$ and $p$, so $x^p-1 = (x+1)\Phi_p(x)$
    and hence
    \begin{equation}
        \Phi_p(x)
        =
        x^{p-1}+x^{p-2}+\cdots+x+1.
    \end{equation}
    We use cyclotomic polynomials only for odd $m$, and in practice only $\Phi_p(x)$ for odd primes $p$.
\end{definition}

\begin{example}
    [Characteristic polynomial of the Bell gate]
    \label{ex:linear_algebra_bell_gate_characteristic_polynomial}
    The two-qubit Bell gate
    \begin{equation}
        M_\mathrm{Bell} = \begin{pmatrix}
            1&1&0&1\\
            0&1&0&1\\
            1&0&1&0\\
            1&1&1&0
        \end{pmatrix}\in\Sp(4,\F_2),
    \end{equation}
    from Ref.~\cite{chakraborty2026nogo} has characteristic polynomial $\chi_{M_\mathrm{Bell}}=\Phi_5(x)=x^4+x^3+x^2+x+1$.
\end{example}

\begin{fact}
    [Irreducible factors of prime cyclotomic polynomials over $\F_2$]
    \label{fact:linear_algebra_cyclotomic_factors_frobenius_orbits}
    Let $p$ be an odd prime and let $\ell=\ord_p(2)$. If $\mu\in\overline{\F}_2$ is a primitive $p^\text{th}$ root of unity, then its Frobenius orbit has length $\ell$, and the corresponding irreducible factor of $\Phi_p(x)$ (\cref{def:polynomials_frobenius}) is
    \begin{equation}
        f_\mu(x)
        =
        \prod_{j=0}^{\ell-1}
        \left(x-\mu^{2^j}\right).
        \label{eq:linear_algebra_cyclotomic_factors_frobenius_orbits}
    \end{equation}
    Every irreducible factor of $\Phi_p(x)$ arises this way, since each has a root among the primitive $p^\text{th}$ roots of unity (\cref{def:linear_algebra_cyclotomic_polynomials}). Hence every irreducible factor of $\Phi_p(x)$ has degree $\ell$ and order $p$, and $\Phi_p(x)$ has $(p-1)/\ell$ distinct irreducible factors~\cite{Lidl_Niederreiter_1996}.
\end{fact}

\begin{remark}
    [Squaring and squarefreeness over $\F_2$]
    \label{rem:polynomials_squaring_squarefree}
    For every $f(x)\in\F_2[x]$, we have $f(x^2)=f(x)^2$. Iterating, if $m=2^a u$ with $u$ odd, then $x^m-1=(x^u-1)^{2^a}$.
    Moreover, $(x^u-1)'=u x^{u-1}=x^{u-1}$ is coprime to $x^u-1$, so $x^u-1$ has no repeated roots, i.e., it is squarefree~\cite{Lidl_Niederreiter_1996}. It follows that every irreducible factor of $x^m-1$ has multiplicity exactly $2^a$; that is, for every irreducible $f(x)\in\F_2[x]$,
    \begin{equation}
        \nu_f(x^m-1)
        \in
        \{0,2^a\}.
    \end{equation}

    In particular, $x^p-1$, and hence $\Phi_p(x)$, is squarefree for every odd prime $p$. Thus the irreducible factors in \cref{fact:linear_algebra_cyclotomic_factors_frobenius_orbits} are distinct.
\end{remark}

\subsubsection{Reciprocal and characteristic polynomials}
\label{app:poly_reciprocal}

\begin{definition}
    [Reciprocal polynomials]
    \label{def:linear_algebra_reciprocal_polynomials}
    Let $f(x)\in\F_2[x]$ have degree $d$ and satisfy $f(0)\neq0$. The \emph{reciprocal polynomial} of $f(x)$ is
    \begin{equation}
        f^*(x)
        =
        x^d f(x^{-1}).
    \end{equation}
    Since $f(0)=1$ over $\F_2$, the polynomial $f^*(x)$ also has degree $d$, and $(f^*)^*=f$. Thus $\alpha\in\overline{\F}_2^\times$ is a root of $f(x)$ if and only if $\alpha^{-1}$ is a root of $f^*(x)$. We call $f(x)$ \emph{self-reciprocal} if $f^*(x)=f(x)$. For irreducible $f(x)$, this is equivalent to its roots being closed under inversion.

    For polynomials with nonzero constant term, $(fg)^*=f^*g^*$. Hence $f(x)$ is irreducible if and only if $f^*(x)$ is irreducible. If $f(x)$ is irreducible, then~\cite{Lidl_Niederreiter_1996}
    \begin{equation}
        \ord(f^*)
        =
        \ord(f).
    \end{equation}
    Moreover, for irreducible $f(x)$ and $g(x)\in\F_2[x]$ with $g(0)\neq0$, applying $*$ to the factorization of $g(x)$ gives
    \begin{equation}
        \nu_{f^*}(g^*)
        =
        \nu_f(g).
    \end{equation}
\end{definition}

\begin{fact}
    [Reciprocal factors of prime cyclotomic polynomials]
    \label{fact:polynomials_cyclotomic_reciprocal_factors}
    Let $p$ be an odd prime, let $\ell=\ord_p(2)$, and let $f_\mu(x)$ be the irreducible factor of $\Phi_p(x)$ corresponding to a primitive $p^\text{th}$ root of unity $\mu$ as in \cref{fact:linear_algebra_cyclotomic_factors_frobenius_orbits}. Taking reciprocals inverts the roots, and $(\mu^{2^j})^{-1}=(\mu^{-1})^{2^j}$. Since $\mu^{-1}$ is again a primitive $p^\text{th}$ root of unity, \cref{def:linear_algebra_reciprocal_polynomials} gives
    \begin{equation}
        f_\mu^*(x)
        =
        f_{\mu^{-1}}(x).
    \end{equation}
    Thus $f_\mu(x)$ is self-reciprocal if and only if $f_{\mu^{-1}}(x)=f_\mu(x)$, which holds if and only if $\mu^{-1}$ lies in the Frobenius orbit of $\mu$. Equivalently, $\mu^{-1}=\mu^{2^j}$ for some $j$, or $2^j\equiv-1\pmod p$, since $\mu$ has order $p$. Hence self-reciprocity depends only on whether $-1\in\langle2\rangle\subseteq(\mathbb{Z}/p\mathbb{Z})^\times$, and not on the choice of $\mu$.
    \begin{itemize}[noitemsep]
        \item If $\ell$ is even, then $2^{\ell/2}\not\equiv1\pmod p$ by the definition of $\ell$, while $(2^{\ell/2})^2\equiv1\pmod p$. The only square roots of $1$ modulo $p$ are $\pm1$, so $2^{\ell/2}\equiv-1\pmod p$. Hence every irreducible factor of $\Phi_p(x)$ is self-reciprocal.
        \item If $\ell$ is odd, then $\langle2\rangle$ has odd order and therefore does not contain $-1$, which has order two. Hence no irreducible factor of $\Phi_p(x)$ is self-reciprocal, and the factors occur in disjoint reciprocal pairs. In particular, $(p-1)/\ell$ is even.
    \end{itemize}
\end{fact}

\begin{example}
    [Small prime cyclotomic factorizations]
    \label{ex:polynomials_small_cyclotomic}
    For $p=5$, we have $\ord_5(2)=4$, so $\Phi_5(x)$ has $(5-1)/4=1$ irreducible factor by \cref{fact:linear_algebra_cyclotomic_factors_frobenius_orbits}. This factor is self-reciprocal by \cref{fact:polynomials_cyclotomic_reciprocal_factors}:
    \begin{equation}
        \Phi_5(x)
        =
        x^4+x^3+x^2+x+1.
    \end{equation}

    For $p=7$, we have $\ord_7(2)=3$, so $\Phi_7(x)$ has $(7-1)/3=2$ irreducible factors of degree $3$. By \cref{fact:polynomials_cyclotomic_reciprocal_factors}, they form a reciprocal pair:
    \begin{equation}
        \Phi_7(x)
        =
        \left(x^3+x+1\right)
        \left(x^3+x^2+1\right).
    \end{equation}
    Indeed, $(x^3+x+1)^*=x^3(x^{-3}+x^{-1}+1)=x^3+x^2+1$.

    For comparison, $\Phi_3(x)=x^2+x+1$ is irreducible and self-reciprocal, with $\ord_3(2)=2$. It is also the minimal polynomial of every nontrivial order-three single-qubit projective Clifford; see \cref{lem:stab_codes_trans_all_or_nothing_physical_c3_cliffords}.
\end{example}

\begin{proposition}
    [Useful characteristic polynomial identities]
    \label{prop:characteristic_poly}
    Here we list several useful characteristic polynomial properties and identities:
    \begin{enumerate}
        \item \label{item:characteristic_poly_1}
        Let $T$ be a linear operator on a finite-dimensional vector space $V$ over $\F_2$, and let $0=V_0\subseteq V_1\subseteq\cdots\subseteq V_r=V$ be a $T$-invariant flag. Then
        \begin{equation}
            \chi_T(x)
            =
            \prod_{i=1}^r \chi_{T,V_i/V_{i-1}}(x),
        \end{equation}
        where $\chi_{T,V_i/V_{i-1}}(x)$ is the characteristic polynomial of the induced action on $V_i/V_{i-1}$.

        \item \label{item:characteristic_poly_2}
        Let $\pi\in S_m$ have cycle lengths $\ell_1,\ldots,\ell_s$, and let $P_\pi$ be its permutation matrix over $\F_2$. Then
        \begin{equation}
            \chi_{P_\pi}(x)
            =
            \prod_{j=1}^s \left(x^{\ell_j}-1\right).
        \end{equation}
        If $f,f'\in\F_2[x]$ are irreducible polynomials with the same odd order, then $\nu_f(x^r-1)=\nu_{f'}(x^r-1)$ for every $r\geq1$. Hence
        \begin{equation}
            \nu_f\!\left(\chi_{P_\pi}\right)
            =
            \nu_{f'}\!\left(\chi_{P_\pi}\right).
        \end{equation}

        \item \label{item:characteristic_poly_3}
        Let $A\in\GL(m,\F_2)$. Then
        \begin{equation}
            \chi_{A^{-1}}(x) = \chi_{A^{-\top}}(x) = \chi_A^*(x).
        \end{equation}
        Hence $\nu_f(\chi_{A^{-\top}})=\nu_{f^*}(\chi_A)$ for every irreducible $f(x)\in\F_2[x]$ with $f(0)\neq0$.

        Furthermore, let $T$ be a linear operator on a finite-dimensional vector space $V$ over $\F_2$, let $\beta$ be a nondegenerate symmetric bilinear form on $V$, let $T$ preserve $\beta$, and let $W\subseteq V$ be $T$-invariant. Then $W^{\perp_\beta}\coloneqq \{v\in V\;:\; \beta(v,w)=0,\;\forall w\in W\}$ is also $T$-invariant. The pairing $([v],u)\mapsto\beta(v,u)$ between $V/W$ and $W^{\perp_\beta}$ is nondegenerate, and the induced actions are contragredient: if the action on $V/W$ is represented by $A$ in some basis, then the action on $W^{\perp_\beta}$ is represented by $A^{-\top}$ in the basis dual to it under this pairing. Therefore,
        \begin{equation}
            \chi_{T,V/W}(x)
            =
            \chi_{T,W^{\perp_\beta}}^*(x).
        \end{equation}

        \item \label{item:characteristic_poly_4}
        If $M\in\Sp(2k,\F_2)$, then
        \begin{equation}
            \chi_M(x)
            =
            \chi_M^*(x).
        \end{equation}
        Hence $\nu_f(\chi_M)=\nu_{f^*}(\chi_M)$ for every irreducible $f(x)\in\F_2[x]$ with $f(0)\neq0$.
    \end{enumerate}
\end{proposition}

\begin{proof}
    We prove the items in sequence.
    \begin{enumerate}
        \item
        Choose a basis adapted to the flag. The matrix of $T$ is block upper triangular, with diagonal blocks given by the induced actions on $V_i/V_{i-1}$. Taking the determinant of $xI-T$ gives the claim.

        \item
        Reorder the basis so that the cycles of $\pi$ form separate blocks. A cycle of length $\ell$ gives a block with characteristic polynomial $x^\ell-1$, which proves the first claim.

        Now let $f$ and $f'$ have the same odd order $d$, and write $r=2^a u$ with $u$ odd. By \cref{def:order_polynomials}, $f\mid x^u-1$ if and only if $d\mid u$, and the same holds for $f'$. By \cref{rem:polynomials_squaring_squarefree}, every irreducible factor of $x^r-1$ has multiplicity $2^a$. Thus $\nu_f(x^r-1)$ and $\nu_{f'}(x^r-1)$ are both $2^a$ if $d\mid u$, and both $0$ otherwise. Applying this equality to each cycle factor of $\chi_{P_\pi}(x)$ gives the final claim.

        \item
        Since $A$ is invertible, $\det(A^{-1})=1$ over $\F_2$. Therefore,
        \begin{equation}
            \begin{split}
                \chi_{A^{-1}}(x)
                =
                \det(xI-A^{-1})
                =
                \det(A^{-1})\det(xA-I)
                =
                \det(I-xA)
                =
                x^m\det(x^{-1}I-A)
                =
                \chi_A^*(x).
            \end{split}
        \end{equation}
        Since transposition does not change the determinant, $\chi_{A^{-\top}}(x)=\chi_{A^{-1}}(x)$. The multiplicity identity follows from \cref{def:linear_algebra_reciprocal_polynomials}. Here $\chi_A(0)=\det(A)=1$, so its reciprocal polynomial is defined as in that definition.

        Now suppose that $T$ preserves $\beta$. Since $\beta$ is nondegenerate, $T$ is injective and hence invertible. Since $W$ is $T$-invariant and finite dimensional, $T(W)=W$. Thus, if $u\in W^{\perp_\beta}$ and $w\in W$, then $T^{-1}w\in W$, so $\beta(Tu,w)=\beta(u,T^{-1}w)=0$. Hence $W^{\perp_\beta}$ is $T$-invariant.

        The pairing $([v],u)\mapsto\beta(v,u)$ is well defined because $\beta(w,u)=0$ for every $w\in W$ and $u\in W^{\perp_\beta}$. It is nondegenerate. Indeed, if $\beta(v,u)=0$ for every $u\in W^{\perp_\beta}$, then $v\in(W^{\perp_\beta})^{\perp_\beta}=W$, so $[v]=0$. Conversely, if $\beta(v,u)=0$ for every $v\in V$, then $u=0$ by nondegeneracy of $\beta$.

        Finally, since $T$ preserves $\beta$, we have $\beta(Tv,Tu)=\beta(v,u)$. In bases of $V/W$ and $W^{\perp_\beta}$ that are dual with respect to this pairing, the two induced matrices are inverse transposes. The first part of this item gives the claim.

        \item
        Since $M\in\Sp(2k,\F_2)$, we have $M^\top\Omega_kM=\Omega_k$ by \cref{def:group_theory_symplectic_group}, and hence $M^{-\top}=\Omega_kM\Omega_k^{-1}$. Thus $M$ and $M^{-\top}$ are similar, so they have the same characteristic polynomial. By Item~\cref{item:characteristic_poly_3}, $\chi_{M^{-\top}}(x)=\chi_M^*(x)$. Hence $\chi_M(x)=\chi_M^*(x)$, and the multiplicity identity follows from \cref{def:linear_algebra_reciprocal_polynomials}.
    \end{enumerate}
\end{proof}

\clearpage

\section{CSS code results}
\label{app:css_codes}

This appendix proves the CSS code results of \cref{sec:css_codes}, using the symplectic conventions and group-theoretic background of \cref{app:preliminaries}. \Cref{app:css_codes/prelims} collects what the later proofs need: the classification of Cliffords into preserving- and exchange-type, and the two notions of self-duality. With these, \cref{app:css_codes/auto} shows that every automorphism gate on an indecomposable CSS code factors into a permutation part and a transversal part, and that its logical group is correspondingly a semidirect product. This reduces the classification to two independent problems: the permutation logical groups (\cref{app:css_codes/perm}) and the transversal logical groups (\cref{app:css_codes/trans}), each treated separately for self-dual and non-self-dual codes. Lastly, \cref{app:css_codes/tCX} considers the addition of transversal interblock CX gates between CSS codeblocks.

\subsection{Definitions and preliminaries}
\label{app:css_codes/prelims}

\subsubsection{Preserving- and exchange-type Cliffords}
\label{app:css_codes/prelims/cliffords}

\begin{definition}
    [Preserving- and exchange-type Cliffords]
    \label{def:preserving_exchange_multi_qubit_cliffords}
    Consider a $k$-qubit Clifford in symplectic representation (\cref{def:stab_codes_symplectic_representation}),
    \begin{equation}\begin{split}
        M = \mqty(
            A & C \\ 
            B & D
        )
        \in 
        \Sp(2k, \mathbb{F}_2).
    \end{split}\end{equation}
    We call the Clifford:
    \begin{itemize}[noitemsep]
        \item \emph{Preserving-type} when $A, D \in \GL(k, \mathbb{F}_2)$, and \emph{pure preserving-type} when additionally $B = C = 0$.
        \item \emph{Exchange-type} when $B, C \in \GL(k, \mathbb{F}_2)$, and \emph{pure exchange-type} when additionally $A = D = 0$.
    \end{itemize}
\end{definition}

\begin{remark}
    [Cliffords on single and multiple qubits]
    \label{rem:preserving_exchange_single_qubit_cliffords}
    A single-qubit Clifford has the symplectic representation
    \begin{equation}\begin{split}
         \mqty(
            a & c \\ 
            b & d
        ) 
        \in 
        \Sp(2, \mathbb{F}_2),
    \end{split}\end{equation}
    and is preserving-type when $bc = 0 \Longleftrightarrow ad = a = d = 1$ and exchange-type when $bc = b = c = 1 \Longleftrightarrow ad = 0$. 
    (Invertibility of symplectic matrices requires $ad + bc = 1$.)
    The six distinct single-qubit projective Cliffords partition into preserving- and exchange-type:
    \begin{equation}\begin{alignedat}{4}
        &\text{Preserving:} \qquad
        & I   &\equiv \mqty(1 & 0 \\ 0 & 1), \qquad
        & S   &\equiv \mqty(1 & 0 \\ 1 & 1), \qquad
        & HSH &\equiv \mqty(1 & 1 \\ 0 & 1), \\
        &\text{Exchange:} \qquad
        & H   &\equiv \mqty(0 & 1 \\ 1 & 0), \qquad
        & HS  &\equiv \mqty(1 & 1 \\ 1 & 0), \qquad
        & SH  &\equiv \mqty(0 & 1 \\ 1 & 1).
    \end{alignedat}\end{equation}
    While single-qubit Cliffords are \emph{either} preserving- or exchange-type, multi-qubit Cliffords can be preserving- or exchange-type, both, or neither.
    An example of the last case is a tensor product of single-qubit Cliffords $\bigotimes_{i=1}^k V_i$ where the $V_i$ factors include both preserving- and exchange-type Cliffords.
\end{remark}

\begin{proposition}
    [Forms and factorizations of preserving- and exchange-type Cliffords]
    \label{prop:preserving_exchange_multi_qubit_cliffords_forms}
    Consider the symplectic representation $M$ of a $k$-qubit Clifford (as in \cref{def:preserving_exchange_multi_qubit_cliffords}). The following holds:
    \begin{enumerate}
        
        \item A preserving-type Clifford admits the factorization
        \begin{equation}\begin{split}
            M 
            = 
            \mqty(
                I & 0 \\ 
                Q & I
            )
            \mqty(
                A & 0 \\ 
                0 & A^{-\top}
            )
            \mqty(
                I & R \\ 
                0 & I
            ),
        \end{split}\end{equation}
        where $A \in \GL(k, \mathbb{F}_2)$ and $Q=BA^{-1}$, $R=A^{-1}C$ are symmetric.
        \label{item:preserving_exchange_multi_qubit_cliffords_forms_1}
        
        \item A pure preserving-type Clifford has the form
        \begin{equation}\begin{split}
            M 
            = 
            \mqty(
                A & 0 \\ 
                0 & A^{-\top}
            ),
        \end{split}\end{equation}
        where $A \in \GL(k, \mathbb{F}_2)$. \label{item:preserving_exchange_multi_qubit_cliffords_forms_2}
        
        \item Multiplying by $\Omega$ on either side swaps preserving- and exchange-type, in both directions:
        \begin{equation}\begin{split}
            \Omega M
            = 
            \mqty( 
                B & D \\ 
                A & C
            ),
            \qquad
            M \Omega
            = 
            \mqty(
                C & A \\ 
                D & B
            ).
        \end{split}\end{equation}
        That is, $\Omega M$ and $M \Omega$ are exchange-type iff $M$ is preserving-type, and they are pure iff $M$ is pure. 
        \label{item:preserving_exchange_multi_qubit_cliffords_forms_3}
        
        \item An exchange-type Clifford admits the factorization
        \begin{equation}\begin{split}
            M 
            =
            \Omega 
            \mqty(
                I & 0 \\ 
                Q & I
            ) 
            \mqty(
                B & 0 \\ 
                0 & B^{-\top}
            ) 
            \mqty(
                I & R \\ 
                0 & I
            ),
        \end{split}\end{equation}
        where $B \in \GL(k, \mathbb{F}_2)$ and $Q=AB^{-1}$, $R=B^{-1}D$ are symmetric. 
        \label{item:preserving_exchange_multi_qubit_cliffords_forms_4}
        
        \item A pure exchange-type Clifford has the form
        \begin{equation}\begin{split}
            M 
            =
            \Omega 
            \mqty(
                B & 0 \\ 
                0 & B^{-\top}
            ) 
            = 
            \mqty(
                0 & B^{-\top} \\ 
                B & 0
            ),
        \end{split}\end{equation}
        where $B \in \GL(k, \mathbb{F}_2)$. 
        \label{item:preserving_exchange_multi_qubit_cliffords_forms_5}
        
    \end{enumerate}
\end{proposition}

\begin{proof}
    $M$ for every $k$-qubit Clifford is symplectic, satisfying $M^{\top} \Omega M = \Omega \Longleftrightarrow M \Omega M^{\top} = \Omega$. Comparing blocks gives $A^\top B = B^\top A$, $A C^\top = C A^\top$, and $A^\top D + B^\top C = I$. Items~\cref{item:preserving_exchange_multi_qubit_cliffords_forms_2,item:preserving_exchange_multi_qubit_cliffords_forms_3,item:preserving_exchange_multi_qubit_cliffords_forms_5} follow immediately. For Item~\cref{item:preserving_exchange_multi_qubit_cliffords_forms_1}, the first two relations give the symmetry of $Q = B A^{-1}$ and $R = A^{-1} C$, and the third gives $D = A^{-\top} + Q A R$, which is the lower-right block of the product. Item~\cref{item:preserving_exchange_multi_qubit_cliffords_forms_4} is Item~\cref{item:preserving_exchange_multi_qubit_cliffords_forms_1} applied to $\Omega M$.
\end{proof}

\begin{lemma}
    [Preserving-type in-sector-identity Clifford subgroups]
    \label{lem:preserving_in_sector_identity_clifford_subgroups}
    Let $G \leq \mathrm{PCl}_k \cong \Sp(2k, \mathbb{F}_2)$ where every element in $G$ is a preserving-type Clifford of the form
    \begin{equation}
        \mqty(I & C \\ B & I):
        \quad
        C = C^\top, 
        \quad
        B = B^\top, 
        \quad
        C B = B C = 0.
    \end{equation}
    Then $G \preceq_{\Sp(2k, \mathbb{F}_2)} \mathcal{U}(2k, \mathbb{F}_2)$.
\end{lemma}

\begin{proof}
    Given any two elements $g, h \in G$, write
    \begin{equation}
        g \coloneqq \mqty(I & C_g \\ B_g & I),
        \qquad
        h \coloneqq \mqty(I & C_h \\ B_h & I),
        \qquad
        g h = \mqty(I + C_g B_h & C_g + C_h \\ B_g + B_h & I + B_g C_h),
    \end{equation}
    so $g h \in G$ forces $C_g B_h = B_g C_h = 0$: every $C$-block annihilates every $B$-block on both sides. Define
    \begin{equation}
        K \coloneqq \bigcap_{g \in G} \ker B_g \subseteq \mathbb{F}_2^k,
        \qquad
        r \coloneqq \dim K.
    \end{equation}
    Then $\im C_h \subseteq K$, because $B_g C_h = 0$ for all $g, h \in G$. Since $C_h$ is symmetric, it follows that $\ker C_h = \ker C_h^\top = (\im C_h)^\perp$. Taking the orthogonal complement reverses the inclusion, hence $K^\perp \subseteq \ker C_h$. This means that each $g \in G$ fixes pointwise (i.e.~every vector in) the subspace $L \coloneqq (K \oplus \{0\}) \oplus (\{0\} \oplus K^\perp) \subseteq \mathbb{F}_2^k \oplus \mathbb{F}_2^k$. If $x \in K$ and $y \in K^\perp$, then
    \begin{equation}
        \mqty(I & C_g \\ B_g & I) \mqty(x \\ y)
        = \mqty(x + C_g y \\  B_g x + y)
        = \mqty(x \\ y).
    \end{equation}
    Since $\dim L=r+(k-r)=k$ and $L$ is isotropic, $L$ is a Lagrangian. Thus every element of $G$ fixes the same Lagrangian pointwise. The rest of the proof chooses a symplectic basis in which $L$ is the standard Lagrangian.

    Choose $R \in \GL(k, \mathbb{F}_2)$ with $R \langle e_1, \ldots, e_r \rangle = K$, where $e_i \in \mathbb{F}_2^k$ is the unit vector with support on the $i^\text{th}$ coordinate. Consequently, $K^\perp = R^{-\top} \langle e_{r + 1}, \ldots, e_k \rangle$.
    Conjugating by the following symplectic matrix
    \begin{equation}
        \mqty(R & 0 \\ 0 & R^{-\top})^{-1}
        \mqty(I & C_g \\ B_g & I)
        \mqty(R & 0 \\ 0 & R^{-\top})
        = \mqty(I & C_g' \\ B_g' & I),
    \end{equation}
    where $C_g' = R^{-1} C_g R^{-\top}$ and $B_g' = R^\top B_g R$. But because 
    $\im C_g \subseteq K$ and $K^\perp \subseteq \ker C_g$, the transformed matrix $C_g'$ has support only on the first $r$ coordinates. Similarly, $K \subseteq \ker B_g$ by definition and $B_g = B_g^\top$, so $\im B_g \subseteq K^\perp$, and the transformed matrix $B_g'$ has support only on the last $k - r$ coordinates. That is, the conjugation gives
    \begin{equation}
        \mqty(R & 0 \\ 0 & R^{-\top})^{-1}
        \mqty(I & C_g \\ B_g & I)
        \mqty(R & 0 \\ 0 & R^{-\top})
        = \mqty(
            I_r & 0 & \alpha_g & 0 \\
            0 & I_{k - r} & 0 & 0 \\
            0 & 0 & I_r & 0 \\
            0 & \beta_g & 0 & I_{k - r}
        ),
        \qquad
        \alpha_g = \alpha_g^\top,
        \qquad
        \beta_g = \beta_g^\top.
    \end{equation}

    Labelling the coordinates $(x_1, \ldots, x_r \, | \, x_{r + 1}, \ldots, x_k \, | \, y_1, \ldots, y_r \, | \, y_{r + 1}, \ldots, y_k)$, we now perform the symplectic coordinate swap $x_i \leftrightarrow y_i$ for $i = r + 1, \ldots, k$. Equivalently, conjugate next by
    \begin{equation}
        \mqty(
            I_r & 0 & 0 & 0 \\
            0 & 0 & 0 & I_{k - r} \\
            0 & 0 & I_r & 0 \\
            0 & I_{k - r} & 0 & 0
        ).
    \end{equation}
    The element $g \in G$ then becomes
    \begin{equation}
        \mqty(
            I_r & 0 & \alpha_g & 0 \\
            0 & I_{k - r} & 0 & \beta_g \\
            0 & 0 & I_r & 0 \\
            0 & 0 & 0 & I_{k - r}
        ).
    \end{equation}
    So we have demonstrated a symplectic change of basis that brings all elements of $G$ into exactly the form of the elements of $\mathcal{U}_\mathrm{x}(2k, \mathbb{F}_2)$. Since $\mathcal{U}_\mathrm{x}(2k, \mathbb{F}_2)$ and $\mathcal{U}(2k, \mathbb{F}_2) = \mathcal{U}_\mathrm{z}(2k, \mathbb{F}_2)$ are conjugate in $\Sp(2k, \mathbb{F}_2)$ (\cref{def:group_theory_siegel_parabolic_unipotent_radicals}), we conclude $G \preceq_{\Sp(2k, \mathbb{F}_2)} \mathcal{U}(2k, \mathbb{F}_2)$.
\end{proof}

\subsubsection{Basics of CSS codes and self-duality} 
\label{app:css_codes/prelims/bases}

\begin{definition}
    [CSS logical bases]
    \label{def:css_logical_basis}
    A \emph{CSS logical basis} on an $\db{n,k}$ CSS code is defined such that the physical representatives of the logical Paulis satisfy $\overline{X}_i \in \{I, X\}^{\otimes n}$ and $\overline{Z}_i \in \{I, Z\}^{\otimes n}$ for all logical qubits $i \in [k]$. 
    Such a logical basis can be described in symplectic form by a pair of binary matrices $L_\mathrm{x}, L_\mathrm{z} \in \mathbb{F}_2^{k \times n}$, such that $\overline{X}_i = X^{(L_\mathrm{x})_i}, \overline{Z}_i = Z^{(L_\mathrm{z})_i}$ for each $i \in [k]$, where $(L_\mathrm{x})_i, (L_\mathrm{z})_i$ are the $i^\text{th}$ row of $L_\mathrm{x}, L_\mathrm{z}$ respectively.
\end{definition}

\begin{definition}
    [CSS logical overlap matrices]
    \label{def:css_logical_overlap_matrices}
    Consider a CSS logical basis $L_\mathrm{x}, L_\mathrm{z} \in \mathbb{F}_2^{k \times n}$ of a CSS code. Then the \emph{CSS logical overlap matrices} are binary matrices
    \begin{equation}
        M_\mathrm{x} \coloneqq L_\mathrm{x} L_\mathrm{x}^\top 
            \in \mathbb{F}_2^{k \times k},
        \qquad
        M_\mathrm{z} \coloneqq L_\mathrm{z} L_\mathrm{z}^\top 
            \in \mathbb{F}_2^{k \times k}.
    \end{equation}
    By their definitions, $M_\mathrm{x}, M_\mathrm{z}$ are symmetric. Moreover, the diagonal entries $(M_\mathrm{x})_{ii}, (M_\mathrm{z})_{ii}$ are nonzero when $\overline{X}_i, \overline{Z}_i$ are odd-weight and zero otherwise, respectively. 
\end{definition}

\begin{definition}
    [Strictly and permutationally self-dual CSS codes]
    \label{def:strictly_and_permutationally_self_dual_css_codes}
    Let $C_\mathrm{x}, C_\mathrm{z}$ be the $X$- and $Z$-type stabilizer spaces of an $\db{n,k}$ CSS code. The code is \emph{permutationally self-dual} (PSD) iff there exists a physical qubit permutation $\pi \in S_n$ such that $\pi(C_\mathrm{x}) = C_\mathrm{z}$ and $\pi(C_\mathrm{z}) = C_\mathrm{x}$; equivalently, iff the physical operator $\overline{W} = U_\pi H^{\otimes n} = H^{\otimes n} U_\pi$ is an automorphism of the code. The code is \emph{strictly self-dual} (SSD) iff $\pi$ can be taken to be the identity; equivalently, iff the physical operator $\overline{W} = H^{\otimes n}$ is an automorphism of the code.
\end{definition}

\begin{remark}
    [Self-duality terminology in literature]
    \label{rem:strictly_and_permutationally_self_dual_terminology_literature}
    We caution that the literature uses several terms for the same notions of CSS self-duality. Codes with $C_\mathrm{x} = C_\mathrm{z}$, which we call SSD codes in  \cref{def:strictly_and_permutationally_self_dual_css_codes}, have also been called \emph{self-dual}~\cite{rengaswamy2020optimality,tansuwannont2025clifford}, \emph{weakly self-dual}~\cite{bravyi2010majorana,haah2017magic}, \emph{self-orthogonal}~\cite{bayanifar2025transversality}, and \emph{homogeneous}~\cite{wang2023construction}.
    Ref.~\cite{calderbank1996good}, following the classical-code terminology of Ref.~\cite{macwilliams1972good}, uses \emph{weakly self-dual} for what we call SSD codes with the added condition that $X^{\otimes n}$, or equivalently $Z^{\otimes n}$, is a stabilizer. Ref.~\cite{haah2017magic} calls these codes \emph{hyperbolic} weakly self-dual CSS codes, and the SSD codes without this additional property \emph{normal}. Our definition of PSD corresponds to the notion of a \emph{self-ZX-dual} CSS code, or a CSS code admitting a ZX-duality, in Ref.~\cite{breuckmann2024fold}.
\end{remark}

\begin{proposition}
    [Guaranteed pure-exchange logical action of a self-dual CSS code]
    \label{prop:self_dual_css_codes_pure_exchange_logical_action}
    Let $\overline{W}$ be any self-duality automorphism, i.e.~any automorphism containing a global physical Hadamard and a qubit permutation ($\overline{W} = U_\pi H^{\otimes n}$), of an $\db{n,k}$ PSD CSS code. Then, in any CSS logical basis, the induced logical action of $\overline{W}$ is a pure exchange-type Clifford.
\end{proposition}

\begin{proof}
    The automorphism $\overline{W}$ has a Hadamard on every physical qubit, so in a CSS logical basis it maps every $X$-type logical operator to a $Z$-type one and vice versa; the accompanying qubit permutation preserves physical Pauli types and so does not affect this. The symplectic representation of the induced logical action therefore has $A = D = 0$. Its off-diagonal blocks must then be invertible: symplecticity (\cref{def:group_theory_symplectic_group}) gives $A^\top D + B^\top C = I$, so $B^\top C = I$ and $C = B^{-\top}$. The induced logical action must therefore be pure exchange-type [cf.~\cref{prop:preserving_exchange_multi_qubit_cliffords_forms}].
\end{proof}

\begin{proposition}
    [Properties of SSD CSS codes]
    \label{prop:strictly_self_dual_css_codes_properties}
    Let $\mathcal{C}$ be an $\db{n,k}$ SSD CSS code. Then:
    \begin{enumerate}[noitemsep]
        \item All stabilizers of $\mathcal{C}$ are even-weight.
        \label{item:strictly_self_dual_css_codes_properties_1}
        \item For any CSS logical basis, the logical overlap matrices $M_\mathrm{x}, M_\mathrm{z}$ of $\mathcal{C}$ are inverses of each other.
        \label{item:strictly_self_dual_css_codes_properties_2}
        \item All pure-$X$- and pure-$Z$-type logicals of $\mathcal{C}$ are even-weight iff $X^{\otimes n}$ (equivalently $Z^{\otimes n}$) is a stabilizer of $\mathcal{C}$.
        \label{item:strictly_self_dual_css_codes_properties_3}
        \item $\mathcal{C}$ has even $X$- and $Z$-distances, hence even distance, if $X^{\otimes n}$ (equivalently $Z^{\otimes n}$) is a stabilizer of $\mathcal{C}$.
        \label{item:strictly_self_dual_css_codes_properties_4}
        \item $k$ is even if $X^{\otimes n}$ (equivalently $Z^{\otimes n}$) is a stabilizer of $\mathcal{C}$. 
        \label{item:strictly_self_dual_css_codes_properties_5}
    \end{enumerate}
\end{proposition}

\begin{proof}
    We prove each property in turn:
    \begin{enumerate}
        \item[\cref{item:strictly_self_dual_css_codes_properties_1}.] To start, choose a full-row-rank stabilizer generator matrix $H \coloneqq H_\mathrm{x} = H_\mathrm{z} \in \mathbb{F}_2^{(n - k) / 2 \times n}$ for $\mathcal{C}$, and let $C \coloneqq \rs(H)$ denote the binary space of the $X$- and $Z$-type stabilizers of $\mathcal{C}$. Thus $X^a$ and $Z^a$ are stabilizers for all $a \in C$. All stabilizers must commute, so $a \cdot b = 0\bmod 2$ for all $a, b \in C$. Taking $a = b$, we get $a \cdot a = \abs{a} \bmod 2 = 0$, so every $a \in C$ is even-weight. That is, every pure $X$- or $Z$-type stabilizer is even-weight. Now consider an arbitrary stabilizer $s = X^a Z^b$ for $a, b \in C$. Its weight is $\abs{a} + \abs{b} - \abs{\supp(a) \cap \supp(b)}$. Modulo two, this is $\abs{a} + \abs{b} + a \cdot b = 0$. So $s$ is even-weight.
        
        \item[\cref{item:strictly_self_dual_css_codes_properties_2}.] Pick any CSS logical basis $L_\mathrm{x}, L_\mathrm{z} \in \mathbb{F}_2^{k \times n}$ of $\mathcal{C}$. Logicals commute with stabilizers, so the rows of $L_\mathrm{x}, L_\mathrm{z}$ lie in $\rs(H)^\perp$, and modulo $\rs(H)$ they each form a basis of the quotient (i.e.~logical) space $\rs(H)^\perp / \rs(H)$. So there exist matrices $R, T$ such that
        \begin{equation}
            L_\mathrm{z} = R H + T L_\mathrm{x}.
            \label{eq:strictly_self_dual_css_codes_logical_overlap_matrices_invertible}
        \end{equation}
        Now use the orthogonality (i.e.~commutation) relations $H L_\mathrm{x}^\top = H L_\mathrm{z}^\top = 0$ and $L_\mathrm{x} L_\mathrm{z}^\top = L_\mathrm{z} L_\mathrm{x}^\top = I$. Multiplying \cref{eq:strictly_self_dual_css_codes_logical_overlap_matrices_invertible} on the right by $L_\mathrm{z}^\top$, we find $L_\mathrm{z} L_\mathrm{z}^\top = T$. Multiplying on the right by $L_\mathrm{x}^\top$ instead, we find $I = T L_\mathrm{x} L_\mathrm{x}^\top$, so $T = (L_\mathrm{x} L_\mathrm{x}^\top)^{-1}$. So we see $M_\mathrm{x} = L_\mathrm{x} L_\mathrm{x}^\top, M_\mathrm{z} = L_\mathrm{z} L_\mathrm{z}^\top$ are inverses of each other.
    
        \item[\cref{item:strictly_self_dual_css_codes_properties_3}.] We show the equivalent statement $\diag(M_\mathrm{x}) = \vb{0} \Longleftrightarrow \vb{1} \in \rs(H)$. Observe that each diagonal entry $(M_\mathrm{x})_{ii} = (L_\mathrm{x})_i \cdot (L_\mathrm{x})_i = (L_\mathrm{x})_i \cdot \vb{1}$, where $(L_\mathrm{x})_i$ is the $i^\text{th}$ row of $L_\mathrm{x}$. So then $\diag(M_\mathrm{x}) = \vb{0} \Longleftrightarrow \vb{1} \in \rs(L_\mathrm{x})^\perp$. As we have shown, all stabilizers of $\mathcal{C}$ are even-weight, so $\vb{1} \in \rs(H)^\perp$ as well. This works out to $\vb{1} \in \rs(H)^\perp \cap \rs(L_\mathrm{x})^\perp = [\rs(H) + \rs(L_\mathrm{x})]^\perp = [\rs(H)^\perp]^\perp = \rs(H)$. The same argument holds for $M_\mathrm{z}$. Because the CSS logical basis $L_\mathrm{x}, L_\mathrm{z} \in \mathbb{F}_2^{k \times n}$ was arbitrary, this two-way implication holds for all pure-$X$ and pure-$Z$-type logicals.
    
        \item[\cref{item:strictly_self_dual_css_codes_properties_4}.] Follows because all pure-$X$- and pure-$Z$-type logicals being even-weight implies that the $X$- and $Z$-distances of $\mathcal{C}$ are even. The distance of $\mathcal{C}$ is the minimum of $X$- and $Z$-distances and is even likewise.
    
        \item[\cref{item:strictly_self_dual_css_codes_properties_5}.] Suppose $\vb{1} \in \rs(H)$ such that $\diag(M_\mathrm{x}) = \vb{0}$.
        We know, moreover, that $M_\mathrm{x}$ is symmetric and invertible. Therefore $M_\mathrm{x}$ defines a nondegenerate alternating symmetric bilinear form on $\mathbb F_2^k$. Such a form exists only in even dimension (see \cref{fact:linear_algebra_albert_classification_f2}). Hence $k$ is even.
    \end{enumerate}
\end{proof}

\clearpage

\subsection{Automorphism gates}
\label{app:css_codes/auto}

\subsubsection{Physical structure of automorphism gates}

\begin{lemma}
    [All-or-nothing physical exchange-type Cliffords on indecomposable CSS codes]
    \label{lem:css_codes_auto_all_or_nothing_exchange_cliffords}
    An automorphism gate on an indecomposable CSS code can have physical exchange-type Cliffords on either none or all of its physical qubits.
\end{lemma}

\begin{proof}
    Consider an indecomposable CSS code defined by stabilizer generator matrices $H_\mathrm{x} \in \mathbb{F}_2^{r_\mathrm{x} \times n}, H_\mathrm{z} \in \mathbb{F}_2^{r_\mathrm{z} \times n}$. Let
    \begin{equation}
        C_\mathrm{x} \coloneqq \rs(H_\mathrm{x}),
        \qquad
        C_\mathrm{z} \coloneqq \rs(H_\mathrm{z}),
        \qquad
        C 
        \coloneqq \rs(H_\mathrm{x} \oplus H_\mathrm{z}),
    \end{equation}
    be the $X$-type, $Z$-type, and overall stabilizer spaces respectively. Suppose there is an automorphism logical gate $\overline U = U_\pi (\bigotimes_{i=1}^n V_i )$ on the code, where $\{V_i\}_{i = 1}^n$ are single-qubit Clifford gates and $U_\pi$ is the unitary corresponding to a qubit permutation $\pi$. This also implies the inverse logical gate $\smash{\smash{\overline{U}}^\dag} = \smash{(\bigotimes_{i = 1}^n V_i^\dag) U_\pi^\dag} = \smash{(\bigotimes_{i = 1}^n V_i^\dag) U_{\pi^{-1}}}$. The Clifford gate $V_i$ and its inverse can be written in symplectic form as
    \begin{equation}
        V_i \equiv \mqty(p_i & r_i \\ q_i & s_i) \in \Sp(2, \mathbb{F}_2),
        \qquad
        V_i^\dag \equiv \mqty(s_i & r_i \\ q_i & p_i) \in \Sp(2, \mathbb{F}_2),
    \end{equation}
    and we associate with each single-qubit Clifford $V_i$ the indicator bit $t_i \coloneqq q_i r_i \in \mathbb{F}_2$, such that $t_i = 0$ indicates that $V_i$ is preserving-type and $t_i=1$ that $V_i$ is exchange-type (see \cref{rem:preserving_exchange_single_qubit_cliffords}). We denote by $T \coloneqq \supp(t) = \{i: t_i = 1\} \subseteq [n]$ the support of exchange-type Cliffords, and write the vectors $\mu \coloneqq (\mu_i)$ for $\mu = p, q, r, s, t$.

    Consider arbitrary stabilizer elements $(a, 0), (0, b) \in C$, where $a \in C_\mathrm{x}, b \in C_\mathrm{z}$. Applying the automorphism logical gate $\overline{U}$, these are mapped to $\pi(p \odot a, q \odot a), \pi(r \odot b, s \odot b)$, respectively, which must remain in the stabilizer space. This means
    \begin{subequations}\begin{align}
        \forall a \in C_\mathrm{x}:
            \qquad
            & \pi(p \odot a) \in C_\mathrm{x},
            \label{eq:css_indecomp_no_add_h_forward_x1}
            \\
            & \pi(q \odot a) \in C_\mathrm{z},
            \label{eq:css_indecomp_no_add_h_forward_x2}
        \\
        \forall b \in C_\mathrm{z}:
            \qquad
            & \pi(r \odot b) \in C_\mathrm{x},
            \label{eq:css_indecomp_no_add_h_forward_z1}
            \\
            & \pi(s \odot b) \in C_\mathrm{z},
            \label{eq:css_indecomp_no_add_h_forward_z2}
    \end{align}\end{subequations}
    where $\odot$ denotes coordinate-wise product. On the other hand, applying the automorphism logical gate $\smash{\overline{U}}^\dag$, the stabilizer elements are mapped to $(s \odot \pi^{-1} a, q \odot \pi^{-1} a), (r \odot \pi^{-1} b, p \odot \pi^{-1} b)$, respectively, which likewise must remain in the stabilizer space. This means
    \begin{subequations}\begin{align}
        \forall a \in C_\mathrm{x}:
            \qquad
            & s \odot \pi^{-1} a \in C_\mathrm{x},
            \label{eq:css_indecomp_no_add_h_inverse_x1}
            \\
            & q \odot \pi^{-1} a \in C_\mathrm{z},
            \label{eq:css_indecomp_no_add_h_inverse_x2}
        \\
        \forall b \in C_\mathrm{z}:
            \qquad
            & r \odot \pi^{-1} b \in C_\mathrm{x},
            \label{eq:css_indecomp_no_add_h_inverse_z1}
            \\
            & p \odot \pi^{-1} b \in C_\mathrm{z},
            \label{eq:css_indecomp_no_add_h_inverse_z2}
    \end{align}\end{subequations}

    From \cref{eq:css_indecomp_no_add_h_forward_x2,eq:css_indecomp_no_add_h_inverse_z1}, we find
    \begin{equation}
        \forall a \in C_\mathrm{x}:
        \qquad
        r \odot \pi^{-1} \left[ \pi(q \odot a) \right]
        = r \odot q \odot a
        = t \odot a
        \in C_\mathrm{x},
    \end{equation}
    and from \cref{eq:css_indecomp_no_add_h_forward_z1,eq:css_indecomp_no_add_h_inverse_x2} we find
    \begin{equation}
        \forall b \in C_\mathrm{z}:
        \qquad
        q \odot \pi^{-1} \left[ \pi(r \odot b) \right]
        = q \odot r \odot b
        = t \odot b
        \in C_\mathrm{z}.
    \end{equation}

    This implies that $\Pi_T(C_\mathrm{x}) \subseteq C_\mathrm{x}$ and $\Pi_T(C_\mathrm{z}) \subseteq C_\mathrm{z}$, where $\Pi_T$ is the projector onto the coordinates in $T$. Unless $T \in \{\varnothing, [n]\}$, by \cref{prop:code_decomposability}, this would mean the code is decomposable across the partition specified by $T$.
\end{proof}

\newcommand{\PEcase}[1]{%
    \ifcase\value{#1}%
    \or P%
    \or E%
    \fi
}
\AddEnumerateCounter{\PEcase}{\PEcase}{E}

\begin{lemma}
    [Physical factorization property of automorphism gates on indecomposable CSS codes]
    \label{lem:css_codes_auto_physical_splitting_property}
    Let $\mathcal{C}$ be an $\db{n,k}$ indecomposable CSS code and $\overline{U} \coloneqq U_\pi (\bigotimes_{i=1}^n V_i)$ be an automorphism gate of $\mathcal{C}$, where $\pi \in S_n$ and $\{V_i\}_{i=1}^n$ are single-qubit physical Cliffords.
    Then exactly one of the following cases occurs:
    \begin{enumerate}[label=(\PEcase*),ref=(\PEcase*)]
        \item \label{item:css_codes_auto_physical_splitting_property_preserving_case} 
        $\{V_i\}_{i=1}^n$ are preserving-type physical Cliffords, and $U_\pi$ and $\bigotimes_{i=1}^n V_i$ are each valid logical gates of $\mathcal{C}$.
        
        \item \label{item:css_codes_auto_physical_splitting_property_exchange_case}
        $\{V_i\}_{i=1}^n$ are exchange-type physical Cliffords, and $U_\pi H^{\otimes n}$ and $\bigotimes_{i=1}^n H_i V_i$ are each valid logical gates of $\mathcal{C}$. 
        This case occurs only when $\mathcal{C}$ is PSD.

        Moreover, let $\overline{W} = U_\sigma H^{\otimes n}$ be any self-duality automorphism of $\mathcal{C}$.
        Then $\overline{U}$ can be factored as $\overline{U} = \overline{W} \overline{V}$, where $\overline{V}=(U_\sigma^{-1}U_\pi) (\bigotimes_{i=1}^n H_iV_i)$ is an automorphism gate of the above preserving type. 

        Additionally, if $\mathcal{C}$ is SSD, then $U_\pi$ and $\bigotimes_{i=1}^n V_i$ are each valid logical gates of $\mathcal{C}$.
    \end{enumerate}
\end{lemma}

\begin{proof}
    By \cref{lem:css_codes_auto_all_or_nothing_exchange_cliffords}, an automorphism gate $\overline{U} \coloneqq U_\pi (\bigotimes_{i=1}^n V_i)$ on an indecomposable CSS code must have $T \in \{\varnothing, [n]\}$, where $T$ is the support of exchange-type physical Cliffords of the automorphism. We consider the two cases:
    \begin{enumerate}[label=(\PEcase*),ref=(\PEcase*)]
        \item $T = \varnothing$, that is, $t_i = q_i r_i = 0$ for all $i \in [n]$. This implies $p_i s_i = 1 \Longrightarrow p_i = s_i = 1$ for all $i \in [n]$, as symplectic matrices are invertible. Then, by \cref{eq:css_indecomp_no_add_h_forward_x1,eq:css_indecomp_no_add_h_forward_z2}, 
        \begin{equation}
            \forall a \in C_\mathrm{x}: \pi a \in C_\mathrm{x} 
            \implies \pi(C_\mathrm{x}) \subseteq C_\mathrm{x},
            \qquad
            \forall b \in C_\mathrm{z}: \pi b \in C_\mathrm{z}
            \implies \pi(C_\mathrm{z}) \subseteq C_\mathrm{z}.
        \end{equation}
        Because $\pi$ is invertible and preserves dimensions, both inclusions are equalities: $\pi(C_\mathrm{x}) = C_\mathrm{x}$ and $\pi(C_\mathrm{z}) = C_\mathrm{z}$. Since qubit permutations preserve the inner product, it follows similarly that $\pi(C_\rmz^\perp)=\pi(C_\rmz)^\perp=C_\rmz^\perp$ and $\pi(C_\rmx^\perp)=\pi(C_\rmx)^\perp=C_\rmx^\perp$. Therefore, $\pi$ induces automorphisms on the quotient spaces $\cL_\rmx\coloneqq C_\rmz^\perp/C_\rmx$ and $\cL_\rmz\coloneqq C_\rmx^\perp/C_\rmz$, namely the $X$- and $Z$-logical spaces, and hence $U_\pi$ is itself a valid logical gate of $\mathcal{C}$. It follows that $\bigotimes_{i=1}^n V_i=U_\pi^\dagger \overline{U}$ is also a valid logical gate.
        
        \item $T = [n]$, that is, $t_i = q_i r_i = 1 \implies q_i = r_i = 1$ for all $i \in [n]$. Then, by \cref{eq:css_indecomp_no_add_h_forward_x2,eq:css_indecomp_no_add_h_forward_z1},
        \begin{equation}
            \forall a \in C_\mathrm{x}: \pi a \in C_\mathrm{z} 
            \implies \pi(C_\mathrm{x}) \subseteq C_\mathrm{z},
            \qquad
            \forall b \in C_\mathrm{z}: \pi b \in C_\mathrm{x}
            \implies \pi(C_\mathrm{z}) \subseteq C_\mathrm{x}.
        \end{equation}
        The two inclusions imply $\dim C_\rmx\le \dim C_\rmz$ and $\dim C_\rmz\le \dim C_\rmx$, respectively. Thus, the dimensions are equal and both inclusions are equalities: $\pi(C_\mathrm{x}) = C_\mathrm{z}$ and $\pi(C_\mathrm{z}) = C_\mathrm{x}$.

        Since qubit permutations preserve the inner product, one similarly has $\pi(C_\rmz^\perp)=\pi(C_\rmz)^\perp=C_\rmx^\perp$ and $\pi(C_\rmx^\perp)=\pi(C_\rmx)^\perp=C_\rmz^\perp$. It follows that $\pi$ induces isomorphisms between the two quotient spaces $\cL_\rmx$ and $\cL_\rmz$. So $U_\pi H^{\otimes n}$ is a self-duality automorphism, and $\bigotimes_{i=1}^n H_iV_i=(U_\pi H^{\otimes n})^\dagger \overline{U}$ is also a valid logical gate; here each $H_iV_i$ is preserving-type.
        
        Lastly, for any self-duality automorphism $\overline{W} = U_\sigma H^{\otimes n}$ of the code, one can expand
        \begin{equation}
            \overline{U} 
            = U_\pi \left(\bigotimes_{i=1}^n V_i\right) 
            = U_\sigma U_{\sigma^{-1}} U_\pi H^{\otimes n} \left(\bigotimes_{i=1}^n H_i V_i\right)
            = \underbrace{U_\sigma H^{\otimes n}}_{\overline{W}}
                \underbrace{U_{\sigma^{-1}} U_\pi \left(\bigotimes_{i=1}^n H_i V_i\right)}_{\overline{V}},
        \end{equation}
        where we have used the fact that $H^{\otimes n}$ commutes with qubit permutations. Since $\overline{W}$ and $\overline{U}$ are automorphism gates, $\overline{V}$ is also an automorphism gate. The automorphism $\overline{V}$ involves preserving-type physical Cliffords $H_iV_i$, and hence, by Case~\cref{item:css_codes_auto_physical_splitting_property_preserving_case}, $U_{\sigma^{-1}} U_\pi$ is itself a permutation gate of the code.
        
        Finally, if $\mathcal{C}$ is SSD, we can choose $\sigma = \id$. Then, because $H^{\otimes n}$ is a transversal gate of the code, $U_\pi$ and $\bigotimes_{i=1}^n V_i$ are themselves a permutation and transversal gate of the code, respectively.
    \end{enumerate} 
\end{proof}

\subsubsection{Logical actions inducible by automorphism gates}

\begin{corollary}
    [Normal forms of automorphism logical actions on indecomposable CSS codes]
    \label{lem:css_codes_auto_logical_splitting_property}
    Let $\mathcal{C}$ be an $\db{n,k}$ indecomposable CSS code and $\mathcal{L}$ be a CSS logical basis of $\mathcal{C}$.
    Let $\overline{U} \coloneqq U_\pi (\bigotimes_{i=1}^n V_i)$ be an automorphism gate of $\mathcal{C}$, where $\pi \in S_n$ and $\{V_i\}_{i=1}^n$ are single-qubit physical Cliffords, and $g \in \Sp(2k,\F_2)$ be the induced logical action of $\overline{U}$ in $\mathcal{L}$.
    Then exactly one of the following cases occurs:
    \begin{enumerate}[label=(\PEcase*),ref=(\PEcase*)]
        \item \label{item:css_codes_auto_logical_splitting_property_preserving_case}
        In Case~\cref{item:css_codes_auto_physical_splitting_property_preserving_case} of \cref{lem:css_codes_auto_physical_splitting_property}, the logical action $g$ is preserving-type and has the form
        \begin{equation}
            g=\mqty(
                A & A R \\ 
                A^{-\top} Q & A^{-\top}
            ):
            \qquad
            A \in \GL(k, \mathbb{F}_2),
            \qquad
            R = R^\top,
            \qquad
            Q = Q^\top,
            \qquad
            RQ = QR = 0.
            \label{eq:css_codes_auto_logical_splitting_property_1}
        \end{equation}
        Indeed, $g$ factorizes as $g=ba$ for some $a \in \mathrm{Trans}_\mathcal{L}(\mathcal{C})$ and $b \in \mathrm{Perm}_\mathcal{L}(\mathcal{C})$.
    
        \item \label{item:css_codes_auto_logical_splitting_property_exchange_case}
        In Case~\cref{item:css_codes_auto_physical_splitting_property_exchange_case} of \cref{lem:css_codes_auto_physical_splitting_property}, the logical action $g$ is exchange-type. This case can occur only if $\mathcal{C}$ is PSD.
        Take any self-duality automorphism of $\mathcal{C}$, whose pure exchange-type induced logical action is $E \in \Sp(2k, \mathbb{F}_2)$.
        Then
        \begin{equation}
            g=Eh,
            \qquad
            h=\begin{pmatrix}A&AR\\A^{-\top}Q & A^{-\top}\end{pmatrix},
        \end{equation}
        for some $A,Q,R$ satisfying the conditions in Case~\cref{item:css_codes_auto_logical_splitting_property_preserving_case} above.
        Equivalently, $g$ factorizes as $g=Eba$ for some $a \in \mathrm{Trans}_\mathcal{L}(\mathcal{C})$ and $b \in \mathrm{Perm}_\mathcal{L}(\mathcal{C})$.
    
        Additionally, if $\mathcal{C}$ is SSD, then $g$ also admits a factorization $g=b\tilde{a}$ for some $\tilde{a} \in \mathrm{Trans}_\mathcal{L}(\mathcal{C})$ and $b \in \mathrm{Perm}_\mathcal{L}(\mathcal{C})$.
    \end{enumerate} 
    
    In particular, an automorphism logical action is \emph{either} preserving or exchange-type (but not both or neither).
\end{corollary}

\begin{proof}
    The factorizations $g=ba$ in Case~\cref{item:css_codes_auto_logical_splitting_property_preserving_case}, and $g=Eh$ and $g=Eba$ in Case~\cref{item:css_codes_auto_logical_splitting_property_exchange_case}, follow directly from \cref{lem:css_codes_auto_physical_splitting_property}.
    Now we examine the two cases more closely:
    \begin{enumerate}[label=(\PEcase*),ref=(\PEcase*)]
        \item By \cref{thm:css_codes_perm_logical_group}, the logical action $b$ of a permutation gate is pure preserving-type and has the form $b=\diag(A, A^{-\top})$. 
        Moreover, $a$ is a logical action induced by a transversal gate where the single-qubit Cliffords are all preserving-type, so by \cref{thm:css_codes_trans_logical_groups_non_ssd_indecomp,thm:css_codes_trans_logical_groups_ssd_indecomp}, its symplectic representation has both diagonal blocks equal to the identity:
        \begin{equation}
            a=\begin{pmatrix}I&R\\Q&I\end{pmatrix}:
            \qquad
            R=R^\top,
            \qquad
            Q=Q^\top,
            \qquad
            RQ=QR=0.
        \end{equation}
        Multiplying these in $g=ba$ gives the form of $g$ in \cref{eq:css_codes_auto_logical_splitting_property_1} as claimed.
        The two diagonal blocks of $g$ are invertible, so $g$ is preserving-type. 
        If $g$ were also exchange-type, then $AR$ and $A^{-\top}Q$ would both be invertible. 
        Since $A$ is invertible, this would imply that both $R$ and $Q$ are invertible, contradicting $RQ=0$ because $k>0$. 
        We conclude $g$ cannot also be exchange-type.

        \item We write the pure exchange-type logical action $E = \smash{\begin{psmallmatrix}0&K^{-\top}\\K&0\end{psmallmatrix}}$ for some $K\in\GL(k,\mathbb{F}_2)$, and $h = \smash{\begin{psmallmatrix}A&AR\\A^{-\top}Q&A^{-\top}\end{psmallmatrix}}$ with $R$ and $Q$ symmetric and $RQ=QR=0$.
        Direct multiplication gives 
        \begin{equation}
            g
            =Eh
            =\begin{pmatrix}K^{-\top}A^{-\top}Q&K^{-\top}A^{-\top}\\KA&KAR\end{pmatrix}.
        \end{equation}

        The two off-diagonal blocks of $g$ are invertible, so $g$ is exchange-type. 
        If $g$ were also preserving-type, then its two diagonal blocks would be invertible. 
        Since $A$ and $K$ are invertible, this would force both $Q$ and $R$ to be invertible, again contradicting $RQ=0$.
        We conclude $g$ cannot also be preserving-type.
    \end{enumerate}

    These two mutually exclusive cases \cref{item:css_codes_auto_logical_splitting_property_preserving_case,item:css_codes_auto_logical_splitting_property_exchange_case} exhaust all automorphism gates by \cref{lem:css_codes_auto_physical_splitting_property}.
\end{proof}

\begin{corollary}
    [No addressable automorphism logical $H$, $SH$, $HS$ gates on indecomposable CSS codes] 
    \label{corr:css_indecomp_no_add_h_sh_hs}
    On an indecomposable CSS code, automorphism gates cannot induce exchange-type single-qubit Clifford logical actions (i.e.~$H$, $SH$, $HS$, or mixture thereof) acting on a nonempty proper subset of the logical qubits, in any CSS logical basis.
\end{corollary}

\begin{proof}
    Such a logical action is a tensor product of single-qubit Cliffords of mixed type, hence neither preserving- nor exchange-type by \cref{rem:preserving_exchange_single_qubit_cliffords}, and so is forbidden by \cref{lem:css_codes_auto_logical_splitting_property}.
\end{proof}

\begin{remark}
    [Decomposability and non-CSS logical bases]
    \Cref{corr:css_indecomp_no_add_h_sh_hs} does not hold on decomposable CSS codes, nor on indecomposable CSS codes when the logical basis is chosen to be non-CSS. 
    To see the former, consider placing $k$ copies of $\db{n,1,d}$ indecomposable SSD codes side-by-side---the overall $\db{nk,k,d}$ code is decomposable. But each SSD codeblock supports transversal $H$, $HS$, and $SH$ gates (see \cref{thm:css_codes_trans_logical_groups_ssd_indecomp}), so the overall code supports every addressable logical $H_i$, $(HS)_i$, and $(SH)_i$ transversally. 
    For the latter, consider an indecomposable CSS code supporting an addressable logical $S_i$ in a CSS logical basis via automorphisms---for example, \cref{cons:concatenated_phantom_clifford,cons:complete_hypergraph_css_t_eq_2}. Then change the $i^\text{th}$ logical basis by $F_i = (HSH)_i$. Since $F_i^{-1} S_i F_i = H_i$, the same physical automorphism implementing $S_i$ in the original logical basis now induces the addressable logical $H_i$ action, up to logical Paulis, in the new basis. This new basis is not CSS because the $F_i$ mixes the $X$- and $Z$-logical sectors: the $\overline{X}_i$ and $\overline{Z}_i$ representatives can no longer be picked to be pure $X$- or $Z$-type at the physical level.
\end{remark}

\begin{corollary}
    [Self-duality required for automorphism logical $H^{\otimes k}$, $(SH)^{\otimes k}$, $(HS)^{\otimes k}$ gates on indecomposable CSS codes]
    \label{corr:css_codes_auto_logical_hadamards_only_on_self_dual_codes}
    An $\db{n,k}$ indecomposable CSS code can host exchange-type logical Clifford actions---which include $H^{\otimes k}$, $(SH)^{\otimes k}$, $(HS)^{\otimes k}$ logical gates, or any mixture thereof, up to composition with logical $S$, CX and CZ circuits---implemented by automorphisms in a CSS logical basis only if the code is PSD.
\end{corollary}

\begin{proof}
    Exchange-type logical actions composed with logical $S$, CX and CZ actions remain exchange-type, and can be induced by automorphisms only on PSD codes by \cref{lem:css_codes_auto_logical_splitting_property}.
\end{proof}

\subsubsection{Automorphism logical groups}

\begin{lemma}
    [Trivial intersection of transversal and permutation logical actions]
    \label{lem:css_codes_trans_perm_logical_actions_trivial_intersection}
    Let $\mathcal{C}$ be an $\db{n,k}$ indecomposable CSS code and $\mathcal{L}$ be any logical basis of $\mathcal{C}$. Then $\mathrm{Trans}_\mathcal{L}(\mathcal{C}) \cap \mathrm{Perm}_\mathcal{L}(\mathcal{C}) = \{I\}$ is the trivial group.
\end{lemma}

\begin{proof}
    First consider $\mathcal{L}$ to be a CSS logical basis of $\mathcal{C}$. For any CSS code $\mathcal{C}$, \cref{thm:css_codes_perm_logical_group} states that every logical action in $\mathrm{Perm}_\mathcal{L}(\mathcal{C})$ must be of the form
    \begin{equation}
        \mqty(C & 0 \\ 0 & C^{-\top}).
    \end{equation}

    For an indecomposable non-SSD $\mathcal{C}$, \cref{thm:css_codes_trans_logical_groups_non_ssd_indecomp} states that every logical action in $\mathrm{Trans}_\mathcal{L}(\mathcal{C})$ must be of the form
    \begin{equation}
        \mqty(I & A \\ B & I).
    \end{equation}

    Therefore a logical action in $\mathrm{Trans}_\mathcal{L}(\mathcal{C}) \cap \mathrm{Perm}_\mathcal{L}(\mathcal{C})$ must have $C = I$ and $A = B = 0$, and is the identity. For an SSD $\mathcal{C}$, \cref{thm:css_codes_trans_logical_groups_ssd_indecomp} states
    \begin{equation}
        \mathrm{Trans}_\mathcal{L}(\mathcal{C})
        = \left\{ \mqty(p I & r M_\mathrm{z} \\ q M_\mathrm{x} & s I): \mqty(p & r \\ q & s) \in \Sp(2, \mathbb{F}_2) \right\},
    \end{equation}
    where $M_\mathrm{x} = L_\mathrm{x} L_\mathrm{x}^\top, M_\mathrm{z} = L_\mathrm{z} L_\mathrm{z}^\top$ are the logical overlap matrices of the code, which by \cref{prop:strictly_self_dual_css_codes_properties}, are invertible. So then a logical action in $\mathrm{Trans}_\mathcal{L}(\mathcal{C}) \cap \mathrm{Perm}_\mathcal{L}(\mathcal{C})$ must have $r M_\mathrm{z} = q M_\mathrm{x} = 0 \Longrightarrow r = q = 0$. Then $p = s = 1$, so the logical action must be the identity. Passing to an arbitrary logical basis from $\mathcal{L}$ amounts to conjugating $\mathrm{Trans}_\mathcal{L}(\mathcal{C})$ and $\mathrm{Perm}_\mathcal{L}(\mathcal{C})$ by the same symplectic rotation, which does not change their trivial intersection.
\end{proof}

\begin{theorem}
    [Automorphism logical groups on indecomposable CSS codes]
    \label{thm:css_codes_auto_logical_groups}
    Let $\mathcal{C}$ be an $\db{n,k}$ indecomposable CSS code and $\mathcal{L}$ be a CSS logical basis of $\mathcal{C}$. Then:
    \begin{itemize}
        
        \item For a SSD $\mathcal{C}$,
        \begin{equation}\begin{split}
            \mathrm{Aut}_\mathcal{L}(\mathcal{C}) 
            &= 
            \mathrm{Trans}_\mathcal{L}(\mathcal{C}) 
            \rtimes 
            \mathrm{Perm}_\mathcal{L}(\mathcal{C}),
            \\
            \abs{\mathrm{Aut}_\mathcal{L}(\mathcal{C})} 
            &<
            \frac{72}{5}
            \cdot
            \frac{1}{2^{k^2}} 
            \cdot
            2^{k(k+1)/2}
            \cdot
            \abs{\GL(k, \mathbb{F}_2)}.
            \label{eq:css_codes_auto_logical_groups_ssd}
        \end{split}\end{equation}
        
        \item For a non-PSD $\mathcal{C}$,
        \begin{equation}\begin{split}
            \mathrm{Aut}_\mathcal{L}(\mathcal{C}) 
            &= 
            \mathrm{Trans}_\mathcal{L}(\mathcal{C}) 
            \rtimes 
            \mathrm{Perm}_\mathcal{L}(\mathcal{C}),
            \\
            \abs{\mathrm{Aut}_\mathcal{L}(\mathcal{C})}
            &\leq 
            2^{k(k+1)/2}
            \cdot
            \abs{\GL(k, \mathbb{F}_2)}.
            \label{eq:css_codes_auto_logical_groups_non_psd}
        \end{split}\end{equation}
    
        \item For a PSD-but-not-SSD $\mathcal{C}$,
        \begin{equation}\begin{split}
            \mathrm{Aut}_\mathcal{L}(\mathcal{C}) 
            &= 
            \mathrm{Trans}_\mathcal{L}(\mathcal{C}) 
            \rtimes 
            \mathrm{Perm}^*_\mathcal{L}(\mathcal{C})
            =
            \mathrm{Trans}_\mathcal{L}(\mathcal{C}) 
            \rtimes 
            \mathrm{Perm}_\mathcal{L}(\mathcal{C}).2,
            \\
            \abs{\mathrm{Aut}_\mathcal{L}(\mathcal{C})}
            &\leq
            2
            \cdot 
            \frac{1}{2^{\ceil{k/2}^2}}
            \cdot 
            2^{k(k+1)/2}
            \cdot
            \abs{\GL(k, \mathbb{F}_2)},
            \label{eq:css_codes_auto_logical_groups_strict_psd}
        \end{split}\end{equation}
        where $\mathrm{Perm}^*_\mathcal{L}(\mathcal{C})$ is the group of logical actions induced allowing permutations of physical qubits and $H^{\otimes n}$ on $\mathcal{C}$. Concretely, let $E$ be the pure exchange-type logical action induced by any self-duality automorphism (see \cref{prop:self_dual_css_codes_pure_exchange_logical_action}) of $\mathcal{C}$ in $\mathcal{L}$. Then $\mathrm{Perm}^*_\mathcal{L}(\mathcal{C}) \coloneqq \langle E \, \mathrm{Perm}_\mathcal{L}(\mathcal{C}), E \rangle = \mathrm{Perm}_\mathcal{L}(\mathcal{C}) \sqcup E \, \mathrm{Perm}_\mathcal{L}(\mathcal{C})$, and is an index-two group extension of $\mathrm{Perm}_\mathcal{L}(\mathcal{C})$. That is, $\mathrm{Perm}_\mathcal{L}(\mathcal{C}) \triangleleft \mathrm{Perm}^*_\mathcal{L}(\mathcal{C})$ and $\mathrm{Perm}^*_\mathcal{L}(\mathcal{C}) / \mathrm{Perm}_\mathcal{L}(\mathcal{C}) \cong C_2$. The group $\mathrm{Perm}^*_\mathcal{L}(\mathcal{C})$ is independent of the choice of $E$.
    \end{itemize}

    The transversal and permutation logical groups can be substituted into \cref{eq:css_codes_auto_logical_groups_ssd,eq:css_codes_auto_logical_groups_non_psd,eq:css_codes_auto_logical_groups_strict_psd} from \cref{thm:css_codes_perm_logical_group,thm:css_codes_perm_logical_group_psd,thm:css_codes_perm_logical_group_ssd,thm:css_codes_trans_logical_groups_ssd_indecomp,thm:css_codes_trans_logical_groups_non_ssd_indecomp} to form $\mathrm{Aut}_\mathcal{L}(\mathcal{C})$.
    Moreover, the maximum $\mathrm{Aut}_\mathcal{L}(\mathcal{C}) \cong S_3$ for $k = 1$, possible on SSD codes, and $\mathrm{Aut}_\mathcal{L}(\mathcal{C}) = \mathcal{P}(2k, \mathbb{F}_2)$ for $k \geq 2$, possible on non-PSD codes, are achievable. 
\end{theorem}

\begin{proof}
    As a preliminary, for any code $\mathcal{C}$, we note that $\mathrm{Perm}_\mathcal{L}(\mathcal{C})$ normalizes $\mathrm{Trans}_\mathcal{L}(\mathcal{C})$, that is, $b \mathrm{Trans}_\mathcal{L}(\mathcal{C}) b^{-1} = \mathrm{Trans}_\mathcal{L}(\mathcal{C})$ for all $b \in \mathrm{Perm}_\mathcal{L}(\mathcal{C})$. This is because the qubit permutation of $b^{-1}$ can be pulled through a transversal gate to cancel with that of $b$, yielding another transversal gate.
    
    For SSD and non-PSD codes, by \cref{lem:css_codes_auto_logical_splitting_property}, every $g \in \mathrm{Aut}_\mathcal{L}(\mathcal{C})$ can be factorized as $g = ba = (b a b^{-1}) b = a' b$ for some $a, a' \in \mathrm{Trans}_\mathcal{L}(\mathcal{C})$ and $b \in \mathrm{Perm}_\mathcal{L}(\mathcal{C})$. Therefore $\mathrm{Aut}_\mathcal{L}(\mathcal{C}) = \mathrm{Trans}_\mathcal{L}(\mathcal{C}) \mathrm{Perm}_\mathcal{L}(\mathcal{C})$. Moreover, by \cref{lem:css_codes_trans_perm_logical_actions_trivial_intersection}, the intersection of $\mathrm{Trans}_\mathcal{L}(\mathcal{C})$ and $\mathrm{Perm}_\mathcal{L}(\mathcal{C})$ is trivial, and $\mathrm{Trans}_\mathcal{L}(\mathcal{C}) \triangleleft \mathrm{Aut}_\mathcal{L}(\mathcal{C})$ since $\mathrm{Perm}_\mathcal{L}(\mathcal{C})$ normalizes $\mathrm{Trans}_\mathcal{L}(\mathcal{C})$. The group product then has the structure of a semidirect product, $\mathrm{Aut}_\mathcal{L}(\mathcal{C}) = \mathrm{Trans}_\mathcal{L}(\mathcal{C}) \rtimes \mathrm{Perm}_\mathcal{L}(\mathcal{C})$.

    For a PSD-but-not-SSD code, fix any self-duality automorphism inducing a pure exchange-type logical action $E$ (see \cref{prop:self_dual_css_codes_pure_exchange_logical_action}). First, we show that $\mathrm{Perm}^*_\mathcal{L}(\mathcal{C}) \coloneqq \langle E \, \mathrm{Perm}_\mathcal{L}(\mathcal{C}), E \rangle = \mathrm{Perm}_\mathcal{L}(\mathcal{C}) \sqcup E \, \mathrm{Perm}_\mathcal{L}(\mathcal{C})$, and is an index-two group extension of $\mathrm{Perm}_\mathcal{L}(\mathcal{C})$. Note that $E$ normalizes $\mathrm{Perm}_\mathcal{L}(\mathcal{C})$, that is, $E \, \mathrm{Perm}_\mathcal{L}(\mathcal{C}) \, E^{-1} = \mathrm{Perm}_\mathcal{L}(\mathcal{C})$. This is because the $H^{\otimes n}$ in the implementation of $E$ and $E^{-1}$ can be pulled through the qubit permutations to cancel, leaving behind a permutation gate. Also, $E^2 \in \mathrm{Perm}_\mathcal{L}(\mathcal{C})$, because the $H^{\otimes n}$ in their implementations likewise cancel. Together, these imply that every element of $\mathrm{Perm}^*_\mathcal{L}(\mathcal{C})$ reduces to either an element of $\mathrm{Perm}_\mathcal{L}(\mathcal{C})$ or $E \, \mathrm{Perm}_\mathcal{L}(\mathcal{C})$. So $\mathrm{Perm}^*_\mathcal{L}(\mathcal{C}) = \mathrm{Perm}_\mathcal{L}(\mathcal{C}) \cup E \, \mathrm{Perm}_\mathcal{L}(\mathcal{C})$. The union is disjoint, because $\mathrm{Perm}_\mathcal{L}(\mathcal{C})$ comprises preserving-type logical actions by \cref{thm:css_codes_perm_logical_group}, while $E \, \mathrm{Perm}_\mathcal{L}(\mathcal{C})$ comprises exchange-type ones. By the preserving- and exchange-type dichotomy of automorphism logical actions in \cref{lem:css_codes_auto_logical_splitting_property}, their intersection vanishes, so $\mathrm{Perm}^*_\mathcal{L}(\mathcal{C}) = \mathrm{Perm}_\mathcal{L}(\mathcal{C}) \sqcup E \, \mathrm{Perm}_\mathcal{L}(\mathcal{C})$. Lastly, since $E \, \mathrm{Perm}_\mathcal{L}(\mathcal{C}) \, E^{-1} = \mathrm{Perm}_\mathcal{L}(\mathcal{C})$, we have that $\mathrm{Perm}_\mathcal{L}(\mathcal{C}) \triangleleft \mathrm{Perm}^*_\mathcal{L}(\mathcal{C})$, and since $\mathrm{Perm}^*_\mathcal{L}(\mathcal{C})$ is the disjoint union of exactly two left cosets of $\mathrm{Perm}_\mathcal{L}(\mathcal{C})$, we have $\mathrm{Perm}^*_\mathcal{L}(\mathcal{C}) / \mathrm{Perm}_\mathcal{L}(\mathcal{C}) \cong C_2$.

    Note that $\mathrm{Perm}^*_\mathcal{L}(\mathcal{C})$ is independent of the choice of $E$. Indeed, if $E_1$ and $E_2$ are induced by two different self-duality automorphisms, then $E_1^{-1}E_2 \in \mathrm{Perm}_\mathcal{L}(\mathcal{C})$, because the two $H^{\otimes n}$ factors in their implementations cancel to leave a permutation. Hence $E_1 \mathrm{Perm}_\mathcal{L}(\mathcal{C}) = E_2 \mathrm{Perm}_\mathcal{L}(\mathcal{C})$, so both choices define the same group $\mathrm{Perm}_\mathcal{L}^*(\mathcal{C})$.

    Now we show that $\mathrm{Aut}_\mathcal{L}(\mathcal{C}) = \mathrm{Trans}_\mathcal{L}(\mathcal{C}) \rtimes \mathrm{Perm}^*_\mathcal{L}(\mathcal{C})$. By \cref{lem:css_codes_auto_logical_splitting_property}, every logical action in $\mathrm{Aut}_\mathcal{L}(\mathcal{C})$ is either preserving- or exchange-type. The preserving-type ones factor as $g = b a$, and the exchange-type ones as $g = E b a' = E (b a' b^{-1}) b = E a b$, for some $a, a' \in \mathrm{Trans}_\mathcal{L}(\mathcal{C})$ and $b \in \mathrm{Perm}_\mathcal{L}(\mathcal{C})$. Note also that $E$ normalizes $\mathrm{Trans}_\mathcal{L}(\mathcal{C})$, since conjugation by its physical implementation maps a transversal gate to another transversal gate. Then we can write $E a b = (E a E^{-1}) E b \in \mathrm{Trans}_\mathcal{L}(\mathcal{C}) E \, \mathrm{Perm}_\mathcal{L}(\mathcal{C})$. Hence $\mathrm{Aut}_\mathcal{L}(\mathcal{C}) = \mathrm{Trans}_\mathcal{L}(\mathcal{C}) \mathrm{Perm}_\mathcal{L}(\mathcal{C}) \sqcup \mathrm{Trans}_\mathcal{L}(\mathcal{C}) E \mathrm{Perm}_\mathcal{L}(\mathcal{C}) = \mathrm{Trans}_\mathcal{L}(\mathcal{C}) \mathrm{Perm}^*_\mathcal{L}(\mathcal{C})$. Now, note that $\mathrm{Perm}^*_\mathcal{L}(\mathcal{C})$ normalizes $\mathrm{Trans}_\mathcal{L}(\mathcal{C})$, because $E$ and $\mathrm{Perm}_\mathcal{L}(\mathcal{C})$ both normalize $\mathrm{Trans}_\mathcal{L}(\mathcal{C})$. So $\mathrm{Trans}_\mathcal{L}(\mathcal{C}) \triangleleft \mathrm{Aut}_\mathcal{L}(\mathcal{C})$. Lastly, we note that the intersection of $\mathrm{Trans}_\mathcal{L}(\mathcal{C})$ and $\mathrm{Perm}^*_\mathcal{L}(\mathcal{C})$ is trivial. This is because, by \cref{lem:css_codes_trans_perm_logical_actions_trivial_intersection}, the intersection of $\mathrm{Trans}_\mathcal{L}(\mathcal{C})$ and $\mathrm{Perm}_\mathcal{L}(\mathcal{C})$ is trivial; and the intersection of $\mathrm{Trans}_\mathcal{L}(\mathcal{C})$ and $E \, \mathrm{Perm}_\mathcal{L}(\mathcal{C})$ vanishes because the former comprises preserving-type logical actions by \cref{thm:css_codes_trans_logical_groups_non_ssd_indecomp} whereas the latter comprises exchange-type ones. So the product is semidirect, $\mathrm{Aut}_\mathcal{L}(\mathcal{C}) = \mathrm{Trans}_\mathcal{L}(\mathcal{C}) \rtimes \mathrm{Perm}^*_\mathcal{L}(\mathcal{C})$.

    The group order bounds in \cref{eq:css_codes_auto_logical_groups_ssd,eq:css_codes_auto_logical_groups_non_psd,eq:css_codes_auto_logical_groups_strict_psd} follow from substituting in bounds for $\abs{\mathrm{Trans}_\mathcal{L}(\mathcal{C})}$ and $\abs{\mathrm{Perm}_\mathcal{L}(\mathcal{C})}$ from \cref{thm:css_codes_perm_logical_group,thm:css_codes_perm_logical_group_psd,thm:css_codes_perm_logical_group_ssd,thm:css_codes_trans_logical_groups_ssd_indecomp,thm:css_codes_trans_logical_groups_non_ssd_indecomp}. 
    In the PSD-but-not-SSD case, for simplicity, we have neglected a factor of $\smash{1/2^{\floor{k^2/4}}}$ arising from the permutation logical group order bound of \cref{thm:css_codes_perm_logical_group_psd}, which applies for $k \geq 3$.
    
    From these group order bounds, we find that the maximum-sized automorphism logical group is $\mathrm{Aut}_\mathcal{L}(\mathcal{C}) \cong S_3$ for $k = 1$, possible on SSD codes, and $\mathrm{Aut}_\mathcal{L}(\mathcal{C}) = \mathcal{P}(2k, \mathbb{F}_2)$ for $k \geq 2$, possible on non-PSD codes. The former is achieved by any $k = 1$ SSD CSS code (see \cref{thm:css_codes_trans_logical_groups_ssd_indecomp}), and the latter is achieved by, for example, \cref{cons:concatenated_phantom_clifford}.
\end{proof}

\begin{remark}
   [Splitting of $\mathrm{Perm}^*_\mathcal{L}(\mathcal{C})$ index-two extension]
   \label{rem:css_codes_auto_logical_group_psd_not_ssd_split_extension}
   In \cref{thm:css_codes_auto_logical_groups}, for a PSD-but-not-SSD indecomposable CSS code $\mathcal{C}$, we stated that $\mathrm{Perm}^*_\mathcal{L}(\mathcal{C})$ is an index-two group extension of $\mathrm{Perm}_\mathcal{L}(\mathcal{C})$, but not $\mathrm{Perm}^*_\mathcal{L}(\mathcal{C}) = \mathrm{Perm}_\mathcal{L}(\mathcal{C}) \rtimes C_2$ specifically. This is because the extension need not be split in general. In general, suppose a group $G$ is an extension of $N \triangleleft G$ by $C_2$. Then the extension splits iff there exists an $F \in G\setminus N$ such that $F^2 = I$. In our context, this amounts to the condition that there exists a self-duality automorphism of $\mathcal{C}$ inducing a logical action $E$ such that $E^2 = I$. On such codes, $\mathrm{Perm}^*_\mathcal{L}(\mathcal{C}) = \mathrm{Perm}_\mathcal{L}(\mathcal{C}) \rtimes C_2$ and accordingly $\mathrm{Aut}_\mathcal{L}(\mathcal{C}) = \mathrm{Trans}_\mathcal{L}(\mathcal{C}) \rtimes (\mathrm{Perm}_\mathcal{L}(\mathcal{C}) \rtimes C_2)$. 
\end{remark}

\subsubsection{Beyond Clifford gates}

Case~\cref{item:css_codes_auto_physical_splitting_property_preserving_case} of \cref{lem:css_codes_auto_physical_splitting_property} shows that, if a product of physical $S$ gates supported on an arbitrary subset of physical qubits, together with a physical qubit permutation, is a valid logical gate, then the product of $S$ gates and the permutation are separately valid logical gates. In fact, more generally, the same conclusion holds for a product of arbitrary single-qubit diagonal gates supported on any subset of the physical qubits.

\begin{lemma}
    [Separation of diagonal physical gates from qubit permutations] 
    \label{lem:css_codes_auto_splitting_property_beyond_clifford}
    Let $\mathcal{C}$ be an $\db{n,k}$ CSS code. 
    Suppose $\overline{U} \coloneqq U_\pi (\bigotimes_{i=1}^n D_i)$ is a valid logical gate on $\mathcal{C}$, where $\pi \in S_n$ is a physical qubit permutation and $\{D_i\}_{i=1}^n$ are single-qubit diagonal gates.
    Then $U_\pi$ and $\bigotimes_{i=1}^n D_i$ must each separately be a valid logical gate.
\end{lemma}

\begin{proof}
    Let $C_\mathrm{x} \subseteq \mathbb{F}_2^n$ and $C_\mathrm{z} \subseteq \mathbb{F}_2^n$ be the $X$- and $Z$-type stabilizer spaces of $\mathcal{C}$ respectively.
    For $u\in C_\rmz^\perp/C_\rmx$, the coset states
    \begin{equation}
        \ket{u}_L=\frac{1}{\sqrt{|C_\rmx|}}\sum_{a\in C_\rmx}\ket{u+a}
    \end{equation}
    form a basis of the codespace.
    We first show that $\pi(C_\rmx)=C_\rmx$ by taking $u=0$.
    Since $\overline{U}$ is a valid logical gate, the state
    \begin{equation}
        \ket{\psi}_L
        =\overline{U}\ket{0}_L
        =U_\pi \left(\bigotimes_{i=1}^n D_i\right) \ket{0}_L
        =\frac{1}{\sqrt{|C_\rmx|}}\sum_{a\in C_\rmx}d(a)\ket{\pi a}
    \end{equation}
    lies in the codespace, where $d(a)\neq 0$ are phases because $D$ is unitary and diagonal. 
    Thus $\supp(\ket{\psi}_L)=\pi(C_\rmx)$.
    On the other hand, $\ket{\psi}_L$ is stabilized by every $X$-type stabilizer $X^c$ for $c\in C_\rmx$. 
    Hence
    \begin{equation}
        X^c\ket{\psi}_L
        =\frac{1}{\sqrt{|C_\rmx|}}\sum_{a\in C_\rmx}d(a)\ket{\pi a+c}
        =\ket{\psi}_L.
    \end{equation}
    Therefore, $\supp(\ket{\psi}_L)+c=\supp(\ket{\psi}_L)$ for every $c\in C_\rmx$. 
    Notice that $0\in C_\rmx$ and $\pi(0)=0\in \pi(C_\rmx)$. 
    Since the coefficient $d(0)$ is nonzero, we have that $c=0+c\in\pi(C_\rmx)$ for every $c\in C_\rmx$. 
    Thus $C_\rmx\subseteq \pi(C_\rmx)$.
    Since $\pi$ is invertible, we have $\dim \pi(C_\rmx)=\dim C_\rmx$ and hence $\pi(C_\rmx)=C_\rmx$.

    Next, we show that $\pi(C_\rmz)=C_\rmz$.
    For each $u\in C_\rmz^\perp$, the support $\pi(u+C_\rmx)=\pi u+C_\rmx$ of the corresponding transformed coset state must lie in $C_\rmz^\perp$. Hence $\pi u\in C_\rmz^\perp$ for every $u\in C_\rmz^\perp$, and therefore $\pi(C_\rmz^\perp)\subseteq C_\rmz^\perp$. Since $\pi$ is invertible and preserves dimensions, this inclusion is an equality: $\pi(C_\rmz^\perp)=C_\rmz^\perp$.
    Permutations preserve the inner product, so we then have $\pi(C_\rmz)=\pi((C_\rmz^\perp)^\perp)=\pi(C_\rmz^\perp)^\perp=(C_\rmz^\perp)^\perp=C_\rmz$.

    Thus $U_\pi$ separately preserves both the $X$- and $Z$-type stabilizer spaces and is itself a valid logical gate. Consequently, $\bigotimes_{i=1}^n D_i = U_\pi^\dagger \overline{U}$ is also a valid logical gate.
\end{proof}

\clearpage

\subsection{Permutation gates}
\label{app:css_codes/perm}

\subsubsection{Permutation logical groups on general codes}

\begin{theorem}
    [Permutation logical groups on CSS codes]
    \label{thm:css_codes_perm_logical_group}
    Let $\mathcal{C}$ be an $\db{n,k}$ CSS code and $\mathcal{L}$ be a CSS logical basis of $\mathcal{C}$. 
    Then every element of $\mathrm{Perm}_\mathcal{L}(\mathcal{C})$ is a pure preserving-type Clifford with symplectic representation
    \begin{equation}
        \mqty(A & 0 \\ 0 & A^{-\top}),
        \qquad
        A \in \GL(k,\mathbb{F}_2).
        \label{eq:css_codes_perm_logical_action_form}
    \end{equation}
    That is, $\mathrm{Perm}_\mathcal{L}(\mathcal{C}) \cong \mathcal{A}_\mathcal{L}(\mathcal{C}) \coloneqq \{A:\diag(A,A^{-\top})\in\mathrm{Perm}_\mathcal{L}(\mathcal{C})\} \leq \GL(k, \mathbb{F}_2)$. 
    The maximum $\mathcal{A}_\mathcal{L}(\mathcal{C}) = \GL(k, \mathbb{F}_2)$ is achievable.
\end{theorem}

\begin{proof}
    Permutations of physical qubits cannot change the type of Pauli operators acting on the physical qubits.
    Hence, in a CSS logical basis, permutation gates map $X$-type (resp.~$Z$-type) logical operators to $X$-type (resp.~$Z$-type) logical operators, so every induced logical action is a pure preserving-type Clifford.
    By \cref{prop:preserving_exchange_multi_qubit_cliffords_forms}, every such Clifford has the form in \cref{eq:css_codes_perm_logical_action_form}.
    This proves $\mathcal A_\mathcal L(\mathcal C)\leq\GL(k,\mathbb F_2)$.
    The maximum $\mathcal A_\mathcal L(\mathcal C)=\GL(k,\mathbb F_2)$ is achieved by phantom codes~\cite{koh2026entangling}.
\end{proof}

\subsubsection{Flagged isometry groups}

\begin{definition}
    [Flagged isometry groups]
    \label{def:flagged_isometry_groups}
    Let $k\geq 1$, $0\leq m\leq\floor{k/2}$, and put $s\coloneqq k-2m$.
    Fix a basis $(e_1,\dots,e_k)$ of $\F_2^k$, and set $E_0\coloneqq\{0\}$ and $E_j\coloneqq\langle e_1,\dots,e_j\rangle$ for $1\leq j\leq k$; then $\{0\}=E_0\subsetneq E_1\subsetneq\cdots\subsetneq E_k=\F_2^k$ forms a flag (see \cref{def:linear_algebra_flags_and_isotropic_flags}).
    Denote by
    \begin{equation}
        P_m(k,\F_2)
        \coloneqq
        \left\{
            g\in\GL(k,\F_2):
            g(E_m)=E_m,\;
            g(E_{k-m})=E_{k-m}
        \right\}
    \end{equation}
    the stabilizer of $E_m\subseteq E_{k-m}\subseteq\F_2^k$.
    Its action on the quotient $Q_m\coloneqq E_{k-m}/E_m$, which has dimension $s$, gives an induced group representation $\pi_m:P_m(k,\F_2)\to\GL(Q_m)\cong\GL(s,\F_2)$.
    
    For a vector space $V$ equipped with a symmetric bilinear form $\beta$, denote by
    $\Isom(V,\beta)$ the group of invertible linear transformations $g:V\to V$ preserving $\beta$, i.e.~$\beta(gv,gw)=\beta(v,w)$ for all $v,w\in V$.
    We define the flagged isometry groups $\mathcal F_m^{\Sp}(k,\F_2)$ and $\mathcal F_m^{O}(k,\F_2)$ as follows:
    \begin{itemize}
        \item If $s$ is even, choose a nondegenerate alternating form $\beta_{\Sp}$ on $Q_m$ and define
        \begin{equation}
            \mathcal F_m^{\Sp}(k,\F_2)
            \coloneqq
            \pi_m^{-1}\!\left(\Isom(Q_m,\beta_{\Sp})\right).
        \end{equation}
        Choosing a basis of $Q_m$ in which $\beta_{\Sp}$ has Gram matrix
        $\Omega_{s/2}$ identifies
        \begin{equation}
            \Isom(Q_m,\beta_{\Sp})
            =
            \Sp(s,\F_2)
            \coloneqq
            \left\{
                M\in\GL(s,\F_2):
                M^\top\Omega_{s/2}M=\Omega_{s/2}
            \right\}.
        \end{equation}
        
        \item If $s\geq1$, choose a nondegenerate nonalternating symmetric form $\beta_O$ on $Q_m$ and define
        \begin{equation}
            \mathcal F_m^O(k,\F_2)
            \coloneqq
            \pi_m^{-1}\!\left(\Isom(Q_m,\beta_O)\right).
        \end{equation}
        Choosing a basis of $Q_m$ in which $\beta_O$ has Gram matrix $I_s$ identifies
        \begin{equation}
            \Isom(Q_m,\beta_O)
            =
            O(s,\F_2)
            \coloneqq
            \left\{
                M\in\GL(s,\F_2):
                M^\top M=I_s
            \right\}.
        \end{equation}
    \end{itemize}

    Explicitly, in a basis adapted to the flag, with block sizes $(m,s,m)$ and with the middle block chosen as above, every element of the flag stabilizer can be written as a block upper-triangular matrix
    \begin{equation}
        \label{eq:flag_block_matrix}
        \begin{pmatrix}
            A & * & *\\
            0 & M & *\\
            0 & 0 & C
        \end{pmatrix},
        \qquad
        A,C\in\GL(m,\F_2),
        \qquad
        M\in\GL(s,\F_2).
    \end{equation}
    The groups $\mathcal F_m^{\Sp}(k,\F_2)$ and $\mathcal F_m^O(k,\F_2)$ are obtained by imposing $M\in\Sp(s,\F_2)$ and $M\in O(s,\F_2)$, respectively.

    For $s=0$, we adopt the formal conventions $\Sp(0,\F_2)=O(0,\F_2)=\{1\}$; the two flagged isometry groups then coincide, $\mathcal F_m^{\Sp}(k,\F_2)=\mathcal F_m^O(k,\F_2)=P_m(k,\F_2)$.
    When $m=0$,
    \begin{equation}
        \mathcal F_0^{\Sp}(k,\F_2)=\Sp(k,\F_2)
        \quad
        (k \,\, \text{even}),
        \qquad
        \mathcal F_0^O(k,\F_2)=O(k,\F_2).
    \end{equation}
\end{definition}

\begin{proposition}
    [Conversion between $\Sp$ and $O$ groups]
    \label{prop:group_conversion_symplectic_orthogonal}
    $O(s,\mathbb F_2)\cong\Sp(s-1,\mathbb F_2)$ when $s$ is odd.
\end{proposition}

\begin{proof}
    Let $s=2r+1$, let $V\coloneqq\F_2^s$, and equip $V$ with the standard dot product $\la x,y\ra\coloneqq x^\top y$.
    Over $\F_2$, one has $\la x,x\ra=\la x,\mathbf{1}\ra$.
    Moreover, every $M\in O(s,\F_2)$ fixes $\mathbf{1}$.
    Indeed, since $M^\top M=I$, for every $x\in V$,
    $\la x,\mathbf{1}\ra=\la x,x\ra=\la Mx,Mx\ra=\la Mx,\mathbf{1}\ra=\la x,M^{-1}\mathbf{1}\ra$.
    Nondegeneracy of the dot product therefore gives $M^{-1}\mathbf{1}=\mathbf{1}$, and hence $M\mathbf{1}=\mathbf{1}$.

    Now consider $W\coloneqq\mathbf{1}^\perp=\{x\in\F_2^s:\sum_i x_i=0\}$.
    This is the even-weight subspace and has dimension $s-1=2r$.
    For every $w\in W$, $\la w,w\ra=\la w,\mathbf{1}\ra=0$, so the restriction of the dot product to $W$ is alternating.
    Since $s$ is odd, $\mathbf{1}\notin W$, and therefore
    $V=\langle\mathbf{1}\rangle\perp W$.
    In particular, the restricted form on $W$ is nondegenerate, so $W$ is a symplectic space of dimension $s-1$.

    Since every $M\in O(s,\F_2)$ fixes $\mathbf{1}$ and preserves $W$, restriction to $W$ gives an injective homomorphism from $O(s,\F_2)$ to the group of invertible linear transformations of $W$ preserving the restricted form.
    Conversely, every isometry $A$ of the restricted form on $W$ extends to an element of $O(s,\F_2)$ by
    $\widetilde A(a\mathbf{1}+w)\coloneqq a\mathbf{1}+Aw$.
    Thus restriction is an isomorphism onto the isometry group of the restricted form on $W$.
    Choosing a symplectic basis of $W$ identifies this latter group with $\Sp(s-1,\F_2)$, and hence
    $O(s,\F_2)\cong\Sp(s-1,\F_2)$.
\end{proof}

\begin{remark}
    [Orders of flagged isometry groups]
    \label{rem:flagged_isometry_group_orders}
    The group orders of the flagged isometry groups $\mathcal F_m^{\Sp}(k,\mathbb F_2)$ and $\mathcal F_m^{O}(k,\mathbb F_2)$ follow from the block matrix forms in \cref{def:flagged_isometry_groups} and are:
    \begin{equation}\begin{split}
        \left|\mathcal F_m^{\Sp}(k,\mathbb F_2)\right|
        =
        2^{m^2+2ms}
        \left|\GL(m,\mathbb F_2)\right|^2
        \left|\Sp(s,\mathbb F_2)\right|,
        \qquad
        \left|\mathcal F_m^{O}(k,\mathbb F_2)\right|
        =
        2^{m^2+2ms}
        \left|\GL(m,\mathbb F_2)\right|^2
        \left|O(s,\mathbb F_2)\right|,
        \label{eq:flagged_isometry_group_orders}
    \end{split}\end{equation}
    where $s \coloneqq k-2m$. Moreover,
    \begin{equation}
        \abs{O(2r+1,\F_2)}
        =
        \abs{\Sp(2r,\F_2)}
        \quad (r\geq0),
        \qquad
        \abs{O(2r,\F_2)}
        =
        2^{2r-1}\abs{\Sp(2r-2,\F_2)}
        \quad (r\geq1).
    \end{equation}
    Together with \cref{eq:group_order_symplectic_upper_bound} of \cref{fact:group_theory_relevant_group_orders}, these give
    $\abs{O(s,\F_2)}\leq2^{s(s-1)/2}$ for all $s\geq0$.
    \Cref{tab:flagged_isometry_group_orders} presents group orders at small $k$.
    \begin{table}[h]
        \centering
        \footnotesize
        \begin{tabular}{
            C{0.6cm}
            R{1.7cm}
            R{1.7cm}
            R{1.7cm}
            R{1.7cm}
            R{1.7cm}
            R{1.7cm}
            R{1.7cm}
            R{1.7cm}
        }
            \toprule
            & \multicolumn{4}{c}{
                \(\abs*{\mathcal F_m^{\Sp}(k,\mathbb F_2)}\)
            }
            & \multicolumn{4}{c}{
                \(\abs*{\mathcal F_m^{O}(k,\mathbb F_2)}\)
            }
            \\
            \cmidrule(lr){2-5}
            \cmidrule(lr){6-9}
            \(k\)
                & \(m=0\)
                & \(m=1\)
                & \(m=2\)
                & \(m=3\)
                & \(m=0\)
                & \(m=1\)
                & \(m=2\)
                & \(m=3\)
            \\
            \midrule
            1
                & -
                & -
                & -
                & -
                & \(1\)
                & -
                & -
                & -
            \\
            2
                & \(6\)
                & \(2\)
                & -
                & -
                & \(2\)
                & \(2\)
                & -
                & -
            \\
            3
                & -
                & -
                & -
                & -
                & \(6\)
                & \(8\)
                & -
                & -
            \\
            4
                & \(720\)
                & \(192\)
                & \(576\)
                & -
                & \(48\)
                & \(64\)
                & \(576\)
                & -
            \\
            5
                & -
                & -
                & -
                & -
                & \(720\)
                & \(768\)
                & \(9\,216\)
                & -
            \\
            6
                & \(1\,451\,520\)
                & \(368\,640\)
                & \(884\,736\)
                & \(14\,450\,688\)
                & \(23\,040\)
                & \(24\,576\)
                & \(294\,912\)
                & \(14\,450\,688\)
            \\
            \bottomrule
        \end{tabular}
        \caption{Orders of the flagged isometry groups $\mathcal F_m^{\Sp}(k,\mathbb F_2)$ and $\mathcal F_m^{O}(k,\mathbb F_2)$ for \(1\leq k\leq 6\).}
        \label{tab:flagged_isometry_group_orders}
    \end{table}
\end{remark}

\begin{remark}
    [Different notions of orthogonal groups]
    Although referred to as orthogonal groups in literature, the $O(s,\mathbb F_2)$ in \cref{def:flagged_isometry_groups} is not to be confused with the $O^\pm(s, \mathbb F_2)$ groups of \cref{def:group_theory_binary_orthogonal_groups}---the former preserves the nonalternating symmetric bilinear form $b(x,y)=x^\top y$ whereas the latter preserve a quadratic form. They are different over $\mathbb F_2$. Coincidental isomorphisms occur when $s \leq 3$: $O(1, \mathbb{F}_2)$ is the trivial group; $O(2, \mathbb{F}_2) = O^+(2, \mathbb{F}_2) = \{I_2, \Omega_1\} \cong S_2$ but is distinct from $O^-(2, \mathbb{F}_2)$; and $O(3, \mathbb{F}_2) \cong \Sp(2, \mathbb{F}_2) \cong S_3$ is the group of $3 \times 3$ permutation matrices. At $s = 4$, the groups diverge. Take, for example,
    \begin{equation}
        P
        =
        \begin{pmatrix}
            0&0&0&1 \\
            0&1&0&0 \\
            0&0&1&0 \\
            1&0&0&0
        \end{pmatrix},
        \qquad
        A^+
        =
        \begin{pmatrix}
            1&1&0&0 \\
            0&1&0&0 \\
            0&0&1&0 \\
            0&0&1&1
        \end{pmatrix},
        \qquad
        A^-
        =
        \begin{pmatrix}
            1&0&1&0 \\
            0&1&0&0 \\
            0&0&1&0 \\
            0&0&0&1
        \end{pmatrix}.
    \end{equation}
    Then $P \in O(4, \mathbb{F}_2)$ but $P \notin O^\pm(4, \mathbb{F}_2)$; $A^+ \in O^+(4, \mathbb{F}_2)$ but $A^+ \notin O(4, \mathbb{F}_2), O^-(4, \mathbb{F}_2)$; and $A^- \in O^-(4, \mathbb{F}_2)$ but $A^- \notin O(4, \mathbb{F}_2), O^+(4, \mathbb{F}_2)$. The difference is also apparent by comparing the group orders in \cref{tab:flagged_isometry_group_orders,tab:group_theory_relevant_group_orders}.
\end{remark}

\subsubsection{Balanced logical space of self-dual codes and constraints}

\begin{proposition}
    [Balanced logical space of PSD CSS codes and its properties]
    \label{prop:psd_css_codes_balanced_space_properties}
    Let $\mathcal C$ be an $\db{n,k}$ PSD CSS code with $X$- and $Z$-stabilizer spaces $C_\mathrm{x},C_\mathrm{z}\subseteq\F_2^n$.
    Define
    \begin{equation}
        R_\mathrm{x}
        \coloneqq C_\mathrm{x}\cap C_\mathrm{x}^{\perp},
        \qquad
        R_\mathrm{z}
        \coloneqq C_\mathrm{z}\cap C_\mathrm{z}^{\perp},
        \qquad
        T
        \coloneqq C_\mathrm{x}\cap C_\mathrm{z}
        = R_\mathrm{x}\cap R_\mathrm{z},
        \qquad
        W
        \coloneqq C_\mathrm{x}^{\perp}\cap C_\mathrm{z}^{\perp}
        = (C_\mathrm{x}+C_\mathrm{z})^\perp.
    \end{equation}

    Then the following statements hold.
    \begin{enumerate}
    
        \item Any self-duality permutation maps $R_\mathrm{x} \leftrightarrow R_\mathrm{z}$ and preserves $T$, so $\dim(R_\mathrm{x}/T)=\dim(R_\mathrm{z}/T) \eqqcolon m$.
        \label{item:psd_css_codes_balanced_space_properties_1}
        
        \item The \emph{balanced logical space}
        \begin{equation}
            B
            \coloneqq \frac{W}{T}
            = \frac{(C_\mathrm{x}+C_\mathrm{z})^\perp}{C_\mathrm{x}\cap C_\mathrm{z}},
            \label{eq:psd_css_codes_balanced_space_properties_2}
        \end{equation}
        has dimension $k$. For $w\in W$, we write $[w]_B\coloneqq w+T$ for its equivalence class in $B$.
        \label{item:psd_css_codes_balanced_space_properties_2}
        
        \item The ordinary dot product induces a symmetric bilinear form on $B$, 
        \begin{equation}
            \beta_B([u]_B,[v]_B)=uv^\top,
            \label{eq:psd_css_codes_balanced_space_properties_3}
        \end{equation}
        whose radical is $\rad(\beta_B)=(R_\mathrm{x}+R_\mathrm{z})/T$.
        In particular, $\rad(\beta_B)$ has dimension $2m$, so $B/\rad(\beta_B)$ has dimension $k-2m$. Moreover, by \cref{def:radical_bilinear_form}, the induced symmetric bilinear form on $B/\rad(\beta_B)$ is nondegenerate.
        \label{item:psd_css_codes_balanced_space_properties_3}
        
        \item The induced nondegenerate symmetric bilinear form on $B/\rad(\beta_B)$ is alternating iff $\mathbf 1\in C_\mathrm{x}+C_\mathrm{z}$.
        \label{item:psd_css_codes_balanced_space_properties_4}
        
        \item If $\mathbf 1\in C_\mathrm{x}+C_\mathrm{z}$, then $n$, $k$, and $k-2m$ are even.
        \label{item:psd_css_codes_balanced_space_properties_5}
        
    \end{enumerate}
\end{proposition}

\begin{proof}
    Since $\mathcal C$ is PSD, there exists a physical qubit permutation $\pi\in S_n$ such that $\pi(C_\mathrm{x})=C_\mathrm{z}$ and $\pi(C_\mathrm{z})=C_\mathrm{x}$.
    In particular, $\dim C_\mathrm{x}=\dim C_\mathrm{z}\eqqcolon r$, and $n=2r+k$.
    Since $C_\mathrm{z}\subseteq C_\mathrm{x}^\perp$ and $C_\mathrm{x}\subseteq C_\mathrm{z}^\perp$, we have $C_\mathrm{x}\cap C_\mathrm{z}\subseteq R_\mathrm{x}\cap R_\mathrm{z}$. 
    The reverse inclusion is immediate, so $T=C_\mathrm{x}\cap C_\mathrm{z}=R_\mathrm{x}\cap R_\mathrm{z}$. Since the permutation preserves the inner product, we have $\pi(C_\mathrm{x}^\perp) = \pi(C_\mathrm{x})^\perp$ and $\pi(C_\mathrm{z}^\perp) = \pi(C_\mathrm{z})^\perp$. The same permutation maps $R_\mathrm{x}\leftrightarrow R_\mathrm{z}$ and preserves $T$, and thus $\dim(R_\mathrm{x}/T)=\dim(R_\mathrm{z}/T)$. 
    This shows Property~\ref{item:psd_css_codes_balanced_space_properties_1}.

    Now, because $C_\mathrm{z}\subseteq C_\mathrm{x}^\perp$ and $C_\mathrm{x}\subseteq C_\mathrm{z}^\perp$, one has $T\subseteq W$, and can therefore define the balanced logical space $B\coloneqq W/T$, as in \cref{eq:psd_css_codes_balanced_space_properties_2}. 
    We show $\dim B=k$ by dimension counting. 
    Call $\ell=\dim(C_\mathrm{x}\cap C_\mathrm{z})$; then $\dim(C_\mathrm{x}+C_\mathrm{z})=\dim C_\mathrm{x}+\dim C_\mathrm{z}-\dim(C_\mathrm{x}\cap C_\mathrm{z})=2r-\ell$. 
    Therefore $\dim B=n-(2r-\ell)-\ell=n-2r=k$, as claimed in Property~\ref{item:psd_css_codes_balanced_space_properties_2}.

    Next, define a symmetric bilinear form on $B$ as in \cref{eq:psd_css_codes_balanced_space_properties_3}. 
    This is well-defined, since changing $u$ or $v$ by an element of $T\subseteq C_\mathrm{x}+C_\mathrm{z}$ does not change the dot product with an element of $W=(C_\mathrm{x}+C_\mathrm{z})^\perp$.
    We then claim $\rad(\beta_B)=(R_\mathrm{x}+R_\mathrm{z})/T$. This is because
    \begin{equation}
        \rad(\beta_B)
        =\frac{W\cap W^\perp}{T}
        =\frac{W\cap(C_\rmx+C_\rmz)}{T}
        =\frac{R_\rmx+R_\rmz}{T},
    \end{equation}
    where $W\cap(C_\rmx+C_\rmz)=R_\rmx+R_\rmz$ since if $w=a+b\in W$ with $a\in C_\rmx$, $b\in C_\rmz$, then $w,b\in C_\rmx^\perp$ gives $a\in R_\rmx$ and similarly $b\in R_\rmz$.
    The reverse inclusion is immediate.
    Now, using the fact that $R_\rmx\cap R_\rmz=T$, $\dim(\rad(\beta_B))=\dim(R_\rmx)+\dim(R_\rmz)-2\dim(T)=\dim(R_\rmx/T)+\dim(R_\rmz/T)=2m$.
    This shows Property~\ref{item:psd_css_codes_balanced_space_properties_3}.

    Finally, we decide whether the quotient form is alternating. 
    For $w\in W$, we have $\beta_B([w]_B,[w]_B)=ww^\top=w\mathbf 1^\top$.
    Therefore $\beta_B$, and equivalently its induced form on $B/\rad(\beta_B)$, is alternating iff $w\mathbf 1^\top=0$ for all $w\in W$.
    This is equivalent to $\mathbf 1\in W^\perp$.
    Since $W^\perp=C_\mathrm{x}+C_\mathrm{z}$, the induced nondegenerate quotient form is alternating iff $\mathbf 1\in C_\mathrm{x}+C_\mathrm{z}$, as claimed in Property~\ref{item:psd_css_codes_balanced_space_properties_4}.
    
    By Property~\ref{item:psd_css_codes_balanced_space_properties_4}, if $\mathbf 1\in C_\mathrm{x}+C_\mathrm{z}$, then the induced quotient form is nondegenerate alternating. 
    Hence $k-2m$ is even. 
    Since $2m$ is even, $k$ is even as well. 
    So $n=2r+k$ is also even. 
    This shows Property~\ref{item:psd_css_codes_balanced_space_properties_5}.
\end{proof}

\begin{lemma}
    [Balanced logical flag of a PSD CSS code]
    \label{lem:psd_css_codes_balanced_logical_flag}
    Let $\mathcal C$ be an $\db{n,k}$ PSD CSS code with $X$- and $Z$-stabilizer spaces $C_\mathrm{x},C_\mathrm{z}\subseteq\F_2^n$.
    Let $T\coloneqq C_\mathrm{x}\cap C_\mathrm{z}$,
    $W\coloneqq C_\mathrm{x}^\perp\cap C_\mathrm{z}^\perp$, the balanced logical space $B\coloneqq W/T$, symmetric bilinear form $\beta_B$, and dimension
    $m$ be as in \cref{prop:psd_css_codes_balanced_space_properties}.
    Denote by $\mathcal L_\mathrm{x}\coloneqq C_\mathrm{z}^{\perp}/C_\mathrm{x}$ and $\mathcal L_\mathrm{z}\coloneqq C_\mathrm{x}^{\perp}/C_\mathrm{z}$ the $X$- and $Z$-logical spaces of $\mathcal{C}$.
    For $x\in C_\mathrm{z}^{\perp}$ and $z\in C_\mathrm{x}^{\perp}$, write their logical equivalence classes $[x]_{\mathcal L_\mathrm{x}}\coloneqq x+C_\mathrm{x}$ and $[z]_{\mathcal L_\mathrm{z}}\coloneqq z+C_\mathrm{z}$; and for $w\in W$, write $[w]_B\coloneqq w+T$.
    Define maps $\phi_\mathrm{x}:B\to\mathcal L_\mathrm{x}$ and $\phi_\mathrm{z}:B\to\mathcal L_\mathrm{z}$ by
    \begin{equation}
        \phi_\mathrm{x}([w]_B) = w+C_\mathrm{x}=:[w]_{\cL_\rmx},
        \qquad
        \phi_\mathrm{z}([w]_B) = w+C_\mathrm{z}=:[w]_{\cL_\rmz}.
        \label{eq:psd_css_codes_balanced_logical_flag_1}
    \end{equation}

    Then the maps $\phi_\mathrm{x}$ and $\phi_\mathrm{z}$ are $\mathrm{PPerm}(\mathcal C)$-equivariant, where $\mathrm{PPerm}(\mathcal{C}) \leq S_n$ is the group of physical implementations of permutation gates on $\mathcal{C}$.
    Moreover, define
    \begin{equation}
        U_\mathrm{x}\coloneqq\im\phi_\mathrm{x},
        \qquad
        U_\mathrm{z}\coloneqq\im\phi_\mathrm{z},
        \qquad
        \Lambda_\mathrm{x}
        \coloneqq U_\mathrm{z}^\perp
        = \left\{
            [x]_{\mathcal L_\mathrm{x}}\in\mathcal L_\mathrm{x}:
            xz^\top=0
            \,\, \forall \,\, [z]_{\mathcal L_\mathrm{z}}\in U_\mathrm{z}
        \right\}
        \subseteq \mathcal L_\mathrm{x}.
    \end{equation}

    Then every permutation logical action preserves the flag
    \begin{equation}
        \Lambda_\mathrm{x}\subseteq U_\mathrm{x}\subseteq\mathcal L_\mathrm{x}, 
        \qquad 
        \dim\Lambda_\mathrm{x}=m,
        \qquad 
        \dim U_\mathrm{x}=k-m.
        \label{eq:flag_in_balanced_logical_space}
    \end{equation}
    
    Later, we will use this flag in the context of \cref{def:flagged_isometry_groups}. 
    In fact, the quotient $U_\rmx/\Lambda_\rmx$ carries a nondegenerate symmetric bilinear form $\overline{\beta}_\rmx$.
    Explicitly, for $w,w'\in W$, this form is given by
    \begin{equation}
        \label{eq:induced_nondegenerate_bilinear_form}
        \overline{\beta}_\mathrm{x}
        \left(
            (w+C_\mathrm{x})+\Lambda_\mathrm{x},
            (w'+C_\mathrm{x})+\Lambda_\mathrm{x}
        \right)
        =
        ww'^\top.
    \end{equation}
\end{lemma}

\begin{proof}
    Recall from \cref{prop:psd_css_codes_balanced_space_properties} that $R_\mathrm{x}\coloneqq C_\mathrm{x}\cap C_\mathrm{x}^\perp$, $R_\mathrm{z}\coloneqq C_\mathrm{z}\cap C_\mathrm{z}^\perp$, and $m\coloneqq\dim(R_\mathrm{x}/T)=\dim(R_\mathrm{z}/T)$.
    We first show that the maps $\phi_\mathrm{x}$ and $\phi_\mathrm{z}$ are well-defined.
    Since $W\subseteq C_\mathrm{z}^\perp$, every $w\in W$ determines a class $w+C_\mathrm{x}\in\mathcal L_\mathrm{x}$.
    Moreover, if $[w_1]_B=[w_2]_B$, then $w_1=w_2+c$ for some $c\in T\subseteq C_\mathrm{x}$, and hence $w_1+C_\mathrm{x}=w_2+C_\mathrm{x}$.
    Thus $\phi_\mathrm{x}$ is well-defined, and the same argument applies to $\phi_\mathrm{z}$.

    Any permutation $\sigma\in\mathrm{PPerm}(\mathcal C)$ preserves $C_\mathrm{x}$ and $C_\mathrm{z}$ separately and induces well-defined logical actions on the $X$- and $Z$-logical spaces $\mathcal L_\mathrm{x}$ and $\mathcal L_\mathrm{z}$; let $\rho_\mathrm{x}$ and $\rho_\mathrm{z}$ denote the corresponding group representations, respectively. We write
    \begin{equation}
        \rho_\rmx:\mathrm{PPerm}(\mathcal{C})\to\GL(\mathcal L_\rmx),
        \qquad
        \rho_\rmz:\mathrm{PPerm}(\mathcal{C})\to\GL(\mathcal L_\rmz),
        \qquad 
        \rho_\mathrm{x}(\sigma)[x]_{\mathcal L_\mathrm{x}} = [xP_\sigma]_{\mathcal L_\mathrm{x}},
        \quad
        \rho_\mathrm{z}(\sigma)[z]_{\mathcal L_\mathrm{z}} = [zP_\sigma]_{\mathcal L_\mathrm{z}},
    \end{equation}
    where $P_\sigma\in\F_2^{n\times n}$ is the permutation matrix corresponding to $\sigma$.
    Every $\sigma\in\mathrm{PPerm}(\mathcal C)$ also preserves $W$ and $T$, so it induces
    \begin{equation}
        \rho_B:\mathrm{PPerm}(\mathcal C)\to\GL(B),
        \qquad
        \rho_B(\sigma)[w]_B=[wP_\sigma]_B.
    \end{equation}

    Observe now that the maps $\phi_\mathrm{x}$ and $\phi_\mathrm{z}$ are $\mathrm{PPerm}(\mathcal C)$-equivariant:
    \begin{equation}
        \phi_\mathrm{x}\circ \rho_B(\sigma)
        =
        \rho_\mathrm{x}(\sigma)\circ \phi_\mathrm{x},
        \qquad
        \phi_\mathrm{z}\circ \rho_B(\sigma)
        =
        \rho_\mathrm{z}(\sigma)\circ \phi_\mathrm{z},
    \end{equation}
    because $\phi_\mathrm{x}(\rho_B(\sigma)[w]_B)=\phi_\mathrm{x}([wP_\sigma]_B)=[wP_\sigma]_{\mathcal L_\mathrm{x}}=\rho_\mathrm{x}(\sigma)[w]_{\mathcal L_\mathrm{x}}=\rho_\mathrm{x}(\sigma)\phi_\mathrm{x}([w]_B)$, and likewise for $\phi_\mathrm{z}$.

    Consequently, $U_\rmx\coloneqq \im\phi_\rmx$ and $U_\rmz\coloneqq \im\phi_\rmz$ are invariant under the induced logical action of any $\sigma\in \mathrm{PPerm}(\mathcal{C})$: for every $[w]_{\cL_\rmx}\in\im\phi_\rmx$ and every $\sigma\in \mathrm{PPerm}(\mathcal{C})$, we have $\rho_\rmx(\sigma)[w]_{\cL_\rmx}\in \im \phi_\rmx$, and likewise for $U_\rmz$.
    We next determine the dimensions of these spaces.
    From the definitions of $U_\mathrm{x},U_\mathrm{z}$ and $\phi_\rmx,\phi_\rmz$, we have 
    \begin{equation}\begin{aligned}
        U_\mathrm{x}
        &=
        \frac{W+C_\mathrm{x}}{C_\mathrm{x}}
        \subseteq \mathcal L_\mathrm{x}, 
        \qquad
        U_\mathrm{z}
        =
        \frac{W+C_\mathrm{z}}{C_\mathrm{z}}
        \subseteq \mathcal L_\mathrm{z},
        \qquad
        \ker\phi_\mathrm{x}
        =
        \frac{W\cap C_\mathrm{x}}{T}
        =
        \frac{R_\mathrm{x}}{T},
        \qquad
        \ker\phi_\mathrm{z}
        =
        \frac{W\cap C_\mathrm{z}}{T}
        =
        \frac{R_\mathrm{z}}{T},
    \end{aligned}\end{equation}
    where we have noted
    \begin{equation}
        W\cap C_\mathrm{x}
        =(C_\mathrm{x}^{\perp}\cap C_\mathrm{z}^{\perp})\cap C_\mathrm{x}
        =C_\mathrm{x}\cap C_\mathrm{x}^{\perp}
        =R_\mathrm{x},
        \qquad
        W\cap C_\mathrm{z}=R_\mathrm{z},
    \end{equation}
    using $C_\mathrm{x}\subseteq C_\mathrm{z}^{\perp}$ and $C_\mathrm{z}\subseteq C_\mathrm{x}^{\perp}$.
    Because $\dim B=k$ and $m=\dim(R_\rmx/T)=\dim(R_\rmz/T)$, rank--nullity gives
    $\dim U_\mathrm{x}=\dim B-\dim\ker\phi_\rmx=k-m$ and similarly $\dim U_\mathrm{z}=k-m$.

    Now we show that $\Lambda_\rmx=U_\rmz^\perp$ is also invariant under all induced logical $X$-actions.
    Consider the logical pairing
    \begin{equation}
        \beta_{\mathrm{xz}}:\mathcal L_\mathrm{x}\times\mathcal L_\mathrm{z}\to\F_2,
        \qquad
        \beta_{\mathrm{xz}}([x]_{\cL_\rmx},[z]_{\cL_\rmz})\coloneqq xz^\top.
    \end{equation}
    This pairing is well-defined and nondegenerate.
    Indeed, changing $x$ by an element of $C_\mathrm{x}$ or $z$ by an element of $C_\mathrm{z}$ does not change the pairing by CSS orthogonality; and if $[x]_{\mathcal L_\mathrm{x}}$ pairs trivially with every element of $\mathcal L_\mathrm{z}$, then $x\in(C_\mathrm{x}^\perp)^\perp=C_\mathrm{x}$, with the analogous statement on the other side.
    Because physical permutations preserve the ordinary dot product, the actions $\rho_\rmx(\sigma)$ and $\rho_\rmz(\sigma)$ jointly preserve the logical pairing $\beta_{\mathrm{xz}}$:
    \begin{equation}
        \beta_{\mathrm{xz}}
        \left(
            \rho_\mathrm{x}(\sigma)[x]_{\mathcal L_\mathrm{x}},
            \rho_\mathrm{z}(\sigma)[z]_{\mathcal L_\mathrm{z}}
        \right)
        =
        \beta_{\mathrm{xz}}
        \left(
            [x]_{\mathcal L_\mathrm{x}},
            [z]_{\mathcal L_\mathrm{z}}
        \right).
    \end{equation}

    Now let $[x]_{\cL_\rmx}\in\Lambda_\rmx$ and $[z]_{\cL_\rmz}\in U_\rmz$.
    Since $U_\rmz$ is invariant and $\rho_\rmz(\sigma)$ is invertible, with inverse induced by $\sigma^{-1}$, there exists $[z']_{\cL_\rmz}\in U_\rmz$ such that
    $[z]_{\cL_\rmz}=\rho_\rmz(\sigma)[z']_{\cL_\rmz}$.
    Therefore
    $\beta_{\mathrm{xz}}\bigl(\rho_\rmx(\sigma)[x]_{\cL_\rmx},[z]_{\cL_\rmz}\bigr)
    =\beta_{\mathrm{xz}}\bigl(\rho_\rmx(\sigma)[x]_{\cL_\rmx},\rho_\rmz(\sigma)[z']_{\cL_\rmz}\bigr)
    =\beta_{\mathrm{xz}}([x]_{\cL_\rmx},[z']_{\cL_\rmz})=0$.
    Thus $\rho_\rmx(\sigma)\Lambda_\rmx\subseteq\Lambda_\rmx$.
    Applying the same argument to $\sigma^{-1}$ gives equality.

    Next, we show that $\dim\Lambda_\rmx=m$.
    Observe that
    \begin{equation}
        \Lambda_\rmx
        =
        U_\rmz^\perp
        =
        \frac{C_\mathrm{x}+R_\mathrm{z}}{C_\mathrm{x}}.
    \end{equation}
    Indeed, if $[x]_{\mathcal L_\mathrm{x}}\in U_\mathrm{z}^{\perp}$, then $x\in C_\mathrm{z}^{\perp}$ and $xw^\top=0$ for all $w\in W$, since every element of $U_\mathrm{z}$ has a representative in $W$.
    Hence $x\in W^\perp=C_\mathrm{x}+C_\mathrm{z}$.
    Write $x=a+b$, where $a\in C_\mathrm{x}$ and $b\in C_\mathrm{z}$.
    Since $x\in C_\mathrm{z}^{\perp}$ and $a\in C_\mathrm{x}\subseteq C_\mathrm{z}^{\perp}$, it follows that $b\in C_\mathrm{z}^{\perp}$.
    Thus $b\in C_\mathrm{z}\cap C_\mathrm{z}^{\perp}=R_\mathrm{z}$, so
    $[x]_{\mathcal L_\mathrm{x}}\in(C_\mathrm{x}+R_\mathrm{z})/C_\mathrm{x}$.
    The reverse inclusion is immediate because every class in $(C_\mathrm{x}+R_\mathrm{z})/C_\mathrm{x}$ has a representative in $R_\mathrm{z}$, and $R_\mathrm{z}$ pairs trivially with $W$.
    Moreover, since $R_\mathrm{z}\subseteq W$, we have $\Lambda_\mathrm{x}\subseteq U_\mathrm{x}$.
    We also have $C_\mathrm{x}\cap R_\mathrm{z}=T$.
    Indeed, $R_\mathrm{z}\subseteq C_\mathrm{z}$ gives $C_\mathrm{x}\cap R_\mathrm{z}\subseteq T$, while $T\subseteq R_\mathrm{z}$ by \cref{prop:psd_css_codes_balanced_space_properties}.
    Therefore $\dim\Lambda_\mathrm{x}=\dim(R_\mathrm{z}/T)=m$.

    Lastly, we prove the existence of a nondegenerate symmetric bilinear form $\overline\beta_\rmx$ on the quotient $U_\rmx/\Lambda_\rmx$.
    To do this, we first show that $\phi_\rmx$ induces an isomorphism
    $B/\rad(\beta_B)\cong U_\mathrm{x}/\Lambda_\mathrm{x}$; the nondegenerate bilinear form on $B/\rad(\beta_B)$ then naturally carries over to $U_\mathrm{x}/\Lambda_\mathrm{x}$.
    Since $U_\rmx=\phi_\rmx(B)$, it remains to show
    $\phi_\rmx^{-1}(\Lambda_\rmx)=\rad(\beta_B)$.
    Indeed, if $w\in W$ and $w+C_\mathrm{x}\in\Lambda_\mathrm{x}$, then $w=a+b$ for some $a\in C_\mathrm{x}$ and $b\in R_\mathrm{z}$.
    Since $w,b\in W$, we have $a=w+b\in W\cap C_\mathrm{x}=R_\mathrm{x}$, and hence $w\in R_\mathrm{x}+R_\mathrm{z}$.
    Thus $[w]_B\in\rad(\beta_B)$.
    Conversely, if $[w]_B\in\rad(\beta_B)$, then we may write $w=a+b$ for some $a\in R_\mathrm{x}$ and $b\in R_\mathrm{z}$.
    Hence
    $\phi_\mathrm{x}([w]_B)=w+C_\mathrm{x}=b+C_\mathrm{x}
    \in(C_\mathrm{x}+R_\mathrm{z})/C_\mathrm{x}
    =\Lambda_\mathrm{x}$.
    Thus $\phi_\mathrm{x}^{-1}(\Lambda_\mathrm{x})=\rad(\beta_B)$.

    Therefore $\phi_\rmx$ induces an isomorphism
    \begin{equation}
        \overline\phi_\rmx:
        B/\rad(\beta_B)
        \longrightarrow
        U_\rmx/\Lambda_\rmx,
        \qquad
        [w]_B+\rad(\beta_B)
        \longmapsto
        \phi_\rmx([w]_B)+\Lambda_\rmx.
    \end{equation}
    Since $\phi_\rmx$ is $\mathrm{PPerm}(\mathcal C)$-equivariant, so is $\overline\phi_\rmx$.

    As a final remark, every $\rho_B(\sigma)$ is a $\beta_B$-isometry, since
    $\beta_B\bigl(\rho_B(\sigma)[w]_B,\rho_B(\sigma)[w']_B\bigr)
    =(wP_\sigma)(w'P_\sigma)^\top=ww'^\top$.
    It therefore preserves $\rad(\beta_B)$ and induces an isometry of $B/\rad(\beta_B)$.
    We can then transport the quotient form through the equivariant isomorphism $\overline\phi_\rmx$ to obtain the nondegenerate symmetric form $\overline\beta_\rmx$ on $U_\rmx/\Lambda_\rmx$ as stated in \cref{eq:induced_nondegenerate_bilinear_form}.
    The induced action of every $\rho_\rmx(\sigma)$ on $U_\rmx/\Lambda_\rmx$ preserves $\overline\beta_\rmx$.
\end{proof}

\begin{lemma}
    [Characteristic-polynomial parity constraint for permutation logical actions on PSD CSS codes]
    \label{lem:psd_css_codes_perm_charpoly_parity}
    Let $\mathcal C$ be a PSD CSS code, $U_\pi H^{\otimes n}$ be a self-duality automorphism of $\mathcal C$ for $\pi\in S_n$, and let $W\coloneqq(C_\mathrm{x}+C_\mathrm{z})^\perp$, $T\coloneqq C_\mathrm{x}\cap C_\mathrm{z}$, and the balanced logical space $B\coloneqq W/T$ be as in \cref{prop:psd_css_codes_balanced_space_properties}.
    For $\tau\in S_n$, let $P_\tau\in\F_2^{n\times n}$ denote its permutation matrix.
    Let $E: B\to B$ denote the invertible linear map induced by $\pi$, with $E([w]_B)\coloneqq[wP_\pi]_B$.
    Let $\rho_B$ denote the action of $\mathrm{PPerm}(\mathcal C)$ on $B$ induced by physical permutations; thus, for every $\sigma\in\mathrm{PPerm}(\mathcal C)$,
    $\rho_B(\sigma):B\to B$ is the invertible linear map given by
    $\rho_B(\sigma)[w]_B\coloneqq[wP_\sigma]_B$.
    Then, for every $\sigma\in\mathrm{PPerm}(\mathcal C)$ and every irreducible self-reciprocal polynomial $f\in\F_2[x]$,
    \begin{equation}
        \nu_f(\chi_{P_{\pi\sigma}})
        \equiv
        \nu_f(\chi_{E\rho_B(\sigma)})
        \pmod 2.
        \label{eq:psd_css_codes_perm_charpoly_parity_1}
    \end{equation}
    Consequently, if $f$ and $f'$ are self-reciprocal irreducible polynomials of the same odd order, then
    \begin{equation}
        \nu_f(\chi_{E\rho_B(\sigma)})
        \equiv
        \nu_{f'}(\chi_{E\rho_B(\sigma)})
        \pmod 2.
        \label{eq:psd_css_codes_perm_charpoly_parity_2}
    \end{equation}
\end{lemma}

\begin{proof}
    The permutation $\pi$ maps $C_\mathrm{x}\leftrightarrow C_\mathrm{z}$, and hence preserves $C_\mathrm{x}+C_\mathrm{z}$, $W$, and $T$.
    Therefore the linear map $E$ on $B$ is well-defined.
    For $\sigma\in\mathrm{PPerm}(\mathcal C)$, the permutation $\pi\sigma$ also maps $C_\mathrm{x}\leftrightarrow C_\mathrm{z}$.
    Its induced action on $B$ is $E\rho_B(\sigma)$, so
    $\smash{\chi_{P_{\pi\sigma},B}=\chi_{E\rho_B(\sigma)}}$.
    Since $T\subseteq W\subseteq\F_2^n$ are invariant under $P_{\pi\sigma}$, the characteristic polynomial of $P_{\pi\sigma}$ factors across these invariant subspaces and quotients as (see Property~\cref{item:characteristic_poly_1} of \cref{prop:characteristic_poly})
    \begin{equation}
        \chi_{P_{\pi\sigma}}
        =
        \chi_{P_{\pi\sigma},T}\,
        \chi_{P_{\pi\sigma},B}\,
        \chi_{P_{\pi\sigma},\F_2^n/W}.
        \label{eq:psd_css_codes_perm_charpoly_parity_3}
    \end{equation}
    The induced action of $P_{\pi\sigma}$ on $\F_2^n/W$ is contragredient to the induced action on $W^\perp$, so by Property~\cref{item:characteristic_poly_3} of \cref{prop:characteristic_poly},
    \begin{equation}
        \chi_{P_{\pi\sigma},\F_2^n/W}=\chi_{P_{\pi\sigma},W^\perp}^*.
    \end{equation}
    
    Moreover, $T\subseteq W^\perp=C_\mathrm{x}+C_\mathrm{z}$, and both spaces are preserved by $P_{\pi\sigma}$.
    Thus
    \begin{equation}
        \chi_{P_{\pi\sigma},W^\perp}
        =\chi_{P_{\pi\sigma},T}\chi_{P_{\pi\sigma},Q},
    \end{equation}
    where $Q \coloneqq W^\perp/T \cong (C_\mathrm{x}/T) \oplus (C_\mathrm{z}/T)$.
    Combining these identities gives
    \begin{equation}
        \chi_{P_{\pi\sigma}}
        =
        \chi_{P_{\pi\sigma},T}\,
        \chi_{E\rho_B(\sigma)}\,
        \chi_{P_{\pi\sigma},Q}^*\,
        \chi_{P_{\pi\sigma},T}^*.
        \label{eq:psd_css_codes_perm_charpoly_parity_4}
    \end{equation}

    Since $P_{\pi\sigma}$ exchanges $C_\mathrm{x}$ and $C_\mathrm{z}$, its induced action on
    $Q\cong(C_\mathrm{x}/T)\oplus(C_\mathrm{z}/T)$ has a $2 \times 2$ block off-diagonal form with off-diagonal blocks $C$ and $D$. 
    Therefore $\chi_{P_{\pi\sigma},Q}(x)=\det(x^2I-CD)=h(x^2)$, where $h(x)\coloneqq\det(xI-CD)$.
    Since $h(x^2)=h(x)^2$ over $\F_2$, every irreducible factor occurs in $\chi_{P_{\pi\sigma},Q}$ with even multiplicity.
    Also, since $f$ is self-reciprocal, by \cref{def:linear_algebra_reciprocal_polynomials}, $\nu_f(\chi_{P_{\pi\sigma},T}) =\nu_f(\chi_{P_{\pi\sigma},T}^*)$. 
    Then \cref{eq:psd_css_codes_perm_charpoly_parity_1} follows from \cref{eq:psd_css_codes_perm_charpoly_parity_4}.
    On the other hand, $P_{\pi\sigma}$ is a permutation of coordinates.
    If $f$ and $f'$ are self-reciprocal irreducible polynomials of the same odd order, then by Property~\cref{item:characteristic_poly_2} of  \cref{prop:characteristic_poly},
    $\nu_f(\chi_{P_{\pi\sigma}})=\nu_{f'}(\chi_{P_{\pi\sigma}})$.
    Applying \cref{eq:psd_css_codes_perm_charpoly_parity_1} to $f$ and $f'$ gives \cref{eq:psd_css_codes_perm_charpoly_parity_2}.
\end{proof}

\begin{lemma}
    [Order-$17$ elements in $\Sp(k,\F_2)$ and $O(k,\F_2)$ groups]
    \label{lem:order_17_witness_binary_classical_groups}
    The cyclotomic polynomial $\Phi_{17}(x)$ splits over $\F_2$ as
    $\Phi_{17}(x)=f_1(x)f_2(x)$, where
    \begin{equation}
        f_1(x)=x^8+x^5+x^4+x^3+1,
        \qquad
        f_2(x)=x^8+x^7+x^6+x^4+x^2+x+1.
    \end{equation}
    Both $f_1$ and $f_2$ are self-reciprocal irreducible polynomials of order $17$.
    Moreover:
    \begin{enumerate}[noitemsep]
        \item For every even $k\ge8$, $\Sp(k,\F_2)$ contains an element with characteristic polynomial $f_1(x)(x+1)^{k-8}$.
        \item For every $k\ge9$, $O(k,\F_2)$ contains an element with characteristic polynomial $f_1(x)(x+1)^{k-8}$.
    \end{enumerate}
\end{lemma}

\begin{proof}
    A direct multiplication gives $\Phi_{17}(x)=f_1(x)f_2(x)$.
    Since $\ord_{17}(2)=8$, \cref{fact:linear_algebra_cyclotomic_factors_frobenius_orbits} implies that every irreducible factor of $\Phi_{17}(x)$ over $\F_2$ has degree $8$ and order $17$.
    Since $f_1$ and $f_2$ both have degree $8$, they are therefore irreducible and have order $17$.
    Moreover, since $\ord_{17}(2)=8$ is even, \cref{fact:polynomials_cyclotomic_reciprocal_factors} implies that both $f_1$ and $f_2$ are self-reciprocal.

    We first construct the required element of $\Sp(k,\F_2)$ with characteristic polynomial $f_1(x)(x+1)^{k-8}$.
    Let $\alpha$ be a root of $f_1$.
    Since $f_1$ has degree $8$, the field extension of $\F_2$ by adjoining $\alpha$ gives $L\coloneqq\F_2(\alpha)=\F_{2^8}$.
    We also have $\alpha^{16}=\alpha^{-1}$ since $f_1$ has order $17$.
    View $L$ as an eight-dimensional vector space over $\F_2$, and let $M_\alpha:L\to L$ be the linear operator $x\mapsto x\alpha$.
    We show that $\chi_{M_\alpha}(x)=f_1(x)$.
    
    Let $\overline{z}=z^{16}$ be the Frobenius automorphism of $L/\F_{16}$; note that $\overline{\overline{z}}=z$.
    Define $\gamma(u,v)=\tr_{L/\F_2}(u\overline{v})$.
    This is a nondegenerate alternating form: it is nondegenerate because the trace pairing is nondegenerate and $v\mapsto\overline{v}$ is invertible; it is alternating because $(u\overline{u})^{16}=u\overline{u}$, so $u\overline{u}\in\F_{16}$, and
    $\tr_{L/\F_2}(u\overline{u})
    =\tr_{\F_{16}/\F_2}(\tr_{L/\F_{16}}(u\overline{u}))=0$,
    since $\tr_{L/\F_{16}}(u\overline{u})=u\overline{u}+(u\overline{u})^{16}=0$.
    Thus $L$ is a symplectic vector space equipped with $\gamma$.
    Since $M_\alpha$ is multiplication by $\alpha$ and $\alpha\overline{\alpha}=\alpha^{17}=1$, we have
    $\gamma(M_\alpha u,M_\alpha v)
    =\tr_{L/\F_2}(\alpha u\overline{\alpha v})
    =\tr_{L/\F_2}(\alpha\overline{\alpha}u\overline{v})
    =\gamma(u,v)$.
    For every $p(x)\in\F_2[x]$, the operator $p(M_\alpha)$ is multiplication by $p(\alpha)$.
    Hence $p(M_\alpha)=0$ if and only if $p(\alpha)=0$.
    Therefore the minimal polynomial of $M_\alpha$ equals the minimal polynomial of $\alpha$, namely $f_1$.
    Since $\deg f_1=8=\dim_{\F_2}L$, the characteristic polynomial of $M_\alpha$ is also $f_1$.
    
    For every even $k\ge8$, $k-8$ is also even.
    Take the direct sum with the identity on a $(k-8)$-dimensional symplectic space: $N\coloneqq M_\alpha\oplus I_{k-8}$.
    Then $N\in\Sp(k,\F_2)$ and
    $\chi_N(x)=f_1(x)\chi_{I_{k-8}}(x)
    =f_1(x)(x-1)^{k-8}
    =f_1(x)(x+1)^{k-8}$.

    Lastly, we construct the required element of $O(k,\F_2)$.
    Consider the standard dot product on $\F_2^9$ and let
    $\mathbf{1}=(1,\ldots,1)^\top$ and $U\coloneqq\mathbf{1}^\perp$.
    Since $\mathbf{1}^\top\mathbf{1}=1$, we have
    $\F_2^9=\langle\mathbf{1}\rangle\perp U$.
    The restriction of the dot product to $U$ is nondegenerate alternating, and $\dim U=8$.
    Hence $U$ is isometric to the symplectic space $(L,\gamma)$ constructed above.
    Transport $M_\alpha$ to an isometry of $U$ and extend it by the identity on $\langle\mathbf{1}\rangle$.
    This gives an element $A_9\in O(9,\F_2)$ with
    $\chi_{A_9}(x)=f_1(x)(x+1)$.
    For every $k\ge9$, set $A_k\coloneqq A_9\oplus I_{k-9}$.
    Then $A_k\in O(k,\F_2)$ and
    $\chi_{A_k}(x)=f_1(x)(x+1)^{k-8}$.
\end{proof}

\subsubsection{Permutation logical groups on self-dual codes}

\begin{theorem}
    [Permutation logical groups on PSD CSS codes]
    \label{thm:css_codes_perm_logical_group_psd}
    Let $\mathcal{C}$ be an $\db{n, k}$ PSD CSS code defined by
    stabilizer generator matrices $H_\mathrm{x},H_\mathrm{z}\in \smash{\mathbb F_2^{(n-k)/2\times n}}$, and $\mathcal L$ be a CSS logical basis of $\mathcal C$.
    Let $C_\mathrm{x}\coloneqq \rs(H_\mathrm{x})$, $C_\mathrm{z}\coloneqq \rs(H_\mathrm{z})$ be the $X$- and $Z$-stabilizer spaces, and $\cL_\rmx\coloneqq C_\rmz^\perp/C_\rmx, \cL_\rmz\coloneqq C_\rmx^\perp/C_\rmz$ be the $X$- and $Z$-logical spaces, of $\mathcal{C}$, respectively.
    By \cref{thm:css_codes_perm_logical_group}, $\mathrm{Perm}_{\mathcal L}(\mathcal C) \cong \mathcal A_{\mathcal L}(\mathcal C) \leq \GL(k,\mathbb F_2)$.
    Define
    \begin{equation}
        R_\mathrm{x}\coloneqq C_\mathrm{x}\cap C_\mathrm{x}^{\perp},
        \qquad
        R_\mathrm{z}\coloneqq C_\mathrm{z}\cap C_\mathrm{z}^{\perp},
        \qquad
        T\coloneqq C_\mathrm{x}\cap C_\mathrm{z}=R_\mathrm{x}\cap R_\mathrm{z},
        \qquad
        m\coloneqq \dim(R_\mathrm{x}/T)=\dim(R_\mathrm{z}/T),
    \end{equation}
    as in \cref{prop:psd_css_codes_balanced_space_properties}.
    Then the following bounds hold.
    \begin{itemize}
        \item If $m>0$, equivalently $T\subsetneq R_\mathrm{x}$ and $T\subsetneq R_\mathrm{z}$, then $\mathcal A_{\mathcal L}(\mathcal C)$ preserves a flag of
        dimensions $m\subseteq k-m$ in $\mathbb F_2^k$. More precisely:    
        \begin{itemize}
            \item If $\mathbf 1\in C_\mathrm{x}+C_\mathrm{z}$,
            then $n$, $k$ and $k-2m$ are even and
            \begin{equation}
                \mathcal A_{\mathcal L}(\mathcal C)
                \preceq_{\GL(k,\mathbb F_2)}
                \mathcal F_m^{\Sp}(k,\mathbb F_2).
                \label{eq:css_codes_perm_logical_group_psd_flagged_symplectic}
            \end{equation}
            \item If $\mathbf 1\notin C_\mathrm{x}+C_\mathrm{z}$, then $k-2m>0$ and
            \begin{equation}
                \mathcal A_{\mathcal L}(\mathcal C)
                \preceq_{\GL(k,\mathbb F_2)}
                \mathcal F_m^O(k,\mathbb F_2).
                \label{eq:css_codes_perm_logical_group_psd_flagged_orthogonal}
            \end{equation}
        \end{itemize}
        
        \item If $m=0$, equivalently $T=R_\mathrm{x}=R_\mathrm{z}$, the bounds reduce to the nondegenerate symplectic or orthogonal cases:
        \begin{itemize}
            \item If $\mathbf 1\in C_\mathrm{x}+C_\mathrm{z}$, then $n$ and $k$ are even and
            \begin{equation}
                \mathcal A_{\mathcal L}(\mathcal C)
                \preceq_{\GL(k,\mathbb F_2)}
                \Sp(k,\mathbb F_2).
                \label{eq:css_codes_perm_logical_group_psd_nondeg_symplectic}
            \end{equation}
            Equality is achievable for $k \le 4$ but not possible for $k\ge 8$.

            \item If $\mathbf 1\notin C_\mathrm{x}+C_\mathrm{z}$, then
            \begin{equation}
                \mathcal A_{\mathcal L}(\mathcal C)
                \preceq_{\GL(k,\mathbb F_2)}
                O(k,\mathbb F_2).
                \label{eq:css_codes_perm_logical_group_psd_nondeg_orthogonal}
            \end{equation}
            Equality is achievable for $k \le 5$ but not possible for $k\ge 9$.
        \end{itemize}
    \end{itemize}

    In particular, these bounds imply
    \begin{equation}
        \abs{\mathrm{Perm}_\mathcal{L}(\mathcal{C})}
        <
        \frac{1}{2^{\floor{k^2/4}}} 
        \cdot 
        \abs{\GL(k, \mathbb{F}_2)},
        \label{eq:css_codes_perm_logical_group_psd_order_bound}
    \end{equation}
    for $k \geq 3$, and $\abs{\mathrm{Perm}_\mathcal{L}(\mathcal{C})} \leq \abs{\GL(k, \mathbb{F}_2)}$ for $k \in \{1, 2\}$.
\end{theorem}

\begin{proof}
    By \cref{lem:psd_css_codes_balanced_logical_flag}, every induced logical $X$-action preserves a flag $\Lambda_\mathrm{x}\subseteq U_\mathrm{x}\subseteq\mathcal L_\mathrm{x}$ with $\dim\Lambda_\mathrm{x}=m$ and $\dim U_\mathrm{x}=k-m$, and the middle quotient $U_\mathrm{x}/\Lambda_\mathrm{x}$ carries a nondegenerate symmetric bilinear form preserved by every permutation logical action.
    By \cref{prop:psd_css_codes_balanced_space_properties}, if $\mathbf 1\in C_\mathrm{x}+C_\mathrm{z}$, this middle quotient form is alternating and $n$, $k$, and $k-2m$ are even.
    Hence the logical permutation group is contained, up to $\GL(k,\mathbb F_2)$ conjugacy, in $\mathcal F_m^{\Sp}(k,\mathbb F_2)$ as per \cref{def:flagged_isometry_groups}.
    If $\mathbf 1\notin C_\mathrm{x}+C_\mathrm{z}$, the middle quotient form is nondegenerate nonalternating, so the logical permutation group is contained, up to $\GL(k,\mathbb F_2)$ conjugacy, in $\mathcal F_m^O(k,\mathbb F_2)$.
    In this nonalternating branch, necessarily $k-2m>0$, since the unique bilinear form on a zero-dimensional space is alternating.
    
    Lastly, when $m=0$, the flag is trivial: $\Lambda_\mathrm{x}=\{0\}$ and $U_\mathrm{x}=\mathcal L_\mathrm{x}$. 
    Therefore the middle quotient is the full logical $X$-space, and the two bounds reduce to
    $\Sp(k,\mathbb F_2)$ in the alternating case and $O(k,\mathbb F_2)$ in the nonalternating case.
    That $n$ and $k$ are even in the former case follows from
    \cref{prop:psd_css_codes_balanced_space_properties}.

    Attainability at small $k$ in the $m=0$ branches is given by \cref{cons:css_codes_maximum_perm_logical_group_ssd}. 
    We now prove the claimed non-attainment statements at larger $k$. 
    Fix a self-duality automorphism $U_\pi H^{\otimes n}$ of $\mathcal C$.
    Let $B\coloneqq(C_\mathrm{x}+C_\mathrm{z})^\perp/(C_\mathrm{x}\cap C_\mathrm{z})$ be the balanced logical space with symmetric bilinear form $\beta_B$ as in \cref{prop:psd_css_codes_balanced_space_properties}.
    Let $\rho_B$ denote the action of $\mathrm{PPerm}(\mathcal C)$ on $B$ induced by physical permutations, and let $E:B\to B$ be the invertible linear map induced by $\pi$, as in \cref{lem:psd_css_codes_perm_charpoly_parity}.
    Since $m=0$, \cref{prop:psd_css_codes_balanced_space_properties} gives $\rad(\beta_B)=\{0\}$, while \cref{lem:psd_css_codes_balanced_logical_flag} gives a $\mathrm{PPerm}(\mathcal C)$-equivariant isomorphism $\phi_\mathrm{x}:B\to\mathcal L_\mathrm{x}$.
    Hence $\rho_B(\mathrm{PPerm}(\mathcal C))$ is conjugate to $\mathcal A_{\mathcal L}(\mathcal C)$.
    The two branches are:
    \begin{itemize}
        \item $\mathbf 1\in C_\mathrm{x}+C_\mathrm{z}$. 
        Then $\beta_B$ is nondegenerate alternating.
        Since the physical permutation $\pi$ preserves the ordinary dot product, $E$ preserves $\beta_B$.
        After choosing a symplectic basis of $B$, $E$ is therefore represented by an element of $\Sp(k,\F_2)$.
        Assume for contradiction that the $\Sp(k,\F_2)$ bound is saturated.
        Then, after the same basis choice, $\rho_B(\mathrm{PPerm}(\mathcal C))=\Sp(k,\F_2)$.
        Since $E\in\Sp(k,\F_2)$, it follows that $E\rho_B(\mathrm{PPerm}(\mathcal C))=\Sp(k,\F_2)$.
        For $k\ge8$, \cref{lem:order_17_witness_binary_classical_groups} gives an element of $\Sp(k,\F_2)$ with characteristic polynomial $f_1(x)(x+1)^{k-8}$, where $f_1$ and $f_2$ are self-reciprocal irreducible polynomials of order $17$.
        For this element, $\nu_{f_1}=1$ and $\nu_{f_2}=0$, contradicting \cref{eq:psd_css_codes_perm_charpoly_parity_2}.
        Therefore the $\Sp(k,\F_2)$ bound cannot be saturated for $k\ge8$.

        \item $\mathbf 1\notin C_\mathrm{x}+C_\mathrm{z}$.
        Then $\beta_B$ is nondegenerate nonalternating.
        Since $E$ preserves $\beta_B$, after choosing a basis of $B$ in which $\beta_B$ has matrix $I$, $E$ is represented by an element of $O(k,\F_2)$.
        Assume for contradiction that the $O(k,\F_2)$ bound is saturated.
        Then, after the same basis choice, $\rho_B(\mathrm{PPerm}(\mathcal C))=O(k,\F_2)$.
        Since $E\in O(k,\F_2)$ in this basis, it follows that $E\rho_B(\mathrm{PPerm}(\mathcal C))=O(k,\F_2)$.
        For $k\ge9$, \cref{lem:order_17_witness_binary_classical_groups} gives an element of $O(k,\F_2)$ with characteristic polynomial $f_1(x)(x+1)^{k-8}$.
        Again $\nu_{f_1}=1$ and $\nu_{f_2}=0$, contradicting \cref{eq:psd_css_codes_perm_charpoly_parity_2}.
        Therefore the $O(k,\F_2)$ bound cannot be saturated for $k\ge9$.
    \end{itemize}

    Finally, we obtain \cref{eq:css_codes_perm_logical_group_psd_order_bound} by comparing the group orders in the four branches [\cref{eq:css_codes_perm_logical_group_psd_flagged_symplectic,eq:css_codes_perm_logical_group_psd_flagged_orthogonal,eq:css_codes_perm_logical_group_psd_nondeg_symplectic,eq:css_codes_perm_logical_group_psd_nondeg_orthogonal}]. 
    First suppose $m>0$, and set $s\coloneqq k-2m$.
    We use the group order formulae for $\mathcal F_m^{\Sp}(k,\mathbb{F}_2)$ and $\mathcal F_m^{O}(k,\mathbb{F}_2)$, together with $\abs{O(s,\F_2)}\leq2^{s(s-1)/2}$, from \cref{rem:flagged_isometry_group_orders}, and the bounds $\abs{\GL(m,\F_2)}\leq 2^{m^2-1}$ for $m\ge1$ and $\abs{\Sp(s,\F_2)}\leq2^{s(s+1)/2}$ from \cref{fact:group_theory_relevant_group_orders}, to obtain
    \begin{equation}
        \abs{\mathcal F_m^{\Sp}(k,\mathbb F_2)}
        \leq
        2^{3m^2+2ms+s(s+1)/2-2}
        \leq
        2^{k^2-\floor{k^2/4}-2},
        \qquad
        \abs{\mathcal F_m^O(k,\mathbb F_2)}
        \leq
        2^{3m^2+2ms+s(s-1)/2-2}
        \leq
        2^{k^2-\floor{k^2/4}-2},
    \end{equation}
    where we have noted $k=2m+s$ and $\floor{k^2/4}=m^2+ms+\floor{s^2/4}$.
    
    Now suppose $m=0$ and $k\ge3$.
    In the $\Sp(k,\mathbb F_2)$ branch, $k$ is even and hence $k\ge4$.
    By \cref{eq:group_order_symplectic_upper_bound},
    $\abs{\Sp(k,\F_2)}<2^{k(k+1)/2}\leq2^{k^2-k^2/4-2}$.
    In the $O(k,\mathbb F_2)$ branch,
    $\abs{O(k,\F_2)}\leq2^{k(k-1)/2}<2^{k^2-\floor{k^2/4}-2}$.
    
    Now recalling $2^{k^2-2}<\abs{\GL(k,\mathbb F_2)}$ from \cref{eq:group_order_general_linear_lower_and_upper_bound}, we obtain \cref{eq:css_codes_perm_logical_group_psd_order_bound} as claimed.
    For $k\in\{1,2\}$, the general containment $\mathcal A_{\mathcal L}(\mathcal C)\leq\GL(k,\F_2)$ from \cref{thm:css_codes_perm_logical_group} gives the stated bound; this is tight as $k=2$ PSD phantom codes exist (see \cref{cons:css_codes_maximum_perm_logical_group_ssd}), while for $k=1$ both groups are trivial.
\end{proof}

\begin{theorem}
    [Permutation logical groups on SSD CSS codes]
    \label{thm:css_codes_perm_logical_group_ssd}
    Let $\mathcal{C}$ be an $\db{n, k, d}$ SSD CSS code defined by
    stabilizer generator matrix $H \coloneqq H_\mathrm{x} = H_\mathrm{z} \in \smash{\mathbb F_2^{(n-k)/2\times n}}$.
    Let $C \coloneqq \rs(H)$ be the binary stabilizer space, and $\mathcal L$ be a CSS logical basis of $\mathcal C$. 
    By \cref{thm:css_codes_perm_logical_group}, $\mathrm{Perm}_{\mathcal L}(\mathcal C) \cong \mathcal A_{\mathcal L}(\mathcal C) \leq \GL(k,\mathbb F_2)$.
    The following bounds hold.
    \begin{itemize}
        \item If $\mathbf 1\in C$, then $n$, $k$ and $d$ are even and
        \begin{equation}
            \mathcal A_{\mathcal L}(\mathcal C)
            \preceq_{\GL(k,\mathbb F_2)}
            \Sp(k,\mathbb F_2).
        \end{equation}
        Equality is achievable for $k \le 4$ but not possible for $k\ge 8$.

        \item If $\mathbf 1\notin C$, then
        \begin{equation}
            \mathcal A_{\mathcal L}(\mathcal C)
            \preceq_{\GL(k,\mathbb F_2)}
            O(k,\mathbb F_2)
        \end{equation}
        Equality is achievable for $k \le 5$ but not possible for $k\ge 9$.
    \end{itemize}

    In particular, these group bounds imply
    \begin{equation}
        \abs{\mathrm{Perm}_\mathcal{L}(\mathcal{C})}
        <
        \frac{12}{5} 
        \cdot 
        \frac{1}{2^{k(k-1)/2}} 
        \cdot
        \abs{\GL(k, \mathbb{F}_2)}.
        \label{eq:css_codes_perm_logical_group_ssd_order_bound}
    \end{equation}
\end{theorem}

\begin{proof}
    SSD codes are a subset of PSD codes, so we invoke \cref{thm:css_codes_perm_logical_group_psd}. 
    Here $C \coloneqq C_\mathrm{x} = C_\mathrm{z}$, so $R_\mathrm{x} = R_\mathrm{z} = C$ and $T = C$. 
    Then $m \coloneqq \dim(R_\mathrm{x}/T) = \dim(R_\mathrm{z}/T) = \dim(C/C) = 0$. 
    That is, only the two cases of \cref{eq:css_codes_perm_logical_group_psd_nondeg_symplectic,eq:css_codes_perm_logical_group_psd_nondeg_orthogonal} can arise. 
    In restating those cases, we simplify $C_\mathrm{x} + C_\mathrm{z} = C$. 
    When $\vb{1} \in C$, the code must have even distance by \cref{prop:strictly_self_dual_css_codes_properties}.
    Attainability of the $\Sp(k,\mathbb F_2)$ and $O(k,\mathbb F_2)$ logical groups at small $k$ is given by \cref{cons:css_codes_maximum_perm_logical_group_ssd}.

    Finally, we prove \cref{eq:css_codes_perm_logical_group_ssd_order_bound}.
    Suppose first that $\mathbf 1\in C$.
    Then $k$ is even; write $k=2r$.
    From the group order formulae in \cref{fact:group_theory_relevant_group_orders},
    \begin{equation}
        2^{k(k-1)/2}
        \cdot
        \frac{\abs{\Sp(k,\F_2)}}{\abs{\GL(k,\F_2)}}
        =
        \prod_{i=1}^r
        \frac{1}{1-2^{-(2i-1)}}
        <
        2
        \left(
            1-\sum_{i=2}^{\infty}2^{-(2i-1)}
        \right)^{-1}
        = 
        \frac{12}{5}.
    \end{equation}
    Now suppose that $\mathbf 1\notin C$.
    Using $\abs{O(k,\F_2)}\leq2^{k(k-1)/2}$, and $2^{k^2-2}<\abs{\GL(k,\F_2)}$ from \cref{fact:group_theory_relevant_group_orders}, we obtain
    \begin{equation}
        2^{k(k-1)/2}
        \cdot
        \frac{\abs{O(k,\F_2)}}{\abs{\GL(k,\F_2)}}
        <
        2^{2-k}
        <
        \frac{12}{5}.
    \end{equation}
    Thus \cref{eq:css_codes_perm_logical_group_ssd_order_bound} holds in both branches.
\end{proof}

\begin{corollary}
    [No $k \geq 3$ PSD CSS phantom codes]
    \label{rem:no_self_dual_phantom_codes_k_geq_3}
    For $m>0$, the flagged isometry groups
    $\mathcal F_m^{\Sp}(k,\mathbb F_2)$ and
    $\mathcal F_m^O(k,\mathbb F_2)$ are proper subgroups of
    $\GL(k,\mathbb F_2)$; and $O(k,\mathbb F_2)<\GL(k,\mathbb F_2)$ for all $k\geq3$ and
    $\Sp(k,\mathbb F_2)<\GL(k,\mathbb F_2)$ for all even $k\geq4$.
    Therefore, \Cref{thm:css_codes_perm_logical_group} implies that no PSD CSS phantom codes can exist at $k \geq 3$. As phantom codes cannot be decomposable, except the potential presence of trivial components encoding no logical qubits, an equivalent statement using \cref{corr:css_codes_auto_logical_hadamards_only_on_self_dual_codes} is that CSS phantom codes at $k \geq 3$ cannot host automorphism logical $H^{\otimes k}$ gates---and addressable automorphism logical $H$ gates are not possible by \cref{corr:css_indecomp_no_add_h_sh_hs}---in any CSS logical basis.
    This answers the question posed in Ref.~\cite[Footnote 9]{koh2026entangling}.
\end{corollary}

\clearpage

\subsection{Transversal gates}
\label{app:css_codes/trans}

\begin{lemma}
    [Uniformity of physical Cliffords in transversal gates on indecomposable SSD CSS codes]
    \label{lem:trans_exchange_uniformity_indecomp_css_codes}
    Suppose $\overline{W} = \bigotimes_{i=1}^n V_i$ is a transversal gate on an indecomposable SSD CSS code. Then all $V_i$ are projectively equal. That is, $\overline{W} \in \{I^{\otimes n}, S^{\otimes n}, (HSH)^{\otimes n}, H^{\otimes n}, (HS)^{\otimes n}, (SH)^{\otimes n} \}$ up to Paulis and global phases.
\end{lemma}

\begin{proof}
    We reference the proof of \cref{lem:css_codes_auto_physical_splitting_property}, specializing to the context of transversal gates by setting the permutations $\pi = \mathrm{id}$. As $\mathcal{C}$ is SSD, we have $C_\mathrm{x} = C_\mathrm{z} \eqqcolon C$. First consider the case of $T = \varnothing$, where $p = s = \vb{1}$. Then from \cref{eq:css_indecomp_no_add_h_forward_x2,eq:css_indecomp_no_add_h_forward_z1}, we have
    \begin{equation}\begin{split}
        \forall (a, 0) \in C \oplus C
        : \,\,
        (a, q \odot a) \in C \oplus C,
        \qquad\quad
        \forall (0, b) \in C \oplus C
        : \,\,
        (r \odot b, b) \in C \oplus C.
    \end{split}\end{equation}
    Because $(a, 0), (0, b) \in C \oplus C$, we have $(0, q \odot a), (r \odot b, 0) \in C \oplus C$. That is, $(0, q \odot C) \subseteq C \oplus C$ and $(r \odot C, 0) \subseteq C \oplus C$, or equivalently stated, $\Pi_{\supp(q)} C \subseteq C$ and $\Pi_{\supp(r)} C \subseteq C$, where $\Pi_{\supp(\mu)}(\cdot)$ denotes projection onto the support of a vector $\mu$.
    Let $Q \coloneqq \supp(q)$. Since $\Pi_Q C \subseteq C$ and $C$ is linear, for every $c \in C$ we also have $\Pi_{Q^c}(c) = c + \Pi_Q(c) \in C$. Hence both coordinate projections associated with $Q \sqcup Q^c = [n]$ preserve the full binary stabilizer space:
    \begin{equation}
        \pi_Q(C \oplus C)
        =
        \Pi_Q C \oplus \Pi_Q C
        \subseteq C \oplus C,
        \qquad\quad
        \pi_{Q^c}(C \oplus C)
        =
        \Pi_{Q^c} C \oplus \Pi_{Q^c} C
        \subseteq C \oplus C.
    \end{equation}
    Therefore, by \cref{prop:code_decomposability}, if $\varnothing \neq Q \neq [n]$, then the code is decomposable. Since the code is indecomposable, $\supp(q)$ must be either $\varnothing$ or $[n]$. The same argument gives $\supp(r) \in \{\varnothing,[n]\}$. Moreover, $q_i r_i = 0$ for every $i \in [n]$, so $\supp(q)$ and $\supp(r)$ cannot both equal $[n]$. Thus the local Cliffords on the physical qubits must be uniform, and $I^{\otimes n}, S^{\otimes n}, (HSH)^{\otimes n}$ are the possible operators.
    
    Now consider the case of $T = [n]$, where $q = r = \vb{1}$. From \cref{eq:css_indecomp_no_add_h_forward_x1,eq:css_indecomp_no_add_h_forward_z2}, we have
    \begin{equation}\begin{split}
        \forall (a, 0) \in C \oplus C
        : \,\,
        (p \odot a, a) \in C \oplus C, 
        \qquad\quad
        \forall (0, b) \in C \oplus C &:
        : \,\,
        (b, s \odot b) \in C \oplus C.
    \end{split}\end{equation}
    So $(p \odot C, 0), (0, s \odot C) \subseteq C \oplus C$, or equivalently, $\Pi_{\supp(p)} C \subseteq C$ and $\Pi_{\supp(s)} C \subseteq C$. Let $P \coloneqq \supp(p)$. Since $\Pi_P C \subseteq C$ and $C$ is linear, for every $c \in C$ we also have $\Pi_{P^c}(c) = c + \Pi_P(c) \in C$. Hence both coordinate projections associated with $P \sqcup P^c = [n]$ preserve the full binary stabilizer space:
    \begin{align}
        \pi_P(C \oplus C)
        =
        \Pi_P C \oplus \Pi_P C
        \subseteq C \oplus C,
        \qquad\quad
        \pi_{P^c}(C \oplus C)
        =
        \Pi_{P^c} C \oplus \Pi_{P^c} C
        \subseteq C \oplus C.
    \end{align}
    Therefore, by \cref{prop:code_decomposability}, if $\varnothing \neq P \neq [n]$, then the code is decomposable. Since the code is indecomposable, $\supp(p)$ must be either $\varnothing$ or $[n]$. The same argument gives $\supp(s) \in \{\varnothing,[n]\}$. In addition, since each $V_i$ is invertible, we have $p_i s_i = 0$, so $\supp(p)$ and $\supp(s)$ cannot both equal $[n]$. Thus again the local Cliffords on the physical qubits must be uniform, and the possible gates are $H^{\otimes n}, (HS)^{\otimes n}$, and $(SH)^{\otimes n}$. 
    
    Lastly, we caution that we have worked only in the projective Clifford picture throughout this analysis, so the notion of uniformity here is only up to local Paulis and phases.
\end{proof}

\begin{theorem}
    [Fixed transversal logical group on indecomposable SSD CSS codes]
    \label{thm:css_codes_trans_logical_groups_ssd_indecomp}
    Let $\mathcal{C}$ be an $\db{n,k}$ indecomposable SSD CSS code. Then, in a CSS logical basis $\mathcal{L} = (L_\mathrm{x}, L_\mathrm{z})$, the projective transversal logical group is exactly
    \begin{equation}
        \mathrm{Trans}_\mathcal{L}(\mathcal{C})
        = \left\{
            \mqty(p I & r M_\mathrm{z} \\ q M_\mathrm{x} & s I)
            :
            \mqty(p & r \\ q & s) \in \Sp(2, \mathbb{F}_2)
        \right\}
        \cong S_3,
        \label{eq:css_codes_trans_logical_groups_ssd_indecomp_1}
    \end{equation}
    where $M_\mathrm{x} \coloneqq L_\mathrm{x} L_\mathrm{x}^\top$ and $M_\mathrm{z} \coloneqq L_\mathrm{z} L_\mathrm{z}^\top$ are inverses of each other. The six projective logical actions are performed by
    \begin{equation}
        I^{\otimes n},\quad
        S^{\otimes n},\quad
        (HSH)^{\otimes n},\quad
        H^{\otimes n},\quad
        (HS)^{\otimes n},\quad
        (SH)^{\otimes n},
    \end{equation}
    possibly after multiplying them by physical Pauli corrections; their induced projective logical actions are given in \cref{eq:css_codes_trans_logical_groups_ssd_indecomp_2}. In any logical basis, $\mathrm{Trans}_\mathcal{L}(\mathcal{C})$ is conjugate in $\Sp(2k,\mathbb{F}_2)$ to the group in \cref{eq:css_codes_trans_logical_groups_ssd_indecomp_1}.
\end{theorem}

\begin{proof}
    By \cref{lem:trans_exchange_uniformity_indecomp_css_codes}, every transversal gate is a uniformly applied single-qubit Clifford with symplectic representation
    \begin{equation}
        c = \mqty(
            p & r \\
            q & s
        )
        \in
        \Sp(2,\mathbb{F}_2).
    \end{equation}
    
    We first compute the induced logical action of such an operation. The $X$- and $Z$-type logical basis representatives, in symplectic form, are $(L_\mathrm{x}, 0), (0, L_\mathrm{z}) \in \mathbb{F}_2^{k \times 2n}$ respectively. The uniform Clifford operation with single-qubit action $c$ sends these to $(p L_\mathrm{x}, q L_\mathrm{x}), (r L_\mathrm{z}, s L_\mathrm{z})$ respectively. Since $\mathcal{L}$ is a CSS logical basis, $L_\mathrm{x} L_\mathrm{z}^\top = L_\mathrm{z} L_\mathrm{x}^\top = I$. Consequently, the $X$-sector logical action block-matrices of $p L_\mathrm{x}$ and $r L_\mathrm{z}$ are $p L_\mathrm{x} L_\mathrm{z}^\top = p I$ and $r L_\mathrm{z} L_\mathrm{z}^\top = r M_\mathrm{z}$, while the $Z$-sector logical action block-matrices of $q L_\mathrm{x}$ and $s L_\mathrm{z}$ are $q L_\mathrm{x} L_\mathrm{x}^\top = q M_\mathrm{x}$ and $s L_\mathrm{z} L_\mathrm{x}^\top = s I$, respectively. Hence the induced logical symplectic action is
    \begin{equation}
        \Phi(c)
        =
        \mqty(
            p I & r M_\mathrm{z}
            \\
            q M_\mathrm{x} & s I
        ).
    \end{equation}

    This gives the containment of $\mathrm{Trans}_\mathcal{L}(\mathcal{C})$ in the displayed set in \cref{eq:css_codes_trans_logical_groups_ssd_indecomp_1}. The six cases are explicitly
    \begin{equation}\begin{aligned}
        I^{\otimes n}      
            &: \; 
            \mqty(I & 0 \\ 0 & I),
            \qquad\qquad
        S^{\otimes n}      
            : \; 
            \underbrace{\mqty(I & 0 \\ M_\mathrm{x} & I)}_{
                S, \mathrm{CZ}},
            \qquad\qquad
        (HSH)^{\otimes n}  
            : \; 
            \underbrace{\mqty(I & M_\mathrm{z} \\ 0 & I)}_{
                HSH, H^{\otimes 2}\mathrm{CZ}H^{\otimes 2}},
            \\[4pt]
        H^{\otimes n}      
            &: \; 
            \mqty(0 & M_\mathrm{z} \\ M_\mathrm{x} & 0)
            = 
            \underbrace{\mqty(0 & I \\ I & 0)}_{
                H^{\otimes k}}
            \underbrace{\mqty(M_\mathrm{x} & 0 \\ 0 & M_\mathrm{z})}_{
                \mathrm{CX} \,\, \text{circuit}}
            = 
            \underbrace{\mqty(M_\mathrm{z} & 0 \\ 0 & M_\mathrm{x})}_{
                \mathrm{CX} \,\, \text{circuit}}
            \underbrace{\mqty(0 & I \\ I & 0)}_{
                H^{\otimes k}},
            \\[4pt]
        (HS)^{\otimes n}   
            &: \; 
            \mqty(I & M_\mathrm{z} \\ M_\mathrm{x} & 0)
            =
            \underbrace{\mqty(I & I \\ I & 0)}_{
                (HS)^{\otimes k}}
            \underbrace{\mqty(I & 0 \\ I + M_\mathrm{z} & I)}_{
                S, \mathrm{CZ}}
            \underbrace{\mqty(M_\mathrm{x} & 0 \\ 0 & M_\mathrm{z})}_{
                \mathrm{CX} \,\, \text{circuit}},
            \\[4pt]
        (SH)^{\otimes n}   
            &: \; 
            \mqty(0 & M_\mathrm{z} \\ M_\mathrm{x} & I)
            =
            \underbrace{\mqty(M_\mathrm{z} & 0 \\ 0 & M_\mathrm{x})}_{
                \mathrm{CX} \,\, \text{circuit}}
            \underbrace{\mqty(I & 0 \\ I + M_\mathrm{z} & I)}_{
                S, \mathrm{CZ}}
            \underbrace{\mqty(0 & I \\ I & I)}_{
                (SH)^{\otimes k}}.
        \label{eq:css_codes_trans_logical_groups_ssd_indecomp_2}
    \end{aligned}\end{equation}

    Conversely, we show that every element of the displayed set in \cref{eq:css_codes_trans_logical_groups_ssd_indecomp_1} is realized projectively. Let $C \coloneqq C_\mathrm{x} = C_\mathrm{z} \subseteq \mathbb{F}_2^n$ be the common $X$- and $Z$-type binary stabilizer space. The full binary stabilizer space is then $C \oplus C$. For any $c$, the corresponding uniform physical Clifford layer sends $(a,b) \mapsto (pa+rb,qa+sb)$. If $(a,b)\in C\oplus C$, then $(pa+rb,qa+sb)\in C\oplus C$ because $C$ is linear. Thus the uniform Clifford operation preserves the binary stabilizer space. The uniform Clifford operation may nevertheless introduce signs on the signed stabilizer group. By \cref{prop:pauli_phase_correction_binary_stabilizer_automorphisms}, there is a physical Pauli correction such that the composition of the uniform Clifford layer and the correction preserves the signed stabilizer group exactly. This Pauli correction does not change the induced projective logical Clifford action. Hence every matrix $\Phi(c)$ in \cref{eq:css_codes_trans_logical_groups_ssd_indecomp_1} is realized by a valid transversal gate.

    It remains to identify the group. Recall that $M_\mathrm{x}$ and $M_\mathrm{z}$ are inverses of each other. Therefore the map
    \begin{equation}
        \Phi:\Sp(2,\mathbb{F}_2)\to\Sp(2k,\mathbb{F}_2),
        \qquad
        \mqty(p&r\\q&s)
        \mapsto
        \mqty(p I & r M_\mathrm{z} \\ q M_\mathrm{x} & s I),
    \end{equation}
    is a homomorphism: for any two elements $c_1, c_2$, the equality $M_\mathrm{x}M_\mathrm{z}=M_\mathrm{z}M_\mathrm{x}=I$ gives $\Phi(c_1)\Phi(c_2)=\Phi(c_1c_2)$.
    The map is injective because $k\geq 1$ and $M_\mathrm{x},M_\mathrm{z}$ are invertible, so the scalars $p,r,q,s$ are recovered from the four blocks of $\Phi(c)$. Hence $\mathrm{Trans}_\mathcal{L}(\mathcal{C})
        \cong
        \Sp(2,\mathbb{F}_2)
        \cong
        S_3$.
    Finally, changing the logical basis conjugates the logical action group by the corresponding element of $\Sp(2k,\mathbb{F}_2)$.
\end{proof}

\begin{corollary}
    [Self-duality required for transversal logical $H^{\otimes k}$, $(SH)^{\otimes k}$, $(HS)^{\otimes k}$ gates on indecomposable CSS codes]
    \label{corr:css_codes_trans_logical_hadamards_only_on_self_dual_codes}
    An $\db{n,k}$ indecomposable CSS code transversally realizes, in a CSS logical basis, exchange-type logical gates $H^{\otimes k}$, $(SH)^{\otimes k}$, and $(HS)^{\otimes k}$, up to composition with logical $S$, $\mathrm{CZ}$, and $\mathrm{CX}$ circuits, iff the code is SSD.
    Moreover, these actions are realized exactly, without such composition, iff the logical overlap matrices $M_\mathrm{x} = M_\mathrm{z} = I$ in the chosen logical basis.
\end{corollary}

\begin{proof}
    \Cref{thm:css_codes_trans_logical_groups_non_ssd_indecomp} restricts transversal logical actions on non-SSD indecomposable codes to be preserving-type.
    The guaranteed exchange-type transversal logical actions on SSD codes can be read off from \cref{eq:css_codes_trans_logical_groups_ssd_indecomp_2} of \cref{thm:css_codes_trans_logical_groups_non_ssd_indecomp}.
    The extraneous $S$, $\mathrm{CZ}$, and $\mathrm{CX}$ circuit components of the logical actions vanish iff $M_\mathrm{x} = M_\mathrm{z} = I$.
\end{proof}

\begin{theorem}
    [Transversal logical groups on indecomposable non-SSD CSS codes]
    \label{thm:css_codes_trans_logical_groups_non_ssd_indecomp}
    Let $\mathcal{C}$ be an $\db{n,k}$ indecomposable non-SSD CSS code. Then, in a CSS logical basis $\mathcal{L}$, transversal gates on $\mathcal{C}$ involve only preserving-type physical Cliffords and induce a group of preserving-type logical actions
    \begin{equation}
        \mathrm{Trans}_\mathcal{L}(\mathcal{C})
        \cong 
        \left\{
            \mqty(I & A \\ B & I)
            = \mqty(I & A \\ 0 & I) \mqty(I & 0 \\ B & I)
            = \mqty(I & 0 \\ B & I) \mqty(I & A \\ 0 & I) 
            :
            A \in \mathcal{A},\;
            B \in \mathcal{B}
        \right\}
        \cong
        \mathcal{A} \oplus \mathcal{B},
        \label{eq:css_codes_trans_logical_groups_non_ssd_indecomp_1}
    \end{equation}
    where $\mathcal{A}, \mathcal{B}$ are linear subspaces of symmetric binary matrices satisfying $\mathcal{A} \mathcal{B} = \mathcal{B} \mathcal{A} = 0$. In particular, $\mathrm{Trans}_\mathcal{L}(\mathcal{C})$ is an elementary abelian unipotent group of order $2^{\ell_\mathcal{A} + \ell_\mathcal{B}}$, where $\ell_\mathcal{A} \coloneqq \dim \mathcal{A},\; \ell_\mathcal{B} \coloneqq \dim \mathcal{B}$, and $\mathrm{Trans}_\mathcal{L}(\mathcal{C}) \preceq_{\Sp(2k, \mathbb{F}_2)} \mathcal{U}(2k, \mathbb{F}_2)$. Moreover:
    \begin{itemize}
        
        \item If $\mathcal{C}$ is PSD, then $\ell \coloneqq \ell_\mathcal{A} = \ell_\mathcal{B}$ and $0 \leq 2\ell \leq \floor{k/2} (\floor{k/2} + 1)$. Hence $\mathrm{Trans}_\mathcal{L}(\mathcal{C}) \preceq_{\Sp(2k, \mathbb{F}_2)} \mathcal{U}(2k, \mathbb{F}_2)$, and
        \begin{equation}
            \abs{\mathrm{Trans}_\mathcal{L}(\mathcal{C})} 
                \leq 2^{\floor{k/2} (\floor{k/2} + 1)}
                = \frac{1}{2^{\ceil{k/2}^2}} \abs{\mathcal{U}(2k, \mathbb{F}_2)}
                \leq \abs{\mathcal{U}(2k, \mathbb{F}_2)}.
            \label{eq:css_codes_trans_logical_groups_non_ssd_indecomp_2}
        \end{equation}

        For $k > 1$, $\mathrm{Trans}_\mathcal{L}(\mathcal{C})$ is a proper subgroup of $\mathcal{U}(2k, \mathbb{F}_2)$.
        The maximum corresponds to $\mathrm{Trans}_\mathcal{L}(\mathcal{C}) \sim_{\Sp(2k, \mathbb{F}_2)} \mathcal{U}_\mathrm{x}(2\floor{k/2}, \mathbb{F}_2) \times \mathcal{U}_\mathrm{z}(2\floor{k/2}, \mathbb{F}_2)$, where the two unipotent radical subgroups are supported on disjoint subsets of $\floor{k/2}$ logical qubits, and is achievable. 
        
        \item Otherwise, the maximum $\mathrm{Trans}_\mathcal{L}(\mathcal{C}) \sim_{\Sp(2k, \mathbb{F}_2)} \mathcal{U}(2k, \mathbb{F}_2)$ is achievable.
        
    \end{itemize}
\end{theorem}

\begin{proof}
    To obtain \cref{eq:css_codes_trans_logical_groups_non_ssd_indecomp_1}, we follow the proof of
    \cref{lem:css_codes_auto_physical_splitting_property} but specialized to transversal gates by setting $\pi=\id$. Since the code is non-SSD, only the $T=\varnothing$ case, where $p=s=\vb{1}$, can occur. Then $X$- and $Z$-type logical representatives are transformed as $(x,0)\mapsto(x,q\odot x)$ and $(0,z)\mapsto(r\odot z,z)$, respectively, so every induced logical action has the form
    \begin{equation}
        \mqty(I&A\\B&I),
    \end{equation}
    where symplecticity requires $A$ and $B$ to be symmetric. Moreover, a transversal Clifford layer preserves the binary stabilizer space precisely when $q\odot C_\mathrm{x}\subseteq C_\mathrm{z}$ and $r\odot C_\mathrm{z}\subseteq C_\mathrm{x}$. Since these conditions are independent, setting $q=\vb{0}$ or $r=\vb{0}$ separately realizes the corresponding pure upper- or lower-triangular logical action. (By \cref{prop:pauli_phase_correction_binary_stabilizer_automorphisms}, physical Pauli corrections may be included to preserve the signed stabilizer group without changing transversality or the induced projective logical action, so these are valid transversal gates.)

    Now, defining the spaces of occurring upper and lower blocks of logical actions in $G \coloneqq \mathrm{Trans}_\mathcal{L}(\mathcal{C})$ by
    \begin{equation}\begin{split}
        \mathcal{A}
        \coloneqq
        \left\{
            A\in\mathbb{F}_2^{k\times k}
            :
            \exists B\in\mathbb{F}_2^{k\times k}
            \text{ s.t. }
            \mqty(I&A\\B&I)\in G
        \right\},
        \qquad
        \mathcal{B}
        \coloneqq
        \left\{
            B\in\mathbb{F}_2^{k\times k}
            :
            \exists A\in\mathbb{F}_2^{k\times k}
            \text{ s.t. }
            \mqty(I&A\\B&I)\in G
        \right\},
    \end{split}\end{equation}
    the independent sector realization above shows that
    \begin{equation}
        \mqty(I&A\\0&I),
        \mqty(I&0\\B&I)
        \in G\;,
        \qquad
        \forall
        \;
        A\in\mathcal{A},
        \,\,
        B\in\mathcal{B}.
    \end{equation}
    Since the composition of two transversal gates is again transversal, $G$ is closed under multiplication. It follows that $\mathcal{A}$ and $\mathcal{B}$ are linear subspaces of symmetric binary matrices. Moreover,
    \begin{equation}
        \mqty(I&A\\0&I)\mqty(I&0\\B&I)
        =
        \mqty(I+AB&A\\B&I),
        \qquad
        \mqty(I&0\\B&I)\mqty(I&A\\0&I)
        =
        \mqty(I&A\\B&I+BA).
    \end{equation}
    Since both products belong to $G$ and must have identity diagonal blocks, $AB=BA=0$. Thus the two pure-sector actions commute and their product realizes every pair $(A,B)\in\mathcal{A}\oplus\mathcal{B}$. This proves \cref{eq:css_codes_trans_logical_groups_non_ssd_indecomp_1}---the elements of $G$ have the displayed form and factorizations, and $G \cong \mathcal{A}\oplus\mathcal{B}$. In particular, $G$ is elementary abelian, and its elements are unipotent. The group order is then $\abs{G} = 2^{\ell_\mathcal{A} + \ell_\mathcal{B}}$. Finally, \cref{lem:preserving_in_sector_identity_clifford_subgroups} gives $G \preceq_{\Sp(2k,\mathbb{F}_2)} \mathcal{U}(2k,\mathbb{F}_2)$. The maximum $\mathrm{Trans}_\mathcal{L}(\mathcal{C}) \sim_{\Sp(2k,\mathbb{F}_2)} \mathcal{U}(2k, \mathbb{F}_2)$ is achievable by, for example, \cref{cons:concatenated_phantom_clifford,cons:complete_hypergraph_css_t_eq_2}.

    We now assume additionally that $\mathcal{C}$ is PSD. Then a self-duality automorphism on $\mathcal{C}$ induces a pure exchange-type logical action (see \cref{prop:self_dual_css_codes_pure_exchange_logical_action})
    \begin{equation}
        E
        =
        \mqty(
            0 & \Gamma^{-\top}
            \\
            \Gamma & 0
        ).
    \end{equation}
    Conjugation by the self-duality automorphism maps transversal gates to transversal gates, and its induced logical action satisfies
    \begin{equation}
        E
        \mqty(I&A\\B&I)
        E^{-1}
        =
        \mqty(
            I & \Gamma^{-\top}B\Gamma^{-1}
            \\
            \Gamma A\Gamma^\top & I
        ).
    \end{equation}
    Since conjugation by the self-duality automorphism is bijective on the transversal logical group, we observe $\mathcal{A} = \Gamma^{-\top}\mathcal{B}\Gamma^{-1}$ and $\mathcal{B} = \Gamma\mathcal{A}\Gamma^\top$. In particular, these subspaces have the same dimension, $\ell \coloneqq \ell_\mathcal{A} = \ell_\mathcal{B}$. 
    
    It remains to bound $\ell$ and identify the maximum-order case. 
    For a matrix $S$, let $\im S=\cs(S)=\{S\bx:\bx\in\F_2^k\}$ denote the column span of $S$.
    Define 
    \begin{equation}
        N_{\mathcal{A}}
        \coloneqq
        \sum_{A\in\mathcal{A}}\im A,
        \qquad
        N_{\mathcal{B}}
        \coloneqq
        \sum_{B\in\mathcal{B}}\im B,
    \end{equation}
    where the summation here means the standard sum of subspaces (i.e.~linear span of all vectors of the summands).
    Because \(\mathcal B=\Gamma\mathcal A\Gamma^\top\), and right multiplication by an invertible matrix does not change a matrix's image, we have \(N_{\mathcal B}=\Gamma N_{\mathcal A}\), and hence the dimensions of the two spaces are identical; we denote them $t \coloneqq \dim N_{\mathcal A}=\dim N_{\mathcal B}$. 
    We next show that these two spaces are orthogonal. For every $A\in\mathcal{A}$ and $B\in\mathcal{B}$, the relation $AB=0$ gives $\im B\subseteq\ker A$. Because $A$ is symmetric, $\ker A=(\im A)^\perp$.
    Taking the sum over $B$ and the intersection over $A$ gives $N_\mathcal{B}\subseteq N_\mathcal{A}^\perp.$
    Consequently $t \leq k - t \Rightarrow 2t \leq k$, equivalently $t \leq \floor{k/2}$.

    We now claim that the space of symmetric matrices whose images are contained in a fixed $t$-dimensional space has dimension $t(t+1)/2$.
    To formalize this, let $N\subseteq\mathbb{F}_2^k$ be a subspace of dimension $t$, and define 
    \begin{equation}
        \operatorname{sym}(N) \coloneqq \{S\in \mathbb{F}_2^{k \times k}: S = \trans{S},\; \im S \subseteq N\}.
    \end{equation}
    In fact, an equivalent way of defining $\operatorname{sym}(N)$ is as follows: choose a matrix $R\in\F_2^{k\times t}$ whose columns form a basis of $N$; then $\operatorname{sym}(N)=\{RMR^\top:M=M^\top\in\F_2^{t\times t}\}$.
    To see this, extend the columns of $R$ to an invertible matrix $F\in\GL(k,\F_2)$. 
    Under the conjugation $S\mapsto F^{-1}SF^{-\top}$, the subspace $N$ becomes the span of the first $t$ coordinate vectors.
    If $\im S\subseteq N$, the last $k-t$ rows of the transformed matrix vanish; symmetry then forces its last $k-t$ columns to vanish as well.
    Therefore, $\dim N=t\Rightarrow \dim\operatorname{sym}(N)=t(t+1)/2$.

    By the definition of \(N_{\mathcal A}\), every \(A\in\mathcal A\) has image contained in \(N_{\mathcal A}\). Thus $\mathcal A\subseteq \operatorname{sym}(N_{\mathcal A})$, and hence $\ell = \dim\mathcal A \leq \dim \operatorname{sym}(N_{\mathcal A}) = t(t+1)/2 \leq \floor{k/2}(\floor{k/2} + 1)/2$. This resolves the maximum value $\ell$ can take. It follows that $\abs{G} = 2^{2\ell} \leq 2^{\floor{k/2}(\floor{k/2}+1)}$, proving \cref{eq:css_codes_trans_logical_groups_non_ssd_indecomp_2}. The containment in $\mathcal{U}(2k,\mathbb{F}_2)$ is proper when $k>1$.

    Equality in the order bound requires $t = \floor{k/2}$ and $\ell = t(t+1)/2$. Since $\mathcal{A} \subseteq \operatorname{sym}(N_{\mathcal{A}})$ and the two spaces have the same dimension, we obtain $\mathcal{A} = \operatorname{sym}(N_{\mathcal{A}})$. Similarly, $\mathcal{B} = \operatorname{sym}(N_{\mathcal{B}})$. 
    
    Lastly, to identify the embedding in $\Sp(2k, \mathbb{F}_2)$ up to conjugacy, define $W_{\mathcal{A}} \coloneqq N_{\mathcal{A}}\oplus \{0\}$ and $W_{\mathcal{B}} \coloneqq \{0\}\oplus N_{\mathcal{B}}$ as subspaces of the logical symplectic space $\mathbb{F}_2^k\oplus\mathbb{F}_2^k$. Both are totally isotropic, and $N_{\mathcal{B}}\subseteq N_{\mathcal{A}}^\perp$ implies that they are mutually orthogonal. Thus $W_{\mathcal{A}}\oplus W_{\mathcal{B}}$ is a totally isotropic subspace of dimension $2t$. Let us choose bases $\{a_1,\ldots,a_t\}$ of $N_{\mathcal{A}}$ and $\{b_1,\ldots,b_t\}$ of $N_{\mathcal{B}}$. 
    
    Since $2t\leq k$, these isotropic bases may be extended to symplectic bases as follows: since $N_\mathcal{B}\subseteq N_\mathcal{A}^\perp$, extend $b_1,\dots,b_t$ to a basis $b_1,\dots,b_t, c_1,\dots, c_{k-2t}$ of $N_\mathcal{A}^\perp$. Choose vectors $\alpha_1,\dots,\alpha_t$ satisfying $\alpha_i^\top a_j=\delta_{ij}$, then $\alpha_1,\dots,\alpha_t,b_1,\dots,b_t,c_1,\dots,c_{k-2t}$ is a basis of $\F_2^k$. Let $P\in\GL(k,\F_2)$ be the matrix having these vectors as its rows. 
    Notice that $PN_\mathcal{A}=\la e_1,\dots,e_t\ra$, $P^{-\top}N_\mathcal{B}=\la e_{t+1},\dots,e_{2t}\ra$.
    Now conjugate the logical group by the change of basis $\begin{psmallmatrix}P&0\\0&P^{-\top}\end{psmallmatrix}\in\Sp(2k,\F_2)$, so that $A\mapsto PAP^\top$, $B\mapsto P^{-\top}BP^{-1}$.
    We have the transformed upper blocks range over all symmetric matrices supported on the first $t$ logical coordinates, while the transformed lower blocks range over all symmetric matrices supported on the next $t$ logical coordinates.
    Thus the two full unipotent factors act on disjoint sets of $t$ logical qubits.
    Equivalently, $G=\mathrm{Trans}_\mathcal{L}(\mathcal{C}) \sim_{\Sp(2k, \mathbb{F}_2)} \mathcal{U}_\mathrm{x}(2\floor{k/2}, \mathbb{F}_2) \times \mathcal{U}_\mathrm{z}(2\floor{k/2}, \mathbb{F}_2)$ as claimed.
    
    This maximum is achieved by, for example, \cref{cons:css_codes_maximum_trans_logical_group_psd}.
\end{proof}

\begin{corollary}
    [No logical CX and SWAP on indecomposable CSS codes from transversal gates]
    \label{corr:css_codes_trans_no_cx_swap}
    On an indecomposable CSS code, transversal gates cannot induce $\mathrm{CX}$ and $\mathrm{SWAP}$ logical actions, or any product thereof, in any CSS logical basis.
\end{corollary}

\begin{proof}
    The symplectic representations of $\mathrm{CX}$ and $\mathrm{SWAP}$, or products thereof, are of the form:
    \begin{equation}
        \begin{pmatrix}
            A & 0 \\
            0 & A^{-\top}
        \end{pmatrix},
        \qquad 
        A \neq I.
    \end{equation}
    This is incompatible with the logical actions of transversal gates in \cref{thm:css_codes_trans_logical_groups_ssd_indecomp} for SSD codes and \cref{thm:css_codes_trans_logical_groups_non_ssd_indecomp} for non-SSD codes.
\end{proof}

\clearpage

\subsection{Augmenting logical groups with transversal interblock CX gates}
\label{app:css_codes/tCX}

\begin{definition}
    [Interblock transversal CX gates on CSS codes]
    \label{def:css_codes_tcxs}
    Let $\mathcal{C}$ be an $\db{n,k}$ CSS code and $\mathcal{L}$ be a CSS logical basis of $\mathcal{C}$. 
    Consider $N>1$ codeblocks of $\mathcal{C}$, each using the logical basis $\mathcal{L}$.
    The \emph{interblock transversal CX gate}, abbreviated as the tCX gate, controlled by codeblock $a$ and targeting codeblock $b \neq a$ is the physical operation $\smash{\prod_{i=1}^n \mathrm{CX}_{i^{(a)} i^{(b)}}}$, where $i^{(a)}$ denotes the $i^\text{th}$ physical qubit on codeblock $a$ and likewise on $b$.
    This gate has the induced logical action $\smash{\prod_{j=1}^k \mathrm{CX}_{j^{(a)} j^{(b)}}}$, where $j^{(a)}$ denotes the $j^\text{th}$ logical qubit on codeblock $a$ and likewise on $b$.
\end{definition}

\begin{definition}
    [Logical groups augmented by tCX gates on CSS codes]
    \label{def:css_codes_logical_groups_with_tcxs}
    Let $\mathcal{C}$ be an $\db{n,k}$ CSS code and $\mathcal{L}$ be a CSS logical basis of $\mathcal{C}$. 
    Consider $N>1$ codeblocks of $\mathcal{C}$, each using the logical basis $\mathcal{L}$.
    We denote by $\mathrm{Trans}^\mathrm{tCX}_\mathcal{L}(\mathcal{C}, N)$, $\mathrm{Perm}^\mathrm{tCX}_\mathcal{L}(\mathcal{C}, N)$, and $\mathrm{Aut}^\mathrm{tCX}_\mathcal{L}(\mathcal{C}, N)$, the group of logical actions induced by transversal, permutation, and automorphism gates on each codeblock, respectively, in conjunction with tCXs between every ordered pair of codeblocks.
\end{definition}

\begin{theorem}
    [Augmentation of logical groups by tCX gates on indecomposable CSS codes]
    \label{thm:css_codes_tcxs_logical_group_promotion}
    Let $\mathcal{C}$ be an $\db{n,k}$ indecomposable CSS code and $\mathcal{L}$ be a CSS logical basis of $\mathcal{C}$. 
    Consider $N>1$ codeblocks of $\mathcal{C}$, each using the logical basis $\mathcal{L}$.
    Then the following hold:
    \begin{enumerate}
        \item $\mathrm{Trans}^\mathrm{tCX}_\mathcal{L}(\mathcal{C}, N)$ is the full projective Clifford group $\Sp(2Nk, \mathbb{F}_2)$ on the $Nk$ total logical qubits iff $\mathcal{C}$ is SSD with $k=1$.
        Non-SSD $\mathcal{C}$, and SSD $\mathcal{C}$ with $k>1$, achieve proper subgroups of $\Sp(2Nk, \mathbb{F}_2)$.
        \label{item:css_codes_tcxs_logical_group_promotion_1}
        \item $\mathrm{Aut}^\mathrm{tCX}_\mathcal{L}(\mathcal{C}, N)$ is the full projective Clifford group $\Sp(2Nk, \mathbb{F}_2)$ on the $Nk$ total logical qubits iff $\mathcal{C}$ is SSD with $k=1$.
        Non-SSD $\mathcal{C}$, and SSD $\mathcal{C}$ with $k>1$, achieve proper subgroups of $\Sp(2Nk, \mathbb{F}_2)$.
        \label{item:css_codes_tcxs_logical_group_promotion_2}
        \item For a non-PSD $\mathcal{C}$, $\mathrm{Perm}^\mathrm{tCX}_\mathcal{L}(\mathcal{C}, N)$ is isomorphic to a subgroup of $\GL(Nk, \mathbb{F}_2)$, and equality is attainable. 
        \label{item:css_codes_tcxs_logical_group_promotion_3}
        \item For a non-PSD $\mathcal{C}$, $\mathrm{Aut}^\mathrm{tCX}_\mathcal{L}(\mathcal{C}, N) \preceq_{\Sp(2k, \mathbb{F}_2)^N} \mathcal{P}(2Nk, \mathbb{F}_2)$, and equality is attainable. 
        \label{item:css_codes_tcxs_logical_group_promotion_4}
    \end{enumerate}
\end{theorem}

\begin{proof}
    Write the overall logical symplectic space of the $N$ codeblocks as $V \coloneqq (\mathbb F_2^k)^{\oplus N}_\mathrm{x} \oplus (\mathbb F_2^k)^{\oplus N}_\mathrm{z}$, where the $\{\mathrm{x}, \mathrm{z}\}$ subscripts denote the logical $X$ and $Z$ sectors for clarity. 
    Elements of this space are written as $(\vb{x}_1,\ldots,\vb{x}_N,\vb{z}_1,\ldots,\vb{z}_N)$, where each $\vb{x}_i,\vb{z}_i \in \mathbb{F}_2^k$ represents the logical Paulis within codeblock $i$.
    The logical action of the tCX gate from codeblock $a$ to codeblock $b$ is
    \begin{equation}
        \vb{x}_b\longmapsto \vb{x}_b+\vb{x}_a,
        \qquad
        \vb{z}_a\longmapsto \vb{z}_a+\vb{z}_b,
        \label{eq:tCX_logical_action_proof}
    \end{equation}
    with all other components unchanged. 
    We use two immediate consequences of \cref{eq:tCX_logical_action_proof}. 
    First, for any subspaces $R,S\leq\mathbb F_2^k$, all tCX logical actions preserve $R^{\oplus N}_{\mathrm{x}}\oplus S^{\oplus N}_{\mathrm{z}}$.
    Second, all tCX logical actions preserve the quadratic form 
    \begin{equation}
        q(\vb{x}_1,\ldots,\vb{x}_N;\vb{z}_1,\ldots,\vb{z}_N)
        \coloneqq
        \sum_{a=1}^N \vb{x}_a^\top \vb{z}_a.
        \label{eq:css_codes_tcxs_logical_group_promotion_quadratic_form}
    \end{equation}
    The full group $\Sp(2Nk,\mathbb F_2)$ preserves no nonzero proper subspace of $V$, since it is transitive on $V\setminus\{0\}$, and it does not preserve $q$, since an addressable logical $S$ gate changes $q$.

    We first prove the positive cases of Items~\cref{item:css_codes_tcxs_logical_group_promotion_1,item:css_codes_tcxs_logical_group_promotion_2}. 
    Suppose that $\mathcal C$ is SSD and $k=1$. 
    By \cref{thm:css_codes_trans_logical_groups_ssd_indecomp}, each codeblock supports the full single-qubit projective Clifford group $\Sp(2,\mathbb F_2)\cong\langle H,S\rangle \cong S_3$ transversally. 
    Moreover, the interblock tCX is then the ordinary logical CX between the
    corresponding logical qubits. 
    Hence the available gates contain $H_a,S_a$ on every logical qubit and $\CX_{ab}$ between every pair, which generate $\Sp(2N,\mathbb F_2)$. 
    Thus both $\mathrm{Trans}^\mathrm{tCX}_\mathcal{L}(\mathcal{C}, N)$ and $\mathrm{Aut}^\mathrm{tCX}_\mathcal{L}(\mathcal{C}, N)$ are the full projective Clifford group on the $Nk = N$ logical qubits.

    We now show that every other code class in Items~\cref{item:css_codes_tcxs_logical_group_promotion_1,item:css_codes_tcxs_logical_group_promotion_2} gives a proper subgroup:
    \begin{itemize}
        
        \item Non-PSD $\mathcal C$.
        Following \cref{thm:css_codes_trans_logical_groups_non_ssd_indecomp,lem:preserving_in_sector_identity_clifford_subgroups},
        the transversal logical group on each codeblock preserves the Lagrangian
        $L \coloneqq K_\mathrm{x} \oplus K^\perp_\mathrm{z}$, where $K\coloneqq\bigcap_{B\in\mathcal B}\ker B$, $K_\mathrm{x} \coloneqq \{(x, 0): x \in K\} = K \oplus \{0\}$, and $K^\perp_\mathrm{z} \coloneqq \{(0, z): z \in K^\perp\} = \{0\} \oplus K^\perp$, and $\mathcal{B}$ is a linear subspace of symmetric binary matrices as defined in \cref{thm:css_codes_trans_logical_groups_non_ssd_indecomp}.
        Moreover, by \cref{thm:css_codes_auto_logical_groups}, $\mathrm{Aut}_{\mathcal L}(\mathcal C)=\mathrm{Trans}_{\mathcal L}(\mathcal C)\rtimes\mathrm{Perm}_{\mathcal L}(\mathcal C)$, with $\mathrm{Perm}_{\mathcal L}(\mathcal C)$ normalizing $\mathrm{Trans}_{\mathcal L}(\mathcal C)$.
        If a permutation logical action has the form $\diag(R, R^{-\top})$ [cf.~\cref{thm:css_codes_perm_logical_group}], this normalization sends $B\mapsto R^{-\top}BR^{-1}$ and therefore implies $RK=K$, and hence $R^{-\top}K^\perp=K^\perp$. 
        We conclude every automorphism logical action also preserves $L$.

        Consequently, on the $N$ codeblocks, automorphism logical actions on each codeblock preserve the Lagrangian $L^{(N)}\coloneqq(K^{\oplus N})_{\mathrm{x}}\oplus((K^\perp)^{\oplus N})_{\mathrm{z}}$.
        The interblock tCX logical actions preserve $L^{(N)}$ as well, since they only add $X$-logical vectors between codeblocks within the common subspace $K$, and separately add $Z$-logical vectors between codeblocks within the common subspace $K^\perp$.
        The generated logical group therefore preserves $L^{(N)}$ and is contained in
        $\Stab_{\Sp(2Nk,\mathbb F_2)}(L^{(N)}) \sim_{\Sp(2k,\mathbb F_2)^N} \mathcal P(2Nk,\mathbb F_2) < \Sp(2Nk,\mathbb F_2)$.
        Incidentally, this proves the containment in Item~\cref{item:css_codes_tcxs_logical_group_promotion_4}; we remark on attainability later.

        \item PSD-but-not-SSD $\mathcal C$.
        Following the proof of \cref{thm:css_codes_trans_logical_groups_non_ssd_indecomp}, define $N_{\mathcal A}\coloneqq\sum_{A\in\mathcal A}\im A$ and $N_{\mathcal B}\coloneqq\sum_{B\in\mathcal B}\im B$.
        If the transversal logical group of $\mathcal C$ is nontrivial, the same theorem shows that $0<\dim N_{\mathcal A}=\dim N_{\mathcal B}\leq \floor{k/2}$ and $N_{\mathcal B}=\Gamma N_{\mathcal A}$, where $\Gamma$ is from the induced logical action of a self-duality automorphism of $\mathcal C$, of the form $E=\begin{psmallmatrix}0&\Gamma^{-\top}\\\Gamma&0\end{psmallmatrix}$.
        The transversal logical group preserves $W \coloneqq (N_{\mathcal A})_\mathrm{x} \oplus (N_{\mathcal B})_\mathrm{z} = (N_{\mathcal A}\oplus\{0\}) \oplus (\{0\}\oplus N_{\mathcal B})$, while the permutation logical group preserves $(N_{\mathcal A})_\mathrm{x}$ and $(N_{\mathcal B})_\mathrm{z}$ separately because it normalizes the transversal logical group, and $E$ exchanges the two spaces.
        Hence, by \cref{thm:css_codes_auto_logical_groups}, every automorphism logical action preserves $W$.

        On the $N$ codeblocks, automorphism logical actions on each codeblock preserve the repeated subspace $W^{(N)}=N_{\mathcal A}^{\oplus N}{}_{\mathrm{x}}\oplus N_{\mathcal B}^{\oplus N}{}_{\mathrm{z}}$.
        The interblock tCX logical actions also preserve $W^{(N)}$, since they only add $X$-logical vectors between codeblocks within the common subspace $N_{\mathcal A}$, and separately add $Z$-logical vectors between codeblocks within the common subspace $N_{\mathcal B}$.
        Since $0<\dim W^{(N)}\leq Nk<2Nk$, this is a nonzero proper invariant subspace, so the generated logical group is a proper subgroup of $\Sp(2Nk,\mathbb F_2)$.

        If instead the transversal logical group of $\mathcal C$ is trivial, then \cref{thm:css_codes_auto_logical_groups} gives $\mathrm{Aut}_{\mathcal L}(\mathcal C)=\mathrm{Perm}_{\mathcal L}^{*}(\mathcal C)$.
        Its logical actions are generated by pure preserving-type actions and a pure exchange-type action induced by a self-duality automorphism.
        Both preserve the quadratic form $(x,z)\mapsto x^\top z$.
        Thus the automorphism logical actions on each codeblock and the tCX logical actions all preserve the quadratic form $q$ as defined in \cref{eq:css_codes_tcxs_logical_group_promotion_quadratic_form}.
        Since the full $\Sp(2Nk,\mathbb F_2)$ does not preserve $q$, the generated group is again proper.

        \item SSD $\mathcal C$ with $k>1$.
        Let $M_\mathrm{x}$ and $M_\mathrm{z}$ be the overlap matrices of $\mathcal C$ in the logical basis $\mathcal L$; by \cref{prop:strictly_self_dual_css_codes_properties}, they are inverses of each other.
        First, transversal gates on $\mathcal C$ induce exactly six projective Clifford logical actions, which are given in \cref{eq:css_codes_trans_logical_groups_ssd_indecomp_1} of \cref{thm:css_codes_trans_logical_groups_ssd_indecomp}.
        By \cref{thm:css_codes_auto_logical_groups}, the automorphism group of $\mathcal C$ is generated by this transversal group and the permutation logical group; for the latter, \cref{thm:css_codes_perm_logical_group_psd,thm:css_codes_perm_logical_group_ssd} shows that every permutation logical action has the form $\diag(R, R^{-\top})$ for some $R\in\GL(k,\mathbb F_2)$ preserving the logical overlap form with Gram matrix $M_\mathrm{x}$, and hence $R^\top M_\mathrm{x}R=M_\mathrm{x}$.
        Now:
        \begin{itemize}
            \item If $M_\mathrm{x}$ is alternating, then the transversal generators $\begin{psmallmatrix}I&0\\M_\mathrm{x}&I\end{psmallmatrix}$ and $\begin{psmallmatrix}0&M_\mathrm{z}\\M_\mathrm{x}&0\end{psmallmatrix}$ from  \cref{eq:css_codes_trans_logical_groups_ssd_indecomp_1} both preserve $(x,z)\mapsto x^\top z$; so do all permutation logical actions.
            Hence all automorphisms and all tCX logical actions preserve $q$ as in \cref{eq:css_codes_tcxs_logical_group_promotion_quadratic_form}, and the generated group is proper.
            \item If $M_\mathrm{x}$ is nonalternating, let $\omega\in\mathbb F_2^k$ be the unique characteristic vector of $M_\mathrm{x}$, defined by $x^\top M_\mathrm{x} x=x^\top M_\mathrm{x}\omega$ for all $x\in\mathbb F_2^k$.
            Since $M_\mathrm{x}$ is nonalternating, $\omega\neq0$. 
            The forms of the transversal logical actions in \cref{eq:css_codes_trans_logical_groups_ssd_indecomp_1} show immediately that the
            transversal logical group preserves $W_\omega\coloneqq \left\langle(\omega,0),(0,M_\mathrm{x}\omega)\right\rangle$.
            Moreover, $R^\top M_\mathrm{x} R=M_\mathrm{x}$ and uniqueness of the characteristic vector imply $R\omega=\omega$, and hence also $R^{-\top}M_\mathrm{x}\omega=M_\mathrm{x}\omega$. 
            Thus every automorphism logical action of $\mathcal{C}$ preserves $W_\omega$. 
            On $N$ codeblocks, automorphism logical actions preserve the repeated subspace $W_\omega^{(N)}=\langle\omega\rangle^{\oplus N}_{\mathrm{x}}\oplus\langle M_\mathrm{x}\omega\rangle^{\oplus N}_{\mathrm{z}}$; moreover tCX logical actions only add the $N$ copies of $\omega$ among themselves, and likewise the $N$ copies of $M_\mathrm{x}\omega$ among themselves, so they also preserve $W_\omega^{(N)}$.
            This preserved subspace is of dimension $2N<2Nk$ because $k>1$, so it is a nonzero proper invariant subspace.
        \end{itemize}

        Thus every non-SSD code and every SSD code with $k>1$ has a proper $\mathrm{Aut}^\mathrm{tCX}_\mathcal{L}(\mathcal{C}, N)$ logical group, and therefore also a proper $\mathrm{Trans}^\mathrm{tCX}_\mathcal{L}(\mathcal{C}, N)$ logical group. Together with the SSD $k=1$ case, this proves Items~\cref{item:css_codes_tcxs_logical_group_promotion_1} and \cref{item:css_codes_tcxs_logical_group_promotion_2}.
    \end{itemize}

    The containment of $\mathrm{Perm}^\mathrm{tCX}_\mathcal{L}(\mathcal{C}, N)$ in Item~\cref{item:css_codes_tcxs_logical_group_promotion_3} follows because permutation logical actions on CSS codes (see \cref{thm:css_codes_perm_logical_group}) and tCXs using CSS logical bases are products of logical CXs; thus logical actions in $\mathrm{Perm}^\mathrm{tCX}_\mathcal{L}(\mathcal{C}, N)$ are CX circuits. Attainability in Items~\cref{item:css_codes_tcxs_logical_group_promotion_3,item:css_codes_tcxs_logical_group_promotion_4} follow from the fact that phantom non-PSD codes exist at any $k$~\cite{koh2026entangling}, and $N$ codeblocks of a phantom code can achieve addressable logical CXs between any ordered pair of the $Nk$ total logical qubits by composing within-block qubit permutations and tCXs~\cite[Thm.~1]{koh2026entangling}. 
    This reaches all logical CX circuits, hence attaining equality in Item~\cref{item:css_codes_tcxs_logical_group_promotion_3}; and using phantom codes that support every addressable $S$ logical gate transversally, for example, \cref{cons:concatenated_phantom_clifford}, we attain $\mathrm{Aut}^\mathrm{tCX}_\mathcal{L}(\mathcal{C}, N) = \langle \mathbf{S}_{Nk}, \mathbf{CX}_{Nk} \rangle \cong \mathcal{P}(2Nk, \mathbb{F}_2)$ as claimed in Item~\cref{item:css_codes_tcxs_logical_group_promotion_4}.
\end{proof}

\begin{remark}
    [Augmentation of logical groups by tCX gates on CSS codes on decomposable CSS codes]
    Item~\cref{item:css_codes_tcxs_logical_group_promotion_2} of \cref{thm:css_codes_tcxs_logical_group_promotion} do not hold when $\mathcal{C}$ is allowed to be decomposable.
    To give a counter-example, consider an indecomposable SSD CSS code $\mathcal{C}_0$ encoding a single logical qubit that supports transversal $H$ and $S$ logical actions (see \cref{thm:css_codes_perm_logical_group_ssd}) in a CSS logical basis $\mathcal{L}_0$; we then take the decomposable $\mathcal{C} = \mathcal{C}_0^{\oplus k}$ and $\mathcal{L} = \mathcal{L}_0^{\oplus k}$ for a number of copies $k>1$.
    Then $\mathcal{C}$ has transversal $S_i$ and $H_i$ logical actions for every $i\in[k]$.
    By composing these addressable $S_i$ and $H_i$ logical gates and tCXs, addressable logical CXs between any pair of logical qubits between the codeblocks can be implemented.
    We illustrate a compilation for a logical CX between the first logical qubits of two $k=3$ codeblocks of $\mathcal{C}$ below.
    Arbitrary automorphism logical SWAPs are available within each $\mathcal{C}$ codeblock by permuting its constituent $\mathcal{C}_0$ blocks, so an addressable logical CX between any pair of logical qubits of the two codeblocks is available by conjugating this logical CX with appropriate SWAPs.
    The addressable $S_i$, $H_i$, and CXs thereby generate the full projective Clifford group on the $Nk$ total logical qubits.
    \begin{center}
        \begin{quantikz}[row sep={0.50cm,between origins}, column sep=0.35cm]
            \lstick[wires=3,braces=right]{$\mathcal{C}$ codeblock}
                & \qw
                & \qw
                & \ctrl{3}
                  \gategroup[wires=6,steps=3,
                    style={draw,dashed,rounded corners,inner xsep=0pt,inner ysep=0pt},
                    background,
                    label style={label position=below,anchor=north,yshift=-0.18cm}]
                    {$\mathrm{tCX}$}
                & \qw
                & \qw
                & \qw
                & \ctrl{3}
                  \gategroup[wires=6,steps=3,
                    style={draw,dashed,rounded corners,inner xsep=0pt,inner ysep=0pt},
                    background,
                    label style={label position=below,anchor=north,yshift=-0.18cm}]
                    {$\mathrm{tCX}$}
                & \qw
                & \qw
                & \gate{S^\dagger}
                & \qw
                \\
                & \qw
                & \qw
                & \qw
                & \ctrl{3}
                & \qw
                & \qw
                & \qw
                & \ctrl{3}
                & \qw
                & \qw
                & \qw
                \\
                & \qw
                & \qw
                & \qw
                & \qw
                & \ctrl{3}
                & \qw
                & \qw
                & \qw
                & \ctrl{3}
                & \qw
                & \qw
                \\
            \lstick[wires=3,braces=right]{$\mathcal{C}$ codeblock}
                & \gate{H}
                & \gate{S^\dagger}
                & \targ{}
                & \qw
                & \qw
                & \gate{S}
                & \targ{}
                & \qw
                & \qw
                & \gate{H}
                & \qw
                \\
                & \qw
                & \qw
                & \qw
                & \targ{}
                & \qw
                & \qw
                & \qw
                & \targ{}
                & \qw
                & \qw
                & \qw
                \\
                & \qw
                & \qw
                & \qw
                & \qw
                & \targ{}
                & \qw
                & \qw
                & \qw
                & \targ{}
                & \qw
                & \qw
        \end{quantikz}
        =
        \begin{quantikz}[row sep={0.50cm,between origins}, column sep=0.35cm]
            \qw
            & \ctrl{3}
            & \qw
            \\
            \qw
            & \qw
            & \qw
            \\
            \qw
            & \qw
            & \qw
            \\
            \qw
            & \targ{}
            & \qw
            \\
            \qw
            & \qw
            & \qw
            \\
            \qw
            & \qw
            & \qw
        \end{quantikz}
    \end{center}
\end{remark}

\begin{remark}
    [Addressable logical $H$, $SH$, $HS$ gates on indecomposable CSS codes with workspace and tCXs]
    \label{rem:css_codes_tcxs_exchange_addressability}
    Recall that \cref{corr:css_indecomp_no_add_h_sh_hs} forbids exchange-type single-qubit Clifford logical actions (i.e.~$H$, $SH$, $HS$, or mixture thereof) induced by automorphisms acting on a nonempty proper subset of the logical qubits of an indecomposable CSS code, in any CSS logical basis.
    On some indecomposable CSS codes, this restriction on addressability can be circumvented by allowing multiple codeblocks and using tCXs between codeblocks in conjunction with automorphisms.
    The intuition is that automorphisms in each codeblock may induce logical CX or SWAP actions, and when composed with the tCXs, may produce subsets of addressable logical SWAPs acting on logical qubits between codeblocks. 
    These can then be composed with $H^{\otimes k}$, $(SH)^{\otimes k}$, $(HS)^{\otimes k}$, or mixtures thereof in each codeblock to produce cancellations that lead to some addressability.

    To give a concrete illustration of this possibility, we consider the $\db{9,3,2}$ indecomposable SSD CSS code of \cref{tab:css_codes_maximum_perm_logical_group_ssd_orthogonal}, but with the logical basis
    \begin{equation}
        L_{\mathrm{x}}
        =
        \begin{pmatrix}
            1&0&0&1&0&0&1&1&1\\
            0&0&0&0&0&0&1&0&1\\
            0&0&0&0&0&0&1&1&0
        \end{pmatrix},
        \quad
        L_{\mathrm{z}}
        =
        \begin{pmatrix}
            1&0&0&1&0&0&1&1&1\\
            0&0&0&0&0&0&1&1&0\\
            0&0&0&0&0&0&1&0&1
        \end{pmatrix},
        \quad
        L_{\mathrm{x}}L_{\mathrm{z}}^\top
        =
        I,
        \quad
        M_{\mathrm{x}}
        =
        M_{\mathrm{z}}
        =
        \begin{pmatrix}
            1&0&0\\
            0&0&1\\
            0&1&0
        \end{pmatrix}.
    \end{equation}

    In this logical basis, the following physical operations induce respective logical actions:
    \begin{equation}
        U_\pi:
        \mathrm{SWAP}_{23},
        \qquad
        U_\sigma:
        \mathrm{CX}_{23},
        \qquad
        U_{\pi\sigma\pi}:
        \mathrm{CX}_{32},
        \qquad
        U_\pi H^{\otimes 9}:
        H^{\otimes 3},
        \qquad
        Z_1 Z_4 Z_8 Z_9 S^{\otimes 9}:
        S_1 \mathrm{CZ}_{23},
    \end{equation}
    where qubit permutations $\pi = (1\,2)(4\,5)(8\,9)$ and $\sigma = (1\,2)(4\,5)(7\,8)$, as written in the convention of \cref{sec:constructions/preliminaries/permutations}.
    With tCXs, we first compile the following addressable logical CXs between logical qubits of two codeblocks:
    \begin{center}
        \begin{quantikz}[row sep={0.50cm,between origins}, column sep=0.35cm]
            \lstick[wires=3,braces=right]{codeblock}
                & \ctrl{3}
                  \gategroup[wires=6,steps=3,
                    style={draw,dashed,rounded corners,inner xsep=0pt,inner ysep=0pt},
                    background,
                    label style={label position=below,anchor=north,yshift=-0.18cm}]
                    {$\mathrm{tCX}$}
                &[-0.25cm] \qw
                &[-0.25cm] \qw
                & \qw
                & \ctrl{3}
                  \gategroup[wires=6,steps=3,
                    style={draw,dashed,rounded corners,inner xsep=0pt,inner ysep=0pt},
                    background,
                    label style={label position=below,anchor=north,yshift=-0.18cm}]
                    {$\mathrm{tCX}$}
                &[-0.25cm] \qw
                &[-0.25cm] \qw
                & \qw
                & \qw
                \\
                & \qw
                & \ctrl{3}
                & \qw
                & \qw
                & \qw
                & \ctrl{3}
                & \qw
                & \qw
                & \qw
                \\
                & \qw
                & \qw
                & \ctrl{3}
                & \qw
                & \qw
                & \qw
                & \ctrl{3}
                & \qw
                & \qw
                \\
            \lstick[wires=3,braces=right]{codeblock}
                & \targ{}
                & \qw
                & \qw
                & \qw
                & \targ{}
                & \qw
                & \qw
                & \qw
                & \qw
                \\
                & \qw
                & \targ{}
                & \qw
                & \targ{}
                & \qw
                & \targ{}
                & \qw
                & \targ{}
                & \qw
                \\
                & \qw
                & \qw
                & \targ{}
                & \ctrl{-1}
                & \qw
                & \qw
                & \targ{}
                & \ctrl{-1}
                & \qw
        \end{quantikz}
        =
        \begin{quantikz}[row sep={0.50cm,between origins}, column sep=0.35cm]
            \qw
            & \qw
            & \qw
            \\
            \qw
            & \qw
            & \qw
            \\
            \qw
            & \ctrl{2}
            & \qw
            \\
            \qw
            & \qw
            & \qw
            \\
            \qw
            & \targ{}
            & \qw
            \\
            \qw
            & \qw
            & \qw
        \end{quantikz},
        \qquad
        \begin{quantikz}[row sep={0.50cm,between origins}, column sep=0.35cm]
            \qw
            & \qw
            & \qw
            & \qw
            & \qw
            \\
            \qw
            & \swap{1}
            & \qw
            & \swap{1}
            & \qw
            \\
            \qw
            & \targX{}
            & \ctrl{2}
            & \targX{}
            & \qw
            \\
            \qw
            & \qw
            & \qw
            & \qw
            & \qw
            \\
            \qw
            & \qw
            & \targ{}
            & \qw
            & \qw
            \\
            \qw
            & \qw
            & \qw
            & \qw
            & \qw
        \end{quantikz}
        =
        \begin{quantikz}[row sep={0.50cm,between origins}, column sep=0.35cm]
            \qw
            & \qw
            & \qw
            \\
            \qw
            & \ctrl{3}
            & \qw
            \\
            \qw
            & \qw
            & \qw
            \\
            \qw
            & \qw
            & \qw
            \\
            \qw
            & \targ{}
            & \qw
            \\
            \qw
            & \qw
            & \qw
        \end{quantikz}.
    \end{center}

    By conjugating the above logical CXs with swapping of the two codeblocks, accomplished by qubit permutation or by composing tCXs in both directions, their reverse directions can be achieved. 
    Consequently, logical SWAPs on those logical qubits can be achieved.
    Then composing with logical $H^{\otimes 3}$ on the first codeblock gives $H_1$ on that codeblock:
    \begin{center}
        \begin{quantikz}[row sep={0.60cm,between origins}, column sep=0.35cm]
            \lstick[wires=3,braces=right]{codeblock}
                & \gate{H}
                & \qw
                & \gate{H}
                & \qw
                & \qw
                & \gate{H}
                & \qw
                & \qw
                \\
                & \gate{H}
                & \swap{3}
                & \gate{H}
                & \swap{3}
                & \qw
                & \gate{H}
                & \qw
                & \qw
                \\
                & \gate{H}
                & \qw
                & \gate{H}
                & \qw
                & \swap{2}
                & \gate{H}
                & \swap{2}
                & \qw
                \\
            \lstick[wires=3,braces=right]{codeblock}
                & \qw
                & \qw
                & \qw
                & \qw
                & \qw
                & \qw
                & \qw
                & \qw
                \\
                & \qw
                & \targX{}
                & \qw
                & \targX{}
                & \targX{}
                & \qw
                & \targX{}
                & \qw
                \\
                & \qw
                & \qw
                & \qw
                & \qw
                & \qw
                & \qw
                & \qw
                & \qw
        \end{quantikz}
        =
        \begin{quantikz}[row sep={0.60cm,between origins}, column sep=0.35cm]
            & \gate{H} & \qw \\
            & \qw      & \qw \\
            & \qw      & \qw \\
            & \qw      & \qw \\
            & \qw      & \qw \\
            & \qw      & \qw
        \end{quantikz},
        \qquad
    \end{center}

    Lastly, composing with logical $S_1 \mathrm{CZ}_{23}$ on that codeblock gives $(SH)_1$ and $(HS)_1$:
    \begin{center}
        \begin{quantikz}[row sep={0.60cm,between origins}, column sep=0.35cm]
            \lstick[wires=3,braces=right]{codeblock}
                & \gate{S}
                & \gate{H}
                & \gate{S}
                & \gate{H}
                & \qw
                \\
                & \ctrl{1}
                & \qw
                & \ctrl{1}
                & \qw
                & \qw
                \\
                & \control{}
                & \qw
                & \control{}
                & \qw
                & \qw
        \end{quantikz}
        = 
        \begin{quantikz}[row sep={0.60cm,between origins}, column sep=0.35cm]
            & \gate{SH} & \qw \\
            & \qw       & \qw \\
            & \qw       & \qw
        \end{quantikz},
        \qquad
        \begin{quantikz}[row sep={0.50cm,between origins}, column sep=0.35cm]
                & \gate{H}
                & \gate{S}
                & \gate{H}
                & \gate{S}
                & \qw
                \\
                & \qw
                & \ctrl{1}
                & \qw
                & \ctrl{1}
                & \qw
                \\
                & \qw
                & \control{}
                & \qw
                & \control{}
                & \qw
        \end{quantikz}
        =
        \begin{quantikz}[row sep={0.50cm,between origins}, column sep=0.35cm]
            & \gate{HS} & \qw \\
            & \qw       & \qw \\
            & \qw       & \qw
        \end{quantikz}.
    \end{center}
\end{remark}

\clearpage

\section{Stabilizer code results}
\label{app:stab_codes}

This appendix proves the general stabilizer code results of \cref{sec:stab_codes}. The factorization of \cref{app:css_codes} does not hold for general stabilizer codes, so we bound each gate class separately. The proof of the automorphism bound proceeds by ruling out every subgroup of $\Sp(2k,\mathbb{F}_2)$ larger than the Siegel parabolic, so we first collect the group theory this requires, including a classification of large maximal subgroups (\cref{app:stab_codes/prelims}). We then bound the permutation and transversal logical groups (\cref{app:stab_codes/perm,app:stab_codes/trans}), and prove the automorphism bound (\cref{app:stab_codes/auto}). Lastly, we determine the number of physical qubits needed to attain each bound (\cref{app:stab_codes/cost}).


We remark that the material in \cref{app:stab_codes/prelims/additional_subgroups,app:stab_codes/prelims/maximal_subgroups_aschbacher,app:stab_codes/prelims/siegel_parabolic_order_uniqueness} concern detailed group-theoretic calculations and are needed only for the automorphism logical group proofs of \cref{app:stab_codes/auto}; readers interested in the broader-level analysis or our other results may skip them.

\subsection{Definitions and preliminaries}
\label{app:stab_codes/prelims}

\subsubsection{Basic definitions}
\label{app:stab_codes/prelims/basic}

\begin{definition}
    [Standard form of a stabilizer code]
    \label{def:stab_code_standard_form}
    The stabilizer generator matrix of an $\db{n,k}$ stabilizer code can always be placed into the following standard form~\cite{gottesman1997stabilizer}:
    \begin{equation}
        H =
        \left(
        \begin{array}{ccc|ccc}
            I & A_1 & A_2 & B & 0 & C
            \\
            0 & 0 & 0 & D & I & E
        \end{array}
        \right),
    \end{equation}
    where the qubit blocks are $(r, s, k)$ in size, with $n = r + s + k$. That is, $A_1 \in \mathbb F_2^{r\times s}, A_2 \in \mathbb F_2^{r\times k}, B \in \mathbb F_2^{r\times r}, C \in \mathbb F_2^{r\times k}, D \in \mathbb F_2^{s\times r}, E \in \mathbb F_2^{s\times k}$. For $H$ to represent a valid stabilizer code, all stabilizers must commute, so $D^\top = A_1 + A_2 E^\top$ and $B+B^\top = A_2 C^\top + C A_2^\top$.

    A standard-form logical basis is given by
    \begin{equation}
        L_\mathrm{x}
        =
        \left(
        \begin{array}{ccc|ccc}
            0 & E^\top & I & C^\top & 0 & 0
        \end{array}
        \right),
        \qquad
        L_\mathrm{z}
        =
        \left(
        \begin{array}{ccc|ccc}
            0 & 0 & 0 & A_2^\top & 0 & I
        \end{array}
        \right),
    \end{equation}
    such that $L_\mathrm{x} \Omega L_\mathrm{z}^\top = I$. Notably, in this standard-form logical basis, the $Z$-type logical operators comprise only $Z$-type physical Paulis (and identities).
\end{definition}

\subsubsection{Additional subgroups of the symplectic binary group}
\label{app:stab_codes/prelims/additional_subgroups}

We make reference to the $G_2(2)$ subgroup of $\Sp(6, \mathbb{F}_2)$ in \cref{app:stab_codes/prelims/maximal_subgroups_aschbacher} and in the proofs for the automorphism logical group results in \cref{app:stab_codes/auto}, and therefore provide definitions and basic properties here.

\begin{definition}
    [The maximal subgroup $G_2(2) < \Sp(6, \mathbb{F}_2)$]
    \label{def:group_theory_g22_subgroup_of_sp62}
    The finite symplectic group $\Sp(6,\mathbb F_2)$ contains a unique conjugacy class of maximal subgroups isomorphic to the Chevalley group $G_2(2)$ of order 12096~\cite{GAP_character_table_library,ATLAS_v3}; see, for example, Ref.~\cite[Chap.~4]{carter1972simple} for the construction and basic theory of Chevalley groups.
    Throughout this appendix, $G_2(2)<\Sp(6,\mathbb F_2)$ denotes an arbitrary representative of this conjugacy class. 
    At an abstract level, $G_2(2)\cong \operatorname{PSU}_3(3)\rtimes C_2,$ and its derived subgroup is $G_2(2)' \coloneqq \comm{G_2(2)}{G_2(2)} \cong\operatorname{PSU}_3(3)$~\cite{ATLAS_v3}. 
    Under the identification
    $\PCl_3\cong\Sp(6,\mathbb F_2)$, one projective Clifford realization is 
    \begin{equation}
        G_2(2)
        \cong
        \langle
        a \coloneqq \mathrm{CZ}_{12} S_3,
        b \coloneqq H_1 \mathrm{CX}_{23} \mathrm{CX}_{12} \mathrm{CX}_{32}
        \rangle
        =
        \langle
        \mathrm{CZ}_{13},
        H_1 \mathrm{CX}_{23},
        S_1 S_3 \mathrm{CX}_{12}
        \rangle.
    \end{equation}
    For the above projective Clifford representative, an order-$14$ dihedral subgroup $D_{14} < G_2(2)$ can be exhibited by selecting the order-$7$ element $g \coloneqq a b \in G_2(2)$ and the involution $t \coloneqq \mathrm{CX}_{12} S_2 \mathrm{CX}_{32} \mathrm{CZ}_{12} S_2 \in G_2(2)$. Noting that $t g t^{-1} = g^{-1}$, we have $\langle g, t \rangle \cong D_{14}$.
\end{definition}

\begin{proposition}
    [No proper subgroup of $G_2(2)$ larger than Siegel parabolic $\mathcal{P}(6, \mathbb{F}_2)$]
    \label{prop:group_theory_no_proper_subgroup_of_g22_larger_than_siegel_parabolic}
    If $K \leq G_2(2)$ with $\abs{K} > \abs{\mathcal{P}(6, \mathbb{F}_2)}$, then $K = G_2(2)$.
\end{proposition}

\begin{proof}
    Every proper subgroup of \(G_2(2)\) has index at least \(2\), and hence order at most $\abs{G_2(2)}/2 = 6048 < 10752 = \abs{\mathcal P(6,\mathbb F_2)}$.
    Therefore any \(K\leq G_2(2)\) satisfying
    \(\abs{K}>\abs{\mathcal P(6,\mathbb F_2)}\) must equal \(G_2(2)\).
\end{proof}

\subsubsection{Maximal subgroups via Aschbacher classification}
\label{app:stab_codes/prelims/maximal_subgroups_aschbacher}

Maximal subgroups of finite classical groups can be classified into Aschbacher collections~\cite{aschbacher1984maximal,kleidman1990subgroup,bray2013maximal}, namely, the collections $\mathcal C_1, \ldots, \mathcal C_8$ and the almost-simple collection $\mathcal S$. 
(These are not to be confused with the cyclic and symmetric groups.) 
Furthermore, there is a refined classification of large maximal subgroups of various finite classical groups by Ref.~\cite{yin2025large}. 
We use this refined classification to show the following preliminaries. 
These preliminaries are needed only for the proofs of the automorphism logical group results in \cref{app:stab_codes/auto}.

\begin{definition}
    [Large subgroups]
    \label{def:group_theory_large_subgroup}
    A proper subgroup $K$ of a group $G$ is called large iff $\abs{K}^3 \geq \abs{G}$.
\end{definition}

\begin{fact}
    [Largeness of the Siegel parabolic]
    \label{fact:siegel_parabolic_subgroup_large}
    $\mathcal{P}(2k, \mathbb{F}_2)$ is a large subgroup of $\Sp(2k, \mathbb{F}_2)$ for all $k \geq 1$.
\end{fact}

\begin{proof}
    From \cref{eq:group_order_symplectic_upper_bound,eq:group_order_siegel_parabolic_lower_bound}, for all $k \geq 1$,
    \begin{equation}\begin{split}
        \frac{\abs{\mathcal{P}(2k,\mathbb{F}_2)}^3}
            {\abs{\Sp(2k,\mathbb{F}_2)}}
        >
        \frac{2^{(9k^2-3k)/2}}{2^{2k^2+k}}
        =
        2^{(5k^2-5k)/2}
        =
        2^{5k(k-1)/2}
        \ge 1.
    \end{split}\end{equation}
\end{proof}

\begin{proposition}
    [List of large maximal subgroups of the binary symplectic group]
    \label{prop:aschbacher_large_symplectic_maximal_subgroups}
    A large maximal subgroup of $\Sp(2k,\F_2)$, where $k\ge 3$, is one of the types in \Cref{tab:aschbacher_large_symplectic_maximal_subgroups}.
    \begin{table}[!h]
        \centering
        \begin{tabular}{p{2cm} p{11cm} p{4cm}}
        \toprule
        Aschbacher & Subgroup $H$ & Parameters
        \\
        \midrule
        $\mathcal C_1$ 
            & Parabolics $P_r(2k, \mathbb{F}_2)$: 
                stabilizers of totally isotropic dimension-$r$ subspaces
            & $1 \le r \le k$ 
        \\
        $\mathcal C_1$ 
            & Nondegenerate subspace stabilizers: 
                $\Sp(2r, \mathbb{F}_2) \times \Sp(2k - 2r, \mathbb{F}_2)$ 
            & $1 \le r < k/2$ 
        \\
        $\mathcal C_2$ 
            & $\Sp(2m, \mathbb{F}_2) \wr S_2$ 
            & $k = 2m$, $m \geq 2$ 
        \\
        $\mathcal C_2$ 
            & $\Sp(2m, \mathbb{F}_2) \wr S_3$ 
            & $k = 3m$, $m \geq 2$
        \\
        $\mathcal C_3$ 
            & $\Sp(2m, \mathbb{F}_4) \rtimes C_2$ 
            & $k = 2m$, $m \geq 2$ 
        \\
        $\mathcal C_3$ 
            & $\Sp(2m, \mathbb{F}_8) \rtimes C_3$ 
            & $k = 3m$, $m \geq 1$ 
        \\
        $\mathcal C_8$ 
            & $O^+(2k, \mathbb{F}_2)$ 
            & all $k \ge 3$ 
        \\
        $\mathcal C_8$ 
            & $O^-(2k, \mathbb{F}_2)$ 
            & all $k \ge 3$ 
        \\
        $\mathcal S$ 
            & $\mathrm{PSU}_3(3) \rtimes C_2 \cong G_2(2)$ 
            & only in $\Sp(6, \mathbb{F}_2)$ 
        \\
        $\mathcal S$ 
            & $S_{10}$ 
            & only in $\Sp(8, \mathbb{F}_2)$ 
        \\
        $\mathcal S$ 
            & $S_{14}$ 
            & only in $\Sp(12, \mathbb{F}_2)$ 
        \\
        $\mathcal S$ 
            & $S_{18}$ 
            & only in $\Sp(16, \mathbb{F}_2)$ 
        \\
        $\mathcal S$ 
            & $S_{22}$ 
            & only in $\Sp(20, \mathbb{F}_2)$ 
        \\
        \bottomrule
        \end{tabular}
        \caption{List of large maximal subgroups of $\Sp(2k, \mathbb{F}_2)$ by subcollections (i.e.~types) of Aschbacher collection for $k \geq 3$.}
        \label{tab:aschbacher_large_symplectic_maximal_subgroups}
    \end{table}
\end{proposition}

\begin{proof}
    We specialize the classification of large maximal subgroups of projective symplectic groups in Ref.~\cite[Thm.~1.3]{yin2025large} to $\mathrm{P}\Sp(2k,\mathbb F_q)$, and then set $q=2$.
    Since $\mathrm{P}\Sp(2k,\mathbb F_2)=\Sp(2k,\mathbb F_2)$, we obtain the candidate maximal subgroups relevant to our case.
    We translate these candidates into our notation and remove those that are not maximal in $\Sp(2k,\mathbb F_2)$.
    The cases are as follows.

    \begin{itemize}
        \item
        The Aschbacher collection $\mathcal C_1$ consists of stabilizers of proper subspaces.
        Stabilizing an $r$-dimensional totally isotropic subspace gives the parabolic subgroup $P_r(2k,\mathbb F_2)$, while stabilizing a nondegenerate symplectic subspace of dimension $2r$ gives $\Sp(2r,\mathbb F_2)\times\Sp(2k-2r,\mathbb F_2)$.
        For the latter, a nondegenerate subspace and its symplectic orthogonal complement have the same stabilizer, so it suffices to take $r<k/2$.
        When $r=k/2$, the two nondegenerate subspaces can be exchanged by a symplectic transformation.
        Thus, $\Sp(k,\mathbb F_2)\times\Sp(k,\mathbb F_2)$ is properly contained in $\Sp(k,\mathbb F_2)\wr S_2$, which is itself properly contained in $\Sp(2k,\mathbb F_2)$.
        Therefore, $\Sp(k,\mathbb F_2)\times\Sp(k,\mathbb F_2)$ is not maximal.

        \item
        There are two types of candidates in the symplectic $\mathcal C_2$ collection.
        The first is the stabilizer of a decomposition $\mathbb F_2^{2k}=V_1\perp\cdots\perp V_t$ into $t$ isometric nondegenerate symplectic subspaces of common dimension $d=2k/t$.
        The subgroup preserving the decomposition as an unordered collection is $\Sp(d,\mathbb F_2)\wr S_t=\Sp(d,\mathbb F_2)^t\rtimes S_t$.
        After setting $q=2$ in Ref.~\cite[Thm.~1.3]{yin2025large}, the large possibilities have $t=2$ or $t=3$, together with the candidate $(2k,t)=(8,4)$.
        The maximality condition excludes $(d,q)=(2,2)$~\cite[Prop.~2.3.6]{bray2013maximal}.
        Thus, for $t=2$, writing $k=2m$ gives $\Sp(2m,\mathbb F_2)\wr S_2$ with $m\geq2$, while for $t=3$, writing $k=3m$ gives $\Sp(2m,\mathbb F_2)\wr S_3$ with $m\geq2$.
        The exceptional $(2k,t)=(8,4)$ case has $d=2$ and is excluded by the same maximality condition.

        The second $\mathcal C_2$ candidate is $\GL(k,\mathbb F_2)\rtimes C_2$, the stabilizer of an unordered pair of complementary Lagrangians.
        In the standard decomposition $\mathbb F_2^{2k}=\mathcal X\oplus\mathcal Z$, its elements preserve the plus-type quadratic form $Q^+(x,z)=x\cdot z$ from \cref{def:standard_quad_form}.
        Hence $\GL(k,\mathbb F_2)\rtimes C_2<O^+(2k,\mathbb F_2)<\Sp(2k,\mathbb F_2)$, where the first inclusion is proper for $k\geq3$.
        Thus, this candidate is not maximal.

        \item
        The $\mathcal C_3$ collection consists of subgroups obtained by regarding the natural binary symplectic space as a vector space over a proper extension of $\mathbb F_2$.
        By Ref.~\cite[Thm.~1.3]{yin2025large}, the large $\mathcal C_3$ subgroups for $\mathrm P\Sp(n,\F_q)$ are of type $\Sp(n/2,\F_{q^2})$, $\Sp(n/3,\F_{q^3})$, and $\Gamma\mathrm{U}_{n/2}(q)$, which is the normalizer of $\GU_k(2)$.
        We specialize to $n=2k$ and $q=2$.

        For the extension of degree two, the natural $2k$-dimensional space over $\mathbb F_2$ becomes a $k$-dimensional space over $\mathbb F_4$.
        Since a nondegenerate symplectic space has even dimension, write $k=2m$.
        The corresponding extension-field subgroup has linear part $\Sp(2m,\mathbb F_4)$, and its normalizer also contains the nontrivial field automorphism $x\mapsto x^2$ of $\mathbb F_4/\mathbb F_2$.
        Hence the full subgroup is $\Sp(2m,\mathbb F_4)\rtimes C_2$, with $m\geq2$ since $k\geq3$.

        For the extension of degree three, the binary space becomes a $2k/3$-dimensional space over $\mathbb F_8$.
        Thus $3$ divides $k$, and writing $k=3m$ gives linear part $\Sp(2m,\mathbb F_8)$.
        Since $x\mapsto x^2$ generates $\operatorname{Gal}(\mathbb F_8/\mathbb F_2)\cong C_3$, the full subgroup is $\Sp(2m,\mathbb F_8)\rtimes C_3$, with $m\geq1$.

        It remains to exclude $\Gamma\mathrm{U}_k(2)=\GU_k(2)\rtimes \la F\ra$, where $F:v\mapsto v^2$, as a maximal subgroup. We show that $\Gamma\mathrm{U}_k(2)<O^\epsilon(2k,\F_2)$ by observing that its elements preserve a nondegenerate binary quadratic form $Q(v)$, $v\in \F_2^{2k}\cong \F_4^k$.
        Let $Q(v)=h(v,v)$, where $h:\mathbb F_4^k\times\mathbb F_4^k\to\mathbb F_4$ is a nondegenerate Hermitian form, i.e., $h(v,u)=\overline{h(u,v)}=h(u,v)^2$.
        By the convention of~\cite{yin2025large}, $\GU_k(2)\coloneqq\{A\in\GL(k,\F_4):h(Au,Av)=h(u,v)\}$.
        Choose an $h$-orthonormal basis, so that $h(u,v)=\sum_{i=1}^k u_i v_i^2$.
        The coordinate-wise Frobenius $F(v_1,\dots,v_k)=(v_1^2,\dots,v_k^2)$ is $\F_2$-linear in each argument.
        In this basis, $Q(v)=h(v,v)=\sum_{i=1}^k v_i^3\in \F_2$, since $v_i^3\in\F_2$ for every $v_i\in\F_4$ and $i\in[k]$.
        Its polar form $B_Q(u,v)=Q(u+v)+Q(u)+Q(v)=h(u,v)+h(v,u)=\tr_{\F_4/\F_2}(h(u,v))$ is nondegenerate because, for every $u\neq 0$, $h(u,\cdot)$ is surjective onto $\F_4$.
        Both parts of $\Gamma\mathrm{U}_k(2)$ preserve $Q$: $Q(Av)=Q(v)$ for all $A\in\GU_k(2)$, and $Q(Fv)=h(Fv,Fv)=h(v,v)^2=Q(v)$.
        Consequently, $\Gamma\mathrm{U}_k(2)$ is not a maximal subgroup of $\Sp(2k,\F_2)$.

        \item
        By Ref.~\cite[Thm.~1.3]{yin2025large}, no large $\mathcal C_4$ or $\mathcal C_7$ subgroups occur.
        Moreover, no large $\mathcal C_5$ subgroup occurs because $\mathbb F_2$ has no proper subfields, while the large $\mathcal C_6$ subgroups occur only in the exceptional cases listed in Ref.~\cite[Thm.~1.3]{yin2025large}, none of which has $q=2$.

        \item
        The $\mathcal C_8$ collection consists of subgroups preserving a nondegenerate quadratic form whose polar form is the ambient symplectic form.
        By \cref{fact:classification_nondegen_bqs}, every nondegenerate quadratic form on a $2k$-dimensional binary symplectic space is, up to isometry, of exactly one of the two types $Q^+$ and $Q^-$.
        By \cref{def:group_theory_binary_orthogonal_groups}, their isometry groups are $O^+(2k,\mathbb F_2)$ and $O^-(2k,\mathbb F_2)$, respectively.
        Since both $Q^+$ and $Q^-$ polarize to the standard symplectic form, we have $O^\pm(2k,\mathbb F_2)<\Sp(2k,\mathbb F_2)$.
        By Ref.~\cite[Thm.~1.3]{yin2025large}, these are large maximal subgroups for all $k\geq3$.

        \item
        It remains to analyze the $\mathcal S$ collection.
        Ref.~\cite[Thm.~1.3]{yin2025large} reduces the large $\mathcal S$ subgroups to the explicit finite lists in Ref.~\cite[Tabs.~3 and 4]{yin2025large}.

        Among the nonalternating cases in Ref.~\cite[Tab.~4]{yin2025large}, the entries whose ambient group is a binary symplectic group of dimension at least six are $\mathrm{PSU}_3(3)\rtimes C_2<\mathrm P\Sp(6,\mathbb F_2)$ and $G_2(2)<\mathrm P\Sp(6,\mathbb F_2)$.
        Since $\mathrm P\Sp(6,\mathbb F_2)=\Sp(6,\mathbb F_2)$ and, by \cref{def:group_theory_g22_subgroup_of_sp62}, $\mathrm{PSU}_3(3)\rtimes C_2\cong G_2(2)$ and there is a unique conjugacy class of such maximal subgroups, these give the single $G_2(2)$ row for $k=3$.

        For the alternating and symmetric cases in Ref.~\cite[Tab.~3]{yin2025large}, the entries explicitly having binary symplectic ambient group and $k\geq3$ are $S_{10}<\Sp(8,\mathbb F_2)$, $S_{14}<\Sp(12,\mathbb F_2)$, $S_{18}<\Sp(16,\mathbb F_2)$, and $S_{22}<\Sp(20,\mathbb F_2)$.
        Ref.~\cite[Tab.~3]{yin2025large} also lists $(\mathrm P\Omega_9(2),A_{10})$, giving an $A_{10}$ candidate in $\Sp(8,\mathbb F_2)$ since $\mathrm P\Omega_9(2)\cong\Sp(8,\mathbb F_2)$.
        However, the maximal-subgroup classification of $\Sp(8,\mathbb F_2)$ gives $S_{10}\cong A_{10}\rtimes C_2$ as a maximal subgroup containing the relevant $A_{10}$~\cite{bray2013maximal,GAP_character_table_library}.
        Hence $A_{10}<S_{10}<\Sp(8,\mathbb F_2)$, so $A_{10}$ is not maximal and contributes no additional row.
        All other entries of Ref.~\cite[Tabs.~3 and 4]{yin2025large} either have a different ambient classical group, a field size different from two, or symplectic dimension less than six.
    \end{itemize}

    These are exactly the rows of \cref{tab:aschbacher_large_symplectic_maximal_subgroups}.
\end{proof}

\begin{proposition}
    [Maximal irreducible subgroups of binary symplectic group no smaller than Siegel parabolic]
    \label{prop:aschbacher_symplectic_maximal_subgroups_at_least_size_of_siegel_parabolic}
    For $k\geq3$, the maximal irreducible subgroups $G$ of $\Sp(2k,\mathbb F_2)$ satisfying $\abs{G}\geq\abs{\mathcal P(2k,\mathbb F_2)}$ are, up to conjugacy, $O^+(2k,\mathbb F_2)$ and $O^-(2k,\mathbb F_2)$ for all $k\geq3$, together with $\mathrm{PSU}_3(3)\rtimes C_2\cong G_2(2)$ for $k=3$.
\end{proposition}

\begin{proof}
    We require $\abs{G} \geq \abs{\mathcal{P}(2k, \mathbb{F}_2)}$ and $\abs{\mathcal{P}(2k, \mathbb{F}_2)}^3 \geq \abs{\Sp(2k, \mathbb{F}_2)}$ by \cref{fact:siegel_parabolic_subgroup_large}, so $G$ must be large. 
    We also require $G$ to be irreducible (see \cref{def:group_theory_binary_reducible_subgroups}). 
    It therefore suffices to examine each row of \cref{tab:aschbacher_large_symplectic_maximal_subgroups} and remove the row if the group is reducible or its order is smaller than the order of the corresponding Siegel parabolic group. 
    First, using \cref{eq:group_order_symplectic_upper_bound,eq:group_order_siegel_parabolic_lower_bound}, we state the useful bounds
    \begin{equation}
        \abs{\Sp(2m,\mathbb F_2)}
        \leq
        2^{2m^2+m},
        \qquad
        \abs{\mathcal{P}(4m,\mathbb F_2)}
        \geq 
        2^{6m^2-m},
        \qquad
        \abs{\mathcal{P}(6m,\mathbb F_2)}
        \geq 
        2^{(27m^2-3m)/2},
    \end{equation}
    for all $m \geq 1$. Then consider each row:
    
    \begin{itemize}
        
        \item
        $\mathcal C_1$ collection, parabolic subgroups \(P_r(2k,\mathbb F_2)\) and stabilizers of nondegenerate \(2r\)-dimensional symplectic subspaces. 
        Both families stabilize a nonzero proper subspace and are therefore reducible.

        \item
        $\mathcal C_2$ collection, $\Sp(2m,\mathbb F_2)\wr S_2$, where $k=2m$ and $m \geq 2$.
        We have
        \begin{equation}
            \abs{\Sp(2m,\mathbb F_2)\wr S_2}
            =
            2 \cdot \abs{\Sp(2m,\mathbb F_2)}^2
            \leq 2^{4m^2+2m+1},
        \end{equation}
        and since $6m^2-m>4m^2+2m+1$ for $m \geq 2$, this subgroup is smaller than $\mathcal{P}(2k, \mathbb{F}_2)$.

        \item
        $\mathcal C_2$ collection, $\Sp(2m,\mathbb F_2)\wr S_3$, where $k=3m$ and $m \geq 2$.
        We have
        \begin{equation}
            \abs{\Sp(2m,\mathbb F_2)\wr S_3}
            =
            6 \cdot \abs{\Sp(2m,\mathbb F_2)}^3
            \leq 2^{6m^2+3m+3},
        \end{equation}
        and since $(27m^2-3m)/2>6m^2+3m+3$ for $m \geq 2$, this subgroup is smaller than $\mathcal{P}(2k, \mathbb{F}_2)$.

        \item
        $\mathcal C_3$ collection, $\Sp(2m, \mathbb{F}_4) \rtimes C_2$, where $k=2m$ and $m \geq 2$.
        Noting that $\abs{\Sp(2m, \mathbb{F}_4)} < 2 \cdot \abs{\Sp(2m, \mathbb{F}_2)}^2$, we have
        \begin{equation}
            \abs{\Sp(2m, \mathbb{F}_4) \rtimes C_2}
            =
            2 \cdot \abs{\Sp(2m, \mathbb{F}_4)}
            <
            4 \cdot \abs{\Sp(2m, \mathbb{F}_2)}^2
            < 2^{4m^2+2m+2},
        \end{equation}
        and since $6m^2-m\geq4m^2+2m+2$ for $m \geq 2$, this subgroup is smaller than $\mathcal{P}(2k, \mathbb{F}_2)$.

        \item
        $\mathcal C_3$ collection, $\Sp(2m, \mathbb{F}_8) \rtimes C_3$, where $k=3m$ and $m \geq 1$.
        Noting that $\abs{\Sp(2m, \mathbb{F}_8)} < 4 \cdot \abs{\Sp(2m, \mathbb{F}_2)}^3$, we have
        \begin{equation}
            \abs{\Sp(2m, \mathbb{F}_8) \rtimes C_3}
            =
            3 \cdot \abs{\Sp(2m, \mathbb{F}_8)}
            <
            12 \cdot \abs{\Sp(2m, \mathbb{F}_2)}^3
            < 2^{6m^2+3m+4},
        \end{equation}
        and $(27m^2-3m)/2>6m^2+3m+4$ for $m \geq 2$.
        At $m = 1$, direct calculation shows $\abs{\Sp(2, \mathbb{F}_8) \rtimes C_3} < \abs{\mathcal{P}(6, \mathbb{F}_2)}$.
        So this subgroup is smaller than $\mathcal{P}(2k, \mathbb{F}_2)$.

        \item
        $\mathcal C_8$ collection, $O^\pm(2k, \mathbb{F}_2)$.
        From \cref{eq:group_order_orthogonal_plus,eq:group_order_siegel_parabolic}, we have for $k \geq 3$,
        \begin{align}
            \frac{|O^+(2k, 2)|}{|\mathcal{P}(2k, 2)|} = 2^{1-k}\prod_{i = 1}^{k-1} (2^i + 1) > 2^{1-k} 2^{1 + 2 + \ldots + (k-1)} = 2^{(k-1)(k-2)/2} > 1.
        \end{align}
        Moreover, from \cref{eq:group_order_orthogonal_plus,eq:group_order_orthogonal_minus}, we have
        \begin{equation}
            \frac{|O^-(2k,2)|}{|O^+(2k,2)|}
            =
            \frac{2^k+1}{2^k-1}
            >
            1.
        \end{equation}
        Thus, both subgroups are irreducible and larger than $\mathcal{P}(2k, \mathbb{F}_2)$ for all $k \geq 3$.

        \item
        $\mathcal S$ collection. The four symmetric-group subgroups are smaller than $\mathcal{P}(2k,\mathbb F_2)$ by direct calculation using $|S_{2k+2}| = (2k + 2)! < |\mathcal{P}(2k, 2)|$ for the $k = 4, 6, 8$, and $10$ cases.
        For $k=3$, however, $\abs{G_2(2)}=12096>10752=\abs{\mathcal{P}(6,\mathbb F_2)}$.
        Moreover, the $\mathcal S$ subgroups are irreducible.
        Thus, $G_2(2)$ gives the sole $\mathcal S$ subgroup satisfying the required conditions.
        
    \end{itemize}
\end{proof}

\begin{proposition}
    [Large irreducible maximal subgroups of binary orthogonal groups are smaller than Siegel parabolic]
    \label{prop:aschbacher_large_irreducible_orthogonal_maximal_subgroups_below_siegel}
    Let \(k\geq 4\) and \(\epsilon\in\{+,-\}\). Then the following holds:
    \begin{enumerate}
        \item If \(K < \Omega^\epsilon(2k,\mathbb F_2)\) is a large maximal subgroup of \(\Omega^\epsilon(2k,\mathbb F_2)\) that is irreducible on the natural symplectic space $V\coloneqq\mathbb{F}_2^{2k}$, then $|K|<|\mathcal{P}(2k, \mathbb{F}_2)|$.
        \label{enum:aschbacher_large_irreducible_orthogonal_maximal_subgroups_below_siegel_case_1}
        \item If \(K' < O^\epsilon(2k,\mathbb F_2)\) is a large maximal subgroup of \(O^\epsilon(2k,\mathbb F_2)\) that is irreducible on the natural symplectic space $V\coloneqq\mathbb{F}_2^{2k}$ and does not
        contain \(\Omega^\epsilon(2k,\mathbb F_2)\), then $|K'|<|\mathcal{P}(2k, \mathbb{F}_2)|$.
        \label{enum:aschbacher_large_irreducible_orthogonal_maximal_subgroups_below_siegel_case_2}
    \end{enumerate}
\end{proposition}

\begin{proof}
    We prove both assertions simultaneously.
    For notational uniformity, let $K$ denote the subgroup in either Part~\cref{enum:aschbacher_large_irreducible_orthogonal_maximal_subgroups_below_siegel_case_1} or Part~\cref{enum:aschbacher_large_irreducible_orthogonal_maximal_subgroups_below_siegel_case_2}.
    Define $K^\Omega\coloneqq K \cap \Omega^\epsilon(2k,\mathbb F_2)$.
    If \(K \leq \Omega^\epsilon(2k,\mathbb F_2)\) as in Part~\cref{enum:aschbacher_large_irreducible_orthogonal_maximal_subgroups_below_siegel_case_1}, then \(K^\Omega = K\).
    For Part~\cref{enum:aschbacher_large_irreducible_orthogonal_maximal_subgroups_below_siegel_case_2}, the maximal subgroup $K<O^\epsilon(2k, \mathbb F_2)$ cannot be contained in \(\Omega^\epsilon(2k,\mathbb F_2)\), since otherwise $K<\Omega^\epsilon(2k,\mathbb F_2)<O^\epsilon(2k,\mathbb F_2)$ would contradict maximality.
    By hypothesis, \(K\) also does not contain \(\Omega^\epsilon(2k,\mathbb F_2)\).
    Hence $K\Omega^\epsilon(2k,\mathbb F_2)=O^\epsilon(2k,\mathbb F_2)$.
    Since \(\Omega^\epsilon(2k,\mathbb F_2)\) has index two in \(O^\epsilon(2k,\mathbb F_2)\), the second isomorphism theorem gives $[K:K\cap\Omega^\epsilon(2k,\mathbb F_2)]=2$, i.e., $|K|=2|K^\Omega|$.
    Thus in every case it is sufficient to show $2|K^\Omega|<|\mathcal{P}(2k, \mathbb{F}_2)|$.
    Note that in Part~\cref{enum:aschbacher_large_irreducible_orthogonal_maximal_subgroups_below_siegel_case_2}, \(K^\Omega\) has index two in \(K\) and is therefore normal.
    Hence its unique nontrivial left and right cosets in \(K\) coincide, both being \(K\setminus K^\Omega\).

    We first eliminate the case in which \(K^\Omega\) is reducible.
    This can only occur in Part~\cref{enum:aschbacher_large_irreducible_orthogonal_maximal_subgroups_below_siegel_case_2}, since in Part~\cref{enum:aschbacher_large_irreducible_orthogonal_maximal_subgroups_below_siegel_case_1} we have \(K^\Omega = K\), which is irreducible by premise.
    Choose an element \(g\in K\setminus K^\Omega\).
    Since \(K^\Omega \triangleleft K\) has index \(2\), the group \(K\) is generated by \(K^\Omega\) together with \(g\).
    Let \(U\leq V\coloneqq\mathbb{F}_2^{2k}\) be a nonzero proper \(K^\Omega\)-invariant subspace, which exists because \(K^\Omega\) is reducible by assumption.

    Let us first state a simple observation that we will invoke repeatedly.
    If \(W\leq V\) is \(K^\Omega\)-invariant, then \(gW\) is also \(K^\Omega\)-invariant.
    Indeed, for every \(a\in K^\Omega\), \(a(gW)=g(g^{-1}ag)W\).
    Since \(K^\Omega \triangleleft K\), we have \(g^{-1}ag\in K^\Omega\), and therefore \(a(gW)=gW\).
    Moreover, \(g\) swaps \(W\) and \(gW\), because \(g^2W=W\) using $g^2\in K^\Omega$ and that $W$ is \(K^\Omega\)-invariant.
    Hence both \(W+gW\) and \(W\cap gW\) are \(K\)-invariant subspaces of \(V\).

    Applying this observation to \(W=U\), irreducibility of \(K\) implies \(U+gU=V\) and \(U\cap gU=\{0\}\).
    Indeed, \(U+gU\) is a nonzero \(K\)-invariant subspace and hence must be \(V\), while \(U\cap gU\leq U<V\) is a \(K\)-invariant subspace and hence must be \(\{0\}\).

    Thus $V=U\oplus gU$ and $\dim U = (\dim V) / 2 = k$.
    Next consider the radical of \(U\) with respect to the ambient symplectic form, $R \coloneqq \rad(U) = U\cap U^\perp$.
    Since \(K^\Omega\leq \Sp(2k,\mathbb F_2)\) preserves the symplectic form and preserves \(U\), it also preserves \(U^\perp\), and therefore preserves $R$.
    That is, \(R\) is \(K^\Omega\)-invariant.
    We apply the observation above to \(W=R\).
    Since \(R\leq U\) and \(gR\leq gU\), and since \(U\cap gU=0\), we have $R\cap gR=\{0\}$.
    Therefore, if \(R\neq \{0\}\), irreducibility of \(K\) forces $R+gR = R \oplus gR = V$, and \(\dim R=k\).
    Since \(R\leq U\) and \(\dim U=k\), this implies $R=U$.
    Thus either $R = \{0\}$ or $R = U$.
    We consider these two cases one at a time.

    \begin{itemize}

        \item $R = \{0\}$.
        Then \(U\) is a nondegenerate \(k\)-dimensional subspace of \(V\).
        Since a nondegenerate alternating form has even dimension, we must have $k=2a$ for some integer \(a\geq 2\).
        Since \(U\) is nondegenerate, we have the symplectic orthogonal decomposition $V=U\perp U^\perp$.

        Recall that the subgroup \(K^\Omega\) preserves both \(U\) and \(U^\perp\).
        Therefore restriction to the two summands gives an injective homomorphism \(K^\Omega \hookrightarrow \Sp(U)\times \Sp(U^\perp) \cong \Sp(2a,\mathbb F_2)\times \Sp(2a,\mathbb F_2)\).
        Since we are in Part~\cref{enum:aschbacher_large_irreducible_orthogonal_maximal_subgroups_below_siegel_case_2}, we have $[K:K^\Omega]=2$.
        Then
        \begin{equation}
            \abs{K}
            = 2 \abs{K^\Omega}
            \leq 2 \abs{\Sp(2a,\mathbb F_2)}^2
            \leq 2^{4a^2+2a+1},
            \qquad
            \abs{\mathcal{P}(2k, \mathbb{F}_2)}
            = \abs{\mathcal{P}(4a, \mathbb{F}_2)}
            \geq 2^{6a^2-a},
        \end{equation}
        where we have used \cref{eq:group_order_symplectic_upper_bound,eq:group_order_siegel_parabolic_lower_bound}.
        For $a \geq 2$, we have $4a^2+2a+1<6a^2-a$, so $\abs{K} < \abs{\mathcal{P}(2k, \mathbb{F}_2)}$.

        \item $R = U$.
        In this case, $U \leq U^\perp$, so the symplectic form vanishes identically on \(U\).
        Since \(\dim U=k\) and \(\dim V=2k\), the subspace \(U\) is a maximally totally isotropic subspace of the symplectic space \(V\).
        Equivalently, \(U\) is a Lagrangian subspace.
        Since \(g\in K\leq O^\epsilon(2k,\mathbb F_2)\leq \Sp(2k,\mathbb F_2)\), the subspace \(gU\) is also Lagrangian.
        Moreover, recall $V=U\oplus gU$.
        So \(U\) and \(gU\) are complementary Lagrangian subspaces.

        The subgroup \(K\) preserves the unordered pair $\{U,gU\}$.
        Indeed, \(K^\Omega\) preserves \(U\) and \(gU\) individually, while \(g\) swaps them.
        Since \(K\) is generated by \(K^\Omega\) and \(g\), every element of \(K\) preserves the unordered pair \(\{U,gU\}\).
        Therefore \(K\) is contained in the stabilizer in \(\Sp(2k,\mathbb F_2)\) of the unordered pair of complementary Lagrangian subspaces $\{U,gU\}$.
        The stabilizer of the ordered pair \((U,gU)\) has order \(|\GL(k,\mathbb F_2)|\).
        To see this, choose any \(A\in \GL(U)\).
        Preservation of the nondegenerate symplectic pairing between \(U\) and \(gU\) forces the action on \(gU\) to be the dual inverse action \(A^{-\top}\).
        Thus an element stabilizing \(U\) and \(gU\) individually is determined by its arbitrary action \(A\) on \(U\).
        Allowing the possible swap of the two Lagrangian summands contributes at most one additional factor of \(2\).
        Hence the unordered-pair stabilizer has order $\le 2|\GL(k,\mathbb F_2)|$.

        Using \cref{eq:group_order_siegel_parabolic,eq:group_order_unipotent_radical}, we find $\abs{K} \leq 2|\GL(k,\mathbb F_2)| < \abs{\mathcal{P}(2k, \mathbb{F}_2)}$ for $k \geq 4$.

    \end{itemize}

    It remains to consider \(K^\Omega\) that is irreducible.
    Here, we apply the refined Aschbacher classification result of Ref.~\cite[Thm.~1.4]{yin2025large}.
    For Part~\cref{enum:aschbacher_large_irreducible_orthogonal_maximal_subgroups_below_siegel_case_1}, we set the ambient group to be \(G_0 = G = P\Omega^\epsilon(2k,\mathbb F_2) = \Omega^\epsilon(2k,\mathbb F_2)\).
    Accordingly, there is no distinction between the subgroups $H$ and $H_0$ in the theorem, and \(H = H_0 = K = K^\Omega\) for our purpose.
    For Part~\cref{enum:aschbacher_large_irreducible_orthogonal_maximal_subgroups_below_siegel_case_2}, we set \(G_0 = \Omega^\epsilon(2k,\mathbb F_2)\) and \(G = O^\epsilon(2k,\mathbb F_2)\), and $K^\Omega$ is then the subgroup denoted $H_0$ in the theorem.
    Notice that in this second case we do not need \(K^\Omega\) itself to be large in \(\Omega^\epsilon(2k,\mathbb F_2)\).
    The largeness hypothesis is on \(K\) as a subgroup of \(O^\epsilon(2k,\mathbb F_2)\), as precisely required for our proposition, and Ref.~\cite[Thm.~1.4]{yin2025large} is formulated precisely to classify the possible intersections \(K^\Omega\).
    Because $\abs{K} = \abs{K^\Omega}$ in Part~\cref{enum:aschbacher_large_irreducible_orthogonal_maximal_subgroups_below_siegel_case_1} and $\abs{K} = 2 \abs{K^\Omega}$ in Part~\cref{enum:aschbacher_large_irreducible_orthogonal_maximal_subgroups_below_siegel_case_2}, it suffices to bound $2 \abs{K^\Omega} < \abs{\mathcal{P}(2k, \mathbb{F}_2)}$ to cover both cases.

    Since \(K^\Omega\) is irreducible, the $\mathcal C_1$ Aschbacher collection is excluded because its subgroups stabilize nonzero proper subspaces.
    In the $\mathcal C_2$ Aschbacher collection, the reducible odd-dimensional orthogonal product case appearing as \(O^\circ(k,\mathbb F_2)^2\) is also excluded, since it preserves the two summands individually.
    Likewise, in the $\mathcal C_3$ Aschbacher collection, the odd-dimensional extension-field type \(\GO^\circ(k,\mathbb F_4)\) is also excluded.
    In characteristic two, the polar form of an odd-dimensional nonsingular quadratic form has a canonical one-dimensional radical over \(\mathbb F_4\).
    This radical is invariant under the orthogonal group and under the associated field automorphism.
    After restriction of scalars to \(\mathbb F_2\), it is a two-dimensional \(K^\Omega\)-invariant binary subspace.
    Thus this case is reducible.
    In addition, we have chosen to omit the Aschbacher collections $\mathcal C_5$ and $\mathcal C_6$ from \cref{tab:aschbacher_large_irreducible_orthogonal_maximal_subgroups_below_siegel}.
    No \(\mathcal C_5\) case occurs after specializing to \(q=2\).
    The \(\mathcal C_5\) possibilities in Ref.~\cite[Thm.~1.4]{yin2025large} require \(q\) to be a square or a cube of a smaller prime power, whereas \(2\) admits no such proper subfield representation.
    No \(\mathcal C_6\) case occurs either since the unique \(\mathcal C_6\) case in Ref.~\cite[Thm.~1.4]{yin2025large} has \(q=3\).
    There are no $\mathcal C_7$ or $\mathcal C_8$ cases in Ref.~\cite[Thm.~1.4]{yin2025large}.

    After specializing the remaining entries of the theorem to \(q=2\), the subgroup \(K^\Omega\), which is the same as $H_0$ in Ref.~\cite[Thm.~1.4]{yin2025large}, is one of the following types listed in \cref{tab:aschbacher_large_irreducible_orthogonal_maximal_subgroups_below_siegel}.
    The fourth column gives an upper bound for \(2|K^\Omega|\).
    We do not require the group displayed in the second column to be a literal ambient container for \(K^\Omega\).

    \begin{table}[!h]
        \centering
        \begin{tabular}{p{1.9cm} p{3.5cm} p{7cm} p{3.5cm}}
            \toprule
            Aschbacher
            & Possible type of \(K^\Omega\)
            & Conditions
            & Upper bound for \(2|K^\Omega|\)
            \\
            \midrule
            \(\mathcal C_2\)
            &
            \(\GL(k,\mathbb F_2).2\)
            &
            \(\epsilon=+\)
            &
            \(4|\GL(k,\mathbb F_2)|\)
            \\
            \(\mathcal C_2\)
            &
            \(O^\eta(k,\mathbb F_2)\wr S_2\)
            &
            \(\epsilon=+\), \(k\) even, \(\eta\in\{+,-\}\)
            &
            \(4|O^\eta(k,\mathbb F_2)|^2\)
            \\
            \(\mathcal C_2\)
            &
            \(3^4.2^3.S_4\)
            &
            \((k,\epsilon)=(4,+)\)
            &
            \(31104\)
            \\
            \(\mathcal C_2\)
            &
            \(3^5.2^4.S_5\)
            &
            \((k,\epsilon)=(5,-)\)
            &
            \(933120\)
            \\
            \(\mathcal C_2\)
            &
            \(A_5^3.2^2.S_3\)
            &
            \((k,\epsilon)=(6,-)\)
            &
            \(10368000\)
            \\
            \(\mathcal C_3\)
            &
            \(O^\epsilon(k,\mathbb F_4)\)
            &
            \(k\) even
            &
            \(4|O^\epsilon(k,\mathbb F_4)|\)
            \\
            \(\mathcal C_3\)
            &
            \(\GU_k(2)\)
            &
            \(\epsilon=+\) if \(k\) is even, and \(\epsilon=-\) if \(k\) is odd
            &
            \(4|\GU_k(2)|\)
            \\
            \(\mathcal C_4\)
            &
            \(\Sp_2(2)\otimes \Sp_k(2)\)
            &
            \((k,\epsilon)=(4,+),(6,+)\)
            &
            \(2|\Sp(2, \mathbb{F}_2)|\,|\Sp(k, \mathbb{F}_2)|\)
            \\
            \(\mathcal S\)
            &
            \(A_d\)
            &
            Some subset of \(d\leq 2k+2\)
            &
            \(d!\)
            \\
            \(\mathcal S\)
            &
            \(P\Omega_7(2)\cong \Sp(6, \mathbb{F}_2)\)
            &
            \((k,\epsilon)=(4,+)\)
            &
            \(2|\Sp(6, \mathbb{F}_2)|\)
            \\
            \bottomrule
        \end{tabular}
        \caption{Possible types of \(K^\Omega\) obtained by specializing Ref.~\cite[Thm.~1.4]{yin2025large}, excluding obvious reducible cases.}
        \label{tab:aschbacher_large_irreducible_orthogonal_maximal_subgroups_below_siegel}
    \end{table}

    We now compare each upper bound for $2 \abs{K^\Omega}$ with \(|\mathcal{P}(2k, \mathbb{F}_2)|\).
    \begin{itemize}

        \item $\mathcal C_2$ rows.
        First, from \cref{eq:group_order_general_linear,eq:group_order_siegel_parabolic}, we find that $4|\GL(k,\mathbb F_2)| < \abs{\mathcal{P}(2k, \mathbb{F}_2)}$ for $k \geq 4$.
        For the next row, using \cref{eq:group_order_orthogonal_upper_bound,eq:group_order_siegel_parabolic_lower_bound}, we have
        \begin{equation}
            4|O^\eta(k,\mathbb F_2)|^2
            <
            2^{k^2-k+6},
            \qquad
            \abs{\mathcal{P}(2k, \mathbb F_2)}
            \geq
            2^{(3k^2-k)/2},
        \end{equation}
        and $k^2-k+6 < (3k^2-k)/2$ for $k \geq 4$.
        For the next three exceptional rows,
        \begin{equation}
            31104<\abs{\mathcal{P}(8, \mathbb{F}_2)},
            \qquad
            933120<\abs{\mathcal{P}(10, \mathbb{F}_2)},
            \qquad
            10368000<\abs{\mathcal{P}(12, \mathbb{F}_2)}.
        \end{equation}

        \item $\mathcal C_3$ rows.
        First, writing $k=2m$ where $m\ge 2$, and using \cref{eq:group_order_orthogonal_plus,eq:group_order_orthogonal_minus}, we have
        \begin{equation}
            |O^\epsilon(2m, 4)| 
            = 2 \cdot 4^{m(m-1)} (4^m \mp 1) \prod_{i = 1}^{m-1} (4^{2i} - 1)
            = 2\cdot 4^{2m^2-m}(1\mp 4^{-m})\prod_{i=1}^{m-1}(1-4^{-2i}).
        \end{equation}
        Therefore, we have $|O^\epsilon(k,\mathbb F_4)|< 2 \cdot 4^{2m^2 - m} < 2^{k^2}$.
        Thus, $4|O^\epsilon(k,\mathbb F_4)| < 2^{k^2+2} < \abs{\mathcal{P}(2k, \mathbb{F}_2)}$ for $k \geq 4$.

        Next, let us analyze the $\GU_k(2)$ type.
        The order of the $\GU_k(2)$ group is given by $2^{k(k-1)/2}\prod_{i=1}^k(2^i-(-1)^i)$~\cite{carter1972simple}.
        In addition, for odd $i \geq 3$, we have $\frac{2^i - 1}{2^i + 1} = 1 - \frac{2}{2^i + 1} > 1 - 2^{1-i}$.
        Therefore, for $k \geq 4$,
        \begin{equation}
            \frac{|\mathcal{P}(2k, \mathbb{F}_2)|}{4|\GU_k(2)|}
            =
            \frac{2^{k(k+1)/2-2}}{3}
            \left(\prod_{\substack{3\leq i\leq k\\ i\ \mathrm{odd}}}
            \frac{2^i-1}{2^i+1}\right)
            \geq
            \frac{2^8}{3}
            \left(1 - \sum_{\substack{i \geq 3 \\ i\ \mathrm{odd}}}2^{1-i}\right)
            =
            2^8 \cdot \frac{2}{9}
            > 1.
        \end{equation}

        \item $\mathcal C_4$ row.
        By direct calculation,
        \begin{equation}
            2|\Sp(2, \mathbb{F}_2)|\,|\Sp(4, \mathbb{F}_2)|
            =
            8640
            <
            \abs{\mathcal{P}(8, \mathbb{F}_2)},
            \qquad
            2|\Sp(2, \mathbb{F}_2)|\,|\Sp(6, \mathbb{F}_2)|
            =
            17418240
            <
            \abs{\mathcal{P}(12, \mathbb{F}_2)}.
        \end{equation}

        \item $\mathcal S$ rows.
        From \cref{eq:group_order_siegel_parabolic_lower_bound_factorial}, we have $d! < \abs{\mathcal{P}(2k, \mathbb{F}_2)}$ for all $d \leq 2k + 2$, in the regime $k \geq 4$.
        And finally, $2|\Sp(6, \mathbb{F}_2)| = 2903040 < |\mathcal{P}(8, \mathbb{F}_2)|$.
    \end{itemize}

    Thus in every possible large irreducible maximal case, $2|K^\Omega|<|\mathcal{P}(2k, \mathbb{F}_2)|$.
    This completes the proof for both assertions.
\end{proof}

\begin{lemma}
    [Orthogonal subgroups larger than Siegel parabolic contain orthogonal cores]
    \label{lem:group_theory_irreducible_orthogonal_subgroups_larger_than_siegel_parabolic_contain_orthogonal_cores}
    Let $k\geq 3$ and $G \leq O^\epsilon(2k, \mathbb{F}_2)$ for $\epsilon\in\{+,-\}$ be a subgroup that acts irreducibly on the natural $2k$-dimensional symplectic space $V \coloneqq \mathbb{F}_2^{2k}$. If $\abs{G} > \abs{\mathcal{P}(2k, \mathbb{F}_2)}$, then $G \geq \Omega^\epsilon(2k,\mathbb{F}_2)$.
\end{lemma}

\begin{proof}
    We first handle the $k=3$ case by direct calculation.
    For $\epsilon = +$, we note the exceptional isomorphism $\Omega^+(6,\mathbb{F}_2)\cong A_8$.
    We have $\abs{A_8} = 20160$ and the largest proper subgroup of $A_8$ has order $2520$~\cite{liebeck1987classification,GAP_character_table_library}.
    Similarly, for $\epsilon = -$, we use the exceptional isomorphism $\Omega^-(6,\mathbb{F}_2)\cong U_4(2)$.
    We have $\abs{U_4(2)} = 25920$ and the largest proper subgroup of $U_4(2)$ has order $960$~\cite{GAP_character_table_library}.
    Hence, for both signs of $\epsilon$, if $G \leq \Omega^\epsilon(6,\mathbb{F}_2)$ and $\abs{G} > \abs{\mathcal{P}(6, \mathbb{F}_2)} = 10752$, then $G = \Omega^\epsilon(6,\mathbb{F}_2)$.
    Otherwise, if $G \not\leq \Omega^\epsilon(6,\mathbb{F}_2)$, then because $[O^\epsilon(6,\mathbb{F}_2):\Omega^\epsilon(6,\mathbb{F}_2)] = 2$, we have $[G:G\cap\Omega^\epsilon(6,\mathbb F_2)]=2$ and hence
    $\abs{G \cap \Omega^\epsilon(6,\mathbb{F}_2)} = \abs{G} / 2 > \abs{\mathcal{P}(6, \mathbb{F}_2)} / 2 = 5376$,
    larger than $2520$ and $960$.
    So $G \cap \Omega^\epsilon(6,\mathbb{F}_2) = \Omega^\epsilon(6,\mathbb{F}_2)$ and accordingly $G \geq \Omega^\epsilon(6,\mathbb{F}_2)$.
    This proves the lemma for \(k=3\).
    
    Now we consider $k \geq 4$. By \cref{fact:siegel_parabolic_subgroup_large}, $\mathcal{P}(2k, \mathbb{F}_2)$ is a large subgroup of $\Sp(2k, \mathbb{F}_2)$. Suppose, toward a contradiction, that $\abs{G} > \abs{\mathcal{P}(2k, \mathbb{F}_2)}$ but $G \ngeq \Omega^\epsilon(2k,\mathbb{F}_2)$. Consider the cases:
    \begin{itemize}
        \item \(G < \Omega^\epsilon(2k,\mathbb{F}_2)\). Choose a maximal subgroup $G\leq K<\Omega^\epsilon(2k,\mathbb{F}_2)$. Since \(G\) is irreducible, \(K\) is irreducible. Also, $|K|^3 \geq |G|^3 > \abs{\mathcal{P}(2k, \mathbb{F}_2)}^3 \geq |\Sp(2k,\mathbb F_2)| > |\Omega^\epsilon(2k,\mathbb{F}_2)|$. Thus \(K\) is a large irreducible maximal subgroup of \(\Omega^\epsilon(2k,\mathbb{F}_2)\). By \cref{prop:aschbacher_large_irreducible_orthogonal_maximal_subgroups_below_siegel}, \(|G|\leq |K|<\abs{\mathcal{P}(2k, \mathbb{F}_2)}\), contradicting $\abs{G} > \abs{\mathcal{P}(2k, \mathbb{F}_2)}$.

        \item \(G\not\leq \Omega^\epsilon(2k,\mathbb{F}_2)\). Choose a maximal subgroup $G\leq K<O^\epsilon(2k,\mathbb{F}_2)$. Since \(G\ngeq \Omega^\epsilon(2k,\mathbb{F}_2)\) by assumption and \(G\not\leq \Omega^\epsilon(2k,\mathbb{F}_2)\), the maximal
        subgroup \(K\) does not contain \(\Omega^\epsilon(2k,\mathbb{F}_2)\). Indeed, if \(K\) contained both \(\Omega^\epsilon(2k,\mathbb{F}_2)\) and an element of \(G\setminus \Omega^\epsilon(2k,\mathbb{F}_2)\), then $K = O^\epsilon(2k,\mathbb{F}_2)$. Also, \(K\) is irreducible because it contains the irreducible subgroup \(G\). Finally, we note $|K|^3 \geq |G|^3 > \abs{\mathcal{P}(2k, \mathbb{F}_2)}^3 \geq |\Sp(2k,\mathbb F_2)| > |O^\epsilon(2k,\mathbb{F}_2)|$. Hence \(K\) is a large irreducible maximal subgroup of \(O^\epsilon(2k,\mathbb{F}_2)\) not
        containing \(\Omega^\epsilon(2k,\mathbb{F}_2)\). By \cref{prop:aschbacher_large_irreducible_orthogonal_maximal_subgroups_below_siegel}, \(|G|\leq |K|<\abs{\mathcal{P}(2k, \mathbb{F}_2)}\), contradicting $\abs{G} > \abs{\mathcal{P}(2k, \mathbb{F}_2)}$.
    \end{itemize}

    Both cases are impossible. Therefore $G$ must contain $\Omega^\epsilon(2k,\mathbb F_2)$.
\end{proof}

\subsubsection{Order-uniqueness of the Siegel parabolic subgroup}
\label{app:stab_codes/prelims/siegel_parabolic_order_uniqueness}

\begin{lemma}
    [Stabilizer of a complete totally isotropic flag and its symplectic overgroups]
    \label{lem:group_theory_complete_totally_isotropic_flag_stabilizer_and_overgroups}
    In the standard symplectic space \(\mathbb F_2^{2k}\), fix a complete totally isotropic flag
    \begin{equation}
        \mathcal F:\quad
        0<E_1<E_2<\cdots<E_k,
        \qquad
        \dim(E_i)=i.
    \end{equation}

    Denote its stabilizer
    \begin{equation}
        U_{\mathcal F}
        \coloneqq
        \Stab_{\Sp(2k,\mathbb F_2)}(\mathcal F)
        =
        \{M\in\Sp(2k,\mathbb F_2):
        M(E_i)=E_i \,\, \forall \, i\in[k]\}.
    \end{equation}
    
    Then \(|U_{\mathcal F}|=2^{k^2}\), and hence \(U_{\mathcal F}\) is a Sylow \(2\)-subgroup of
    \(\Sp(2k,\mathbb F_2)\). Moreover, if \(U_{\mathcal F}\leq K\leq \Sp(2k,\mathbb F_2)\) for some subgroup \(K\), then there exists a subset \(I=\{d_1<\cdots<d_r\}\subseteq[k]\) such that \(K = \Stab_{\Sp(2k,\mathbb F_2)} (E_{d_1}<\cdots<E_{d_r})\). The case \(I=\varnothing\) gives \(K=\Sp(2k,\mathbb F_2)\).
\end{lemma}

\begin{proof}
    Choose a symplectic basis \((e_1,\ldots,e_k,f_1,\ldots,f_k)\) adapted to \(\mathcal F\), so that \(E_i=\spn_{\mathbb F_2}(e_1,\ldots,e_i)\) for every $i \in [k]$. With respect to this basis, the elements of \(U_{\mathcal F}\) have the form
    \begin{equation}
        \begin{pmatrix}
            A & AS\\
            0 & A^{-\top}
        \end{pmatrix},
        \label{eq:standard_sylow_two_symplectic_form}
    \end{equation}
    where \(A\) is upper unitriangular and \(S\) is symmetric. Indeed, stabilizing \(E_k\) gives the upper block-triangular form, the symplectic condition gives the lower-right block \(A^{-\top}\) and, upon writing the upper-right block as \(AS\), requires \(S\) to be symmetric, and stabilizing every \(E_i\) requires \(A\) to be upper triangular. Over \(\mathbb F_2\), every invertible upper triangular matrix has all diagonal entries equal to \(1\). There are \(k(k-1)/2\) freely chosen entries in an upper unitriangular \(k\times k\) binary matrix and \(k(k+1)/2\) freely chosen entries in a symmetric \(k\times k\) binary matrix. Therefore \(
        |U_{\mathcal F}|
        =
        2^{k(k-1)/2}2^{k(k+1)/2}
        =
        2^{k^2}
    \). From the order formulae in \cref{fact:group_theory_relevant_group_orders}, the full \(2\)-part of \(|\Sp(2k,\mathbb F_2)|\) is \(2^{k^2}\). Hence \(U_{\mathcal F}\) is a Sylow \(2\)-subgroup.

    The remaining assertion is the standard classification of overgroups of a complete isotropic-flag stabilizer. In the symplectic setting, \cite[Prop.~3.2.3 and Ex.~3.2.6]{digne2020representations} shows that every subgroup containing the stabilizer of a complete isotropic flag is the stabilizer of one of its subflags; see also \cite[Chap.~14]{aschbacher2000finite}. Therefore, if \(U_{\mathcal F}\leq K\leq\Sp(2k,\mathbb F_2)\), there exists a subset \(I=\{d_1<\cdots<d_r\}\subseteq[k]\) such that \(
        K
        =
        \Stab_{\Sp(2k,\mathbb F_2)}
        (E_{d_1}<\cdots<E_{d_r})
    \) as claimed. If \(I=\varnothing\), no proper subspace is required to be stabilized, and the stabilizer is all of \(\Sp(2k,\mathbb F_2)\).
\end{proof}

\begin{lemma}
    [Order-uniqueness of the Siegel parabolic subgroup]
    \label{lem:order_uniqueness_siegel_parabolic}
    Suppose $G < \Sp(2k, \mathbb{F}_2)$ and $|G| = |\mathcal{P}(2k, \mathbb{F}_2)|$, for $k \neq 2$. Then $G$ is conjugate to $\mathcal{P}(2k, \mathbb{F}_2)$ inside $\Sp(2k, \mathbb{F}_2)$. The $k = 2$ exception is discussed in \cref{rem:order_uniqueness_siegel_parabolic_k_eq_2}.
\end{lemma}

\begin{proof}
    By the order formula in \cref{fact:group_theory_relevant_group_orders}, the largest power of \(2\) dividing each of
    \(\abs{\Sp(2k,\mathbb F_2)}\) and \(\abs{G}=\abs{\mathcal P(2k,\mathbb F_2)}\) is \(2^{k^2}\).
    Let \(S\) be a Sylow \(2\)-subgroup of \(G\). Then \(\abs{S}=2^{k^2}\), so \(S\) is also a Sylow \(2\)-subgroup of \(\Sp(2k,\mathbb F_2)\). Since all Sylow \(2\)-subgroups of a finite group are conjugate, after conjugating \(G\) inside \(\Sp(2k,\mathbb F_2)\), we may assume that \(U_{\mathcal F}\leq G\), where \(U_{\mathcal F}\) is the Sylow \(2\)-subgroup defined in \cref{lem:group_theory_complete_totally_isotropic_flag_stabilizer_and_overgroups}.

    By \cref{lem:group_theory_complete_totally_isotropic_flag_stabilizer_and_overgroups}, there is a subset \(I=\{d_1<\cdots<d_r\}\subseteq[k]\) such that \(
        G
        =
        \smash{\Stab_{\Sp(2k,\mathbb F_2)}
        (E_{d_1}<\cdots<E_{d_r})}
    \).
    The set \(I\) is nonempty. Indeed, if \(I=\varnothing\), then \(G=\Sp(2k,\mathbb F_2)\), whereas
    \(
        \abs{\Sp(2k,\mathbb F_2)}/
        \abs{\mathcal P(2k,\mathbb F_2)}
        =
        \prod_{i=1}^{k}(2^i+1)
        >1,
    \)
    contradicting \(\abs{G}=\abs{\mathcal P(2k,\mathbb F_2)}\).
    Set \(d_0\coloneqq0\), \(a_j\coloneqq d_j-d_{j-1}\) for \(1\leq j\leq r\), and \(m\coloneqq k-d_r\). Then every \(a_j\) is positive and \(a_1+\cdots+a_r+m=k\).

    By the standard Levi decomposition of symplectic parabolic subgroups~\cite[Prop.~3.4.1]{digne2020representations}, \(G\) has a normal \(2\)-subgroup \(R\) such that
    \(
        G/R
        \cong
        \GL(a_1,\mathbb F_2)
        \times\cdots\times
        \GL(a_r,\mathbb F_2)
        \times
        \Sp(2m,\mathbb F_2).
    \)
    Concretely, with respect to a symplectic basis adapted to the partial flag, the corresponding Levi subgroup consists of the block-diagonal transformations
    \(
        \operatorname{diag}
        (A_1,\ldots,A_r,T,A_r^{-\top},\ldots,A_1^{-\top}),
    \)
    with \(A_j\in\GL(a_j,\mathbb F_2)\) and \(T\in\Sp(2m,\mathbb F_2)\). Hence
    \begin{equation}
        \abs{G}
        =
        \abs{R}
        \left(
            \prod_{j=1}^{r}
            \abs{\GL(a_j,\mathbb F_2)}
        \right)
        \abs{\Sp(2m,\mathbb F_2)},
    \end{equation}
    where \(\abs{R}\) is a power of \(2\).
    Define
    \begin{equation}
        p(t)
        \coloneqq
        \prod_{i=1}^{t}(2^i-1),
        \qquad
        q(t)
        \coloneqq
        \prod_{i=1}^{t}(2^{2i}-1),
    \end{equation}
    with \(p(0)=q(0)=1\). By \cref{fact:group_theory_relevant_group_orders}, \(
        \abs{\GL(t,\mathbb F_2)}
        =
        2^{t(t-1)/2}p(t)
    \) and \(
        \abs{\Sp(2t,\mathbb F_2)}
        =
        2^{t^2}q(t)
    \). Since \(\abs{R}\) is a power of \(2\), the odd part of \(\abs{G}\) is therefore \(
        \smash{\left(\prod_{j=1}^{r}p(a_j)\right)q(m)}
    \).
    Moreover, \(G\) contains \(U_{\mathcal F}\), whose order is \(2^{k^2}\), while the largest power of \(2\) dividing \(\abs{\Sp(2k,\mathbb F_2)}\) is also \(2^{k^2}\). Hence the \(2\)-part of \(\abs{G}\) is exactly \(2^{k^2}\), and thus
    \begin{equation}
        \abs{G}
        =
        2^{k^2}
        \left(\prod_{j=1}^{r}p(a_j)\right)q(m).
    \end{equation}
    On the other hand, \(
        \abs{\mathcal P(2k,\mathbb F_2)}
        =
        2^{k^2}p(k)
    \). Therefore the assumed order equality is equivalent to
    \begin{equation}
        \left(\prod_{j=1}^{r}p(a_j)\right)q(m)
        =
        p(k).
        \label{eq:order_uniqueness_siegel_parabolic_reduced_equation}
    \end{equation}

    We now solve \cref{eq:order_uniqueness_siegel_parabolic_reduced_equation}. First observe that, for all \(a,b\geq0\), \(p(a)p(b)\leq p(a+b)\), with equality if and only if \(a=0\) or \(b=0\). Indeed, \(
        p(a+b)/p(a)
        =
        \prod_{i=1}^{b}(2^{a+i}-1)
        \geq
        \prod_{i=1}^{b}(2^i-1)
        =
        p(b),
    \)
    and every factorwise inequality is strict when \(a,b>0\). Iterating this inequality gives \(
        \prod_{j=1}^{r}p(a_j)
        \leq
        p(a_1+\cdots+a_r),
    \)
    and, since every \(a_j\) is positive, equality holds if and only if \(r=1\).
    We consider three cases:
    \begin{itemize}
        \item Suppose first that \(m=0\). Then \(a_1+\cdots+a_r=k\), and \(
            \prod_{j=1}^{r}p(a_j)\leq p(k)
        \). Equality in \cref{eq:order_uniqueness_siegel_parabolic_reduced_equation} therefore forces \(r=1\) and \(a_1=k\).

        \item Next suppose that \(0<m\leq k/2\). Since \(a_1+\cdots+a_r=k-m\), we have \(
            \prod_{j=1}^{r}p(a_j)
            \leq
            p(k-m)
        \), and
        \begin{equation}
            \left(\prod_{j=1}^r p(a_j)\right)q(m)
            \leq
            p(k-m)q(m)
            =
            p(k-m)\prod_{i=1}^m(2^{2i}-1),
            \qquad
            p(k)
            =
            p(k-m)\prod_{i=1}^m(2^{k-m+i}-1).
        \end{equation}
        Since \(m\leq k-m\), for every \(1\leq i\leq m\) we have \(2i\leq k-m+i\). Therefore \(p(k-m)q(m)\leq p(k)\). Equality can hold only if \(2i=k-m+i\) for every \(1\leq i\leq m\), equivalently \(i=k-m\) for every \(1\leq i\leq m\). This is impossible when \(m>1\), while for \(m=1\) it forces \(k=2\). Thus, for \(k\neq2\), equality in \cref{eq:order_uniqueness_siegel_parabolic_reduced_equation} is impossible in this case. The exceptional case \(k=2\) is described in \cref{rem:order_uniqueness_siegel_parabolic_k_eq_2}.

        \item Finally, suppose that \(m>k/2\). If \cref{eq:order_uniqueness_siegel_parabolic_reduced_equation} held, then \(q(m)\) would divide \(p(k)\), and in particular
        \(
            2^{2m}-1
            \mid
            p(k).
        \)
        Suppose first that \(2m\neq6\). By Zsigmondy's theorem~\cite{zsigmondy1892theorie}, there exists a prime \(\ell\) dividing \(2^{2m}-1\) which does not divide \(2^s-1\) for any \(1\leq s<2m\). Since \(k<2m\), \(\ell\) divides none of the factors of \(p(k)=\prod_{s=1}^{k}(2^s-1)\), a contradiction. It remains to consider \(2m=6\), so that \(m=3\). Since \(I\neq\varnothing\), we have \(m\leq k-1\), while \(m>k/2\); hence \(k\in\{4,5\}\). Direct substitution into \cref{eq:order_uniqueness_siegel_parabolic_reduced_equation} rules out both cases.
    \end{itemize}

    For \(k\neq2\), the only solution of \cref{eq:order_uniqueness_siegel_parabolic_reduced_equation} is \(m=0\), \(r=1\), and \(a_1=k\). It follows that \(d_1=k\). Therefore
    \(
        G
        =
        \Stab_{\Sp(2k,\mathbb F_2)}(E_k),
    \)
    which is the standard Siegel parabolic subgroup \(\mathcal P(2k,\mathbb F_2)\). Undoing the initial conjugation, the original subgroup \(G\) is conjugate inside \(\Sp(2k,\mathbb F_2)\) to \(\mathcal P(2k,\mathbb F_2)\).
\end{proof}

\begin{remark}
    [Exception to order-uniqueness of the Siegel parabolic subgroup at $k = 2$]
    \label{rem:order_uniqueness_siegel_parabolic_k_eq_2}
    The exception to \cref{lem:order_uniqueness_siegel_parabolic} at $k=2$ arises because, in $\Sp(4,\mathbb{F}_2)$, both the stabilizer of a Lagrangian plane (i.e.~the Siegel parabolic subgroup) and the stabilizer of an isotropic line have order $2^4(2^2-1)=48$, and they are not conjugate within $\Sp(4,\mathbb{F}_2)$. The latter has the symplectic representation
    \begin{equation}
        \left\{
        \begin{pmatrix}
            1&v^\top \Omega_1 M & c\\0&M&v\\0&0&1
        \end{pmatrix}
        :
        M\in\Sp(2,\F_2),\;v\in \F_2^2,\;c\in\F_2
        \right\}
        \le
        \Sp(4,\F_2).
    \end{equation}
\end{remark}

\clearpage

\subsection{Permutation gates}
\label{app:stab_codes/perm}

\begin{theorem}
    [Permutation logical groups on stabilizer codes]
    \label{thm:stab_codes_perm_logical_groups}
    Let $\mathcal{C}$ be an $\db{n,k}$ stabilizer code. Then, in any logical basis $\mathcal{L}$, $\mathrm{Perm}_\mathcal{L}(\mathcal{C}) \preceq_{\Sp(2k,\mathbb F_2)} \mathcal{P}(2k, \mathbb{F}_2)$. The maximum $\mathrm{Perm}_\mathcal{L}(\mathcal{C}) \sim_{\Sp(2k,\mathbb F_2)} \mathcal{P}(2k, \mathbb{F}_2)$ is achievable.
\end{theorem}

\begin{proof}
    A stabilizer code can be placed into standard form (see \cref{def:stab_code_standard_form}), which reveals a logical basis in which all $Z$-type logicals are of the form $\{I, Z\}^{\otimes n}$. A permutation gate involves only a physical qubit permutation but no physical Clifford gates on the qubits, and so cannot change the type of Paulis on the physical qubits. Then every $Z$-type logical remains a $Z$-type logical under the action of any transversal gate. Equivalently stated, the subspace of $Z$-type logicals, a $k$-dimensional Lagrangian subspace within the $2k$-dimensional logical space, is preserved. The group of Lagrangian-preserving transformations in $\Sp(2k,\mathbb F_2)$ is precisely the Siegel parabolic subgroup $\mathcal{P}(2k, \mathbb{F}_2)$. The maximum $\mathrm{Perm}_\mathcal{L}(\mathcal{C}) \sim_{\Sp(2k,\mathbb F_2)} \mathcal{P}(2k, \mathbb{F}_2)$ is achieved by, for example, \cref{cons:non_css_code_permutation_all_logical_s_and_cx}.
\end{proof}

\clearpage

\subsection{Transversal gates}
\label{app:stab_codes/trans}

\begin{lemma}
    [All-or-nothing physical order-three Cliffords on indecomposable stabilizer codes]
    \label{lem:stab_codes_trans_all_or_nothing_physical_c3_cliffords}
    Transversal gates on indecomposable stabilizer codes can have physical order-three Cliffords (i.e.~$HS, SH$) on either none or all of its physical qubits.
\end{lemma}

\begin{proof}
    Consider an indecomposable $\db{n,k}$ stabilizer code. Write the physical binary symplectic space as \(W=\bigoplus_{j=1}^n W_j\) where \(W_j\cong \mathbb F_2^2\), and let \(S\leq W\) be the stabilizer space of the code. Consider any physical implementation of a transversal gate $\gamma \in \Sp(2, \mathbb{F}_2)^n$, which satisfies $\gamma(S) = S$. We denote the support of $\gamma$ that has order-three elements as 
    \(
        K
        \coloneqq
        K_\gamma
        =
        \{j\in[n]: \gamma_j\text{ has order three}\}
    \),
    where $\gamma_j$ is the single-qubit Clifford of $\gamma$ on the $j^\text{th}$ physical qubit.
    
    We construct a coordinate projection onto \(W_K\coloneqq \bigoplus_{j\in K}W_j\) as a polynomial in $\gamma$. In particular, consider \(q(x)\coloneqq (x+1)^3\in\mathbb F_2[x]\). For coordinate \(j\), we claim that \(q(\gamma_j) = I\) when $j \in K$, and \(q(\gamma_j) = 0\) when $j \notin K$. Indeed:
    \begin{itemize}[noitemsep]
        \item If \(\gamma_j=I\), then \(q(\gamma_j)=(I+I)^3=0\). 
        \item If \(\gamma_j\) has order \(2\), then, in characteristic two, \((\gamma_j+I)^2=\gamma_j^2+I=I+I=0\). So again \(q(\gamma_j)=(\gamma_j+I)^3=0\).
        \item If $\gamma_j$ has order $3$, then its minimal polynomial divides
        $x^3+1=(x+1)(x^2+x+1)$. Since $\gamma_j\neq I$ and $\gamma_j$
        acts on the two-dimensional space $W_j$, its minimal polynomial is
        $x^2+x+1$. Over $\mathbb F_2$, $(x+1)^3+1=x(x^2+x+1)$, so $(x+1)^3\equiv 1\pmod{x^2+x+1}$. Therefore $q(\gamma_j)=I$.
    \end{itemize}
    
    It follows that \(\Pi\coloneqq q(\gamma)=(\gamma+I)^3\) is exactly the coordinate projection \(W\rightarrow W_K\) desired. Since \(\gamma(S)=S\) and $S$ is an $\mathbb F_2$-linear subspace, every polynomial in \(\gamma\) preserves \(S\). Hence \(\Pi(S)\subseteq S\). Also, \(I+\Pi\) is the complementary coordinate projection onto \(W_{K^\mathrm{c}}\coloneqq \bigoplus_{j\notin K}W_j\), and likewise \((I+\Pi)(S)\subseteq S\). Together, for every $s \in S$, we have \(s=\Pi(s)+(I+\Pi)(s)\) with \(\Pi(s)\in W_K\) and \((I+\Pi)(s)\in W_{K^\mathrm{c}}\). Therefore \(S=\Pi(S)\oplus (I+\Pi)(S)\), where the two summands are supported on the disjoint qubit sets \(K\) and \(K^\mathrm{c}\). Now, if \(K\) were a proper nonempty subset of \([n]\), this would split the stabilizer space across the nontrivial qubit partition \([n]=K\sqcup K^\mathrm{c}\), contradicting indecomposability (see \cref{prop:code_decomposability}). Therefore either $K = \varnothing$ or $K = [n]$.
\end{proof}

\begin{theorem}
    [Transversal logical groups on indecomposable stabilizer codes]
    \label{thm:stab_codes_trans_logical_groups}
    Let $\mathcal{C}$ be an $\db{n,k}$ indecomposable stabilizer code. Then, in any logical basis $\mathcal{L}$, either $\mathrm{Trans}_\mathcal{L}(\mathcal{C}) \preceq_{\Sp(2k,\mathbb F_2)} \mathcal{U}(2k, \mathbb{F}_2)$, or $\mathrm{Trans}_\mathcal{L}(\mathcal{C}) \cong C_3$, or $\mathrm{Trans}_\mathcal{L}(\mathcal{C}) \cong S_3$. The maxima $\mathrm{Trans}_\mathcal{L}(\mathcal{C}) \sim_{\Sp(2k,\mathbb F_2)} \mathcal{U}(2k, \mathbb{F}_2)$ for $k \geq 2$ and $\mathrm{Trans}_\mathcal{L}(\mathcal{C}) \cong S_3$ for $k = 1$ are each achievable. In addition, $\mathrm{Trans}_\mathcal{L}(\mathcal{C}) \cong C_3$ is also attainable.
\end{theorem}

\begin{proof}
    Let \(W=\mathbb F_2^{2n}\) be the physical binary symplectic space, written \(W=\bigoplus_{j=1}^n W_j\) where each \(W_j\cong \mathbb F_2^2\), and \(S\leq W\) be the stabilizer space of the code. That is, \(\dim S=n-k\) and \(S \subseteq S^\perp\), and the logical binary symplectic space is \(S^\perp/S\) of dimension \(2k\). We denote by \(\Gamma \coloneqq \mathrm{PTrans}(\mathcal{C}) \leq \Sp(2,\mathbb F_2)^n\cong S_3^n\) the group of physical implementations of transversal gates of the code, and let \(\rho: \Gamma \rightarrow \Sp(S^\perp/S)\cong \Sp(2k,\mathbb F_2)\) be the homomorphism mapping physical implementations to their logical actions. In this notation, \(G \coloneqq \mathrm{Trans}_\mathcal{L}(\mathcal{C}) = \rho(\Gamma)\).

    For each physical qubit, let \(C_3\triangleleft S_3\) denote the unique normal order-three subgroup. Define \(N \coloneqq C_3^n\triangleleft S_3^n\), and let \(\Gamma_A \coloneqq \Gamma\cap N\) and \(A \coloneqq \rho(\Gamma_A)\). Since \(N\triangleleft S_3^n\), we have \(\Gamma_A\triangleleft \Gamma\); and since \(\rho\) is surjective onto \(\mathrm{Trans}_\mathcal{L}(\mathcal{C})\) and the image of a normal subgroup is normal in the image of the whole group, we have \(A \triangleleft G\). Moreover,
    \begin{equation}
        \Gamma/\Gamma_A
        =
        \Gamma/(\Gamma\cap N)
        \cong
        \Gamma N/N
        \leq
        S_3^n/N
        \cong
        C_2^n.
    \end{equation} 
    Thus \(\Gamma/\Gamma_A\) is an elementary abelian \(2\)-group. Since \(G/A\) is an image of \(\Gamma/\Gamma_A\), it follows that \(G/A\) is also an elementary abelian \(2\)-group.

    We first show that \(\Gamma_A\) has rank at most one. Let \(t\in \Gamma_A\) be a nonidentity element. Since \(t\in C_3^n\), on each physical qubit \(t\) acts either trivially or by a nontrivial physical order-\(3\) Clifford. By \cref{lem:stab_codes_trans_all_or_nothing_physical_c3_cliffords}, the physical order-\(3\) Cliffords must be on either none or all of the physical qubits. Since $t$ is nonidentity, the none case is impossible; \(t\) must act nontrivially on every physical qubit.

    Suppose \(\Gamma_A\) has rank at least two. For any coordinate \(j\), the projection \(\Gamma_A \rightarrow C_3\) to the \(j\)-th factor cannot be injective, because \(\Gamma_A\) would have order at least \(3^2 = 9\), whereas \(C_3\) has order \(3\). Hence there exists a nonidentity \(t\in\Gamma_A\) acting trivially on coordinate \(j\), contradicting our conclusion in the previous paragraph. Therefore \(\Gamma_A\) has rank at most one, so either $\Gamma_A = \{I\}$ or $\Gamma_A \cong C_3$. Consequently, $A = \{I\}$ or $A \cong C_3$. We consider each case.
    
    \begin{itemize}
        
        \item $A = \{I\}$. Then \(G/A=G\), so \(G\) is an elementary abelian \(2\)-group. Now, since \(\Gamma_A\triangleleft\Gamma\), since \(\Gamma_A\) has order \(1\) or \(3\), and since \(\Gamma/\Gamma_A\) is a \(2\)-group, any Sylow \(2\)-subgroup \(P\leq\Gamma\) satisfies \(\Gamma=P\Gamma_A\). Then, because \(\rho(\Gamma_A)=A=\{1\}\), we have \(\rho(P)=\rho(\Gamma)=G\).
        
        For each physical qubit \(j\), let \(P_j\leq S_3\) be the projection of \(P\) to the \(j^\text{th}\) local factor. Since \(P\) is a \(2\)-group, \(P_j\) is a \(2\)-subgroup of \(S_3\). Hence \(P_j\) is either trivial or of order \(2\). In either case, \(P_j\) fixes at least one line (i.e.~dimension-one subspace) \(\ell_j\leq W_j\). Over \(\mathbb F_2\), a line contains a single nonzero vector, so this line is fixed pointwise (i.e.~the zero and nonzero vectors are individually fixed).

        Consider \(L \coloneqq \bigoplus_{j=1}^n \ell_j\). Then \(L\) is a physical Lagrangian in \(W\): it has dimension \(n\), and it is isotropic because it is the direct sum of one-dimensional subspaces, one in each mutually orthogonal local summand \(W_j\). By construction, \(P\) fixes \(L\) pointwise. Now define the induced logical subspace \(\Lambda 
            \coloneqq
            (L\cap S^\perp)/(L\cap S)
            \leq
            S^\perp/S
        \). Equivalently, \(\Lambda=(L\cap S^\perp+S)/S\). Since \(P\) fixes \(L\) pointwise and preserves both \(S\) and \(S^\perp\), the group \(\mathrm{Trans}_\mathcal{L}(\mathcal{C})=\rho(P)\) fixes \(\Lambda\) pointwise. We now compute the dimension of \(\Lambda\). Let \(a \coloneqq \dim(L\cap S)\), and note that since \(L\) is Lagrangian, \(L^\perp=L\). Therefore
        \begin{equation}\begin{split}
            \dim(L\cap S^\perp)
            =\dim(L^\perp\cap S^\perp)
            =\dim(L+S)^\perp
            &=2n-\dim(L+S) \\
            &=2n-\left(\dim L+\dim S-\dim(L\cap S)\right) \\
            &=2n-\left(n+(n-k)-a\right)
            =k+a.
        \end{split}\end{equation}

        Therefore \(
            \dim\Lambda
            =
            \dim(L\cap S^\perp)-\dim(L\cap S)
            =
            (k+a)-a
            =
            k
        \). Moreover, \(\Lambda\) is represented by vectors in the isotropic space
        \(L\), so \(\Lambda\) is isotropic. Thus \(\Lambda\) is a \(k\)-dimensional
        isotropic subspace of the \(2k\)-dimensional logical symplectic space
        \(S^\perp/S\), hence \(\Lambda\) is a logical Lagrangian. We conclude \(\mathrm{Trans}_\mathcal{L}(\mathcal{C})\) is an elementary abelian 2-group that fixes a logical Lagrangian pointwise. This identifies \(G \preceq_{\Sp(2k,\mathbb F_2)}\mathcal U(2k,\mathbb F_2)\).

        \item \(A\cong C_3\). Here we must have \(\Gamma_A\cong C_3\), and the restriction \(\rho|_{\Gamma_A}:\Gamma_A\rightarrow A\) is an isomorphism. We claim that the centralizer of $A$ inside $G$, denoted $C_G(A) = \{g \in G: gag^{-1} = a \, \forall a \in A\}$, is precisely $A$. 

        To see this, first note that the inclusion \(A\leq C_G(A)\) is immediate because \(A\cong C_3\) is abelian. Conversely, let \(g\in C_G(A)\), and choose \(\gamma\in\Gamma\) such that \(\rho(\gamma)=g\). Let \(t\in\Gamma_A\) be any nonidentity element. Since \(g\) centralizes \(A\), we have 
        \(
            \rho([\gamma,t])
            =
            [\rho(\gamma),\rho(t)]
            =
            [g,\rho(t)]
            =
            1
        \).
        Also, because \(\Gamma_A\triangleleft\Gamma\), we have \([\gamma,t]\in\Gamma_A\). Lastly, since \(\rho|_{\Gamma_A}\) is injective, it follows that \([\gamma,t]=1\). Thus \(\gamma\) centralizes \(t\). Next, by \cref{lem:stab_codes_trans_all_or_nothing_physical_c3_cliffords}, \(t\in\Gamma_A\) must act as a nontrivial order-\(3\) Clifford on every physical qubit. In a single local factor \(S_3\), the centralizer of a nontrivial order-\(3\) element is the local \(C_3\). Therefore every local component of \(\gamma\) lies in the corresponding local \(C_3\), and hence \(\gamma\in N=C_3^n\). Since also \(\gamma\in\Gamma\), we get \(\gamma\in\Gamma\cap N=\Gamma_A\). Therefore \(g=\rho(\gamma)\in \rho(\Gamma_A)=A\), and \(C_G(A)\leq A\).

        Now the conjugation actions of elements of $G$ on $A \triangleleft G$ gives a homomorphism between $G$ and the automorphism group $\mathrm{Aut}(A)$, whose kernel is \(C_G(A)=A\). Therefore it induces an injection \(G/A\hookrightarrow \Aut(A)\). As \(A\cong C_3\), we have \(\mathrm{Aut}(A)\cong C_2\), so either \(G/A=\{1\}\) or \(G/A\cong C_2\). In the former case, \(G=A\cong C_3\). In the latter case, \(\abs{G} = 6\). Moreover \(C_G(A)=A\neq G\), so \(G\) is nonabelian. The unique nonabelian group of order \(6\) is \(S_3\).
    \end{itemize}

    Finally, $\mathrm{Trans}_\mathcal{L}(\mathcal{C}) \sim_{\Sp(2k,\mathbb F_2)} \mathcal{U}(2k, \mathbb{F}_2)$ is achieved by, for example, the indecomposable code in~\cref{cons:complete_hypergraph_css_t_eq_2}; $\mathrm{Trans}_\mathcal{L}(\mathcal{C}) \cong S_3$ is achieved by all indecomposable SSD CSS codes by \cref{thm:css_codes_trans_logical_groups_ssd_indecomp}; and $\mathrm{Trans}_\mathcal{L}(\mathcal{C}) \cong C_3$ is achieved by, for example, the $\db{5,1,3}$ perfect code~\cite{gottesman1997stabilizer}.
\end{proof}

\begin{remark}
    [Comparison to indecomposable CSS codes]
    \label{rem:stab_codes_trans_logical_groups_comparison_to_css}
    First, at the physical level, \cref{lem:trans_exchange_uniformity_indecomp_css_codes} for indecomposable CSS codes is more constraining result than \cref{lem:stab_codes_trans_all_or_nothing_physical_c3_cliffords} for indecomposable stabilizer codes. Indeed, an indecomposable CSS code must be SSD for order-three single-qubit physical Cliffords, which are all exchange-type, to be admitted in transversal gates (see \cref{lem:css_codes_auto_physical_splitting_property}), so \cref{lem:trans_exchange_uniformity_indecomp_css_codes} applies; but \cref{lem:trans_exchange_uniformity_indecomp_css_codes} states more strictly that $(HS)^{\otimes n}$ and $(SH)^{\otimes n}$ are both transversal gates of the code while no other combination of order-three physical Cliffords are. This restriction is not present for indecomposable stabilizer codes in \cref{lem:stab_codes_trans_all_or_nothing_physical_c3_cliffords}: the local order-three physical Cliffords need not all be equal.

    Second, at the logical level, the maximum-sized transversal logical groups on indecomposable stabilizer codes from \cref{thm:stab_codes_trans_logical_groups} are $\mathrm{Trans}_\mathcal{L}(\mathcal{C}) \cong S_3$ at $k = 1$ and $\mathrm{Trans}_\mathcal{L}(\mathcal{C}) \preceq_{\Sp(2k,\mathbb F_2)} \mathcal{U}(2k, \mathbb{F}_2)$ for $k \geq 2$. These are identical to those on indecomposable CSS codes by \cref{thm:css_codes_trans_logical_groups_ssd_indecomp,thm:css_codes_trans_logical_groups_non_ssd_indecomp}. In this sense, stabilizer codes do not confer more power than CSS codes. However, the non-maximum $\mathrm{Trans}_\mathcal{L}(\mathcal{C}) \cong C_3$ is achievable on indecomposable stabilizer codes but not on indecomposable CSS codes. Indeed, the $\db{5,1,3}$ perfect code hosts $\mathrm{Trans}_\mathcal{L}(\mathcal{C}) \cong C_3$~\cite{gottesman1997stabilizer}. But on indecomposable CSS codes, \cref{thm:css_codes_trans_logical_groups_ssd_indecomp,thm:css_codes_trans_logical_groups_non_ssd_indecomp} constrains $\mathrm{Trans}_\mathcal{L}(\mathcal{C}) \preceq_{\Sp(2k,\mathbb F_2)} \mathcal{U}(2k, \mathbb{F}_2)$ or $\mathrm{Trans}_\mathcal{L}(\mathcal{C}) \cong S_3$. Note that $C_3$ is not contained in $\mathcal{U}(2k, \mathbb{F}_2)$, since $\abs{C_3} = 3$ but $3 \nmid \abs{\mathcal{U}(2k, \mathbb{F}_2)}$.
\end{remark}

\clearpage

\subsection{Automorphism gates}
\label{app:stab_codes/auto}

\begin{theorem}
    [Automorphism logical groups on $k \leq 2$ stabilizer codes]
    \label{thm:stab_codes_auto_logical_max_order_k_leq_2}
    Let $\mathcal{C}$ be an $\db{n, k \leq 2}$ stabilizer code and $\mathcal{L}$ be an arbitrary logical basis of $\mathcal{C}$. 
    Then:
    \begin{itemize}
        \item For $k = 1$, $\mathrm{Aut}_\mathcal{L}(\mathcal{C}) \leq \Sp(2, \mathbb{F}_2) \cong \mathrm{PCl}_1$, so $\abs{\mathrm{Aut}_\mathcal{L}(\mathcal{C})} \leq 6$. 
        Equality is achievable.
        \item For $k = 2$, $\abs{\mathrm{Aut}_\mathcal{L}(\mathcal{C})} \leq 72$. 
        Equality holds iff $\mathrm{Aut}_\mathcal{L}(\mathcal{C})$ is conjugate in $\Sp(4,\mathbb{F}_2)$ to $O^+(4,\mathbb{F}_2)$, and this maximum is achievable.
        Note that $O^+(4, \mathbb{F}_2) \cong S_3 \wr S_2 \cong \langle H_1, S_1, \mathrm{SWAP}_{12} \rangle \cong \langle H_1, H_2, \mathrm{CZ}_{12} \rangle$.
    \end{itemize}
\end{theorem}

\begin{proof}
    We consider the $k = 2$ case first. By \cref{lem:stab_codes_auto_prime_factor_nogo_1}, no automorphism gate on $\mathcal{C}$ can induce an order-$5$ logical action, so $\mathrm{Aut}_\mathcal{L}(\mathcal{C})$ must be a proper subgroup of $\Sp(4,\mathbb F_2)$ and must be contained in some maximal proper subgroup $M < \Sp(4,\mathbb F_2)$, and by Cauchy's theorem, $5 \nmid \abs{\mathrm{Aut}_\mathcal{L}(\mathcal{C})}$. The maximal proper subgroups~\cite{GAP_character_table_library,ATLAS_v3} of $\Sp(4,\mathbb F_2) \cong S_6$ of order at least $72$ are isomorphic to $A_6, S_5, S_3 \wr S_2$, with respective orders $360, 120, 72$. We consider these cases:
    \begin{itemize}[noitemsep]
        \item $M \cong A_6$. Then $\mathrm{Aut}_\mathcal{L}(\mathcal{C})$ is a proper subgroup of $A_6$, since $5 \nmid \abs{\mathrm{Aut}_\mathcal{L}(\mathcal{C})}$ but $5\mid \abs{A_6}$. The maximal proper subgroups~\cite{GAP_character_table_library,ATLAS_v3} of $A_6$ have orders at most $60$, so $\abs{\mathrm{Aut}_\mathcal{L}(\mathcal{C})} \leq 60$.
        \item $M \cong S_5$. Then $\abs{M}=120=2^3\cdot 3\cdot 5$, and since $\abs{\mathrm{Aut}_\mathcal{L}(\mathcal{C})} \mid \abs{M}$ and $5 \nmid \abs{\mathrm{Aut}_\mathcal{L}(\mathcal{C})}$, we have $\abs{\mathrm{Aut}_\mathcal{L}(\mathcal{C})} \leq 2^3\cdot 3=24$.
        \item $M\cong S_3\wr S_2$. Then $\abs{\mathrm{Aut}_\mathcal{L}(\mathcal{C})} \leq \abs{M} = 72$.
    \end{itemize}

    The unique case giving $\abs{\mathrm{Aut}_\mathcal{L}(\mathcal{C})} = 72$ is $\mathrm{Aut}_\mathcal{L}(\mathcal{C}) \cong S_3 \wr S_2$, and there is only one conjugacy class of isomorphism type $S_3 \wr S_2$ in $\Sp(4, \mathbb{F}_2)$~\cite{GAP_character_table_library,ATLAS_v3}. We therefore conclude that $\abs{\mathrm{Aut}_\mathcal{L}(\mathcal{C})} = 72$ is achieved uniquely by this subgroup up to conjugacy in $\Sp(4, \mathbb{F}_2)$. Simple projective Clifford realizations of this subgroup are $\langle H_1, S_1, \mathrm{SWAP}_{12} \rangle \cong \langle H_1, H_2, \mathrm{CZ}_{12} \rangle$ up to conjugacy in the two-qubit projective Clifford group. This maximum is achieved, for example, by \cref{cons:non_css_code_auto_k_eq_2_h1_h2_cz}, or by two copies of any $k=1$ stabilizer code each achieving the full $k=1$ Clifford group by automorphisms.

    The $k = 1$ case of the full projective Clifford group is achieved by, for example, the $\db{5,1,3}$ perfect code~\cite{sayginel2025fault,chakraborty2026nogo}, or SSD CSS codes (see \cref{thm:css_codes_trans_logical_groups_ssd_indecomp}). Any other logical group must be a subgroup of the full projective Clifford group.
\end{proof}

\begin{theorem}
    [Automorphism logical groups on $k \geq 3$ stabilizer codes]
    \label{thm:stab_codes_auto_logical_max_order_k_geq_3}
    Let $\mathcal{C}$ be a stabilizer code encoding $k \geq 3$ logical qubits and $\mathcal{L}$ be an arbitrary logical basis of $\mathcal{C}$. Then $\abs{\mathrm{Aut}_\mathcal{L}(\mathcal{C})} \leq \abs{\mathcal{P}(2k, \mathbb{F}_2)}$. The maximum $\abs{\mathrm{Aut}_\mathcal{L}(\mathcal{C})} = \abs{\mathcal{P}(2k, \mathbb{F}_2)}$ occurs iff $\mathrm{Aut}_\mathcal{L}(\mathcal{C})$ is conjugate in $\Sp(2k, \mathbb{F}_2)$ to $\mathcal{P}(2k, \mathbb{F}_2)$, and this maximum is achievable.
\end{theorem}

\begin{proof}
    Set $\Gamma\coloneqq\mathrm{Aut}_\mathcal{L}(\mathcal C)$. 
    We prove the bound by induction on $k\geq3$, with the induction hypothesis for $3\leq j<k$ understood to be vacuous when $k=3$.

    If $\Gamma$ is reducible, \cref{lem:stab_codes_auto_reducible_logical_group_order_bound} gives $\abs{\Gamma}\leq\abs{\mathcal P(2k,\F_2)}$, using the induction hypothesis when $k>3$. 
    Assume henceforth that $\Gamma$ is irreducible and, for contradiction, that $\abs{\Gamma}>\abs{\mathcal P(2k,\F_2)}$.

    The equality $\Gamma=\Sp(2k,\F_2)$ is impossible because $\Sp(2k,\F_2)$ contains $\Omega^\pm(2k,\F_2)$, contradicting \cref{lem:stab_codes_auto_logical_groups_cannot_contain_orthogonal_cores}. 
    Choose a maximal subgroup $\Gamma\leq M<\Sp(2k,\F_2)$. 
    Every $M$-invariant subspace is $\Gamma$-invariant, so $M$ is irreducible; moreover, $\abs{M}>\abs{\mathcal P(2k,\F_2)}$. 
    By \cref{prop:aschbacher_symplectic_maximal_subgroups_at_least_size_of_siegel_parabolic}, after a change of logical basis, $M$ is one of $O^+(2k,\F_2)$ and $O^-(2k,\F_2)$, or, only when $k=3$, $G_2(2)$.
    If $M=O^\epsilon(2k,\F_2)$, then \cref{lem:group_theory_irreducible_orthogonal_subgroups_larger_than_siegel_parabolic_contain_orthogonal_cores} gives $\Omega^\epsilon(2k,\F_2)\leq\Gamma$, again a contradiction. 
    If $k=3$ and $M=G_2(2)$, then \cref{prop:group_theory_no_proper_subgroup_of_g22_larger_than_siegel_parabolic} forces $\Gamma=G_2(2)$, contradicting \cref{corr:stab_codes_auto_logical_group_k_eq_3_cannot_contain_g22}. 
    Thus $\abs{\Gamma}\leq\abs{\mathcal P(2k,\F_2)}$ for every $k\geq3$.

    Finally, \cref{lem:order_uniqueness_siegel_parabolic} shows that equality holds iff $\Gamma$ is conjugate to $\mathcal P(2k,\F_2)$, while \cref{cons:concatenated_phantom_clifford}, for example, achieves equality.
\end{proof}

\subsubsection{Logical symplectic maps and stabilizer extensions}

We recall that all physical and logical Clifford actions and orders are understood through their binary (i.e.~projective) symplectic images---that is, modulo Pauli operators and global phases.
The physical symplectic action of an automorphism gate lies in $\Sp(2,\F_2)^n\rtimes S_n\cong S_3^n\rtimes S_n\leq\Sp(2n,\F_2)$: the $n$ local Clifford gates change the Pauli frames of the individual qubits, while $S_n$ permutes their single-qubit Pauli spaces.

An elementary fact formalized in \cref{lem:stab_codes_auto_logical_symplectic_action_deformation} is that a physical Clifford transformation carrying one stabilizer code to another automatically carries the encoded Pauli operators of the first code to the encoded Pauli operators of the second code, while preserving all their commutation relations. 
Moreover, if one deforms a stabilizer code by local Clifford gates, then conjugating a physical automorphism by that same deformation does not change the logical action it implements, provided one transports the logical basis along with the code.

\begin{lemma}
    [Logical symplectic maps induced by physical Clifford isomorphisms]
    \label{lem:stab_codes_auto_logical_symplectic_action_deformation}
    Let $S,S'\subseteq\F_2^{2n}$ be binary stabilizer spaces, and let $M\in\Sp(2n,\F_2)$ relate the two spaces by $MS=S'$. 
    Then $M$ induces a logical space (i.e.~symplectic) isomorphism $\overline M:S^\perp/S\to(S')^\perp/S'$, $[v]\mapsto[Mv]$.

    In particular, if $MS=S$, then $\overline M$ is the logical symplectic action induced by the automorphism gate $M$. 
    Moreover, if $D\in\Sp(2,\F_2)^n$ is a local Clifford deformation, write $S_D\coloneqq DS$ and $M_D\coloneqq DMD^{-1}$.
    Then $M_DS_D=S_D$ and $\overline{M_D}\circ\overline D=\overline D\circ\overline M$. 
    Thus $M$ and $M_D$ have the same induced logical actions in a logical basis $\mathcal L$ of $S^\perp/S$ and its transported basis $\overline D(\mathcal L)$ of $S_D^\perp/S_D$.
\end{lemma}

\begin{proof}
    Since $M$ is symplectic, $(MS)^\perp=MS^\perp$. 
    Indeed, for every $v\in S^\perp$ and $s\in S$, $(Mv)^\top\Omega_n(Ms)=v^\top(M^\top\Omega_nM)s=v^\top\Omega_ns=0$, and equality follows by dimension counting.
    Hence $MS^\perp=(S')^\perp$ and $[v]\mapsto[Mv]$ is well-defined, invertible, and preserves the symplectic form induced on the quotient. 
    For the deformation statement, $M_DS_D=DMD^{-1}DS=DMS=DS=S_D$, and for every $v\in S^\perp$, $\overline{M_D}(\overline D([v]))=\overline{M_D}([Dv])=[M_DDv]=[DMv]=\overline D([Mv])=\overline D(\overline M([v]))$.
\end{proof}

\begin{lemma}
    [Bound on reducible automorphism logical group order on stabilizer codes]
    \label{lem:stab_codes_auto_reducible_logical_group_order_bound}
    Let $\mathcal{C}$ be an $\db{n, k \geq 3}$ stabilizer code and $\mathcal{L}$ be an arbitrary logical basis of $\mathcal{C}$. Assume that $\mathrm{Aut}_\mathcal{L}(\mathcal{C}) \leq \Sp(2k, \mathbb{F}_2)$ is reducible, that is, it stabilizes a nonzero proper subspace $U \subset \mathbb{F}_2^{2k}$. For $k > 3$, additionally assume that every stabilizer code encoding $3 \leq j < k$ logical qubits has an automorphism logical group of order at most $\abs{\mathcal{P}(2j, \mathbb{F}_2)}$. Then $\abs{\mathrm{Aut}_\mathcal{L}(\mathcal{C})} \leq \abs{\mathcal{P}(2k, \mathbb{F}_2)}$.
\end{lemma}

\begin{proof}
    As shorthand, set $\Gamma \coloneqq \mathrm{Aut}_\mathcal{L}(\mathcal{C}) \leq \Sp(2k, \mathbb{F}_2)$ to be the automorphism logical group of the code, and by premise $\Gamma$ preserves some nonzero proper subspace $U$ of the logical symplectic space $V \coloneqq \mathbb{F}_2^{2k}$. Consider $\rad(U) \coloneqq U \cap U^\perp$. 
    Since $\Gamma$ preserves $U$ and the symplectic form, it also preserves $U^\perp$, and hence preserves $\rad(U)$. 
    
    We use symplectic bases $(e_1,\dots,e_k,f_1,\dots,f_k)$ with $e_i$ and $f_i$ interpreted as logical $X_i$ and $Z_i$, respectively.
    The two cases below have different physical meanings: a nonzero radical is a preserved collection of commuting logical operators, whereas a nondegenerate invariant subspace is a complete logical subsystem.
        
    \begin{itemize}
        \item $\rad(U) \neq \{0\}$. For brevity, denote $R \coloneqq \rad(U) \subset V$, so $R$ is a nonzero totally isotropic subspace stabilized by $\Gamma$. 
        Write $r \coloneqq \dim R \geq 1$. 
        The form induced on $U/R$ is nondegenerate; choose a symplectic basis there, lift it to $U$, and extend a basis of $R$ to a symplectic basis of $V$.
        We may therefore arrange that, for some $m\ge r$, $U=\la e_1,\dots,e_m,f_{r+1},\dots,f_m\ra$ and $R=\la e_1,\dots,e_r\ra$.
        If $r = k$, then $R$ is a Lagrangian. 
        The stabilizer of a Lagrangian is precisely a Siegel parabolic, so $|\Gamma| \leq |\mathcal{P}(2k, \mathbb{F}_2)|$.

        Otherwise, suppose $1 \leq r < k$. 
        Put $W\coloneqq\la e_{r+1},\ldots,e_k,f_{r+1},\ldots,f_k\ra$ and $R^\vee\coloneqq\langle f_1,\ldots,f_r\rangle$. 
        Then $V=R\oplus W\oplus R^\vee$ and $W\cong R^\perp/R$. 
        Every element of $\Stab_{\Sp(2k,\F_2)}(R)$ has the form
        \begin{equation}
            \begin{pmatrix}
                A&0&0\\
                0&B&0\\
                0&0&A^{-\top}
            \end{pmatrix}
            \begin{pmatrix}
                I&P^\top\Omega_{k-r}&Q\\
                0&I&P\\
                0&0&I
            \end{pmatrix},
            \qquad
            Q+Q^\top=P^\top\Omega_{k-r}P,
            \label{eq:stab_codes_auto_isotropic_stabilizer_matrix}
        \end{equation}
        where $A\in\GL(r,\F_2)$, $B\in\Sp(W)\cong \Sp(2(k-r),\F_2)$, and $P\in\F_2^{2(k-r)\times r}$.
        Thus $\Stab(R)=\mathcal U\rtimes\bigl(\GL(r,\F_2)\times\Sp(W)\bigr)$, with $\mathcal U$ given by the second factor in \cref{eq:stab_codes_auto_isotropic_stabilizer_matrix}.
        
        Let $\mathcal B\leq\Sp(W)$ be the image of $\Gamma$ on $R^\perp/R$. The image of $\Gamma$ in the Levi quotient $\Stab(R)/\mathcal U$ is contained in $\GL(r,\F_2)\times\mathcal B$, and its kernel is contained in $\mathcal U$. Therefore
        $\abs{\Gamma}\leq\abs{\mathcal U}\abs{\GL(r,\F_2)}\abs{\mathcal B}$.

        First we compute $\abs{\mathcal U}$. 
        The matrix $P$ contributes $2r(k-r)$ binary parameters. Once $P$ is fixed, the solutions for $Q$ form an affine space over the symmetric $r\times r$ matrices, contributing $r(r+1)/2$ further parameters. Hence
        \begin{equation}
            \abs{\mathcal{U}} = 2^{u(k,r)},
            \qquad
            u(k,r)=2r(k-r)+\frac{r(r+1)}{2}.
        \end{equation}
        Next, we bound $\abs{\mathcal B}$. 
        Gauge-fix the first $r$ logical qubits to $\ket{+}^{\otimes r}$, i.e., add $R=\la e_1,\dots,e_r\ra$ to $X$-type stabilizers.
        The gauge-fixed code has logical space $R^\perp/R\cong W$ and encodes $b\coloneqq k-r$ qubits.
        The matrix factors in \cref{eq:stab_codes_auto_isotropic_stabilizer_matrix} now have a direct interpretation. On the gauge-fixed code, the factor $A$ only changes the generating basis of the newly added stabilizer subspace $R$ and is invisible on $R^\perp/R$. The factor $B$ gives the action in $\mathcal B$ on the surviving logical qubits. The unipotent factor changes surviving logical representatives by elements of $R$ or changes the discarded conjugates $f_1,\ldots,f_r$, and is therefore invisible on $R^\perp/R$. For example, a logical $\mathrm{CX}_{j1}$ with $j>r$ is invisible after logical qubit $1$ is fixed in $\ket{+}$.
        
        Consequently, $\mathcal B$ is contained in the automorphism logical group of a stabilizer code encoding $b$ qubits. Let $B_b$ denote the largest possible order of such a group. 
        By \cref{thm:stab_codes_auto_logical_max_order_k_leq_2}, $B_1 = 6$ and $B_2 = 72$, and by assumption, $B_b \leq \abs{\mathcal{P}(2b, \mathbb{F}_2)}$ for all $3 \leq b < k$.
        
        We now compare orders.
        \begin{itemize}
            \item For $3 \leq b < k$,
            \begin{equation}\begin{split}
                \frac{
                    \abs{\mathcal U}\,
                    \abs{\GL(r,\mathbb F_2)}\,
                    \abs{\mathcal P(2b,\mathbb F_2)}
                }{
                    \abs{\mathcal P(2k,\mathbb F_2)}
                }
                &= 
                \frac{
                    2^{2rb+r(r+1)/2}
                    \cdot
                    2^{r(r-1)/2}\prod_{i=1}^r(2^i-1)
                    \cdot
                    2^{b^2}\prod_{i=1}^b(2^i-1)
                }{
                    \abs{\mathcal P(2k,\mathbb F_2)}
                }
                \\
                &=
                \frac{
                    2^{(r+b)^2}
                    \left(\prod_{i=1}^r(2^i-1)\right)
                    \left(\prod_{i=1}^b(2^i-1)\right)
                }{
                    2^{k^2}\prod_{i=1}^k(2^i-1)
                }
                \\
                &= \frac{
                    \prod_{i=1}^r(2^i-1)
                }{
                    \prod_{i=b+1}^{b+r}(2^i-1)
                }
                < 1.
            \end{split}\end{equation}

            \item For $b = 2$, $\abs{\mathcal B}\leq B_2=72$ and $\abs{\mathcal P(4,\mathbb F_2)} = 48$. Then
            \begin{equation}\begin{split}
                \frac{
                    \abs{\mathcal U}
                    \abs{\GL(r,\mathbb F_2)}
                    B_2
                }{
                    \abs{\mathcal P(2k,\mathbb F_2)}
                }
                &= \frac{72}{48} \cdot \frac{
                    \abs{\mathcal U}
                    \abs{\GL(r,\mathbb F_2)}\abs{\mathcal P(4,\mathbb F_2)}}{
                    \abs{\mathcal P(2k,\mathbb F_2)}
                } 
                \\
                & = \frac{72}{48}
                \cdot
                \frac{
                    \prod_{i=1}^{k-2}(2^i-1)
                }{
                    \prod_{i=3}^{k}(2^i-1)
                } 
                \\
                &=
                \frac{9}{2(2^{k-1}-1)(2^k-1)}
                <1.
            \end{split}\end{equation}

            \item For $b = 1$, $\abs{\mathcal B}\leq B_1=6$ and $\abs{\mathcal P(2,\mathbb F_2)} = 2$. Then, noting that $k \geq 3$,
            \begin{equation}\begin{split}
                \frac{
                    \abs{\mathcal U}
                    \abs{\GL(r,\mathbb F_2)}
                    B_1
                }{
                    \abs{\mathcal P(2k,\mathbb F_2)}
                }
                =
                3
                \cdot
                \frac{
                    \prod_{i=1}^{k-1}(2^i-1)
                }{
                    \prod_{i=2}^{k}(2^i-1)
                }
                =
                \frac{3}{2^k-1}
                <1.
            \end{split}\end{equation}

        \end{itemize}
        
        In summary, unless $R$ is Lagrangian in which case $\Gamma \leq \mathcal{P}(2k, \mathbb{F}_2)$, we have $\abs{\Gamma} < \abs{\mathcal{P}(2k, \mathbb{F}_2)}$.

        \item $\rad(U) = \{0\}$. In this case, $U$ is a nonzero nondegenerate symplectic subspace of $V$. Since $\Gamma$ preserves $U$ and is symplectic, it also preserves $U^\perp$, and $V=U\oplus U^\perp$. Write $2a \coloneqq \dim U \geq 2$. Replacing $U$ by $U^\perp$ if necessary, we may assume $1\leq a\leq k/2$. 
        Without loss of generality, we may write $U=\la e_1,\dots,e_a,f_1,\dots,f_a\ra$ and $U^\perp=\la e_{a+1},\dots,e_k,f_{a+1},\dots,f_k\ra$.
        Thus $U$ defines an $a$-qubit logical subspace, and $U^\perp$ its complement. The two block actions need not, however, be independent.
        
        Let $K \coloneqq \ker\left(\Gamma\to \Sp(U) \cong \Sp(2a, \mathbb{F}_2)\right)$, which is normal in $\Gamma$ as a kernel. Then $\abs{\Gamma} \leq \abs{\Sp(2a,\mathbb F_2)}\,\abs{K}$. 
        $K$ can be interpreted as follows. Choose the Lagrangian $A=\la e_1,\dots,e_a\ra\subset U$ and add its logical $X$ operators to the stabilizer -- gauge fixing the first $a$ logical qubits in $\ket{+}^{\otimes a}$.
        The resulting code encodes $b\coloneqq k-a$ logical qubits with the logical space $A^\perp/A\cong U^\perp$.
        A general element of $\Gamma$ may rotate $A$ to another Lagrangian in $U$, e.g. $\la f_1,\dots,f_a\ra$, and hence map this gauge-fixed code to a different one.
        The subgroup $K$, by contrast, fixes $U$ pointwise and therefore preserves the gauge-fixed code, so for any $g\in K$, its induced logical action on $U^\perp$ is well-defined. Moreover, the action of $K$ on $U^\perp$ is faithful: if $g\in K$ is such that $gx=x$ for all $x\in U^\perp$, then $g=\text{id}$; this is because an element trivial on both $U$ and $U^\perp$ is trivial on $V=U\oplus U^\perp$.
        Thus $K$ embeds into the automorphism logical group of the gauge-fixed code, and therefore $\abs{K} \leq B_b$. 
        As before, by \cref{thm:stab_codes_auto_logical_max_order_k_leq_2}, $B_1 = 6$ and $B_2 = 72$, and by assumption, $B_b \leq \abs{\mathcal{P}(2b, \mathbb{F}_2)}$ for all $3 \leq b < k$. Moreover, because $a\leq k/2$, we have $a\leq b$.

        We now compare orders.
        \begin{itemize}
            \item For $3 \leq b < k$. Then
            \begin{equation}
                \frac{
                    \abs{\Sp(2a,\mathbb F_2)}
                    \abs{\mathcal P(2b,\mathbb F_2)}
                }{
                    \abs{\mathcal P(2k,\mathbb F_2)}
                }
                =
                \frac{
                    2^{a^2}\prod_{i=1}^a(2^{2i}-1)
                    \cdot
                    2^{b^2}\prod_{i=1}^b(2^i-1)
                }{
                    2^{k^2}\prod_{i=1}^k(2^i-1)
                }
                =
                2^{-2ab}
                \prod_{i=1}^a
                \frac{2^{2i}-1}{2^{b+i}-1}
                < 1,
            \end{equation}
            where we have noted $b+i\geq 2i$ for every $1\leq i\leq a$ since $a \leq b$.

            \item For $b = 2$, $a\leq b$ gives $a=1$ or $a=2$. Directly,
            \begin{equation}\begin{split}
                \abs{\Sp(2,\mathbb F_2)} B_2
                =
                6\cdot 72
                =
                432
                &<
                \abs{\mathcal P(6,\mathbb F_2)}
                =
                10752,
                \\
                \abs{\Sp(4,\mathbb F_2)}B_2
                =
                720\cdot 72
                =
                51840
                &<
                \abs{\mathcal P(8,\mathbb F_2)}
                =
                20643840.
            \end{split}\end{equation}

            \item The case $b=1$ cannot occur with $a\leq b$ and $k\geq 3$.
        \end{itemize}

        In summary, we obtain $\abs{\Gamma} < \abs{\mathcal{P}(2k, \mathbb{F}_2)}$ for all cases here as well.
    \end{itemize}
\end{proof}

\subsubsection{Gate-dependent automorphism-to-permutation reduction}

\begin{lemma}
    [Reduction of coprime-to-six physical automorphism groups to permutation groups]
    \label{lem:stab_codes_auto_perm_reduction_cyclic_group_logical_action}
    Let $Q\leq\Sp(2,\F_2)^n\rtimes S_n$ be a group of physical automorphism gates of a stabilizer code $\mathcal C$ with binary stabilizer space $S \subseteq \mathbb F_2^{2n}$. 
    If $\abs{Q}$ is coprime to $6$, then there is a local Clifford deformation $D\in\Sp(2,\F_2)^n$ such that $DQD^{-1}\leq S_n$ consists only of qubit permutations on the deformed code with binary stabilizer space $S_D\coloneqq DS$ (cf.~\cref{lem:stab_codes_auto_logical_symplectic_action_deformation}). 
    Conjugation by $D$ leaves the permutation part of every element of $Q$ unchanged and, in the transported logical basis on the deformed code, $DQD^{-1}$ induces the same logical group as $Q$.
\end{lemma}

\begin{proof}
    Let $E_i\cong\F_2^2$ be the Pauli space of physical qubit $i$, and for each $u\in Q$, let $\pi_u\in S_n$ denote the permutation part of $u$. Thus $Q$ acts on the qubit labels by $i\mapsto\pi_u(i)$. Consider one $Q$-orbit $\mathcal O$ and choose qubit $j\in\mathcal O$. Define the point stabilizer $Q_j\coloneqq\{u\in Q:\pi_u(j)=j\}\le Q$, consisting of those automorphism gates in $Q$ whose permutation parts fix qubit $j$. Restriction to $E_j$ gives a homomorphism $Q_j\to\Sp(E_j)\cong S_3$. Its image has order dividing both $\abs{Q_j}$ (and thus $\abs{Q}$) and $6$, and is therefore trivial.

    For each $i\in\mathcal O$, choose $q_i\in Q$ whose permutation part sends $j$ to $i$, and transport a chosen symplectic basis of $E_j$ to $E_i$ using $q_i$. This basis is independent of the choice of $q_i$: if $q_i'$ is another choice, then $q_i^{-1}q_i'\in Q_j$, which acts trivially on $E_j$. Moreover, for $q\in Q$, we have $q_{\pi_q(i)}^{-1}qq_i\in Q_j$, so $qq_i$ and $q_{\pi_q(i)}$ induce the same map $E_j\to E_{\pi_q(i)}$. Thus $q$ sends the transported basis of $E_i$ to that of $E_{\pi_q(i)}$ without any additional local Clifford action.

    Repeating the construction on every orbit gives compatible Pauli frames for all qubits. Their change from the original frames is implemented by some $D\in\Sp(2,\F_2)^n$, and every element of $DQD^{-1}$ is a pure permutation. Since $D$ is a tensor product of single-qubit Cliffords, conjugation by $D$ does not change permutation parts. The claim about the logical action follows from \cref{lem:stab_codes_auto_logical_symplectic_action_deformation}.
\end{proof}

\begin{remark}
    [Reduction of a prime-order physical automorphism gate to a permutation]
    \label{rem:stab_codes_auto_subgroup_to_permutation_reduction_literature}
    The prime-order single-gate case of \cref{lem:stab_codes_auto_perm_reduction_cyclic_group_logical_action} appeared in Ref.~\cite[Lem.~6]{chakraborty2026nogo}. Our lemma extends that reduction to every \emph{group} of automorphism gates of order coprime to $6$.
\end{remark}

\begin{remark}
    [Implementation of a coprime-to-six cyclic automorphism logical group by permutations]
    \label{rem:stab_codes_auto_perm_reduction_cyclic_group_logical_action}
    As a special case of \cref{lem:stab_codes_auto_perm_reduction_cyclic_group_logical_action}, suppose that the induced logical group is $\la \Sigma\ra$ for some logical Clifford $\Sigma$ of order $N$ coprime to $6$. Then the local Clifford deformation can be chosen such that a pure qubit permutation induces exactly $\Sigma$, rather than merely some element of $\la \Sigma\ra$.

    Indeed, let $g$ be an automorphism gate of the original stabilizer code that induces $\Sigma$, and write $\ord(g)=2^a3^bm$, where $\gcd(m,6)=1$. Set $q\coloneqq g^{2^a3^b}$. Then $\ord(q)=m$, while $q$ induces $\Sigma^r$ with $r\coloneqq2^a3^b$. Since $\gcd(r,N)=1$, choose $s$ such that $rs\equiv1\!\pmod{N}$. Applying \cref{lem:stab_codes_auto_perm_reduction_cyclic_group_logical_action} to $Q=\la q\ra$, we may locally Clifford deform the code so that $q$ becomes a pure qubit permutation, without changing its induced logical action in the transported logical basis. Therefore $q^s$ is also a pure qubit permutation and induces $\Sigma^{rs}=\Sigma$.
\end{remark}

\subsubsection{Invariant Lagrangians and characteristic-polynomial constraints}

\begin{lemma}
    [Coprime-to-six automorphism logical actions preserve a logical Lagrangian]
    \label{lem:stab_codes_auto_coprime_six_preserves_lagrangian}
    Let $\Sigma$ be a logical action of order $N$ coprime to six induced by an automorphism gate on an $\db{n,k}$ stabilizer code. Then $\Sigma$ preserves a logical Lagrangian subspace of dimension $k$.
\end{lemma}

\begin{proof}
    Let $\mathcal{C}_0$ be a stabilizer code encoding $k$ logical qubits on which an automorphism gate induces the logical action $\Sigma$ in some logical basis. By \cref{rem:stab_codes_auto_perm_reduction_cyclic_group_logical_action}, there exists an equivalent stabilizer code $\mathcal{C}$, obtained from $\mathcal{C}_0$ by single-qubit Clifford deformation, and a corresponding deformed logical basis such that a physical qubit permutation $\sigma\in S_n$ induces the same logical action $\Sigma$ on $\mathcal C$.

    Let $S\subseteq\mathbb F_2^{2n}$ be the binary stabilizer space of $\mathcal C$, so that its logical symplectic space is $L\coloneqq S^\perp/S$. Let $L_Z\subseteq L$ be the subspace of logical Pauli classes having representatives consisting only of physical $Z$ operators (and identities). An explicit basis for $L_Z$ can be obtained by placing the stabilizer code into standard form~\cite{gottesman1997stabilizer}, which indicates $\dim L_Z = k$ and that $L_Z$ is a Lagrangian. Because physical qubit permutations preserve the $X$/$Z$ type of physical Pauli operators, the permutation $\sigma$ preserves $L_Z$. Therefore the induced logical action $\Sigma$ preserves the logical Lagrangian $L_Z$ on $\mathcal C$.

    Finally, the single-qubit Clifford deformation relating $\mathcal C_0$ and $\mathcal C$ induces a symplectic isomorphism between their logical spaces. Pulling $L_Z$ back through this isomorphism gives a Lagrangian subspace of the logical space of $\mathcal C_0$ preserved by the original logical action $\Sigma$. Hence $\Sigma$ preserves a logical Lagrangian subspace of dimension $k$.
\end{proof}

\begin{lemma}
    [Reciprocal factors of prime-order automorphism logical actions]
    \label{lem:stab_codes_auto_prime_order_reciprocal_factor_constraint}
    Let $p \geq 5$ be prime, and $\Sigma$ be an order-$p$ logical action induced by an automorphism gate on a stabilizer code encoding $k$ logical qubits. Let $\chi_\Sigma(x)$ denote the characteristic polynomial of $\Sigma$ on the $2k$-dimensional logical space of the code. 
    Then, for every irreducible factor $f(x)$ of $x^p-1$ over $\mathbb{F}_2$,
    \begin{equation}
        \nu_f(\chi_\Sigma)
        =
        \nu_{f^*}(\chi_\Sigma),
        \label{eq:stab_codes_auto_prime_order_reciprocal_factor_constraint_1}
    \end{equation}
    where $\nu_f(\chi_\Sigma)$ denotes the multiplicity of $f(x)$ in $\chi_\Sigma(x)$ and $f^*(x)$ denotes the reciprocal polynomial of $f(x)$---see \cref{def:linear_algebra_reciprocal_polynomials}.
    Moreover, if $f = f^*$, then $\nu_f(\chi_\Sigma)$ is even.
\end{lemma}

\begin{proof}
    By \cref{lem:stab_codes_auto_coprime_six_preserves_lagrangian}, the order-$p$ logical action $\Sigma$ preserves a logical Lagrangian $\Lambda \subset \mathbb{F}_2^{2k}$. Choose a symplectic basis adapted to $\Lambda$. In such a basis, $\Sigma$ has a block-upper-triangular symplectic representation
    \begin{equation}
        \Sigma
        =
        \begin{pmatrix}
            A & B\\
            0 & A^{-\top}
        \end{pmatrix},
    \end{equation}
    with $A\in \GL(k,\mathbb{F}_2)$, and the lower-right block is $A^{-\top}$ because $\Sigma$ is symplectic. Since $\Sigma$ is block upper triangular,
    \begin{equation}
        \chi_\Sigma(x)
        =
        \chi_A(x)\chi_{A^{-\top}}(x)
        =
        \chi_A(x)\chi_A^*(x),
    \end{equation}
    where the second equality follows from Property~\cref{item:characteristic_poly_3} of \cref{prop:characteristic_poly}.

    Let $f(x)$ be an irreducible factor of $x^p-1$. Since $f(0)\neq0$, its reciprocal $f^*(x)$ is defined as in \cref{def:linear_algebra_reciprocal_polynomials}. By \cref{def:linear_algebra_reciprocal_polynomials}, $\nu_f(\chi_A^*)=\nu_{f^*}(\chi_A)$ and $\nu_{f^*}(\chi_A^*)=\nu_f(\chi_A)$. Hence
    \begin{equation}\begin{split}
        \nu_f(\chi_\Sigma)
        &=
        \nu_f(\chi_A)+\nu_f(\chi_A^*)
        =
        \nu_f(\chi_A)+\nu_{f^*}(\chi_A),\\
        \nu_{f^*}(\chi_\Sigma)
        &=
        \nu_{f^*}(\chi_A)+\nu_{f^*}(\chi_A^*)
        =
        \nu_{f^*}(\chi_A)+\nu_f(\chi_A).
    \end{split}\end{equation}
    Hence $\nu_f(\chi_\Sigma)=\nu_{f^*}(\chi_\Sigma)$. Finally, if $f=f^*$, then $\nu_f(\chi_\Sigma)=2\nu_f(\chi_A)$ and is therefore even.
\end{proof}

\begin{corollary}
    [Even exponent of factors of prime-order automorphism logical actions]
    \label{corr:stab_codes_auto_prime_order_even_factor_constraint}
    Let $p \geq 5$ be a prime and suppose $\ord_p(2)$ is even, and let $\Sigma$ be an order-$p$ logical action induced by an automorphism gate on a stabilizer code encoding $k$ logical qubits. Let $\chi_\Sigma(x)$ denote the characteristic polynomial of $\Sigma$ on the $2k$-dimensional logical space of the code. Then every irreducible factor of the cyclotomic polynomial $\Phi_p(x)$ over $\mathbb{F}_2$ occurs in $\chi_\Sigma(x)$ with even exponent.
\end{corollary}

\begin{proof}
    Since $\ord_p(2)$ is even, \cref{fact:polynomials_cyclotomic_reciprocal_factors} tells us that every irreducible factor $f(x)$ of $\Phi_p(x)$ over $\mathbb F_2$ is self-reciprocal, $f=f^*$.
    By \cref{lem:stab_codes_auto_prime_order_reciprocal_factor_constraint}, every self-reciprocal irreducible factor of $x^p-1$ occurs in $\chi_\Sigma(x)$ with even exponent. Since $\Phi_p(x)\mid x^p-1$, every irreducible factor of $\Phi_p(x)$ therefore occurs in $\chi_\Sigma(x)$ with even exponent.
\end{proof}

\subsubsection{Orthogonal core obstruction}

\begin{lemma}
    [Automorphism logical groups cannot contain orthogonal cores]
    \label{lem:stab_codes_auto_logical_groups_cannot_contain_orthogonal_cores}
    Let $\mathcal{C}$ be a stabilizer code encoding $k \geq 3$ logical qubits and $\mathcal{L}$ be an arbitrary logical basis of $\mathcal{C}$. Then $\mathrm{Aut}_\mathcal{L}(\mathcal{C})$ cannot contain either orthogonal core $\Omega^\pm(2k, \mathbb{F}_2)$, as defined in \cref{def:group_theory_binary_orthogonal_cores}, up to conjugacy in $\Sp(2k, \mathbb{F}_2)$.
\end{lemma}

\begin{proof}
    Conjugation by an element of $\Sp(2k, \mathbb{F}_2)$ amounts to a change of logical basis, so without loss of generality, we directly compare $\mathrm{Aut}_\mathcal{L}(\mathcal{C})$ and $\Omega^\pm(2k, \mathbb{F}_2)$. Suppose, for contradiction, that $\Omega^\epsilon(2k,\mathbb{F}_2)\leq \mathrm{Aut}_\mathcal{L}(\mathcal{C})$ for some $\epsilon\in\{+,-\}$. We will construct an order-$5$ element of $\Omega^\epsilon(2k,\mathbb{F}_2)$ whose characteristic polynomial has one nontrivial order-$5$ block.

    Let $A$ be a $4$-dimensional nondegenerate quadratic space of minus type over $\mathbb{F}_2$. Its full orthogonal group has order $\abs{O^-(4,2)} = 2 \cdot 2^2(2^2+1)(2^2-1) = 120$, so by Cauchy's theorem, $O^-(4,2)$ contains an element $c$ of order $5$. 
    A concrete order-five element of $\Sp(4,\F_2)$ is the Bell gate~\cite{chakraborty2026nogo}, which has characteristic polynomial $\chi_{M_{\mathrm{Bell}}}(x)=\Phi_5(x)$---see \cref{ex:linear_algebra_bell_gate_characteristic_polynomial}. Since $\Sp(4,\F_2)\cong S_6$ has a single conjugacy class of order-five elements, the element $c$ above must be symplectically conjugate to $M_{\mathrm{Bell}}$, and hence $\chi_c(x)=\Phi_5(x)$.

    Now choose a nondegenerate quadratic space $B$ of dimension $2k-4$ over $\mathbb{F}_2$ such that the orthogonal direct sum $A\perp B$ has type $\epsilon$. This is possible because $k\geq 3$, so $2k-4\geq 2$, and both $\{+,-\}$ types occur in every positive even dimension over $\mathbb{F}_2$. Define $h \coloneqq c \oplus I_B$. Then $h \in O^\epsilon(2k,\mathbb{F}_2)$, and $h$ has order $5$. Its characteristic polynomial on the natural $2k$-dimensional space is $\chi_h(x) = \chi_c(x)\chi_{I_B}(x) = \Phi_5(x)(x+1)^{2k-4}$.

    For all $k \geq 3$, the subgroup $\Omega^\epsilon(2k,\mathbb{F}_2)$ has index $2$ in $O^\epsilon(2k,\mathbb{F}_2)$. Therefore every odd-order element of $O^\epsilon(2k,\mathbb{F}_2)$ lies in $\Omega^\epsilon(2k,\mathbb{F}_2)$. Since $h$ has order $5$, we have $h \in \Omega^\epsilon(2k,\mathbb{F}_2)$, and hence $h \in \mathrm{Aut}_\mathcal{L}(\mathcal{C})$.
    
    On the other hand, $h$ is an order-$5$ logical action induced by an automorphism gate and $\ord_5(2)=4$ is even. By \cref{corr:stab_codes_auto_prime_order_even_factor_constraint}, applied with $p=5$, every irreducible factor of $\Phi_5(x)$ must occur in $\chi_h(x)$ with even exponent. This contradicts the fact that $\Phi_5(x)$ occurs with exponent $1$ in $\chi_h(x) = (x+1)^{2k-4}\Phi_5(x)$. Therefore $\mathrm{Aut}_\mathcal{L}(\mathcal{C})$ cannot contain $\Omega^\epsilon(2k,\mathbb{F}_2)$. Since $\epsilon\in\{+,-\}$ was arbitrary, neither orthogonal core can be contained in $\mathrm{Aut}_\mathcal{L}(\mathcal{C})$.
\end{proof}

\subsubsection{Dihedral subgroup obstruction}

\begin{lemma}
    [Physical Sylow subgroup compatible with logical inversion]
    \label{lem:stab_codes_auto_sylow_lift_inverter}
    Let $p\geq5$ be prime, $\mathrm{PAut}(\mathcal C)$ be the group of physical automorphism gates of a stabilizer code $\mathcal{C}$, and $\phi:\mathrm{PAut}(\mathcal C)\to\Sp(2k,\F_2)$ be the homomorphism mapping physical gates to their logical actions.
    Suppose $g,t\in\im\phi$ satisfy $\ord(g)=p$ and $tgt^{-1}=g^{-1}$. 
    Set $G\coloneqq\langle g\rangle$ and $\overline G\coloneqq\phi^{-1}(G)$. 
    Then there are a Sylow $p$-subgroup $P\leq\overline G$ and a physical automorphism $\widehat t \in \mathrm{PAut}(\mathcal C)$ implementing a logical action in the coset $Gt$ such that:
    \begin{enumerate}[noitemsep]
        \item $\phi(P)=G$;
        \item $\widehat t$ normalizes $P$ and induces inversion on $P/(P\cap\ker\phi)\cong G$;
        \item After one local Clifford deformation of the code and corresponding transport of the logical basis, $P$ acts by pure qubit permutations, while the preceding two properties remain valid.
    \end{enumerate}
\end{lemma}

\begin{proof}
    Choose $\overline g\in\overline G$ with $\phi(\overline g)=g$, and write $\ord(\overline g)=p^s m$ with $s\geq1$ and $\gcd(p,m)=1$. Then $Q\coloneqq\langle\overline g^m\rangle$ has order $p^s$, and $\phi(Q)=G$ because $g^m$ still generates $G$. Choose a Sylow $p$-subgroup $P\leq\overline G$ containing $Q$. Then $\phi(P)=G$. Define $P_0\coloneqq P\cap\ker\phi$. Then $P/P_0\cong G$.

    Choose any physical automorphism $\overline t\in\mathrm{PAut}(\mathcal C)$ implementing $t$, so that $\phi(\overline t)=t$. Since $t$ normalizes $G$, conjugation by $\overline t$ normalizes $\overline G$. Hence $\overline tP\overline t^{-1}$ is another Sylow $p$-subgroup of $\overline G$. By Sylow conjugacy, choose $b\in\overline G$ such that $b(\overline tP\overline t^{-1})b^{-1}=P$, and set $\widehat t\coloneqq b\overline t$. Then $\widehat tP\widehat t^{-1}=P$. Since $\phi(b)\in G$, the logical action $\phi(\widehat t)$ belongs to $Gt$. Since $\widehat t(\ker\phi)\widehat t^{-1}=\ker\phi$, the relation $\widehat tP\widehat t^{-1}=P$ implies that $\widehat tP_0\widehat t^{-1}=P_0$. Moreover, for every $u\in P$, write $\phi(u)=g^a$ for some non-negative integer $a$. Then
    \begin{equation}
        \phi(\widehat t u\widehat t^{-1})
        =
        \phi(b)tg^at^{-1}\phi(b)^{-1}
        =
        \phi(b)g^{-a}\phi(b)^{-1}
        =
        g^{-a}
        =
        \phi(u)^{-1},
    \end{equation}
    where the penultimate equality uses that $\phi(b)\in G=\la g\ra$ and hence commutes with $g^{-a}$. Thus $\widehat t$ induces inversion on $P/P_0\cong G$.

    Finally, $\gcd(\abs{P},6)=1$, so by \cref{lem:stab_codes_auto_perm_reduction_cyclic_group_logical_action}, there is a local Clifford deformation under which $P$ becomes a pure permutation group. Conjugating $\widehat t$ by the same deformation and transporting the logical basis accordingly preserves all the stated relations by \cref{lem:stab_codes_auto_logical_symplectic_action_deformation}.
\end{proof}

\begin{lemma}
    [Square characteristic polynomial for inverted prime-order automorphism logical actions]
    \label{lem:stab_codes_auto_square_characteristic_polynomial_inverted_actions}
    Let $p\geq5$ be prime, and $\mathcal C$ be an $\db{n,k}$ stabilizer code with stabilizer space $S\subseteq\F_2^{2n}$ and logical space $L=S^\perp/S$. Suppose that $g$ and $t$ are logical actions induced by automorphism gates on $\mathcal C$, such that $g$ has order $p$ and $tgt^{-1}=g^{-1}$. Then $\chi_{g,L}(x)$ is a square in $\F_2[x]$. Equivalently, every irreducible factor of $\chi_{g,L}(x)$ occurs with even exponent.
\end{lemma}

\begin{proof}
    Let $V\coloneqq\F_2^{2n}$ be the physical Pauli space and $\phi:\mathrm{PAut}(\mathcal C)\to\Sp(2k,\F_2)$ the logical-action homomorphism. By \cref{lem:stab_codes_auto_sylow_lift_inverter}, after one local Clifford deformation there are a $p$-group $P\leq\phi^{-1}(\langle g\rangle)$ and a physical automorphism $\widehat t$ such that $\phi(P)=\langle g\rangle$, the group $P$ acts by pure qubit permutations, and $\widehat t$ normalizes $P$ and induces inversion on $P/P_0$, where $P_0\coloneqq P\cap\ker\phi$. We work with the deformed code and transported logical basis without changing notation. Choose $u\in P$ with $\phi(u)=g$.

    For a group $G$, call an $\F_2$-vector space $E$ a $G$-module if $E$ is equipped with a homomorphism $G\to \GL(E)$. Examples of $P$-modules include the stabilizer space $S$, its symplectic orthogonal $S^\perp$ and dual space $S^\vee=\text{Hom}_{\F_2}(S,\F_2)$ (i.e.~the space of $\F_2$-linear maps from $S$ to $\F_2$), and the logical space $L\coloneqq S^\perp/S$; they are also $P_0$-modules.
    
    For every $P_0$-module $E$, write $E_0\coloneqq E^{P_0}=\{e\in E: a\cdot e=e\ \forall a\in P_0\}$ for its fixed subspace.
    Define $\Pi_E\coloneqq\sum_{a\in P_0}a$. Then $\Pi_E$ is a projection of $E$ onto $E_0$ because for every $e\in E$ and $b\in P_0$, $b\cdot \Pi_E e=\sum_{a\in P_0} (ba)\cdot e=\Pi_E e$. 
    Since $P_0$ is a subgroup of the $p$-group $P$ and $p\geq5$, $\abs{P_0}$ is odd and therefore equals $1$ as a scalar in $\F_2$. Hence $\Pi_E$ is surjective onto $E_0$: for every $e\in E_0$, $\Pi_E e=\abs{P_0}e=e$.
    Moreover, $P_0\lhd P$ implies that $\Pi_E$ commutes with the action of $P$. Indeed, for all $r\in P$ and $e\in E$, $\Pi_E(r\cdot e)=\sum_{a\in P_0}a\cdot (r\cdot e)=r\cdot \left(\sum_{a\in P_0}(r^{-1}ar)\cdot e\right)=r\cdot \Pi_E e$.
    Normality also implies that $E_0$ is preserved by $P$, and since $P_0$ acts trivially on $E_0$, this action factors through $P/P_0$. Indeed, for every $r\in P$, $e_0\in E_0$, and $a\in P_0$, $a\cdot(r\cdot e_0)=r\cdot((r^{-1}ar)\cdot e_0)=r\cdot e_0$, because $r^{-1}ar\in P_0$.
    Therefore, $P/P_0$ acts on and preserves the flag $S_0\subseteq (S^\perp)_0\subseteq V_0$.
    
    We next identify the two successive quotients of this flag as $P/P_0$-modules: $(S^\perp)_0/S_0\cong L$ and $V_0/(S^\perp)_0\cong (S^\vee)_0$. 
    For the former, the natural map $(S^\perp)_0\to L$ is surjective: if $\ell\in L$ and $x\in S^\perp$ represents $\ell$, then $[\Pi_{S^\perp} x]=\sum_{a\in P_0}[a\cdot x]=\abs{P_0}\ell=\ell$, where $[a\cdot x]=\ell$ because $P_0\leq\ker\phi$ acts trivially on $L$. Its kernel is $S\cap(S^\perp)_0=S_0$, hence $(S^\perp)_0/S_0\cong L$.
    For the latter, let $\beta:V\to S^\vee$ be defined by $\beta(v)(s)\coloneqq \la v,s\ra$. This map is $P$-equivariant, surjective, and has kernel $S^\perp$. Its restriction $\beta|_{V_0}:V_0\to(S^\vee)_0$ is also surjective: if $\lambda\in(S^\vee)_0$ and $\beta(v)=\lambda$, then $\beta(\Pi_Vv)=\Pi_{S^\vee}\lambda=\lambda$. The kernel of this restricted map is $(S^\perp)_0$. Hence $V_0/(S^\perp)_0\cong(S^\vee)_0$.
    
    By item \cref{item:characteristic_poly_1} of \cref{prop:characteristic_poly}, we can then factor the characteristic polynomial of the automorphism gate $u\in P$ that induces logical action $g$ as
    \begin{equation}
        \chi_{u,V_0}(x)
        =
        \chi_{u,S_0}(x)\chi_{g,L}(x)\chi_{u,(S^\vee)_0}(x).
        \label{eq:stab_codes_auto_fixed_point_characteristic_filtration}
    \end{equation}

    We next argue that $\chi_{u,(S^\vee)_0}(x)=\chi_{u,(S_0)^\vee}(x)=\chi_{u,S_0}^*(x)$ by showing $(S^\vee)_0\cong(S_0)^\vee$ as $P/P_0$-modules.
    The forward and backward maps are given by
    \begin{equation}
        (S^\vee)_0\xrightleftharpoons[J]{R} (S_0)^\vee,
        \qquad
        R(\lambda)=\lambda|_{S_0},
        \qquad
        J(\mu)=\mu\circ \Pi_S.
    \end{equation}
    
    The contragredient action of $r\in P$ on $\lambda\in S^\vee$ is $(r\cdot \lambda)(s)\coloneqq \lambda(r^{-1}\cdot s)$ for every $s\in S$. 
    The map $J$ indeed takes values in $(S^\vee)_0$: for every $a\in P_0$ and $s\in S$, $(a\cdot J(\mu))(s)=J(\mu)(a^{-1}\cdot s)=\mu(\Pi_S(a^{-1}\cdot s))=\mu(\Pi_S s)=J(\mu)(s)$.
    Moreover, $RJ=\id_{(S_0)^\vee}$ and $JR=\id_{(S^\vee)_0}$ because $JR(\lambda)(s)=\lambda(\Pi_S s)=\sum_{a\in P_0} \lambda(a\cdot s)=\abs{P_0}\lambda(s)=\lambda(s)$, and similarly for $RJ$.
    The maps $J,R$ are $P/P_0$-equivariant. Indeed, for every $r\in P$ and $s\in S$, $J((rP_0)\cdot\mu)(s)=((rP_0)\cdot \mu)(\Pi_S s)=\mu(r^{-1}\cdot \Pi_S s)=\mu(\Pi_S(r^{-1}\cdot s))=(r\cdot J(\mu))(s)$; for $s_0\in S_0$, $R(r\cdot \lambda)(s_0)=(r\cdot \lambda)(s_0)=\lambda(r^{-1}\cdot s_0)=((rP_0)\cdot R(\lambda))(s_0)$.

    Moreover, $\widehat t$ normalizes $P_0$ and therefore preserves $S_0$. Since it induces inversion on $P/P_0$, there is some $v_0\in P_0$ such that $\widehat t u\widehat t^{-1}=u^{-1}v_0$. The element $v_0$ acts trivially on $S_0$, so $u|_{S_0}$ is conjugate to $u^{-1}|_{S_0}$. Consequently,
    $\chi_{u,S_0}(x)=\chi_{u,S_0}^*(x)$. Substitution into \cref{eq:stab_codes_auto_fixed_point_characteristic_filtration} gives
    \begin{equation}
        \chi_{u,V_0}(x)
        =
        \chi_{g,L}(x)\chi_{u,S_0}(x)^2.
        \label{eq:stab_codes_auto_fixed_point_stabilizer_square}
    \end{equation}

    Finally, write $V=V_{\mathrm x}\oplus V_{\mathrm z}$, where $V_\rmx=\F_2^n\oplus\{0\}$ and $V_\rmz=\{0\}\oplus \F_2^n$ denote the physical $X$- and $Z$-coordinate spaces. Because $P$ acts by pure qubit permutations, it has identical actions on $V_\rmx$ and $V_\rmz$. Therefore $V_0=(V_{\mathrm x})_0\oplus(V_{\mathrm z})_0$, and the two summands are isomorphic as $\langle u\rangle$-modules. Hence $\chi_{u,V_0}(x)=\chi_{u,(V_{\mathrm x})_0}(x)^2$. Together with \cref{eq:stab_codes_auto_fixed_point_stabilizer_square}, this gives the identity
    \begin{equation}
        \chi_{u,(V_{\mathrm x})_0}(x)^2
        =
        \chi_{g,L}(x)\chi_{u,S_0}(x)^2.
        \label{eq:stab_codes_auto_logical_characteristic_square_identity}
    \end{equation}
    Taking the multiplicity $\nu_f$ of any irreducible $f(x)\in\F_2[x]$ in this identity gives
    $2\nu_f(\chi_{u,(V_{\mathrm x})_0})=\nu_f(\chi_{g,L})+2\nu_f(\chi_{u,S_0})$. Thus $\nu_f(\chi_{g,L})$ is even for every $f$, which is equivalent by unique factorization to $\chi_{g,L}(x)$ being a square.
\end{proof}

\begin{corollary}
    [Automorphism logical groups at $k = 3$ cannot contain $D_{14}$ and $G_2(2)$]
    \label{corr:stab_codes_auto_logical_group_k_eq_3_cannot_contain_g22}
    Let $\mathcal{C}$ be an $\db{n, 3}$ stabilizer code and $\mathcal{L}$ be an arbitrary logical basis of $\mathcal{C}$. Then the dihedral group $D_{14} \coloneqq \langle g,t:g^7=t^2=1,\ tgt^{-1}=g^{-1}\rangle \nleq \mathrm{Aut}_\mathcal{L}(\mathcal{C})$.
    Since $D_{14} < G_2(2)$, consequently, $G_2(2) \nleq \mathrm{Aut}_\mathcal{L}(\mathcal{C})$.
\end{corollary}

\begin{proof}
    Suppose, for contradiction, that $D_{14}\leq\mathrm{Aut}_{\mathcal L}(\mathcal C)$, with generators $g,t$ as in the statement.
    By \cref{ex:polynomials_small_cyclotomic}, $\Phi_7(x)=f(x)f^*(x)$, where $f(x)\coloneqq x^3+x+1$ and $f^*(x)=x^3+x^2+1$ are reciprocal irreducible cubic polynomials.
    Since $g$ has order $7$, its minimal polynomial (see \cref{def:linear_algebra_minimal_and_characteristic_polynomials,def:linear_algebra_cyclotomic_polynomials}) satisfies $m_g(x)\mid x^7-1=(x+1)\Phi_7(x)$.
    Since $g\neq I$, we have $m_g(x)\neq x+1$, so at least one irreducible factor of $\Phi_7(x)$ occurs in $m_g(x)$, and hence in $\chi_g(x)$.
    On the other hand, by Property~\cref{item:characteristic_poly_4} of \cref{prop:characteristic_poly}, $\chi_g(x)$ is self-reciprocal, so $f(x)$ and $f^*(x)$ occur in $\chi_g(x)$ with equal multiplicity.
    Since both have degree $3$ and $\deg\chi_g=6$, they must each occur with multiplicity exactly $1$.
    Hence $\chi_g(x)=f(x)f^*(x)=\Phi_7(x)$, so each irreducible factor of $\Phi_7(x)$ occurs in $\chi_g(x)$ with exponent $1$.
    But \cref{lem:stab_codes_auto_square_characteristic_polynomial_inverted_actions} requires $\chi_g(x)$ to be a square polynomial, a contradiction.
    Therefore $D_{14}\nleq\mathrm{Aut}_{\mathcal L}(\mathcal C)$.
    Since $D_{14}<G_2(2)$, it follows immediately that $G_2(2)\nleq\mathrm{Aut}_{\mathcal L}(\mathcal C)$.

    For concreteness, one realization of the defining relations of $D_{14}$ is
    $g=\CX_{12}\CX_{23}\SWAP_{13}$ and $t=H^{\otimes 3}$. 
    One can verify that $g^7=t^2=I$ and $tgt^{-1}=g^{-1}$, so $\la g,t\ra\cong D_{14}$.
    The symplectic matrices of the two are
    \begin{equation}
        M_g
        =
        \begin{pmatrix}
            A&0\\
            0&A^{-\top}
        \end{pmatrix}
        =
        \begin{pmatrix}
            A&0\\
            0&A^{-1}
        \end{pmatrix},
        \qquad
        A=
        \begin{pmatrix}
            0&0&1\\
            0&1&1\\
            1&1&0
        \end{pmatrix}
        \in\GL(3,\F_2),
        \qquad
        M_t=
        \begin{pmatrix}
            0&I_3\\
            I_3&0
        \end{pmatrix},
    \end{equation}
    where $A^{-\top}=A^{-1}$ because $A^\top=A$.
    One can calculate the characteristic polynomial of $M_g$:
    \begin{equation}
        \chi_A(x)=x^3+x^2+1,
        \qquad
        \chi_{A^{-1}}(x)=\chi^*_A(x)=x^3+x+1,
        \qquad
        \chi_{M_g}(x)=\chi_A(x)\chi_{A^{-1}}(x)=\Phi_7(x).
    \end{equation}
\end{proof}

\subsubsection{Forbidden prime divisors of automorphism logical group orders}
\label{sec:stab_codes/auto/prime_divisors}

\begin{lemma}
    [Forbidden prime-order automorphism logical actions]
    \label{lem:stab_codes_auto_prime_factor_nogo_1}
    Let $p \geq 5$ be a prime factor of $\abs{\Sp(2k, \mathbb{F}_2)}$ but not of $\abs{\GL(k, \mathbb{F}_2)}$. Then automorphism gates on stabilizer codes encoding $k$ logical qubits cannot be order-$p$ in logical action. 
    (Admissible primes $p$ are listed in \cref{prop:group_theory_prime_factors_of_sp_not_gl} for small $k$.)
\end{lemma}

\begin{proof}
    By \cref{lem:stab_codes_auto_coprime_six_preserves_lagrangian}, the order-$p$ logical action $\Sigma$ preserves a logical Lagrangian $\Lambda \subset \mathbb{F}_2^{2k}$. Choose a symplectic basis adapted to $\Lambda$. In such a basis, $\Sigma$ has block-upper-triangular symplectic form
    \begin{equation}
        \Sigma
        =
        \begin{pmatrix}
            A & B\\
            0 & A^{-\top}
        \end{pmatrix},
    \end{equation}
    with $A\in \GL(k,\mathbb{F}_2)$. Here the lower-right block is $A^{-\top}$ because $\Sigma$ is symplectic. We note
    \begin{equation}
        \Sigma^p = \mqty(A^p & \cdot \\ 0 & (A^{-\top})^p).
    \end{equation}
    Because $\Sigma^p = I$, $A$ must have order dividing $p$. But $p$ is prime and $p$ is not a divisor of $\abs{\GL(k, \mathbb{F}_2)}$, so by Lagrange's theorem, $\GL(k, \mathbb{F}_2)$ contains no element of order $p$. The only choice is then $A = I$, and
        \begin{equation}
             \Sigma^p = \mqty(I & B \\ 0 & I)^p = \mqty(I & B \\ 0 & I) = I
             \Longrightarrow B = 0.
        \end{equation}
    This implies $\Sigma$ is the identity, but $\Sigma$ was assumed to have order $p$. We conclude the impossibility of the logical action $\Sigma$, a contradiction.
\end{proof}

\begin{corollary}
    [Forbidden prime-order automorphism logical actions on subsets of logical qubits]
    \label{lem:stab_codes_auto_prime_factor_nogo_2}
    Let $p \geq 5$ be a prime factor of $\abs{\Sp(2k_0, \mathbb{F}_2)}$ but not of $\abs{\GL(k_0, \mathbb{F}_2)}$. Then automorphism gates on stabilizer codes encoding $k \geq k_0$ logical qubits cannot induce an order-$p$ logical action of the form $h (g \otimes I) h^{-1}$ where $g$ acts on any subset of $k_0$ logical qubits, and $h$ is any $k$-qubit Clifford.
    (Admissible primes $p$ are listed in \cref{prop:group_theory_prime_factors_of_sp_not_gl} for small $k$.)
\end{corollary}

\begin{proof}
    Suppose there is a stabilizer code encoding $k \geq k_0$ logical qubits supporting the logical action $h (g \otimes I) h^{-1}$ in some logical basis, where $g$ acts on a subset of $k_0$ logical qubits. Then, performing the logical basis rotation $h$, the logical action becomes just $g \otimes I$. The order of a group element is preserved by conjugation, so $g$ is order-$p$. Fix the states of the $k - k_0$ remaining logical qubits to either $\ket{0}$ or $\ket{+}$, so that those logical operators become stabilizers. This gives a stabilizer code encoding $k_0$ logical qubits supporting the logical action $g$, and \cref{lem:stab_codes_auto_prime_factor_nogo_1} forbids this.
\end{proof}

\begin{proposition}
    [Forbidden prime factors]
    \label{prop:group_theory_prime_factors_of_sp_not_gl}
    For a positive integer $m$, let $\pi(m)$ denote the set of prime divisors of $m$.
    Then the prime factors of $\abs{\Sp(2k, \mathbb{F}_2)}$ which do not divide $\abs{\GL(k, \mathbb{F}_2)}$ are
    \begin{equation}
        \pi\left(\abs{\Sp(2k, \mathbb{F}_2)}\right)
        \setminus \pi\left(\abs{\GL(k, \mathbb{F}_2)}\right)
        =
        \left\{
            p \,\, \text{odd prime}:
            \ord\nolimits_p(2) \,\, \text{is even and} \,\,
            k < \ord\nolimits_p(2) \leq 2k
        \right\},
        \label{eq:group_theory_prime_factors_of_sp_not_gl}
    \end{equation}
    except that for $k=1$, the prime factor $2$ must additionally be included.
    Thus, apart from this exception when $k=1$, all prime factors in the set difference are odd.
    \Cref{tab:group_theory_prime_factors_of_sp_not_gl} gives a list of such prime factors for small $k$.
    Note that, for the settings of \cref{lem:stab_codes_auto_prime_factor_nogo_1,lem:stab_codes_auto_prime_factor_nogo_2}, only primes $p \geq 5$ should be considered;
    for every $k\geq2$, the set difference in \cref{eq:group_theory_prime_factors_of_sp_not_gl} contains at least one such prime.

    \begin{table}[ht]
        \centering
        \begin{tabular}{p{1cm} p{6cm}}
            \toprule
            \(k\) 
                & \(\pi(|\Sp(2k,\mathbb{F}_2)|) \setminus \pi(|\GL(k,\mathbb{F}_2)|)\)
            \\
            \midrule
            \(1\) & \(\{2,3\}\) \\
            \(2\) & \(\{5\}\) \\
            \(3\) & \(\{5\}\) \\
            \(4\) & \(\{17\}\) \\
            \(5\) & \(\{11,17\}\) \\
            \(6\) & \(\{11,13,17\}\) \\
            \(7\) & \(\{11,13,17,43\}\) \\
            \(8\) & \(\{11,13,43,257\}\) \\
            \bottomrule
        \end{tabular}
        \caption{Prime factors of \( |\Sp(2k,\mathbb{F}_2)| \) which do not divide
        \( |\GL(k,\mathbb{F}_2)| \), for \(1 \leq k \leq 8\).}
        \label{tab:group_theory_prime_factors_of_sp_not_gl}
    \end{table}
\end{proposition}

\begin{proof}
    We use the standard order formulas for $\Sp(2k, \mathbb{F}_2)$ and $\GL(k, \mathbb{F}_2)$ in \cref{eq:group_order_symplectic,eq:group_order_general_linear}.
    Let $p$ be an odd prime, and set $f \coloneqq \ord_p(2)$.
    Thus $f$ is the least positive integer such that $2^f \equiv 1 \pmod p$.
    Therefore, for any positive integer $m$, $p \mid 2^m-1 \Longleftrightarrow f \mid m$.
    
    First consider $\GL(k,\mathbb{F}_2)$.
    Since $p$ is odd, the power of two in the order formula contributes no factor of $p$.
    Hence $p \mid |\GL(k,\mathbb{F}_2)| \Longleftrightarrow p \mid 2^\ell-1$ for some $1 \leq \ell \leq k$.
    By the defining property of $f$, this is equivalent to $f \mid \ell$ for some $1 \leq \ell \leq k$.
    Such an $\ell$ exists if and only if $f \leq k$.
    Thus $p \in \pi(|\GL(k,\mathbb{F}_2)|) \Longleftrightarrow \ord_p(2) \leq k$.

    Next consider $\Sp(2k,\mathbb{F}_2)$.
    Again, since $p$ is odd, the power of two in the order formula contributes no factor of $p$.
    Hence $p \mid |\Sp(2k,\mathbb{F}_2)| \Longleftrightarrow p \mid 2^{2j}-1$ for some $1 \leq j \leq k$.
    Equivalently, $f \mid 2j$ for some $1 \leq j \leq k$.

    We now characterize those odd primes $p$ which divide $|\Sp(2k,\mathbb{F}_2)|$ but not $|\GL(k,\mathbb{F}_2)|$.
    The condition $p \nmid |\GL(k,\mathbb{F}_2)|$ is equivalent to $f > k$.
    On the other hand, the condition $p \mid |\Sp(2k,\mathbb{F}_2)|$ is equivalent to the existence of some $1 \leq j \leq k$ such that $f \mid 2j$.
    Suppose first that such a $j$ exists and that $f>k$.
    If $f$ were odd, then $f \mid 2j$ would imply $f \mid j$, since $\gcd(f,2)=1$.
    But then $j \geq f > k$, contradicting $j \leq k$.
    Hence $f$ must be even.
    Writing $f=2a$, the divisibility condition $f \mid 2j$ becomes $2a \mid 2j$, or equivalently, $a \mid j$.
    Since $j \leq k$, this implies $a \leq k$, and therefore $f = 2a \leq 2k$.
    Thus any such prime satisfies $\ord_p(2)$ is even and $k < \ord_p(2) \leq 2k$.

    Conversely, suppose that $p$ is an odd prime such that $f=\ord_p(2)$ is even and $k < f \leq 2k$.
    Write $f=2a$.
    Then $a \leq k$.
    Taking $j=a$, we have $1 \leq j \leq k$ and $f = 2a = 2j$, so in particular $f \mid 2j$.
    Hence $p \mid 2^{2j}-1$, and therefore $p \mid |\Sp(2k,\mathbb{F}_2)|$.
    Meanwhile, since $f>k$, there is no $1 \leq \ell \leq k$ with $f \mid \ell$, so $p \nmid |\GL(k,\mathbb{F}_2)|$.

    Combining the two directions gives \cref{eq:group_theory_prime_factors_of_sp_not_gl}.
    For all $k \geq 2$, the prime factor $2$ appears in both $|\Sp(2k,\mathbb{F}_2)|$ and $|\GL(k,\mathbb{F}_2)|$.
    But for $k = 1$, $|\Sp(2,\mathbb{F}_2)| = 6$ and $|\GL(1,\mathbb{F}_2)| = 1$, so the prime factor $2$ must additionally be included in the set difference.

    Finally, we show that for every $k\geq2$ the set difference contains a prime $p\geq5$.
    Suppose first that $k\neq3$.
    By Zsigmondy's theorem~\cite{zsigmondy1892theorie}, $2^{2k}-1$ has a primitive prime divisor $p$, meaning that $p\mid2^{2k}-1$ but $p\nmid2^j-1$ for every $1\leq j<2k$.
    Hence $\ord_p(2)=2k$, so by \cref{eq:group_theory_prime_factors_of_sp_not_gl}, $p$ lies in the desired set difference.
    Moreover, $\ord_p(2)=2k\geq4$ divides $p-1$, so $p\geq2k+1\geq5$.
    The sole relevant Zsigmondy exception is $2^6-1$, corresponding to $k=3$.
    In this case, $p=5$ has $\ord_5(2)=4$, so $3<4\leq6$, and therefore $5$ lies in the set difference by \cref{eq:group_theory_prime_factors_of_sp_not_gl}.
    Thus such a prime $p\geq5$ exists for every $k\geq2$.
\end{proof}

\clearpage

\subsection{Block length bounds}
\label{app:stab_codes/cost}

\subsubsection{Transversal logical groups}

\begin{lemma}
    [Size of transversal logical group bounded by block length on stabilizer codes]
    \label{lem:stab_codes_trans_logical_group_bound_by_n}
    Let $\mathcal{C}$ be an $\db{n,k}$ stabilizer code and $\mathcal{L}$ be a logical basis of $\mathcal{C}$. 
    Then the largest power of $2$ dividing $\abs{\mathrm{Trans}_\mathcal{L}(\mathcal{C})}$ is at most $2^n$.
    Moreover, suppose $\mathrm{Trans}_\mathcal{L}(\mathcal{C})$ contains no subgroup isomorphic to $C_3$. 
    Then $\mathrm{Trans}_\mathcal{L}(\mathcal{C}) \preceq_{\Sp(2k,\mathbb{F}_2)} \mathcal{U}(2k,\mathbb{F}_2)$, and in particular, $\mathrm{Trans}_\mathcal{L}(\mathcal{C}) \cong C_2^r$ with order $2^r$ for some $0\leq r\leq \min\{n,k(k+1)/2\}$. 
    Consequently, if $n<k(k+1)/2$, then the containment is proper, $\mathrm{Trans}_\mathcal{L}(\mathcal{C}) \prec_{\Sp(2k,\mathbb{F}_2)} \mathcal{U}(2k,\mathbb{F}_2)$.
\end{lemma}

\begin{proof}
    We have $\mathrm{PTrans}(\mathcal{C}) \leq \Sp(2, \mathbb{F}_2)^n \simeq S_3^n$ which is the group of physical binary symplectic actions of transversal gates. Since $\mathrm{Trans}_\mathcal{L}(\mathcal{C})$ is an image and hence a quotient of $\mathrm{PTrans}(\mathcal{C})$, $\abs{\mathrm{Trans}_\mathcal{L}(\mathcal{C})} \mid \abs{\mathrm{PTrans}(\mathcal{C})} \mid 6^n$. Hence, the largest power of $2$ dividing $\abs{\mathrm{Trans}_\mathcal{L}(\mathcal{C})}$ is at most $2^n$.

    We make no assumption on the decomposability of $\mathcal{C}$, so $\mathcal{C}$ could comprise multiple indecomposable stabilizer codes. Write them as $\{\mathcal{C}_1, \ldots, \mathcal{C}_m\}$, where $\mathcal{C}_i$ has parameters $\db{n_i,k_i}$, so that $n=\sum_i n_i$ and $k=\sum_i k_i$. In a logical basis adapted to (i.e.~separable with respect to) this decomposition, transversal gates act component-wise, and hence $\mathrm{Trans}(\mathcal{C}) \cong \prod_i \mathrm{Trans}(\mathcal{C}_i)$. Components with $k_i=0$ have trivial transversal logical group and may be ignored.

    Suppose $\mathrm{Trans}_\mathcal{L}(\mathcal{C})$ contains no subgroup isomorphic to $C_3$. Then no $k_i > 0$ component can have a transversal logical group containing $C_3$. By \cref{thm:stab_codes_trans_logical_groups}, each such component therefore satisfies $\mathrm{Trans}(\mathcal{C}_i)\preceq_{\Sp(2k_i,\mathbb F_2)}\mathcal{U}(2k_i,\mathbb F_2)$. Choosing the component logical bases suitably, their direct product is contained in $\mathcal{U}(2k,\mathbb F_2)$: indeed, $\prod_i\mathcal{U}(2k_i,\mathbb F_2)$ identifies with the subgroup of $\mathcal{U}(2k,\mathbb F_2)$ whose symmetric upper-right block is block diagonal with blocks of sizes $k_i$. Thus, in the original logical basis, $\mathrm{Trans}_\mathcal{L}(\mathcal{C})\preceq_{\Sp(2k,\mathbb F_2)}\mathcal{U}(2k,\mathbb F_2)$.

    Since $\mathcal{U}(2k,\mathbb F_2)\cong \smash{C_2^{k(k+1)/2}}$, it follows that $\mathrm{Trans}_\mathcal{L}(\mathcal{C})\cong C_2^r$ for some $0\leq r\leq k(k+1)/2$. We also showed that $r \leq n$, and therefore $0\leq r\leq\min\{n,k(k+1)/2\}$. If $n < k(k+1)/2$, then we have $r\leq n <  k(k+1)/2$, hence $C_2^r \cong \mathrm{Trans}_\mathcal{L}(\mathcal{C})\prec_{\Sp(2k,\mathbb{F}_2)} \mathcal{U}(2k,\mathbb{F}_2)$ and the containment is proper.
\end{proof}

\begin{theorem}
    [Block lengths of stabilizer codes achieving \(\mathcal{U}(2k, \mathbb{F}_2)\) transversal logical group]
    \label{thm:stab_codes_trans_saturating_n_scaling}
    Let $\mathcal{C}$ be an $\db{n,k,d}$ stabilizer code that supports $\mathrm{Trans}_\mathcal{L}(\mathcal{C}) \succeq_{\Sp(2k, \mathbb{F}_2)} \mathcal{U}(2k, \mathbb{F}_2) \cong \langle \mathbf{S}_k, \mathbf{CZ}_k \rangle$ in some logical basis $\mathcal{L}$. Then $n \geq k(k+1)/2$ holds. Equality can be achieved at $d = 1$.
\end{theorem}

\begin{proof}
    Since $\mathcal{U}(2k,\mathbb{F}_2) \cong C_2^{k(k+1)/2}$, the hypothesis implies that $2^{k(k+1)/2}$ divides $\abs{\mathrm{Trans}_\mathcal{L}(\mathcal{C})}$. On the other hand, by \cref{lem:stab_codes_trans_logical_group_bound_by_n}, the largest power of $2$ dividing $\abs{\mathrm{Trans}_\mathcal{L}(\mathcal{C})}$ is at most $2^n$. Therefore $k(k+1)/2\leq n$. Equality is achieved at $d=1$ by \cref{cons:complete_hypergraph_css_t_eq_2}.
\end{proof}

\subsubsection{Permutation logical groups}

\begin{theorem}
    [Block lengths of stabilizer codes achieving \(\GL(k, \mathbb{F}_2)\) permutation logical group]
    \label{thm:stab_codes_perm_saturating_general_linear_n_scaling}
    Let $\mathcal{C}$ be an $\db{n,k,d}$ stabilizer code hosting $\mathrm{Perm}_\mathcal{L}(\mathcal{C}) \succeq_{\Sp(2k, \mathbb{F}_2)} \GL(k, \mathbb{F}_2) \cong \langle \mathbf{CX}_k \rangle$ in some logical basis $\mathcal{L}$. Then $n \geq 2^k - 1$ holds. Equality can be achieved at $d = 1$.
\end{theorem}

\begin{proof}
    Because permutation gates are a subgroup of automorphism gates, \cref{thm:stab_codes_auto_saturating_general_linear_n_scaling} holds. 
    The $k = 2$ exception does not occur: a $\db{2, 2}$ code realizing a $\GL(2, \mathbb{F}_2)$ or larger permutation logical group is impossible because the group of permutations on two physical qubits is $S_2$ and $\abs{S_2} < \abs{\GL(2, \mathbb{F}_2)}$.
    The $n = 2^k - 1$ equality is achieved at $d = 1$ by phantom codes~\cite{koh2026entangling}---see \cref{cons:concatenated_phantom_clifford}.
\end{proof}

\begin{theorem}
    [Block lengths of stabilizer codes achieving \(\mathcal{P}(k, \mathbb{F}_2)\) permutation logical group]
    \label{thm:stab_codes_perm_saturating_siegel_parabolic_n_scaling}
    Let $\mathcal{C}$ be an $\db{n,k}$ stabilizer code hosting $\mathrm{Perm}_\mathcal{L}(\mathcal{C}) \succeq_{\Sp(2k, \mathbb{F}_2)} \mathcal{P}(2k, \mathbb{F}_2) \cong \langle \mathbf{S}_k, \mathbf{CX}_k \rangle$ in some logical basis $\mathcal{L}$. Then $n \geq 2^k - 1$ holds.
\end{theorem}

\begin{proof}
    Because permutation gates are a subgroup of automorphism gates, \cref{thm:stab_codes_auto_saturating_siegel_parabolic_n_scaling} holds. The $k = 2$ exception does not occur: a $\db{2, 2}$ code realizing a $\mathcal{P}(4, \mathbb{F}_2)$ or larger permutation logical group is impossible because the group of permutations on two physical qubits is $S_2$ and $\abs{S_2} < \abs{\mathcal{P}(4, \mathbb{F}_2)}$.
\end{proof}

\subsubsection{Automorphism logical groups}

\begin{theorem}
    [Block lengths of stabilizer codes achieving \(\GL(k, \mathbb{F}_2)\) automorphism logical group]
    \label{thm:stab_codes_auto_saturating_general_linear_n_scaling}
    Let $\mathcal{C}$ be an $\db{n,k,d}$ stabilizer code hosting $\mathrm{Aut}_\mathcal{L}(\mathcal{C}) \succeq_{\Sp(2k, \mathbb{F}_2)} \GL(k, \mathbb{F}_2) \cong \langle \mathbf{CX}_k \rangle$ in some logical basis $\mathcal{L}$. Then either $n \geq 2^k - 1$ or $(n, k, d) = (2, 2, 1)$. Equality in the bound can be achieved at $d = 1$.
\end{theorem}

\begin{proof}
    For $k=1$, the result is equivalent to the trivial inequality $n \geq k$. We now consider $d \ge 2$. For $k\geq2$ and $k\neq4$, the result is precisely the phantom-LU bound of Ref.~\cite[Thm.~5]{morris2026constraints}. For $k=4$, we computationally verify that no $\db{\le 14, 4, \ge 2}$ stabilizer code exists that support all addressable logical CXs by automorphisms, allowing any logical basis, by constructing SAT instances following Ref.~\cite[App.~D]{koh2026entangling} and confirming UNSAT outputs over one month of solving time. Compared to Ref.~\cite{koh2026entangling}, which developed the SAT solver for permutation gates, we include Boolean variables for the $\Sp(2, \mathbb{F}_2)$ matrices representing single-qubit Cliffords on the $n$ physical qubits, so as to accommodate automorphism gates.

    Lastly we consider $d=1$. There is then a nonzero logical Pauli class $\ell = (x \mid z) \in \mathbb{F}_2^{2k}$ having a weight-one physical representative $\overline{\ell}$. Every automorphism gate maps weight-one physical Pauli operators to weight-one Pauli operators. Therefore every element in the $\langle \mathbf{CX}_k \rangle$-orbit of $\ell$ also has a weight-one representative. Recall the CX logical action,
    \begin{equation}
        (x\mid z)
        \longmapsto
        (Ax\mid A^{-\top}z),
        \qquad
        A\in\GL(k,\mathbb F_2).
        \label{eq:stab_codes_auto_saturating_general_linear_n_scaling_1}
    \end{equation}

    Suppose first that $x=0$, that is, $\ell$ is a $Z$-type logical operator. Then the orbit of $\ell$ under the $\langle \mathbf{CX}_k \rangle$ logical actions is
    \(\left\{
            (0\mid z'):
            z'\in\mathbb F_2^k\setminus\{0\}
        \right\}
    \), which has size $2^k-1$. All of these logical Pauli classes commute. At most one of them can have a weight-one representative on any fixed physical qubit: two different nonidentity single-qubit Pauli labels on the same qubit anticommute, while the same physical Pauli label represents the same logical class. Hence \(n\geq2^k-1\). The case $z=0$ is identical.

    It remains to consider $x\neq0$ and $z\neq0$. Observe that \(\epsilon \coloneqq x^\top z\in\mathbb F_2\) is preserved by \cref{eq:stab_codes_auto_saturating_general_linear_n_scaling_1}. Conversely, any two pairs \((x,z)\) and \((x',z')\) with \(x,z,x',z'\neq0\) and \(x^\top z=x'^\top z'\) are related by some \(A\in\GL(k,\mathbb F_2)\). Hence \(\ell\) form exactly two orbits according to the binary value of \(\epsilon\). Then the orbit size of $\ell$ is
    \begin{equation}
        \begin{cases}
            (2^k-1)(2^{k-1}-1)
            &
            \epsilon=0,
            \\
            (2^k-1)2^{k-1}
            &
            \epsilon=1.
        \end{cases}
        \label{eq:stab_codes_auto_saturating_general_linear_n_scaling_2}
    \end{equation}
    
    Indeed, there are $2^k-1$ choices of nonzero $x'$. For each such $x'$, there are $2^{k-1}-1$ nonzero vectors $z'$ satisfying $x'^\top z'=0$, and $2^{k-1}$ vectors satisfying $x'^\top z'=1$. But there are only $3n$ nonidentity weight-one physical Pauli strings, and distinct logical classes require distinct physical representatives. Thus \(3n \ge (2^k-1)(2^{k-1}-1)\). If $k\geq3$, then \(2^{k-1}-1\geq3\), so we have \(3n \ge 3(2^k-1) \Longrightarrow n \ge 2^k-1\). At $k=2$, the $\db{2,2,1}$ trivial decomposable code hosts $\langle \mathbf{CX}_2 \rangle$ through automorphisms---see \cref{cons:non_css_code_auto_22_general_linear}.

    The $n = 2^k - 1$ equality is achieved at $d = 1$ by phantom codes~\cite{koh2026entangling}---see \cref{cons:concatenated_phantom_clifford}.
\end{proof}

\begin{theorem}
    [Block lengths of stabilizer codes achieving \(\mathcal{P}(2k, \mathbb{F}_2)\) automorphism logical group]
    \label{thm:stab_codes_auto_saturating_siegel_parabolic_n_scaling}
    Let $\mathcal{C}$ be an $\db{n,k,d}$ stabilizer code hosting $\mathrm{Aut}_\mathcal{L}(\mathcal{C}) \succeq_{\Sp(2k, \mathbb{F}_2)} \mathcal{P}(2k, \mathbb{F}_2) \cong \langle \mathbf{S}_k, \mathbf{CX}_k \rangle$ in some logical basis $\mathcal{L}$. Then either $n \geq 2^k - 1$ or $(n, k, d) = (2, 2, 1)$. Equality can be achieved at $d = 1$.
\end{theorem}

\begin{proof}
    Since all Siegel parabolic subgroups are conjugate in $\Sp(2k, \mathbb{F}_2)$, we may choose the logical basis so that \(\mathrm{Aut}_\mathcal{L}(\mathcal{C}) \geq \mathcal{P}(2k,\mathbb{F}_2)\) as a subgroup of the logical binary symplectic group. In this basis, $\mathcal{P}(2k, \mathbb{F}_2)$ contains the Levi subgroup \(\GL(k, \mathbb{F}_2) \cong \langle \mathbf{CX}_k \rangle\). So the code must host the \(\langle \mathbf{CX}_k \rangle\) logical subgroup by automorphisms, and \cref{thm:stab_codes_auto_saturating_general_linear_n_scaling} applies. 
    The $\db{2,2,1}$ exception does not arise because the code has automorphism logical group $O^+(4, \mathbb{F}_2)$ distinct from $\mathcal{P}(4, \mathbb{F}_2)$---see \cref{cons:non_css_code_auto_22_general_linear}.
    The $n = 2^k - 1$ equality is achieved at $d = 1$ by phantom codes~\cite{koh2026entangling}---see \cref{cons:concatenated_phantom_clifford}.
\end{proof}

\clearpage

\section{Code constructions}

This appendix constructs codes attaining the bounds of \cref{app:css_codes,app:stab_codes}. A convenient formalism for diagonal gates at the Clifford level and

Several constructions implement diagonal logical gates through physical qubits that store parities of logical qubits, which the phase-polynomial formalism describes, so we review it first (\cref{app:constructions/prelims}). We then give the CSS constructions (\cref{app:constructions/css}): concatenated phantom codes attaining the automorphism, permutation, and transversal bounds, and complete-hypergraph codes attaining the transversal bound at every level of the Clifford hierarchy. Lastly, we give the non-CSS constructions (\cref{app:constructions/stab}): codes implementing logical $S$, $H$, and the full Siegel parabolic by qubit permutations, and codes attaining the $k \leq 2$ maxima.

\subsection{Definitions and preliminaries}
\label{app:constructions/prelims}

\subsubsection{Clifford hierarchy}
\label{app:constructions/preliminaries/clifford_hierarchy}

\begin{definition}
    [Clifford hierarchy and diagonal gates]
    \label{def:clifford_hierarchy_diagonal_gates}
    We identify unitaries that differ only by a global phase.
    The \(m\)-qubit Clifford hierarchy is defined recursively.
    Its first level \(\mathcal{C}_m^{(1)}\) is the \(m\)-qubit Pauli group modulo global phases, and for \(t \geq 2\),
    \begin{equation}
        \mathcal{C}_m^{(t)}
        \coloneqq
        \left\{
            U
            :
            U P U^\dagger \in \mathcal{C}_m^{(t-1)}
            \text{ for every \(m\)-qubit Pauli operator \(P\)}
        \right\}.
    \end{equation}
    In particular, \(\mathcal{C}_m^{(2)} = \mathrm{Cl}_m\).
    For \(t \geq 3\), the sets \(\mathcal{C}_m^{(t)}\) are not groups~\cite{cui2017diagonal}.
    Moreover, we define
    \begin{equation}
        \widetilde{\mathbf D}_m^{(t)}
        \coloneqq
        \left\{
            U\in\mathcal C_m^{(t)}
            :
            U \text{ is diagonal}
        \right\},
        \qquad
        \mathbf Z_m
        \coloneqq
        \langle Z_1,\ldots,Z_m\rangle,
        \qquad
        \mathbf D_m^{(t)}
        \coloneqq
        \widetilde{\mathbf D}_m^{(t)}/\mathbf Z_m.
    \end{equation}
    Thus, \(\widetilde{\mathbf D}_m^{(t)}\) is the set of all diagonal unitaries in \(\mathcal C_m^{(t)}\), while \(\mathbf D_m^{(t)}\) is the same modulo \(Z\)-type Pauli operators.
    For every diagonal unitary, we call the unique representative of its global-phase equivalence class whose phase on the computational-basis state \(\ket{0^m}\) is \(1\) its \emph{normalized representative}.
    Let
    \begin{equation}
        \omega_t
        \coloneqq
        e^{2\pi i/2^t},
        \qquad
        Z^{(t)}
        \coloneqq
        \diag(1,\omega_t)
        =
        Z^{1/2^{t-1}}.
    \end{equation}
    Then the normalized representatives of the elements of \(\widetilde{\mathbf D}_1^{(t)}\) are exactly the powers \((Z^{(t)})^a\), with \(a\in\mathbb Z_{2^t}\).
    Consequently, \(\widetilde{\mathbf D}_1^{(t)}\cong\mathbb Z_{2^t}\) and \(\mathbf D_1^{(t)}\cong\mathbb Z_{2^{t-1}}\).
    Note that, although the higher Clifford hierarchy levels are not generally groups, \(\smash{\widetilde{\mathbf D}_m^{(t)}}\) and \(\smash{\mathbf D_m^{(t)}}\) are finite abelian groups (see \cref{fact:phase_polynomial_diagonal_clifford_hierarchy}).
    At \(t=2\), the symplectic representation identifies \(\smash{\mathbf D_m^{(2)}}\) with the unipotent radical \(\mathcal U(2m,\mathbb F_2)\).
\end{definition}

\begin{definition}
    [Higher-level transversal and automorphism logical groups]
    \label{def:higher_level_logical_groups}
    Let \(\mathcal{C}\) be an \(\db{n,k}\) stabilizer code, \(\mathcal{L}\) be a logical basis of \(\mathcal{C}\), and \(t \geq 1\). 
    A \emph{level-\(t\) transversal gate} on \(\mathcal{C}\) is a code-preserving physical gate of the form
    $
        \overline{U}
        =
        \bigotimes_{i=1}^n
        U_i
    $,
    where each $U_i$ is a single-qubit gate in \(\smash{\mathcal{C}_m^{(t)}}\).
    The \emph{level-\(t\) transversal logical group} \(\smash{\mathrm{Trans}^{(t)}_\mathcal{L}(\mathcal{C})}\) is the group of logical actions, modulo logical Paulis and global phases, induced by all such gates
    
    Likewise, a \emph{level-\(t\) automorphism gate} on \(\mathcal{C}\) is a code-preserving physical gate of the form
    $
        \overline{U}
        =
        U_\pi
        \bigotimes_{i=1}^n
        U_i
    $, where $\pi \in S_n$ is a permutation of the physical qubits and $U_\pi$ the corresponding unitary, and each $U_i$ is a single-qubit gate in \(\smash{\mathcal{C}_m^{(t)}}\).
    The \emph{level-\(t\) automorphism logical group} \(\smash{\mathrm{Aut}^{(t)}_\mathcal{L}(\mathcal{C})}\) is the group of logical actions, modulo logical Paulis and global phases, induced by all such gates.
    Taking \(\pi=\mathrm{id}\) gives
    \begin{equation}
        \mathrm{Trans}^{(t)}_\mathcal{L}(\mathcal{C})
        \leq
        \mathrm{Aut}^{(t)}_\mathcal{L}(\mathcal{C}).
    \end{equation}

    At \(t=2\), the forms of these physical gates are exactly the Clifford transversal and automorphism gates previously defined in \cref{def:transversal_permutation_automorphism_gate,def:transversal_permutation_automorphism_logical_groups} and considered in much of this paper. As a result,
    \begin{equation}
        \mathrm{Trans}^{(2)}_\mathcal{L}(\mathcal{C})
        =
        \mathrm{Trans}_\mathcal{L}(\mathcal{C}),
        \qquad
        \mathrm{Aut}^{(2)}_\mathcal{L}(\mathcal{C})
        =
        \mathrm{Aut}_\mathcal{L}(\mathcal{C}).
    \end{equation}
\end{definition}

\subsubsection{Phase polynomials}
\label{app:constructions/preliminaries/phase_polynomial}

\begin{definition}
    [Phase polynomials]
    \label{def:phase_polynomials}
    Let \(U\) be an \(m\)-qubit diagonal unitary whose normalized representative (see \cref{def:clifford_hierarchy_diagonal_gates}) has computational-basis phases that are \((2^t)^\text{th}\) roots of unity.
    Writing \(U\) for this normalized representative, it can be expressed uniquely as
    \begin{equation}
        U
        =
        \sum_{x\in\mathbb{F}_2^m}
        \omega_t^{P(x)}
        \ket{x}\!\bra{x},
        \label{eq:phase_polynomial_diagonal_unitary}
    \end{equation}
    where \(P:\mathbb{F}_2^m\to\mathbb{Z}_{2^t}\) is the \emph{phase polynomial} of \(U\), with \(P(0)=0\), and \(\omega_t\) is as defined in \cref{def:clifford_hierarchy_diagonal_gates}.
    The phase polynomial has a unique multilinear expansion in the monomial basis,
    \begin{equation}
        P(x_1,\ldots,x_m)
        =
        \sum_{S\subseteq[m]}
        c_S
        \prod_{i\in S}x_i
        \pmod{2^t},
        \label{eq:phase_polynomial_monomial_basis}
    \end{equation}
    with \(c_S\in\mathbb{Z}_{2^t}\).
    Here \(\prod_{i\in S}x_i\) is the Boolean monomial, valued in \(\{0,1\}\subseteq\mathbb{Z}_{2^t}\), which equals \(1\) exactly when all bits indexed by \(S\) are \(1\).
    The normalization \(P(0)=0\) is equivalent to \(c_\varnothing=0\).

    Given the phase exponents \(p_x\coloneqq P(x)\), the coefficients are recovered by Möbius inversion:
    \begin{equation}
        c_S
        =
        \sum_{T\subseteq S}
        (-1)^{|S|-|T|}
        p_{\mathbf{1}_T}
        \pmod{2^t},
        \label{eq:boolean_mobius_phase_polynomial}
    \end{equation}
    where \(\mathbf{1}_T\in\mathbb{F}_2^m\) is the bit string whose \(i^\text{th}\) entry is \(1\) iff \(i\in T\).
    See Refs.~\cite{koh2026entangling,amy2014polynomial,campbell2017unified} for further background on phase polynomials.
\end{definition}

\begin{fact}
    [Phase-polynomial characterization of diagonal Clifford-hierarchy gates]
    \label{fact:phase_polynomial_diagonal_clifford_hierarchy}
    Every \(U\in\widetilde{\mathbf D}_m^{(t)}\) has a normalized representative whose computational-basis phases are \((2^t)^\text{th}\) roots of unity.
    Conversely, let \(U\) be an \(m\)-qubit diagonal unitary whose normalized representative has such phases, with phase polynomial \(P\) and monomial coefficients \(c_S\) as in \cref{def:phase_polynomials}.
    Specializing Ref.~\cite[Thm.~3]{cui2017diagonal} to qubits, \(U\in\widetilde{\mathbf D}_m^{(t)}\) iff
    \begin{equation}
        c_S
        \in
        2^{|S|-1}\mathbb{Z}_{2^t}
        \qquad
        \text{for every nonempty }S\subseteq[m].
        \label{eq:phase_polynomial_ch_divisibility}
    \end{equation}
    Equivalently, a degree-\(\ell\) monomial can appear only with coefficient divisible by \(2^{\ell-1}\) modulo \(2^t\).
    For \(\ell\leq t\), its corresponding phase is therefore an integer multiple of \(\pi/2^{t-\ell}\), while for \(\ell>t\) its coefficient necessarily vanishes modulo \(2^t\).
    Since multiplication and inversion of diagonal gates correspond to addition and negation of phase polynomials modulo \(2^t\), and \cref{eq:phase_polynomial_ch_divisibility} is preserved under both operations, \(\widetilde{\mathbf D}_m^{(t)}\) and consequently \(\mathbf D_m^{(t)}\) are abelian groups.
\end{fact}

\begin{corollary}
    [Structure of the diagonal Clifford-hierarchy groups]
    \label{cor:diagonal_clifford_hierarchy_group_structure}
    For \(m,t\geq1\),
    \begin{equation}
        \widetilde{\mathbf D}_m^{(t)}
        \cong
        \prod_{r=1}^{t}
        \left(
            \mathbb Z_{2^{t-r+1}}
        \right)^{\binom{m}{r}},
        \qquad
        \mathbf D_m^{(t)}
        \cong
        \left(
            \mathbb Z_{2^{t-1}}
        \right)^m
        \times
        \prod_{r=2}^{t}
        \left(
            \mathbb Z_{2^{t-r+1}}
        \right)^{\binom{m}{r}}.
        \label{eq:diagonal_clifford_hierarchy_group_structure}
    \end{equation}
    
    In particular, denoting by \(\Delta(G)\) the minimum number of generators of a finite group \(G\), we have
    \begin{equation}
        \Delta\left(\widetilde{\mathbf D}_m^{(t)}\right)
        =
        \sum_{r=1}^{t}
        \binom{m}{r},
        \qquad
        \Delta\left(\mathbf D_m^{(t)}\right)
        =
        \begin{cases}
            0, & t=1, \\
            \sum_{r=1}^{t}\binom{m}{r}, & t\geq2.
        \end{cases}
        \label{eq:diagonal_clifford_hierarchy_minimum_generators}
    \end{equation}
\end{corollary}

\begin{proof}
    By \cref{def:phase_polynomials,fact:phase_polynomial_diagonal_clifford_hierarchy}, an element of \(\widetilde{\mathbf D}_m^{(t)}\) is uniquely specified by the monomial coefficients \(c_S\) of the phase polynomial of its normalized representative.
    Coefficients with \(|S|>t\) vanish, while for \(|S|=r\leq t\),
    \begin{equation}
        c_S
        =
        2^{r-1}a_S,
        \qquad
        a_S\in\mathbb Z_{2^{t-r+1}}.
    \end{equation}
    Conversely, by \cref{fact:phase_polynomial_diagonal_clifford_hierarchy}, every such choice of coefficients is realized by an element of \(\widetilde{\mathbf D}_m^{(t)}\).
    Multiplication adds the coefficients \(c_S\) independently modulo \(2^t\), or equivalently adds each \(a_S\) modulo \(2^{t-r+1}\).
    Since there are \(\binom{m}{r}\) subsets \(S\subseteq[m]\) of size \(r\), this gives the group structure of \(\widetilde{\mathbf D}_m^{(t)}\) in \cref{eq:diagonal_clifford_hierarchy_group_structure}.
    Quotienting by \(\mathbf Z_m\) reduces each degree-one factor \(\mathbb Z_{2^t}\) to \(\mathbb Z_{2^{t-1}}\), since \(Z_j\) has phase polynomial \(2^{t-1}x_j\), while leaving all higher-degree factors unchanged.
    This gives the group structure of \(\mathbf D_m^{(t)}\).

    For a finite abelian \(2\)-group \(G\), \(\Delta(G)=\dim_{\mathbb F_2}G/G^2\), where \(G^2\coloneqq\{g^2:g\in G\}\).
    Each nontrivial cyclic factor in the decompositions above contributes one copy of \(\mathbb Z_2\) to this quotient.
    Thus the formula in \cref{eq:diagonal_clifford_hierarchy_minimum_generators} for \(\Delta(\widetilde{\mathbf D}_m^{(t)})\) follows.
    For \(\mathbf D_m^{(t)}\), all factors are trivial when \(t=1\), while for \(t\geq2\) every displayed cyclic factor is nontrivial, giving the formula for \(\Delta(\mathbf D_m^{(t)})\).
\end{proof}

\begin{lemma}
    [Parity-phase representation]
    \label{lem:phase_polynomial_parity_representation}
    Let \(U\in\widetilde{\mathbf D}_m^{(t)}\) have phase polynomial \(P\) as in \cref{def:phase_polynomials}.
    Then \(P\) can be written as
    \begin{equation}
        P(x)
        =
        \sum_{\varnothing\neq S\subseteq[m]}
        b_S
        \bigoplus_{i\in S}x_i
        \pmod{2^t},
        \qquad
        b_S\in\mathbb{Z}_{2^t},
        \label{eq:phase_polynomial_parity_form}
    \end{equation}
    where \(\bigoplus_{i\in S}x_i\in\{0,1\}\) is the parity of the bits indexed by \(S\).
    Unlike the monomial coefficients \(c_S\), the parity coefficients \(b_S\) need not be unique.
\end{lemma}

\begin{proof}
    For every nonempty \(S\subseteq[m]\), the following integer-valued identity holds on Boolean inputs~\cite[Prop.~A.2]{webster2023transversal}:
    \begin{equation}
        2^{|S|-1}
        \prod_{i\in S}x_i
        =
        \sum_{\varnothing\neq T\subseteq S}
        (-1)^{|T|-1}
        \bigoplus_{i\in T}x_i.
        \label{eq:monomial_to_parity_phase_identity}
    \end{equation}
    By \cref{fact:phase_polynomial_diagonal_clifford_hierarchy}, each nonzero monomial coefficient of \(P\), necessarily with \(|S|\leq t\), has the form \(c_S=2^{|S|-1}a_S\) for some \(a_S\in\mathbb{Z}_{2^{t-|S|+1}}\).
    Applying \cref{eq:monomial_to_parity_phase_identity} to each monomial and collecting the coefficients of each parity gives \cref{eq:phase_polynomial_parity_form}.
    
    The parity coefficients need not be unique because the parity functions can satisfy nontrivial \(\mathbb{Z}_{2^t}\)-linear relations.
    For \(m\geq2\), for example,
    \begin{equation}
        2^{t-1}(x_1\oplus x_2)
        =
        2^{t-1}x_1
        +
        2^{t-1}x_2
        \pmod{2^t},
    \end{equation}
    since \(x_1\oplus x_2=x_1+x_2-2x_1x_2\) and \(2^t x_1x_2\equiv0\pmod{2^t}\).
\end{proof}

\begin{example}
    [Phase polynomials of standard diagonal gates]
    \label{ex:standard_diagonal_gate_phase_polynomials}
    We give the phase polynomials of some standard diagonal gates.
    \begin{itemize}[noitemsep]
        \item The gate \(Z^{(t)}\) on qubit \(j\) has \(P(x)=x_j \pmod{2^t}\). More generally, \((Z^{(t)})^a\) on qubit \(j\) has \(P(x)=a x_j \pmod{2^t}\).

        \item At \(t=2\), we have \(\omega_2=i\); the gate \(Z^{(2)}=S\) has \(P(x)=x \pmod{4}\), while \(Z=S^2\) has \(P(x)=2x \pmod{4}\).

        \item At \(t=3\), we have \(\omega_3=e^{i\pi/4}\); the gate \(Z^{(3)}=T\) has \(P(x)=x \pmod{8}\), \(S=T^2\) has \(P(x)=2x \pmod{8}\), and \(Z=T^4\) has \(P(x)=4x \pmod{8}\).

        \item The \(\mathrm{CZ}\) gate satisfies \(\mathrm{CZ}\ket{x_1,x_2}=(-1)^{x_1x_2}\ket{x_1,x_2}\). At \(t=2\), using \cref{eq:monomial_to_parity_phase_identity},
        \begin{equation}
            P(x_1,x_2)
            =
            2x_1x_2
            =
            x_1+x_2-(x_1\oplus x_2)
            \pmod{4}.
        \end{equation}
        The coefficient \(2\) is divisible by \(2^{2-1}=2\), as required by \cref{eq:phase_polynomial_ch_divisibility}.

        \item The \(\mathrm{CS}\) gate satisfies \(\mathrm{CS}\ket{x_1,x_2}=i^{x_1x_2}\ket{x_1,x_2}\). At \(t=3\),
        \begin{equation}
            P(x_1,x_2)
            =
            2x_1x_2
            =
            x_1+x_2-(x_1\oplus x_2)
            \pmod{8}.
        \end{equation}

        \item The \(\mathrm{CCZ}\) gate satisfies \(\mathrm{CCZ}\ket{x_1,x_2,x_3}=(-1)^{x_1x_2x_3}\ket{x_1,x_2,x_3}\). At \(t=3\),
        \begin{equation}
            \begin{split}
                P(x_1,x_2,x_3)
                &=
                4x_1x_2x_3 \\
                &=
                x_1+x_2+x_3
                -(x_1\oplus x_2)
                -(x_1\oplus x_3)
                -(x_2\oplus x_3)
                +(x_1\oplus x_2\oplus x_3)
                \pmod{8}.
            \end{split}
        \end{equation}
    \end{itemize}
\end{example}

\begin{proposition}
    [Circuit compilation for diagonal unitaries with a trivial CX skeleton]
    \label{prop:phase_polynomial_circuit_implementation_diagonal_unitary}
    Let \(U\in\widetilde{\mathbf D}_m^{(t)}\).
    Then the normalized representative of \(U\) can be implemented using \(\mathrm{CX}\) gates and single-qubit powers of \(Z^{(t)}\) as defined in \cref{def:clifford_hierarchy_diagonal_gates}.
    Moreover, the circuit can be chosen so that if all powers of \(Z^{(t)}\) are deleted, the remaining CX circuit, which we call the \emph{CX skeleton}, multiplies to the identity.
\end{proposition}

\begin{proof}
    By \cref{def:phase_polynomials,lem:phase_polynomial_parity_representation}, the phase polynomial \(P:\mathbb{F}_2^m\to\mathbb{Z}_{2^t}\) of the normalized representative of \(U\) can be written in the parity form \cref{eq:phase_polynomial_parity_form}.
    It therefore suffices to implement a single parity-phase term
    \begin{equation}
        V_S\ket{x}
        =
        \omega_t^{b_S\left(\bigoplus_{i\in S}x_i\right)}
        \ket{x},
        \qquad
        \varnothing\neq S\subseteq[m],
        \qquad
        b_S\in\mathbb{Z}_{2^t},
    \end{equation}
    with a CX skeleton that multiplies to the identity.
    To do so, select any qubit \(j\in S\).
    Let \(L_S\coloneqq\smash{\prod_{i\in S\setminus\{j\}}\mathrm{CX}_{ij}}\) be the CX circuit that computes the parity \(\bigoplus_{i\in S}x_i\) into qubit \(j\).
    Indeed, on computational-basis states, \(L_S\) maps \(x_j\mapsto x_j\oplus\smash{\bigoplus_{i\in S\setminus\{j\}}x_i}=\smash{\bigoplus_{i\in S}x_i}\).
    Then the compute--phase--uncompute circuit \(\smash{L_S^{-1}(Z^{(t)})_j^{b_S}L_S}\) imparts the correct phases on computational-basis states to implement \(V_S\).
    Moreover, if the middle \(\smash{(Z^{(t)})_j^{b_S}}\) gate is deleted from this block, the remaining CX skeleton is \(L_S^{-1}L_S=I\).
    Placing these parity-phase blocks back-to-back over all nonempty \(S\subseteq[m]\) therefore implements the normalized representative of \(U\), with a CX skeleton that multiplies to the identity.
\end{proof}

\subsubsection{Conventions for code constructions}
\label{sec:constructions/preliminaries/permutations}
\label{sec:constructions/preliminaries/weights}

\begin{definition}
    [Convention for qubit permutations]
    \label{def:constructions_qubit_permutations}
    For \(\lambda\in S_n\), let \(U_\lambda\) denote the qubit-permutation gate that moves physical qubit \(i\) to position \(\lambda(i)\). 
    The bit at position \(j\) after permuting therefore came from position \(\lambda^{-1}(j)\), so
    \begin{equation}
        U_\lambda\ket{x_1,\ldots,x_n}
        =
        \ket{x_{\lambda^{-1}(1)},\ldots,x_{\lambda^{-1}(n)}},
        \qquad
        U_\lambda
        \left(
            \bigotimes_{i=1}^n A_i
        \right)
        U_\lambda^\dagger
        =
        \bigotimes_{i=1}^n
        A_{\lambda^{-1}(i)}.
        \label{eq:qubit_permutation_convention}
    \end{equation}
    For example, the cycle \(\lambda=(123)\) moves qubits \(1\to2\), \(2\to3\), and \(3\to1\), so
    \begin{equation}
        U_{(123)}\ket{x_1,x_2,x_3}
        =
        \ket{x_3,x_1,x_2},
        \qquad
        U_{(123)}
        \left(A_1\otimes A_2\otimes A_3\right)
        U_{(123)}^\dagger
        =
        A_3\otimes A_1\otimes A_2.
    \end{equation}
    We compose permutations from right to left, as functions; with this convention, \(U_\lambda U_\mu=U_{\lambda\circ\mu}\).
\end{definition}

\begin{definition}
    [Convention for stabilizer weights and qubit degrees on CSS codes]
    \label{def:constructions_stabilizer_weights_degrees}
    Let a CSS code be specified by stabilizer generator matrices \(H_\mathrm{x}\in\mathbb{F}_2^{r_\mathrm{x}\times n}\) and \(H_\mathrm{z}\in\mathbb{F}_2^{r_\mathrm{z}\times n}\).
    For \(\mu\in\{\mathrm{x},\mathrm{z}\}\), define the maximum stabilizer weight and maximum qubit degree in this presentation by
    \begin{equation}
        w_\mu
        \coloneqq
        \max_{i\in[r_\mu]}
        \abs{(H_\mu)_{i:}},
        \qquad
        \delta_\mu
        \coloneqq
        \max_{i\in[n]}
        \abs{(H_\mu)_{:i}},
    \end{equation}
    where \(\abs{v}\) denotes the Hamming weight of a binary vector \(v\), and \((H_\mu)_{i:}\) and \((H_\mu)_{:i}\) denote the \(i^\text{th}\) row and column of \(H_\mu\), respectively. 
    The overall maximum stabilizer weight and qubit degree are
    \begin{equation}
        w
        \coloneqq
        \max(w_\mathrm{x},w_\mathrm{z}),
        \qquad
        \delta
        \coloneqq
        \max_{i\in[n]}
        \left(
            \abs{(H_\mathrm{x})_{:i}}
            +
            \abs{(H_\mathrm{z})_{:i}}
        \right)
        \leq
        \delta_\mathrm{x}+\delta_\mathrm{z}.
    \end{equation}
    The maximum stabilizer weights and qubit degrees bound the minimum depth of a bare-ancilla syndrome-extraction (i.e.~stabilizer measurement) circuit. 
    These quantities depend on the chosen stabilizer generators. 
    We use natural generating sets in our constructions and, unless specifically stated, do not claim they are optimal on the codes.
\end{definition}

\clearpage

\subsection{CSS codes}
\label{app:constructions/css}

\subsubsection{Concatenated phantom codes at any level of the Clifford hierarchy}

\begin{construction}
    [Phantom-as-outer concatenated CSS code at any Clifford hierarchy level]
    \label{cons:concatenated_phantom_as_outer_css_generalized}
    Take $t \geq 1$, and consider the following code concatenation:
    \begin{enumerate}
        \item \textit{Outer code $\mathcal{C}^\mathrm{p}$.} Take a $\db{n^\mathrm{p}, k, (d^\mathrm{p}_\mathrm{x}, d^\mathrm{p}_\mathrm{z} = 1)}$ CSS phantom code that hosts a CSS logical basis $\mathcal{L}^\mathrm{p} \coloneqq \{\overline{X}_j, \overline{Z}_j\}_{j=1}^k$ in which all $\overline{Z}_j$ logicals are weight-one.
        
        \item \textit{Inner code $\mathcal{C}^\mathrm{s}$.} Take a $\db{n^\mathrm{s}, 1, (d^\mathrm{s}_\mathrm{x}, d^\mathrm{s}_\mathrm{z})}$ CSS code, that hosts a CSS logical basis $\mathcal{L}^\mathrm{s}$ in which a tensor product of physical single-qubit gates in the $t^\text{th}$ level of the Clifford hierarchy implements a logical $Z^{(t)}$ gate. (A bare physical qubit having parameters $\db{1, 1, (1, 1)}$ could also be used.)
    \end{enumerate}

    The concatenation here is standard: we encode each physical qubit of the outer code into a codeblock of the inner code. The resulting code $\mathcal{C}$ has parameters
    \begin{equation}
        \db{n \coloneqq n^\mathrm{p} n^\mathrm{s}, k, (d^\mathrm{p}_\mathrm{x} d^\mathrm{s}_\mathrm{x}, d^\mathrm{s}_\mathrm{z})}.
        \label{eq:concatenated_phantom_as_outer_css_parameters}
    \end{equation}
    
    Explicitly, let the outer (resp.~inner) code be described by binary stabilizer generator matrices $H^\mathrm{p}_\mathrm{x}, H^\mathrm{p}_\mathrm{z}$ (resp.~$H^\mathrm{s}_\mathrm{x}, H^\mathrm{s}_\mathrm{z}$) and the logical basis $\mathcal{L}^\mathrm{p}$ (resp.~$\mathcal{L}^\mathrm{s}$) be described by $L^\mathrm{p}_\mathrm{x}, L^\mathrm{p}_\mathrm{z}$ (resp.~$L^\mathrm{s}_\mathrm{x}, L^\mathrm{s}_\mathrm{z}$). Then the stabilizer generators and logical basis of $\mathcal{C}$ are
    \begin{equation}
        H_\mathrm{x} \coloneqq \mqty(
            I_{n^\mathrm{p}} \otimes H^\mathrm{s}_\mathrm{x}
            \\
            H^\mathrm{p}_\mathrm{x} \otimes L^\mathrm{s}_\mathrm{x}
        ),
        \qquad
        H_\mathrm{z} \coloneqq \mqty(
            I_{n^\mathrm{p}} \otimes H^\mathrm{s}_\mathrm{z}
            \\
            H^\mathrm{p}_\mathrm{z} \otimes L^\mathrm{s}_\mathrm{z}
        ),
        \qquad
        L_\mathrm{x} \coloneqq 
            L^\mathrm{p}_\mathrm{x} \otimes L^\mathrm{s}_\mathrm{x},
        \qquad
        L_\mathrm{z} \coloneqq 
            L^\mathrm{p}_\mathrm{z} \otimes L^\mathrm{s}_\mathrm{z}.
        \label{eq:concatenated_phantom_as_outer_css_matrices}
    \end{equation}

    Moreover, let $w_\mathrm{\mu}^\mathrm{\nu}$ and $\delta_\mathrm{\mu}^\mathrm{\nu}$ be the maximum stabilizer weights and qubit degrees, $\mu \in \{\mathrm{x}, \mathrm{z}\}$ and $\nu \in \{\mathrm{p}, \mathrm{s}\}$, of the outer and inner codes (see \cref{sec:constructions/preliminaries/weights}); and $\ell^\mathrm{s}_\mathrm{x} \coloneqq \max_{i \in [k]} \abs{(L^\mathrm{s}_\mathrm{x})_{i:}}$ be the logical weights of the inner code, in the basis presented. Then the maximum stabilizer weights and qubit degrees of $\mathcal{C}$ are bounded by
    \begin{equation}
        w_\mathrm{x} 
        \leq 
        \max(
            w^\mathrm{s}_\mathrm{x}, 
            w^\mathrm{p}_\mathrm{x} \ell^\mathrm{s}_\mathrm{x}
        ),
        \qquad
        w_\mathrm{z} 
        \leq 
        \max(
            w^\mathrm{s}_\mathrm{z}, 
            w^\mathrm{p}_\mathrm{z}
            \ell^\mathrm{s}_\mathrm{z}
        ),
        \qquad
        \delta_\mathrm{x} 
        \leq 
        \delta^\mathrm{p}_\mathrm{x} + \delta^\mathrm{s}_\mathrm{x},
        \qquad
        \delta_\mathrm{z} 
        \leq 
        \delta^\mathrm{p}_\mathrm{z} + \delta^\mathrm{s}_\mathrm{z},
        \qquad
        \delta
        \leq 
        \delta^\mathrm{p} + \delta^\mathrm{s}.
        \label{eq:concatenated_phantom_as_outer_css_weights}
    \end{equation}

    In the CSS logical basis $\mathcal{L} = (L_\mathrm{x}, L_\mathrm{z})$ written in \cref{eq:concatenated_phantom_as_outer_css_matrices}, the logical groups of the resulting code are:
    \begin{itemize}
        
        \item $\mathrm{Aut}^{(t)}_\mathcal{L}(\mathcal{C}) \geq \langle \mathbf{D}^{(t)}_k, \mathbf{CX}_k \rangle$. 
        
        \item $\mathrm{Perm}_\mathcal{L}(\mathcal{C}) = \langle \mathbf{CX}_k \rangle \cong \GL(k, \mathbb{F}_2)$. Inherited from the phantom outer code. Indeed, if $\pi \in S_{n^\mathrm{p}}$ is a physical qubit permutation that implements a logical $\mathrm{CX}_{ab}$ on the phantom code, then $\pi \otimes \id_{n^\mathrm{s}} \in S_n$---that is, permuting the $n^\mathrm{p}$ codeblocks of the inner code according to $\pi$ but doing nothing within each codeblock---is a physical qubit permutation that implements a logical $\mathrm{CX}_{ab}$ on $\mathcal{C}$.

        \item $\mathrm{Trans}^{(t)}_\mathcal{L}(\mathcal{C}) \geq \mathbf{D}^{(t)}_k$. First, because all $\overline{Z}_j$ logicals are weight-one on the outer code, a physical $Z^{(t)}$ gate on the support of each $\overline{Z}_j$ implements a logical $Z^{(t)}$. Upon concatenation, each of these physical $Z^{(t)}$ gates is promoted to a logical $Z^{(t)}$ on the corresponding inner code, which is implementable by a tensor product of physical single-qubit gates by premise. Therefore the resulting code supports transversal $Z^{(t)}$ logical gates on every logical qubit. 
        
        Next, by \cref{prop:phase_polynomial_circuit_implementation_diagonal_unitary}, all diagonal unitaries at the $t^\text{th}$ level of the Clifford hierarchy can be compiled using $Z^{(t)}$ rotations and CX gates, such that the CX skeleton of the circuit multiplies to the identity. Consider the compilation of any such logical diagonal unitary. As $\mathrm{Perm}_\mathcal{L}(\mathcal{C}) = \langle \mathbf{CX}_k \rangle$, the required logical CXs can all be implemented by qubit permutations; and as the logical CX skeleton is trivial, the qubit permutations can be chosen so that their composition is trivial. All logical $Z^{(t)}$ rotations can be pulled through the qubit permutations into a single layer, and the qubit permutations cancel, therefore producing a transversal implementation.

        To give an example, at the $t = 2$ level, consider the compilation of a logical $\mathrm{CZ}$:
        \begin{center}
            \begin{quantikz}[row sep={0.7cm,between origins}]
                & \ctrl{1} & \qw \\
                & \ctrl{0}  & \qw
            \end{quantikz}
            =
            \begin{quantikz}[row sep={0.7cm,between origins}]
                & \gate{S} & \ctrl{1} & \qw      & \ctrl{1} & \qw \\
                & \gate{S} & \targ{}  & \gate{S^\dagger} & \targ{} & \qw
            \end{quantikz}
        \end{center}

        Let a physical permutation $\pi \in S_n$ implement the first logical $\mathrm{CX}$ above. Then, as $\mathrm{CX}$ is involutory, $\pi^{-1} \in S_n$ can implement the second logical $\mathrm{CX}$. The tensor product of physical single-qubit gates implementing the logical $S, S^\dag$ gates can be consolidated into a single layer and the qubit permutations cancel.

        Similarly, at the $t = 3$ level, we have:
        \begin{center}
            \begin{quantikz}[row sep={0.7cm,between origins},column sep=0.25cm]
                & \ctrl{1} & \qw \\
                & \gate{S}  & \qw
            \end{quantikz}
            =
            \begin{quantikz}[row sep={0.7cm,between origins},column sep=0.25cm]
                & \gate{T} & \ctrl{1} & \qw      & \ctrl{1} & \qw \\
                & \gate{T} & \targ{}  & \gate{T^\dagger} & \targ{} & \qw
            \end{quantikz}
            \qquad\quad
            \begin{quantikz}[row sep={0.7cm,between origins},column sep=0.25cm]
                & \ctrl{1} & \qw \\
                & \ctrl{1} & \qw \\
                & \ctrl{0} & \qw
            \end{quantikz}
            =
            \begin{quantikz}[row sep={0.7cm,between origins},column sep=0.25cm]
                & \gate{T} & \qw      & \qw              & \ctrl{2} & \qw      & \qw & \qw              & \ctrl{2} & \ctrl{1} & \qw              & \ctrl{1} & \qw \\
                & \gate{T} & \ctrl{1} & \qw              & \qw      & \qw      & \ctrl{1} & \qw        & \qw      & \targ{}  & \gate{T^\dagger} & \targ{}  & \qw \\
                & \gate{T} & \targ{}  & \gate{T^\dagger} & \targ{}  & \gate{T} & \targ{}          & \gate{T^\dagger} & \targ{} & \qw & \qw & \qw & \qw
            \end{quantikz}
        \end{center}

        The qubit permutations implementing the $\mathrm{CX}$s can likewise be arranged to cancel to the identity.
        
    \end{itemize}

    See \cref{cons:concatenated_phantom_concrete} for concrete examples of admissible $\mathcal{C}^\mathrm{p}$ and $\mathcal{C}^\mathrm{s}$ codes.
\end{construction}

\begin{construction}
    [Phantom-as-inner concatenated CSS code at any Clifford hierarchy level]
    \label{cons:concatenated_phantom_as_inner_css_generalized}
    This is a version of \cref{cons:concatenated_phantom_as_outer_css_generalized} but with the outer and inner codes reversed in roles. Take $t \geq 1$, and consider the following code concatenation:
    \begin{enumerate}
        \item \textit{Outer code $\mathcal{C}^\mathrm{s}$.} Take $k$ codeblocks of a $\db{n^\mathrm{s}, 1, (d^\mathrm{s}_\mathrm{x}, d^\mathrm{s}_\mathrm{z})}$ CSS code, that hosts a CSS logical basis $\mathcal{L}^\mathrm{s}$ in which a tensor product of (powers of) physical $Z^{(t)}$ gate implements a logical $Z^{(t)}$ gate. (A bare physical qubit having parameters $\db{1, 1, (1, 1)}$ could also be used.)
        \item \textit{Inner code $\mathcal{C}^\mathrm{p}$.} Take a $\db{n^\mathrm{p}, k, (d^\mathrm{p}_\mathrm{x}, d^\mathrm{p}_\mathrm{z} = 1)}$ CSS phantom code that hosts a CSS logical basis $\mathcal{L}^\mathrm{p} \coloneqq \{\overline{X}_j, \overline{Z}_j\}_j$ in which all $\overline{Z}_j$ logicals are weight-one.
    \end{enumerate}

    We impose a particular structure to the code concatenation. In particular, for each $i \in [n^\mathrm{s}]$, we encode the corresponding $k$-tuple of $i^\text{th}$ qubits of the $k$ outer codeblocks into a single codeblock of the inner code. The resulting code $\mathcal{C}$ has identical parameters to \cref{eq:concatenated_phantom_as_outer_css_parameters}. Explicitly, let the outer (resp.~inner) code be described by binary stabilizer generator matrices $H^\mathrm{s}_\mathrm{x}, H^\mathrm{s}_\mathrm{z}$ (resp.~$H^\mathrm{p}_\mathrm{x}, H^\mathrm{p}_\mathrm{z}$) and the logical basis $\mathcal{L}^\mathrm{s}$ (resp.~$\mathcal{L}^\mathrm{p}$) be described by $L^\mathrm{s}_\mathrm{x}, L^\mathrm{s}_\mathrm{z}$ (resp.~$L^\mathrm{p}_\mathrm{x}, L^\mathrm{p}_\mathrm{z}$). Then the stabilizer generators and logical basis of $\mathcal{C}$ are
    \begin{equation}
        H_\mathrm{x} \coloneqq \mqty(
            I_{n^\mathrm{s}} \otimes H^\mathrm{p}_\mathrm{x}
            \\
            H^\mathrm{s}_\mathrm{x} \otimes L^\mathrm{p}_\mathrm{x}
        ),
        \qquad
        H_\mathrm{z} \coloneqq \mqty(
            I_{n^\mathrm{s}} \otimes H^\mathrm{p}_\mathrm{z}
            \\
            H^\mathrm{s}_\mathrm{z} \otimes L^\mathrm{p}_\mathrm{z}
        ),
        \qquad
        L_\mathrm{x} \coloneqq 
            L^\mathrm{s}_\mathrm{x} \otimes L^\mathrm{p}_\mathrm{x},
        \qquad
        L_\mathrm{z} \coloneqq 
            L^\mathrm{s}_\mathrm{z} \otimes L^\mathrm{p}_\mathrm{z}.
        \label{eq:concatenated_phantom_as_inner_css_matrices}
    \end{equation}

    Moreover, let $w_\mathrm{\mu}^\mathrm{\nu}$ and $\delta^\nu, \delta_\mathrm{\mu}^\mathrm{\nu}$ be the maximum stabilizer weights and qubit degrees, $\mu \in \{\mathrm{x}, \mathrm{z}\}$ and $\nu \in \{\mathrm{p}, \mathrm{s}\}$, of the inner and outer codes (see \cref{sec:constructions/preliminaries/weights}); and $\ell^\mathrm{p}_\mathrm{\mu} \coloneqq \max_{i \in [k]} \abs{(L^\mathrm{p}_\mathrm{\mu})_{i:}}$ be the logical weights of the inner code, $\mu \in \{\mathrm{x}, \mathrm{z}\}$, in the basis presented. Then the maximum stabilizer weights and qubit degrees of $\mathcal{C}$ are bounded by
    \begin{equation}
        w_\mathrm{x} \leq \max(
            w^\mathrm{p}_\mathrm{x}, 
            w^\mathrm{s}_\mathrm{x} \ell^\mathrm{p}_\mathrm{x}
        ),
        \qquad
        w_\mathrm{z} \leq \max(
            w^\mathrm{p}_\mathrm{z}, 
            w^\mathrm{s}_\mathrm{z} \ell^\mathrm{p}_\mathrm{z}
        ),
        \qquad
        \delta_\mathrm{x} 
        \leq 
        \delta^\mathrm{p}_\mathrm{x} + k \delta^\mathrm{s}_\mathrm{x},
        \qquad
        \delta_\mathrm{z} 
        \leq 
        \delta^\mathrm{p}_\mathrm{z} + k \delta^\mathrm{s}_\mathrm{z},
        \qquad
        \delta
        \leq 
        \delta^\mathrm{p} + k \delta^\mathrm{s},
        \label{eq:concatenated_phantom_as_inner_css_weights}
    \end{equation}

    Notice that \cref{eq:concatenated_phantom_as_inner_css_weights} is different from \cref{eq:concatenated_phantom_as_outer_css_weights}, which differentiates this construction from \cref{cons:concatenated_phantom_as_outer_css_generalized}. In the CSS logical basis $\mathcal{L} = (L_\mathrm{x}, L_\mathrm{z})$ written in \cref{eq:concatenated_phantom_as_inner_css_matrices}, the logical groups of the resulting code are:
    \begin{itemize}
        
        \item $\mathrm{Aut}^{(t)}_\mathcal{L}(\mathcal{C}) \geq \langle \mathbf{D}^{(t)}_k, \mathbf{CX}_k \rangle$.
        
        \item $\mathrm{Perm}_\mathcal{L}(\mathcal{C}) = \langle \mathbf{CX}_k \rangle \cong \GL(k, \mathbb{F}_2)$. These derive from the transversal logical $\mathrm{CX}$s between codeblocks of the single-logical-qubit CSS outer code. Upon concatenation, the physical $\mathrm{CX}$s are promoted into logical $\mathrm{CX}$s within each phantom inner code, which are implementable by physical qubit permutations.

        \item $\mathrm{Trans}^{(t)}_\mathcal{L}(\mathcal{C}) \geq \mathbf{D}^{(t)}_k$. Because there exists a weight-one representative for all $\overline{Z}_j$ logicals on the inner phantom code, a physical $Z^{(t)}$ gate on the support of each $\overline{Z}_j$ implements a logical $Z^{(t)}$ action. These then enable the transversal implementation of logical $Z^{(t)}$ gates on $\mathcal{C}$. Together with $\mathrm{Perm}_\mathcal{L}(\mathcal{C}) = \langle \mathbf{CX}_k \rangle$, this enables the transversal implementation of any diagonal unitary on the $t^\text{th}$ level of the Clifford hierarchy, as discussed in \cref{cons:concatenated_phantom_as_outer_css_generalized}.
        
    \end{itemize}

    See \cref{cons:concatenated_phantom_concrete} for concrete examples of admissible $\mathcal{C}^\mathrm{p}$ and $\mathcal{C}^\mathrm{c}$ codes.
\end{construction}

\begin{corollary}
    [Concatenated phantom CSS codes at second Clifford hierarchy level]
    \label{cons:concatenated_phantom_clifford}
    Take $t = 2$ in \cref{cons:concatenated_phantom_as_outer_css_generalized} or \cref{cons:concatenated_phantom_as_inner_css_generalized}.
    Then the resulting codes achieve $\mathrm{Aut}_\mathcal{L}(\mathcal{C}) = \langle \mathbf{S}_k, \mathbf{CX}_k \rangle \cong \mathcal{P}(2k, \mathbb{F}_2)$, $\mathrm{Perm}_\mathcal{L}(\mathcal{C}) = \langle \mathbf{CX}_k \rangle \cong \GL(k, \mathbb{F}_2)$, and $\mathrm{Trans}_\mathcal{L}(\mathcal{C}) = \langle \mathbf{S}_k, \mathbf{CZ}_k \rangle \cong \mathcal{U}(2k, \mathbb{F}_2)$. That is, the code simultaneously achieves the maximum-sized automorphism, transversal, and permutation gates logical groups possible on indecomposable CSS codes (see \cref{thm:css_codes_auto_logical_groups,thm:css_codes_perm_logical_group,thm:css_codes_trans_logical_groups_ssd_indecomp,thm:css_codes_trans_logical_groups_non_ssd_indecomp}) for all $k \geq 2$.
\end{corollary}

\begin{remark}
    [Concrete admissible outer and inner codes for concatenated phantom CSS codes]
    \label{cons:concatenated_phantom_concrete}
    Various code families are admissible for the input codes, $\mathcal{C}^\mathrm{p}$ and $\mathcal{C}^\mathrm{s}$, in \cref{cons:concatenated_phantom_as_outer_css_generalized,cons:concatenated_phantom_as_inner_css_generalized,cons:concatenated_phantom_clifford}. For example, simplex codes and their variants are natural for the phantom code $\mathcal{C}^\mathrm{p}$:
    \begin{itemize}
        
        \item \emph{Simplex codes.} 
        For dimension $D \geq 2$, let $H_D \in \mathbb{F}_2^{D \times (2^D-1)}$ be the binary matrix whose columns are the nonzero vectors of $\mathbb{F}_2^D$. The classical simplex code is then $\rs(H_D)$, with parameters $[2^D-1,D,2^{D-1}]$, and every nonzero codeword has weight $2^{D-1}$~\cite{ding2018minimal}. Its dual, $\ker H_D$, is the classical Hamming code~\cite{falcone2021binary}. The associated simplex phantom code~\cite{koh2026entangling} is the CSS code with $X$- and $Z$-stabilizer spaces
        \begin{equation}
            C_\mathrm{x} = \{0\},
            \qquad
            C_\mathrm{z} = \ker H_D,
        \end{equation}
        such that the $X$-logical subspace is exactly the classical simplex code. The code has parameters
        \begin{equation}
            \db{2^D-1,D,(2^{D-1},1)},
            \qquad
            w_\mathrm{x} = 0,
            \qquad
            w_\mathrm{z} = 3,
            \qquad
            \delta_\mathrm{x} = 0,
            \qquad
            \delta_\mathrm{z} \leq 4.
        \end{equation}
        The maximum stabilizer weight and qubit degrees follow from Ref.~\cite{tromp1997small}, which determines low-weight bases for the binary Hamming code; for $D \le 500$, a refinement of $\delta_\mathrm{z} = 3$ applies.

        The simplex phantom codes can also be obtained from the $\db{2^D,D,(2^{D-1},2)}$ hypercube codes~\cite{hangleiter2025fault}. These codes have a unique $X$-type stabilizer supported on all $2^D$ vertices (i.e.~physical qubits), while their $Z$-stabilizer space is the extended binary Hamming code. Deleting any qubit punctures the $Z$-stabilizer space to the binary Hamming code $\ker H_D$; upon additionally discarding the $X$-stabilizer, one obtains the simplex phantom code.

        \item \emph{Replicated simplex codes.}
        For $r \geq 2$, concatenate the $D$-dimensional simplex phantom code with the $\db{r,1,(r,1)}$ repetition code. Equivalently, the $X$-logical subspace is the $r$-fold replication of the classical simplex code, 
        $
            \left\{
                (\vb{s},\ldots,\vb{s})
                :
                \vb{s}\in\rs(H_D)
            \right\}
        $.
        Since concatenation with a single-logical-qubit inner code preserves phantomness~\cite{koh2026entangling}, the resulting CSS code is phantom and has parameters
        \begin{equation}
            \db{r(2^D-1),D,(r2^{D-1},1)},
            \qquad
            w_\mathrm{x} = 0,
            \qquad
            w_\mathrm{z} = 3,
            \qquad
            \delta_\mathrm{x} = 0,
            \qquad
            \delta_\mathrm{z} \leq 3.
        \end{equation}
        
       Using the degree-$\leq 4$ Hamming-code basis of Ref.~\cite{tromp1997small}, the incidences of each outer stabilizer can be distributed among the $r$ physical qubits of its repetition block so that, together with a path basis for the inner repetition code stabilizers, every qubit is involved in at most three $Z$-stabilizers.
    \end{itemize}

    A broad range of codes are admissible for $\mathcal{C}^\mathrm{s}$. For example, for $t = 2$, any $k=1$ SSD CSS code is admissible since their transversal logical groups contain an $S$ logical action (see \cref{thm:css_codes_trans_logical_groups_ssd_indecomp}). As the role of $\mathcal{C}^\mathrm{s}$ is to confer distance in \cref{cons:concatenated_phantom_as_outer_css_generalized,cons:concatenated_phantom_as_inner_css_generalized,cons:concatenated_phantom_clifford}, their distance scalings are of interest. 
    Broadly, some concrete options are:
    \begin{itemize}
        
        \item \emph{Colour codes.} 
        Specifically, the one-logical-qubit $\mathrm{CC}_t(0,t-2)$ colour codes in the terminology of Ref.~\cite{kubica2015universal}, which are defined on $t$-dimensional lattices, support transversal $Z^{(t)}$ logical rotations. On the standard higher-dimensional generalizations of the triangular lattice, these codes have parameters
        \begin{equation}
            \db*{\Theta(d^t), 1, \left(\Theta(d^{t-1}), d\right)},
            \qquad
            w_\mathrm{x} \leq (t+1)!,
            \qquad
            w_\mathrm{z} \leq 6,
            \qquad
            \delta_\mathrm{x} \leq t+1,
            \qquad
            \delta_\mathrm{z} \leq \binom{t+1}{2}.
        \end{equation}

        Above, the bounds on the stabilizer weights follow by taking the bulk $t$-cells to be $t$-dimensional permutohedra, which have $(t+1)!$ vertices and whose two-dimensional faces have at most six vertices~\cite{ceballos2023generalized}. At $t=3$, these codes coincide with the quantum Reed--Muller codes $\mathrm{QRM}(t+1)$~\cite{kubica2015universal,anderson2014fault}. For $t=2$, a standard example is the triangular colour code with $6.6.6$ tiling~\cite{landahl2011fault} and for $t=3$, a standard example is the tetrahedral colour code with triangular $6.6.6$ boundaries~\cite{butt2026complementary}.
        
        \item \emph{Self-concatenated colour codes.} The $n=\Theta(d^t)$ scaling of colour codes can be broken by self-concatenating their $d=3$ instances. After $r$ levels of concatenation, the resulting code has parameters
        \begin{equation}
            \db*{
                (2^{t+1}-1)^r,
                1,
                \left((2^t-1)^r,3^r\right)
            },
            \quad
            w_\mathrm{x}
            \leq
            2^t(2^t-1)^{r-1},
            \quad
            w_\mathrm{z}
            \leq
            4 \cdot 3^{r-1},
            \quad
            \delta_\mathrm{x}
            \leq
            r(t+1),
            \quad
            \delta_\mathrm{z}
            \leq
            r\binom{t+1}{2}.
        \end{equation}
        
        That is, we have $n = d^{\log_3(2^{t+1}-1)} < d^t$, amounting to $n \approx d^{1.77}$ for $t = 2$ and $n \approx d^{2.47}$ for $t = 3$.

        \item \emph{Asymptotically good codes for $t=2$.}
        Asymptotically good families of binary self-dual classical codes are known to exist, including both Type~I (singly-even) and Type~II (doubly-even) families~\cite{macwilliams1972good,shi2017good}. A puncturing construction~\cite{jain2025transversal} can be used to map a classical self-dual code $[n,n/2,d]$ to a one-logical-qubit SSD code with parameters $\db{n-1,1,\geq d-1}$. Consequently, applying this construction to an asymptotically good self-dual classical code family gives
        \begin{equation}
            \db*{\Theta(d),1,(d,d)},
            \qquad
            w_\mathrm{x}=w_\mathrm{z}=O(d),
            \qquad
            \delta_\mathrm{x}=\delta_\mathrm{z}=O(d).
        \end{equation}

        The existence of a transversal $S$ logical gate is guaranteed by \cref{thm:css_codes_trans_logical_groups_ssd_indecomp}, and also by \cite[Corr.~1]{tansuwannont2025clifford}.

        \item \emph{Generalized divisible-code towers.}
        For every fixed $t \geq 3$, Ref.~\cite[Sec.~VI~A]{haah2018towers} gives families of generalized divisible CSS codes supporting transversal $Z^{(t)}$ logical rotations, with code parameters $\db{O(d^{t-1}),\Omega(d),d}$. Promoting the additional logical qubits to $Z$-stabilizers gives
        \begin{equation}
            \db*{O(d^{t-1}),1,\geq d},
            \qquad
            w_\mathrm{x}=w_\mathrm{z}=O(n),
            \qquad
            \delta_\mathrm{x}=\delta_\mathrm{z}=O(n).
            \label{eq:gen_divisible_code_towers}
        \end{equation}

        Thus their block-length scaling improves by one power of $d$ over the $t$-dimensional colour-code; but, unlike colour codes, these constructions do not furnish bounded stabilizer weights or qubit degrees.

        \item \emph{Asymptotically good higher-level codes.}
        For every fixed $t \geq 3$, the recent Ref.~\cite{sanjose2026asymptotically} constructs explicit asymptotically good families of CSS qubit codes, with parameters $\db*{n,\Theta(n),\Theta(n)}$, supporting transversal $Z^{(t)}$ logical rotations.
        Specializing the construction to a single retained logical qubit gives families with
        \begin{equation}
            \db*{\Theta(d),1,\geq d},
            \qquad
            w_\mathrm{x}=w_\mathrm{z}=O(n),
            \qquad
            \delta_\mathrm{x}=\delta_\mathrm{z}=O(n).
        \end{equation}
        
        These families improve the $O(d^{t-1})$ block-length scaling of generalized divisible-code towers to the asymptotically optimal linear scaling $n=\Theta(d)$.
        However, the construction likewise does not provide bounded stabilizer weights or qubit degrees, and its presently proved constants deteriorate rapidly with $t$.
    \end{itemize}

\end{remark}

\subsubsection{Complete-hypergraph codes at any level of the Clifford hierarchy}

\begin{construction}
    [Complete-hypergraph CSS codes at any Clifford hierarchy level]
    \label{cons:complete_hypergraph_css_generalized}
    Choose a level of the Clifford hierarchy $t \geq 1$ and a number of logical qubits $k \geq t$. Then consider the rank-$t$ complete hypergraph $G = (V, E)$ defined by vertex and hyperedge sets
    \begin{equation}
        V = [k], \qquad
        E = \{A \subseteq [k]: 1 \leq \abs{A} \leq t\}.
    \end{equation}

    For convenience, we denote non-vertex hyperedges $E^* = \{A \subseteq [k]: 2 \leq \abs{A} \leq t\} \subseteq E$. We define an $\db{n,k}$ CSS code $\mathcal{C}$ by placing one physical qubit on every hyperedge \(A\in E\). Thus $n = \abs{E} = \sum_{s=1}^t \binom{k}{s}$. We take no \(X\)-type stabilizers, and impose the \(Z\)-type parity constraints $z_A = \bigoplus_{i\in A} z_{\{i\}}$ for all $A \in E^*$, where $z_A \in \mathbb{F}_2$ is the computational-basis bit value of the physical qubit indexed by \(A\). Equivalently, we choose for every \(A \in E^*\) an element \(i_A\in A\), for example $i_A = \min A$, and include the weight-three \(Z\)-type stabilizer $Z_A Z_{A\setminus\{i_A\}} Z_{\{i_A\}}$. Indeed, then for each \(A \in E^*\), this weight-three \(Z\)-type stabilizer enforces the parity constraint $z_A = z_{A\setminus\{i_A\}} \oplus z_{\{i_A\}}$, and by induction, the stabilizers together enforce $z_A = \bigoplus_{i\in A} z_{\{i\}}$ as desired.

    To see that the code encodes $k$ logical qubits, we note that the \(Z\)-type stabilizers added above are all independent: ordering the hyperedges by increasing size, the stabilizer indexed by \(A\in E^*\) contains the qubit \(A\), whereas all other qubits in that generator have smaller size. Hence the \(Z\)-type stabilizer generator matrix has one pivot for each \(A\in E^*\). Therefore $r_\mathrm{z} = \abs{E^*} = \abs{E} - k$, and $r_\mathrm{x} = 0$. Then the number of logical qubits is $n - r_\mathrm{x} - r_\mathrm{z} = \abs{E} - (\abs{E} - k) = k$.

    The $k$ logical qubits can, in fact, be associated with the singleton bits $z_{\{i\}}$ as they are unconstrained. Imposing the \(Z\)-type parity constraints $z_A = \bigoplus_{i\in A} z_{\{i\}}$ for all $A \in E^*$ fixes the encoded computational basis states
    \begin{equation}
        \ket{\overline{x_1, \ldots, x_k}}
        =
        \bigotimes_{A \in E}
        \left|
            \bigoplus_{i\in A} x_i
        \right\rangle_A,
    \end{equation}
    where $\bigoplus_i x_i$ is the parity of the $x_i$ input bits and $\ket{x}_A$ denotes the computational basis state $\ket{x}$ on the qubit $A$. A corresponding CSS logical basis $\mathcal{L}$ is
    \begin{equation}
        \overline Z_i = Z_{\{i\}},
        \qquad
        \overline X_i =
        \prod_{A \in E: i \in A} X_A .
    \end{equation}

    Clearly, $d_\mathrm{z} = 1$. To see $d_\mathrm{x}$, it suffices to consider products of the $\overline X_i$ logical operators as there are no $X$-type stabilizers. For a nonempty \(R\subseteq [k]\), we consider
    \begin{equation}\begin{split}
        \overline X_R
        &\coloneqq
        \prod_{i\in R}\overline X_i
        =
        \prod_{i\in R}
        \prod_{\substack{A\in E\\ i\in A}} X_A
        = \prod_{A\in E} X_A^{|A\cap R|\bmod 2},
        \qquad
        d_\mathrm{x}
        =
        \min_{\varnothing \neq R \subseteq [k]} 
        \abs*{\underbrace{\left\{A\in E: |A\cap R|\equiv 1 \pmod 2\right\}
            }_{\supp(\overline X_R)}}.
    \end{split}\end{equation}

    Now choose an \(i\in R\), and consider all subsets $B \subseteq [k] \setminus\{i\}$ with $\abs{B} \leq t - 1$. For each such $B$, the two subsets $B$ and $B \cup \{i\}$ have opposite parity of intersection with \(R\), since \(i\in R\). That is, $|(B\cup\{i\})\cap R| \equiv |B\cap R|+1 \pmod 2$. Hence exactly one of \(B\) and \(B\cup\{i\}\) belongs to
    \(\supp(\overline X_R)\). When \(B=\varnothing\), the empty set is not a
    physical-qubit label, but it has even intersection with \(R\), so the
    contributing label is \(\{i\}\in E\). The number of such choices of $B$ is $m \coloneqq \sum_{s=0}^{t-1} \binom{k-1}{s}$, so every nontrivial \(X\)-logical has weight at least $m$. Furthermore, this lower bound is achieved by taking \(R=\{i\}\) for any $i \in [k]$. That is, in summary, the code has the following parameters for $t \geq 2$
    \begin{equation}
        \db*{
            n = \sum_{s=1}^{t} \binom{k}{s} = \Theta(k^t), \,\,
            k, \,\,
            \left(
                d_\mathrm{x} = \sum_{s=0}^{t-1} \binom{k-1}{s} = \Theta(k^{t-1}),
                d_\mathrm{z} = 1
            \right)
        },
        \quad
        w_\mathrm{x} = 0,
        \quad
        w_\mathrm{z} = 3,
        \quad
        \delta_\mathrm{x} = 0,
        \quad
        \delta_\mathrm{z} = d_\mathrm{x} - 1.
    \end{equation}

    We claim that the entire $\smash{\mathbf{D}^{(t)}_k}$ logical group is implemented transversally on $\mathcal{C}$. The intuition is that a diagonal level-$t$ diagonal gate is a sum of parity phases on the subset of qubits the gate acts on, and the code has one physical qubit storing exactly each such parity; so the usual compute--phase--uncompute parity circuit (see e.g.~\cref{prop:phase_polynomial_circuit_implementation_diagonal_unitary}) amounts to a simple transversal implementation.

    To see this more rigorously, let \(J\subseteq [k]\) be any subset of logical qubits and let \(U\) be a diagonal unitary in the \(t^\text{th}\) level of the Clifford hierarchy acting on the logical qubits in \(J\). Up to global phase, we may write $U$ in the form of \cref{eq:phase_polynomial_diagonal_unitary}, where the phase polynomial \(P\) depends only on the variables \(\{x_j:j\in J\}\). Converting to parity-phase form, as in \cref{eq:phase_polynomial_parity_form}, we may equivalently express
    \begin{equation}
        P(x)
        =
        \sum_{\substack{\varnothing\neq B\subseteq J\\ |B|\leq t}}
        b_B
        \bigoplus_{i\in B}x_i
        \pmod{2^t},
    \end{equation}
    for suitable coefficients \(b_B\in\mathbb Z_{2^t}\). Now, every subset \(B\) in the above sum is a hyperedge of the complete hypergraph, and hence labels a physical qubit of the code. Define the physical transversal operator
    \begin{equation}
        \overline U
        \coloneqq
        \prod_{\substack{\varnothing\neq B\subseteq J\\ |B|\leq t}}
        (Z^{(t)}_B)^{b_B},
    \end{equation}
    where \(Z^{(t)}_B\) denotes the physical \(Z^{(t)} \coloneqq Z^{1/2^{t-1}}\) gate (see \cref{def:clifford_hierarchy_diagonal_gates}) on the qubit indexed by \(B\). Now evaluate this operator on an encoded computational-basis state. Since the physical qubit \(B\) stores the parity \(x_B=\bigoplus_{i\in B}x_i\), we have
    \begin{equation}
        \overline U \ket{\overline{x_1, \ldots, x_k}}
        =
        \prod_{\substack{\varnothing\neq B\subseteq J\\ |B|\leq t}}
        \omega_t^{b_B x_B}
        \ket{\overline{x_1, \ldots, x_k}}
        =
        \omega_t^{P(x)}
        \ket{\overline{x_1, \ldots, x_k}}.
    \end{equation}

    That is, \(\overline U\) preserves the code space and implements \(U\) on the logical qubits up to global phase. Since \(\overline U\) is a tensor product of single-qubit powers of physical \(Z^{(t)}\) gates, the implementation is transversal. 
    Lastly, we note that these codes attain exactly the minimum-possible $n$ for stabilizer codes hosting the $\smash{\mathbf{D}^{(t)}_k}$ logical group transversally (see \cref{thm:stab_codes_trans_any_clifford_hierarchy_level_n_scaling}).
\end{construction}

\begin{remark}
    [Concatenating complete-hypergraph CSS codes for distance]
    \label{rem:complete_hypergraph_css_generalized_concatenation}
    The distance of codes from \cref{cons:complete_hypergraph_css_generalized} can be made nontrivial ($d \geq 2$) by concatenating with any $k = 1$ nontrivial-distance CSS code that supports a transversal logical $Z^{(t)}$ gate. 
    See \cref{cons:concatenated_phantom_concrete} for examples of such codes.
\end{remark}

\begin{corollary}
    [Complete-graph CSS codes at the second Clifford hierarchy level]
    \label{cons:complete_hypergraph_css_t_eq_2}
    Take $t = 2$ and $k \geq 2$ in \cref{cons:complete_hypergraph_css_generalized}. Then the code $\mathcal{C}$ has parameters
    \begin{equation}
        \db*{
            n = k + \binom{k}{2} = \frac{k(k+1)}{2}, \,\,
            k, \,\,
            (d_\mathrm{x} = k, d_\mathrm{z} = 1)
        }.
    \end{equation}

    The code $\mathcal{C}$ is defined on a complete graph of $k$ vertices: the physical qubits are labelled by all singletons $\{i\}$ and unordered pairs $\{i, j\}$ for $1 \leq i < j \leq k$, and we include a $Z$-type ``edge'' stabilizer $\smash{Z_{\{i, j\}} Z_{\{i\}} Z_{\{j\}}}$ for every $\{i, j\}$.
    Moreover, $\mathrm{Trans}_\mathcal{L}(\mathcal{C}) = \smash{\mathbf{D}^{(2)}_k} = \langle \mathbf{S}_k, \mathbf{CZ}_k \rangle \cong \mathcal{U}(2k, \mathbb{F}_2)$. Each of the logical gates in $\langle \mathbf{S}_k, \mathbf{CZ}_k \rangle$ is implemented by a tensor product of powers of physical $S$ gates. Indeed:
    \begin{itemize}[noitemsep]
        \item A logical $S_i$ for any $i \in [k]$ is implemented by physical \(\smash{S_{\{i\}}}\) at the physical level.
        \item A logical $\mathrm{CZ}_{ij}$ for any distinct $i, j \in [k]$ is implemented by physical $\smash{S_{\{i\}}S_{\{j\}}S_{\{i,j\}}^\dagger}$. 
    \end{itemize}

    \Cref{rem:complete_hypergraph_css_generalized_concatenation} applies to increase the distance of the code; $k = 1$ codes that support transversal logical $S$ gates include the colour code and families with asymptotically good distance (see \cref{cons:concatenated_phantom_concrete} for details).
\end{corollary}

\begin{corollary}
    [Complete-hypergraph CSS codes at the third Clifford hierarchy level]
    \label{cons:complete_hypergraph_css_t_eq_3}
    Take $t = 3$ and $k \geq 3$ in \cref{cons:complete_hypergraph_css_generalized}. Then the code $\mathcal{C}$ has parameters
    \begin{equation}
        \db*{
            n = k + \binom{k}{2} + \binom{k}{3} = \frac{k(k^2+5)}{6}, \,\,
            k, \,\,
            \left(
                d_\mathrm{x} = \frac{k(k-1)}{2} + 1, 
                d_\mathrm{z} = 1
            \right)
        }.
    \end{equation}

    The code $\mathcal{C}$ is defined on a rank-$3$ complete hypergraph of $k$ vertices: the physical qubits are labelled by all singletons $\{i\}$, unordered pairs $\{i, j\}$, and unordered triplets $\{i, j, l\}$ for $1 \leq i < j < l \leq k$, and we include a $Z$-type ``edge'' stabilizer $\smash{Z_{\{i, j\}} Z_{\{i\}} Z_{\{j\}}}$ for every $\{i, j\}$ and ``triangle'' stabilizer $\smash{Z_{\{i, j, l\}} Z_{\{j, l\}} Z_{\{i\}}}$ for every $\{i, j, l\}$. 
    Moreover, $\smash{\mathbf{D}^{(3)}_k} = \langle \mathbf{T}_k, \mathbf{CS}_k, \mathbf{CCZ}_k \rangle$ is transversal on $\mathcal{C}$. Each of the logical gates is implemented by a tensor product of powers of physical $T$ gates. Indeed:
    \begin{itemize}[noitemsep]
        \item A logical $T_i$ for any $i \in [k]$ is implemented by physical \(\smash{T_{\{i\}}}\).
        \item A logical $\mathrm{CS}_{ij}$ for any distinct $i, j \in [k]$ is implemented by physical \(\smash{T_{\{i\}}T_{\{j\}}T_{\{i,j\}}^\dagger}\).
        \item A logical $\mathrm{CCZ}_{ijl}$ for any distinct $i, j, l \in [k]$ is implemented by physical $\smash{
            T_{\{i\}}T_{\{j\}}T_{\{l\}}
            T_{\{i,j\}}^\dagger
            T_{\{i,l\}}^\dagger
            T_{\{j,l\}}^\dagger
            T_{\{i,j,l\}}}$.
    \end{itemize}

    \Cref{rem:complete_hypergraph_css_generalized_concatenation} applies to increase the distance of the code; $k = 1$ codes that support transversal logical $T$ gates include the three-dimensional colour codes (see \cref{cons:concatenated_phantom_concrete} for details).
\end{corollary}

\begin{remark}
    [Addressable within-block logical CXs on CSS codes]
    \label{rem:complete_hypergraph_css_generalized_addressable_cx_gates}
    Arbitrary within-block logical $\CX_{ij}$ gates can be performed on $t \geq 2$ codes from \cref{cons:concatenated_phantom_as_inner_css_generalized,cons:concatenated_phantom_as_outer_css_generalized,cons:complete_hypergraph_css_generalized}. 
    In fact, it can be performed on any CSS code with within-block addressable logical $S$ and $\CZ$ gates. 
    We show an elementary approach here.
    To perform a within-block logical $\CX$ gate on a data codeblock, we employ Hadamard teleportation into the codeblock, which can be done with interblock $\CZ^{\otimes k}$ logical gates between the data codeblock and an ancillary codeblock that is initialized in the logical $\ket{+}$ state~(see~\cite[Fig.~13]{koh2026entangling} for an illustration).
    To perform an interblock logical $\CZ_{AB}^{\otimes k}$ on codeblocks $A$ and $B$, we can use the identity $\CX_{AB}^{\otimes k} (S^\dagger)^{\otimes k}_B \CX_{AB}^{\otimes k} S_A^{\otimes k} S_B^{\otimes k}$. 
    In other words, an interblock transversal logical $\CZ^{\otimes k}$ gate can be decomposed into within-block parallel logical $S$ and interblock transversal logical $\CX$ gates. 
    The former is available in the code when $t \geq 2$ and the latter is always available because the code is CSS. 
    Then we can obtain a within-block $\CX_{ij}$ via the following circuit identity, where the Hadamards in the boxes come from the gate-teleported $H^{\otimes k}$. 
    \begin{equation*}
        \begin{quantikz}[row sep={0.65cm,between origins}, column sep={0.75cm,between origins}]
            & \ctrl{1} & \qw \\
            & \targ{}      & \qw \\
            & &
        \end{quantikz}
        \;=\;
        \begin{quantikz}[row sep={0.65cm,between origins}, column sep={0.8cm,between origins}]
            & \qw & \gate{H} 
                \gategroup[wires=3,steps=1,style={draw,rounded corners,inner xsep=0pt,inner ysep=0pt},label style={label position=above,anchor=south,yshift=-6pt}]{tele}
                & \qw & \gate{H} 
                \gategroup[wires=3,steps=1,style={draw,rounded corners,inner xsep=0pt,inner ysep=0pt},label style={label position=above,anchor=south,yshift=-6pt}]{tele}
                & \qw & \ctrl{1}
                & \qw & \gate{H} 
                \gategroup[wires=3,steps=1,style={draw,rounded corners,inner xsep=0pt,inner ysep=0pt},label style={label position=above,anchor=south,yshift=-6pt}]{tele}
                & \qw & \gate{H} 
                \gategroup[wires=3,steps=1,style={draw,rounded corners,inner xsep=0pt,inner ysep=0pt},label style={label position=above,anchor=south,yshift=-6pt}]{tele}
                & \qw & \qw \\
            & \gate{S} & \gate{H} & \gate{S} & \gate{H} & \gate{S} & \control{} & \gate{S} & \gate{H} & \gate{S} & \gate{H} & \gate{S} & \qw \\
            & & \gate{H} & & \gate{H} & & & & \gate{H} & & \gate{H} & &
        \end{quantikz}
    \end{equation*}
    The gates before and after the within-block $\CZ$ conjugates the within-block $\CZ$ with a Hadamard operation on the second qubit up to a global phase difference.
\end{remark}

\subsubsection{Self-dual codes saturating permutation logical groups}

\begin{construction}
    [Minimum-$n$ examples of SSD CSS codes achieving $\Sp(k, \mathbb{F}_2)$ and $O(k, \mathbb{F}_2)$ permutation logical groups]
    \label{cons:css_codes_maximum_perm_logical_group_ssd}
    First, we give in \cref{tab:css_codes_maximum_perm_logical_group_ssd_symplectic} examples of $\db{n, k, d}$ SSD CSS codes, each defined by a stabilizer generator matrix $H \coloneqq H_\mathrm{x} = H_\mathrm{z} \in \smash{\mathbb{F}_2^{(n-k)/2 \times n}}$, that saturate the $\mathrm{Perm}_\mathcal{L}(\mathcal{C}) \cong \Sp(k, \mathbb{F}_2)$ branch of \cref{thm:css_codes_perm_logical_group_ssd} for a CSS logical basis $\mathcal{L}$. These codes must have $\vb{1} \in \rs(H)$, and $n,k,d$ are even.
    The $\db{4,2,2}$ and $\db{16,2,4}$ codes are phantom~\cite{koh2026entangling}; they do not contradict \cref{thm:css_codes_perm_logical_group_ssd} because of the exceptional isomorphism $\Sp(2, \mathbb{F}_2) \cong \GL(2, \mathbb{F}_2)$.
    \begin{table}[h]
        \centering
        \scriptsize
        \setlength{\arraycolsep}{0pt}
        \begin{tabular}{p{1.5cm} p{2cm} p{2.8cm} p{2.8cm} p{2.8cm} p{5.5cm}}
            \toprule
            Code 
            & $\mathrm{Perm}_\mathcal{L}(\mathcal{C})$ 
            & $H$ 
            & $L_\mathrm{x}$ 
            & $L_\mathrm{z}$ 
            & Gates
            \\
            \midrule
            $\db{4,2,2}$
            & $\begin{aligned}
                & \Sp(2, \mathbb{F}_2) \\[-2pt]
                & \cong \GL(2, \mathbb{F}_2)
            \end{aligned}$
            & $\begin{matrix}
                1&1&1&1
            \end{matrix}$
            & $\begin{matrix}
                1&1&0&0\\
                0&1&1&0
            \end{matrix}$
            & $\begin{matrix}
                0&1&1&0\\
                1&1&0&0
            \end{matrix}$
            & $\begin{aligned}
                & \mathrm{CX}_{12}: 
                    (2\,3) 
                \\[-2pt]
                & \mathrm{CX}_{21}: 
                    (1\,2)
            \end{aligned}$
            \\
            \midrule
            $\db{16,2,4}$
            & $\begin{aligned}
                & \Sp(2,\mathbb F_2)\\[-2pt]
                & \cong \GL(2,\mathbb F_2)
            \end{aligned}$
            & $\begin{matrix}
                1&1&1&1&1&1&1&1&1&1&1&1&1&1&1&1\\
                0&0&0&0&0&0&0&0&1&1&1&1&1&1&1&1\\
                0&0&0&0&1&1&1&1&0&0&0&0&1&1&1&1\\
                0&0&1&1&0&0&1&1&0&0&1&1&0&0&1&1\\
                0&1&0&1&0&1&0&1&0&1&0&1&0&1&0&1\\
                0&0&0&0&0&0&0&0&0&0&0&0&1&1&1&1\\
                0&0&0&0&0&0&1&1&0&1&0&1&0&1&1&0
            \end{matrix}$
            & $\begin{matrix}
                0&0&0&0&0&0&0&0&0&0&1&1&0&0&1&1\\
                0&0&0&0&0&1&0&1&0&0&0&0&0&1&0&1
            \end{matrix}$
            & $\begin{matrix}
                0&0&0&0&0&1&0&1&0&0&0&0&0&1&0&1\\
                0&0&0&0&0&0&0&0&0&0&1&1&0&0&1&1
            \end{matrix}$
            & $\begin{aligned}
                & \mathrm{CX}_{12}:
                    (2\,4)(5\,13)(6\,16)
                    (7\,15)(8\,14)(10\,12)
                \\[-2pt]
                & \mathrm{CX}_{21}:
                    (3\,4)(7\,8)(9\,13)
                    (10\,14)(11\,16)(12\,15)
            \end{aligned}$
            \\
            \midrule
            $\db{6,4,2}$
            & $\Sp(4, \mathbb{F}_2)$
            & $\begin{matrix}
                1&1&1&1&1&1
            \end{matrix}$
            & $\begin{matrix}
                1&1&0&0&0&0\\
                0&1&1&0&0&0\\
                1&1&1&1&0&0\\
                0&0&0&1&1&0
            \end{matrix}$
            & $\begin{matrix}
                0&1&1&0&0&0\\
                1&1&0&0&0&0\\
                0&0&0&1&1&0\\
                1&1&1&1&0&0
            \end{matrix}$
            & $\begin{aligned}
                \mathrm{CX}_{12} &:
                    (2\,3) 
                \\[-2pt]
                \mathrm{CX}_{21} &:
                    (1\,2) 
                \\[-2pt]
                \mathrm{CX}_{34} &:
                    (4\,5) 
                \\[-2pt]
                \mathrm{CX}_{43} &:
                    (5\,6) 
                \\[-2pt]
                \mathrm{CX}_{23}\mathrm{CX}_{41} &:
                    (1\,2)(3\,4)(5\,6)
            \end{aligned}$
            \\
            \midrule
            $\db{16,4,4}$
            & $\Sp(4,\mathbb F_2)$
            & $\begin{matrix}
                1&1&1&1&1&1&1&1&1&1&1&1&1&1&1&1\\
                0&0&0&0&0&0&0&0&1&1&1&1&1&1&1&1\\
                0&0&0&0&1&1&1&1&0&0&0&0&1&1&1&1\\
                0&0&1&1&0&0&1&1&0&0&1&1&0&0&1&1\\
                0&1&0&1&0&1&0&1&0&1&0&1&0&1&0&1\\
                0&0&0&1&0&0&0&1&0&0&0&1&1&1&1&0
            \end{matrix}$
            & $\begin{matrix}
                1&0&1&0&0&0&0&0&1&0&1&0&0&0&0&0\\
                1&1&0&0&1&1&0&0&0&0&0&0&0&0&0&0\\
                1&1&0&0&0&0&0&0&1&1&0&0&0&0&0&0\\
                1&0&1&0&1&0&1&0&0&0&0&0&0&0&0&0
            \end{matrix}$
            & $\begin{matrix}
                1&1&0&0&1&1&0&0&0&0&0&0&0&0&0&0\\
                1&0&1&0&0&0&0&0&1&0&1&0&0&0&0&0\\
                1&0&1&0&1&0&1&0&0&0&0&0&0&0&0&0\\
                1&1&0&0&0&0&0&0&1&1&0&0&0&0&0&0
            \end{matrix}$
            & $\begin{aligned}
                \mathrm{CX}_{12} &:
                    (3\,7)(4\,8)(9\,10)
                    (11\,16)(12\,15)(13\,14)
                \\[-2pt]
                \mathrm{CX}_{21} &:
                    (2\,10)(4\,12)(5\,7)
                    (6\,16)(8\,14)(13\,15)
                \\[-2pt]
                \mathrm{CX}_{34} &:
                    (2\,6)(4\,8)(9\,11)
                    (10\,16)(12\,14)(13\,15)
                \\[-2pt]
                \mathrm{CX}_{43} &:
                    (3\,11)(4\,12)(5\,6)
                    (7\,16)(8\,15)(13\,14)
                \\[-2pt]
                \mathrm{CX}_{23}\mathrm{CX}_{41} &:
                    (5\,13)(6\,14)(7\,15)(8\,16)
            \end{aligned}$
            \\
            \bottomrule
        \end{tabular}
        \caption{
            Minimum-$n$ examples of SSD CSS codes achieving $\Sp(k,\mathbb{F}_2)$ permutation logical group. Convention for qubit permutations follows \cref{sec:constructions/preliminaries/permutations}.
            All codes shown are indecomposable.
        }
        \label{tab:css_codes_maximum_perm_logical_group_ssd_symplectic}
    \end{table}
  
    Next, we give in \cref{tab:css_codes_maximum_perm_logical_group_ssd_orthogonal} examples of $\db{n, k,d}$ SSD CSS codes that saturate the $\mathrm{Perm}_\mathcal{L}(\mathcal{C}) \cong O(k, \mathbb{F}_2)$ branch of \cref{thm:css_codes_perm_logical_group_ssd}. These codes must have $\vb{1} \notin \rs(H)$, and $n,k,d$ need not be even.
    \begin{table}[ht]
        \centering
        \scriptsize
        \setlength{\arraycolsep}{0pt}
        \begin{tabular}{p{1.5cm} p{3cm} p{3cm} p{3cm} p{6cm}}
            \toprule
            Code 
            & $\mathrm{Perm}_\mathcal{L}(\mathcal{C})$ 
            & $H$ 
            & $L_\mathrm{x} = L_\mathrm{z}$ 
            & Gates 
            \\
            \midrule
            $\db{2,2,1}$
            & $O(2, \mathbb{F}_2) \cong S_2$
            & Empty
            & $\begin{matrix}
                1&0\\
                0&1
            \end{matrix}$
            & $\begin{aligned}
                \mathrm{SWAP}_{12} &: 
                    (1\,2)
            \end{aligned}$
            \\
            \midrule
            $\db{6,2,2}$
            & $O(2, \mathbb{F}_2) \cong S_2$
            & $\begin{matrix}
                1&1&1&1&0&0\\
                1&1&0&0&1&1
            \end{matrix}$
            & $\begin{matrix}
                1&0&1&0&1&0\\
                1&0&0&1&1&0
            \end{matrix}$
            & $\begin{aligned}
                \mathrm{SWAP}_{12} &: 
                    (3\,4)
            \end{aligned}$
            \\
            \midrule
            $\db{12,2,3}$
            & $O(2, \mathbb{F}_2) \cong S_2$
            & $\begin{matrix}
                0&1&0&0&1&0&0&1&0&0&1&0\\
                1&1&0&1&0&1&1&1&0&1&0&1\\
                0&0&1&1&0&0&0&0&1&1&0&0\\
                0&0&0&0&1&1&0&0&0&0&1&1\\
                0&0&0&0&0&0&1&1&1&1&1&1
            \end{matrix}$
            & $\begin{matrix}
                1&0&1&1&0&0&0&0&0&0&0&0\\
                0&1&0&0&1&1&0&0&0&0&0&0
            \end{matrix}$
            & $\begin{aligned}
                \mathrm{SWAP}_{12} &:
                    (1\,2)(3\,5)(4\,6)
                    (7\,8)(9\,11)(10\,12)
            \end{aligned}$
            \\
            \midrule
            $\db{3,3,1}$
            & $O(3, \mathbb{F}_2) \cong S_3$
            & Empty
            & $\begin{matrix}
                1&0&0\\
                0&1&0\\
                0&0&1
            \end{matrix}$
            & $\begin{aligned}
                \mathrm{SWAP}_{12} &: 
                    (1\,2) 
                \\[-2pt]
                \mathrm{SWAP}_{23} &: 
                    (2\,3)
            \end{aligned}$
            \\
            \midrule
            $\db{9,3,2}$
            & $O(3,\mathbb F_2)\cong S_3$
            & $\begin{matrix}
                1&0&0&0&1&1&1&1&1\\
                0&1&0&1&0&1&1&1&1\\
                0&0&1&1&1&0&1&1&1
            \end{matrix}$
            & $\begin{matrix}
                1&0&0&1&0&0&1&0&0\\
                0&1&0&0&1&0&0&1&0\\
                0&0&1&0&0&1&0&0&1
            \end{matrix}$
            & $\begin{aligned}
                \mathrm{SWAP}_{12} &: 
                    (7\,8) 
                \\[-2pt]
                \mathrm{SWAP}_{23} &: 
                    (8\,9)
            \end{aligned}$
            \\
            \midrule
            $\db{15,3,3}$
            & $O(3,\mathbb F_2)\cong S_3$
            & $\begin{matrix}
                1&0&1&0&0&0&0&0&0&0&1&1&1&0&1\\
                0&1&1&0&0&0&0&1&1&0&1&1&1&1&0\\
                0&0&0&1&0&1&0&0&0&0&1&0&1&0&0\\
                0&0&0&0&1&1&0&1&1&0&0&1&1&0&0\\
                0&0&0&0&0&0&1&0&1&0&0&1&0&0&1\\
                0&0&0&0&0&0&0&0&0&1&1&1&1&1&1
            \end{matrix}$
            & $\begin{matrix}
                1&1&1&0&0&0&0&0&0&0&0&0&0&0&0\\
                0&0&0&1&1&1&0&0&0&0&0&0&0&0&0\\
                0&0&0&0&0&0&1&1&1&0&0&0&0&0&0
            \end{matrix}$
            & $\begin{aligned}
                \mathrm{SWAP}_{12} &:
                    (1\,4)(2\,5)(3\,6)
                    (10\,13)(11\,14)(12\,15)
                \\[-2pt]
                \mathrm{SWAP}_{23} &:
                    (4\,7)(5\,8)(6\,9)
                    (10\,14)(11\,15)(12\,13)
            \end{aligned}$
            \\
            \midrule
            $\db{8,4,2}$
            & $O(4, \mathbb{F}_2)$
            & $\begin{matrix}
                1&1&1&1&1&1&0&0\\
                1&1&0&0&0&0&1&1
            \end{matrix}$
            & $\begin{matrix}
                1&0&1&0&0&0&1&0\\
                1&0&0&1&0&0&1&0\\
                1&0&0&0&1&0&1&0\\
                1&0&0&0&0&1&1&0
            \end{matrix}$
            & $\begin{aligned}
                \mathrm{SWAP}_{12} &: 
                    (34) 
                \\[-2pt]
                \mathrm{SWAP}_{23} &: 
                    (45) 
                \\[-2pt]
                \mathrm{SWAP}_{34} &: 
                    (56) 
                \\[-2pt]
                \mathrm{TVEC}_{1234} &: 
                    (12)
            \end{aligned}$
            \\
            \midrule
            $\db{16,4,3}$
            & $O(4,\mathbb F_2)$
            & $\begin{matrix}
                1&0&0&0&0&0&1&1&0&0&1&0&1&0&1&0\\
                0&1&0&0&0&0&1&0&0&1&1&0&0&1&0&1\\
                0&0&1&0&0&0&0&1&0&1&1&0&0&1&0&1\\
                0&0&0&1&0&0&0&0&1&1&1&1&1&1&0&1\\
                0&0&0&0&1&0&1&1&1&1&1&0&0&0&1&1\\
                0&0&0&0&0&1&0&0&0&0&0&1&1&1&1&1
            \end{matrix}$
            & $\begin{matrix}
                0&0&0&1&1&0&0&0&1&0&0&0&0&0&0&0\\
                1&0&0&0&0&0&0&0&0&0&0&1&1&0&0&0\\
                0&0&0&0&0&1&0&0&0&1&0&0&0&0&0&1\\
                0&0&0&0&0&0&0&0&0&0&1&0&0&1&1&0
            \end{matrix}$
            & $\begin{aligned}
                \mathrm{SWAP}_{12} &:
                    (1\,5)(3\,8)(4\,12)(9\,13)(10\,16)(11\,14)
                \\[-2pt]
                \mathrm{SWAP}_{23} &:
                    (1\,10)(3\,7)(5\,9)(6\,12)(13\,16)(14\,15)
                \\[-2pt]
                \mathrm{SWAP}_{34} &:
                    (3\,8)(4\,9)(6\,15)(10\,11)(12\,13)(14\,16)
                \\[-2pt]
                \mathrm{TVEC}_{1234} &:
                    (1\,5)(4\,11)(6\,15)(9\,10)(12\,14)(13\,16)
            \end{aligned}$
            \\
            \midrule
            $\db{7,5,1}$
            & $O(5, \mathbb{F}_2) \cong \Sp(4, \mathbb{F}_2)$
            & \multicolumn{3}{l}{$\db{6,4,2} \oplus \db{1,1,1}$}
            \\
            \midrule
            $\db{13,5,2}$
            & $O(5, \mathbb{F}_2) \cong \Sp(4, \mathbb{F}_2)$
            & \multicolumn{3}{l}{$\db{6,4,2} \oplus \db{7,1,3}$}
            \\
            \midrule
            $\db{15,5,3}$
            & $O(5,\mathbb F_2)\cong \Sp(4,\mathbb F_2)$
            & $\begin{matrix}
                0&0&0&0&1&1&0&1&1&0&1&1&0&1&1\\
                0&1&1&0&1&1&0&0&0&1&0&1&1&0&1\\
                0&0&0&1&0&1&1&0&1&1&1&0&1&0&1\\
                1&0&1&1&0&1&0&0&0&0&1&1&1&1&0\\
                0&0&0&1&1&0&0&0&0&0&1&1&1&0&1
            \end{matrix}$
            & $\begin{matrix}
                1&1&1&0&0&0&0&0&0&0&0&0&0&0&0\\
                0&0&0&1&1&1&0&0&0&0&0&0&0&0&0\\
                0&0&0&0&0&0&1&1&1&0&0&0&0&0&0\\
                0&0&0&0&0&0&0&0&0&1&1&1&0&0&0\\
                0&0&0&0&0&0&0&0&0&0&0&0&1&1&1
            \end{matrix}$
            & $\begin{aligned}
                \mathrm{SWAP}_{12} &:
                    (1\,5)(2\,4)(3\,6)
                    (7\,8)(10\,11)(14\,15)
                \\[-2pt]
                \mathrm{SWAP}_{23} &:
                    (1\,2)(4\,8)(5\,7)
                    (6\,9)(10\,12)(13\,14)
                \\[-2pt]
                \mathrm{SWAP}_{34} &:
                    (1\,3)(5\,6)(7\,10)
                    (8\,11)(9\,12)(14\,15)
                \\[-2pt]
                \mathrm{SWAP}_{45} &:
                    (1\,2)(4\,5)(7\,8)
                    (10\,14)(11\,15)(12\,13)
                \\[-2pt]
                \mathrm{TVEC}_{1234} &:
                    (1\,8)(2\,4)(3\,11)
                    (5\,7)(6\,10)(9\,12)
            \end{aligned}$
            \\
            \bottomrule
        \end{tabular}
        \caption{
            Minimum-$n$ examples of SSD CSS codes achieving $O(k,\mathbb{F}_2)$ permutation logical group. 
            The $\mathrm{TVEC}_{1234}$ gate is a CX circuit defined by $X$-type Pauli propagation under conjugation 
            $
                X_1 \mapsto X_2 X_3 X_4, \,
                X_2 \mapsto X_1 X_3 X_4, \,
                X_3 \mapsto X_1 X_2 X_4, \,
                X_4 \mapsto X_1 X_2 X_3
            $, 
            or equivalently stated, 
            $
                \mathrm{TVEC}_{1234} 
                = 
                \mathrm{CX}_{21}
                \mathrm{CX}_{31}
                \mathrm{CX}_{41}
                \mathrm{CX}_{12}
                \mathrm{CX}_{13}
                \mathrm{CX}_{14}
                \mathrm{CX}_{21}
                \mathrm{CX}_{31}
                \mathrm{CX}_{41}
            $. 
            Convention for qubit permutations follows \cref{sec:constructions/preliminaries/permutations}.
            The $\db{2,2,1}$ and $\db{3,3,1}$ trivial codes, and $\db{7,5,1}$ and $\db{13,5,2}$ codes, are decomposable; all others are not. 
            The $\db{6,4,2}$ code is from \cref{tab:css_codes_maximum_perm_logical_group_ssd_symplectic};
            $\db{1,1,1}$ is a bare physical qubit and $\db{7,1,3}$ is the Steane code.
        }
        \label{tab:css_codes_maximum_perm_logical_group_ssd_orthogonal}
    \end{table}

    Note that, for odd $k$, because $O(k, \mathbb{F}_2) \cong \Sp(k-1, \mathbb{F}_2)$, an $\db{n,k,d}$ SSD code hosting $\mathrm{Perm}_\mathcal{L}(\mathcal{C}) \cong O(k, \mathbb{F}_2)$ can trivially be constructed as the direct sum of any $\db{n_1,k-1,d_1}$ SSD code hosting $\mathrm{Perm}_\mathcal{L}(\mathcal{C}) \cong \Sp(k-1, \mathbb{F}_2)$ and any $\db{n_2,1,d_2}$ SSD code, where $n=n_1+n_2$ and $d=\min(d_1,d_2)$; codes constructed this way are decomposable. The $\db{7,5,1}$ and $\db{13,5,2}$ codes in \cref{tab:css_codes_maximum_perm_logical_group_ssd_orthogonal} are of this form.

    All codes shown are minimum-$n$ SSD examples achieving the $O(k, \mathbb{F}_2)$ and $\Sp(k, \mathbb{F}_2)$ permutation logical groups at their distances, as verified by exhaustive search.
\end{construction}

\clearpage

\subsubsection{Self-dual codes saturating transversal logical groups}

\begin{construction}
    [PSD CSS codes achieving $\mathcal{U}_\mathrm{x}(2\floor{k/2}, \mathbb{F}_2) \times \mathcal{U}_\mathrm{z}(2\floor{k/2}, \mathbb{F}_2)$ transversal logical group]
    \label{cons:css_codes_maximum_trans_logical_group_psd}
    Let \(k \geq 2\), $s \coloneqq \floor{k/2}$, and $\epsilon \coloneqq k-2s \in \{0,1\}$. Let \(\{e_1,\ldots,e_k\}\) denote the standard basis of \(\mathbb F_2^k\), where the vector $e_i$ has unit entry at coordinate $i$ and is zero everywhere else, and define $u_i \coloneqq e_{2i-1}+e_{2i}$ for $i \in [s]$. Then $U \coloneqq \spn \{u_1,\ldots,u_s\} \subseteq \mathbb F_2^k$ is a totally isotropic subspace, since $u_i\cdot u_j=0$ for all $i,j \in [s]$. We now define a set of active row labels
    \begin{equation}
        \mathcal I_0
        \coloneqq
        \{(i,i):i\in[s]\}
        \cup
        \{(i,j):1\leq i<j\leq s\},
        \qquad\qquad
        \abs{\mathcal I_0} = s(s+1)/2.
    \end{equation}
    
    For each active row label, we define an active row vector \(f_\alpha\in\mathbb F_2^k\) by $f_{(i,i)} \coloneqq u_i$ and $f_{(i,j)} \coloneqq u_i+u_j$ for $i < j$. Hence all active row vectors lie in the totally isotropic subspace \(U\), and $f_\alpha \cdot f_\beta=0$ for all $\alpha,\beta\in\mathcal I_0$. If \(k\) is odd, we add one connector row label \(\star\) and define $f_\star \coloneqq e_1+e_k$. Then let
    \begin{equation}
        \mathcal I
        \coloneqq
        \begin{cases}
            \mathcal I_0, & k\text{ even},\\
            \mathcal I_0\cup\{\star\}, & k\text{ odd},
        \end{cases}
        \qquad\qquad
        m\coloneqq |\mathcal I|=s(s+1)/2+\epsilon.
    \end{equation}
    
    Now we assemble a matrix \(F\in\mathbb F_2^{m\times k}\) whose rows are the vectors \(f_\alpha\), \(\alpha\in\mathcal I\), in any fixed order. We also assemble a matrix \(P\in\mathbb F_2^{m\times m}\) in the following way: if \(k\) is even, we take $P = 0$, and if $k$ is odd, we set its entries by $P_{\alpha,\star} \coloneqq f_\alpha\cdot f_\star$ for each $\alpha\in\mathcal I_0$, and all other entries are zero. That is, in the odd-$k$ case, \(P\) has possible nonzero entries only in the column indexed by \(\star\), and its \(\star\)-row is zero.

    Now we define an \(\db{n \coloneqq 2m + k, k}\) CSS code $\mathcal{C}$ with the stabilizer generator matrices
    \begin{equation}
        H_\mathrm{x}
        =
        \mqty[I & P & F],
        \qquad
        H_\mathrm{z}
        =
        \mqty[P & I & F].
        \label{eq:css_codes_maximum_trans_logical_group_psd_check_matrices}
    \end{equation}

    The physical qubits are thus partitioned into three blocks of sizes $(m, m, k)$. We will denote the first-block qubit labelled by \(\alpha\in\mathcal I\) by \(x_\alpha\), the second-block qubit labelled by \(\alpha\in\mathcal I\) by \(z_\alpha\), and the final \(k\) physical qubits by \(\ell_1,\ldots,\ell_k\). The code is manifestly PSD (but not SSD) from the forms of $H_\mathrm{x}$ and $H_\mathrm{z}$ above. One can also verify that the code is indecomposable. We claim that in any CSS logical basis $\mathcal{L}$,
    \begin{equation}
        \mathrm{Trans}_\mathcal{L}(\mathcal{C})
        \cong
        \mathcal{U}_\mathrm{z}(2s, \mathbb{F}_2)
        \times
        \mathcal{U}_\mathrm{x}(2s, \mathbb{F}_2).
    \end{equation}

    To see this logical group structure, we explicitly describe the transversal implementations of $\mathrm{CZ}$ and $S$ logical gates on the first clique of $s$ logical qubits, which generate $\mathcal{U}_\mathrm{z}(2s, \mathbb{F}_2)$, and $H^{\otimes 2} \mathrm{CZ} H^{\otimes 2}$ and $H S H$ logical gates on the second clique of $s$ logical qubits, which generate $\mathcal{U}_\mathrm{x}(2s, \mathbb{F}_2)$. A valid CSS logical basis $\mathcal{L}' = (L_\mathrm{x}', L_\mathrm{z}')$ is
    \begin{equation}
        L_\mathrm{x}'
        =
        \mqty[0 & F^\top & I],
        \qquad
        L_\mathrm{z}'
        =
        \mqty[F^\top & 0 & I].
        \label{eq:css_codes_maximum_trans_logical_group_psd_logical_basis_1}
    \end{equation}

    We consider a logical change-of-basis rotation and the corresponding rotated CSS logical basis $\mathcal{L} = (L_\mathrm{x}, L_\mathrm{z})$:
    \begin{equation}
        R
        \coloneqq
        \bigoplus_{i=1}^s
        \begin{pmatrix}
            1&0\\
            1&1
        \end{pmatrix}
        \oplus I_\epsilon
        \in \GL(k,\mathbb F_2),
        \qquad
        L_\mathrm{x} \coloneqq R L_\mathrm{x}',
        \qquad
        L_\mathrm{z} \coloneqq R^{-\top} L_\mathrm{z}'.
        \label{eq:css_codes_maximum_trans_logical_group_psd_logical_basis_2}
    \end{equation}

    Above, we have chosen $R$ such that $R u_i=e_{2i-1}$ and $R^{-\top}u_i=e_{2i}$. In the CSS logical basis $\mathcal{L}$, the lower-triangular $\mathcal{U}_\mathrm{z}(2s, \mathbb{F}_2)$ sector is supported on the even logical coordinates $\mathcal C_\mathrm{z} = \{2,4,\ldots,2s\}$, while the upper-triangular $\mathcal{U}_\mathrm{x}(2s, \mathbb{F}_2)$ sector is supported on the odd logical coordinates $\mathcal C_\mathrm{x} = \{1,3,\ldots,2s-1\}$. If \(k\) is odd, the remaining logical qubit \(2s+1\) is inert under transversal Clifford gates. Indeed, let us first consider $\mathcal C_\mathrm{z}$. One can check, by multiplying out the symplectic matrices of the gates with the stabilizer generator matrices and logical matrices, that the following hold:
    \begin{itemize}
        \item For each \(1 \leq i \leq s\), applying a physical \(\smash{S_{z_{(i,i)}}}\) gate induces the logical action
        \begin{equation}
            \begin{pmatrix}
                I & 0 \\
                A' & I
            \end{pmatrix} 
            \in \Sp(2k, \mathbb{F}_2),
            \qquad            
            A'=u_i u_i^\top,
        \end{equation}
        in the logical basis $\mathcal{L}'$. After the logical basis rotation, this becomes $A = e_{2i}e_{2i}^\top$ in $\mathcal{L}$. Hence this physical gate implements the $S_{2i}$ logical gate.
        
        \item For \(1 \leq i < j \leq s\), applying physical \(\smash{S_{z_{(i,j)}}S_{z_{(i,i)}}S_{z_{(j,j)}}}\) induces the logical action
        \begin{equation}
            A'
            =
            (u_i+u_j)(u_i+u_j)^\top
            +
            u_i u_i^\top
            +
            u_j u_j^\top
            =
            u_i u_j^\top+u_j u_i^\top,
        \end{equation}
        in the logical basis $\mathcal{L}'$. After the logical basis rotation, this becomes $A = e_{2i}e_{2j}^\top+e_{2j}e_{2i}^\top$ in $\mathcal{L}$. Hence this physical gate implements the $\mathrm{CZ}_{(2i)(2j)}$ logical gate.
    \end{itemize}

    Similarly, one can check that the following holds on $\mathcal C_\mathrm{x}$.
    \begin{itemize}
        \item For each \(1 \leq i \leq s\), applying a physical \(\smash{(HSH)_{x_{(i,i)}}}\) gate induces the logical action
        \begin{equation}
            \begin{pmatrix}
                I & B' \\
                0 & I
            \end{pmatrix}
             \in \Sp(2k, \mathbb{F}_2),
            \qquad            
            B'=u_i u_i^\top,
        \end{equation}
        in the logical basis $\mathcal{L}'$. After the logical basis rotation, this becomes $B = e_{2i-1}e_{2i-1}^\top$ in $\mathcal{L}$. Hence this physical gate implements the $(HSH)_{2i-1}$ logical gate.
        
        \item For \(1 \leq i < j \leq s\), applying physical \(\smash{(HSH)_{x_{(i,j)}}(HSH)_{x_{(i,i)}}(HSH)_{x_{(j,j)}}}\) induces the logical action
        \begin{equation}
            B'
            =
            (u_i+u_j)(u_i+u_j)^\top
            +
            u_i u_i^\top
            +
            u_j u_j^\top
            =
            u_i u_j^\top+u_j u_i^\top,
        \end{equation}
        in the logical basis $\mathcal{L}'$. After the logical basis rotation, this becomes $B = e_{2i-1}e_{2j-1}^\top+e_{2j-1}e_{2i-1}^\top$ in $\mathcal{L}$. Hence this physical gate implements the $(H^{\otimes 2} \mathrm{CZ} H^{\otimes 2})_{(2i-1)(2j-1)}$ logical gate.
    \end{itemize}
\end{construction}

\clearpage

\subsection{Stabilizer codes}
\label{app:constructions/stab}

\subsubsection{Small non-CSS codes}

\begin{construction}
    [Unique minimum-$n$ code hosting permutation logical $S$]
    \label{cons:non_css_code_permutation_logical_s}
    Consider the $\db{4,1,2}$ non-CSS stabilizer code with stabilizer group\footnote{For readability, Pauli strings in this section are typeset in a monospaced font.}
    \begin{equation}
        \mathcal{S}
        \coloneqq
        \left\langle
        s_1 \coloneqq  \texttt{IXYZ},
        \,\,
        s_2 \coloneqq  \texttt{XYZI},
        \,\,
        s_3 \coloneqq  \texttt{YZIX}
        \right\rangle.
    \end{equation}
    
    The fourth cyclic shift is linearly dependent on the above generators and is given by $s \coloneqq  \texttt{ZIXY} = s_1s_2s_3$. Consider the logical basis
    \begin{equation}
        \overline{X} \coloneqq \texttt{IIZX},
        \qquad
        \overline{Z} \coloneqq \texttt{ZZZZ}.
    \end{equation}

    Then the following permutation and automorphism gates induce $S$ and $H$ logical gates, respectively:
    \begin{enumerate}
        \item $U_\lambda$ where $\lambda = (4321)$. The conjugation action on the logical Paulis are:
        \begin{equation}\begin{split}
            \overline{X} = \texttt{IIZX}
            \longrightarrow 
            \texttt{IZXI} = s\overline{Y}, 
            \qquad
            \overline{Z} = \texttt{ZZZZ} 
            \longrightarrow 
            \texttt{ZZZZ} = \overline{Z},
        \end{split}\end{equation}
        Hence, $U_\lambda$ implements a logical $S$ gate.
    
        \item $U_\tau (SH\otimes S^\dagger\otimes HS^\dagger H\otimes HS^\dagger)$ where $\tau = (1243)$. The conjugation action on the logical Paulis are:
            \begin{equation}
            \begin{split}
                \overline{X} = \texttt{IIZX} 
                \longrightarrow 
                \texttt{YIYI} = (s_1s_2)\overline{Z},
                \qquad
                \overline{Z} = \texttt{ZZZZ} 
                \longrightarrow 
                \texttt{YYXZ} = (s_1s_3)\overline{X}.
            \end{split}            
            \end{equation}
        Hence, $U_\tau (SH\otimes S^\dagger\otimes HS^\dagger H\otimes HS^\dagger)$ implements a logical $H$ gate.
    \end{enumerate}
    
    This code is unique in the following ways. No other $\db{\leq 4,1}$ stabilizer code up to $\mathrm{LC\Pi}$ code equivalence (i.e.~allowing single-qubit Clifford deformations and qubit permutations) has either the property that a qubit permutation implements a logical $S$ gate up to global phase, or has nontrivial distance ($d>1$) and the property that automorphisms implement the full logical Clifford group. Moreover, no $\db{4, \geq 2}$ stabilizer code exists that host addressable permutation logical $S$ gates. These claims are verified by exhaustive search.
\end{construction}

\begin{example}
    [Unique minimum-$n$ code with trivial transversal and permutation but nontrivial automorphism logical groups]
    \label{ex:stab_code_auto_semidirect_product_counterexample}
    Consider Clifford-deforming the non-CSS stabilizer code in \cref{cons:non_css_code_permutation_logical_s} by $HS^\dagger HZ\otimes HS^\dagger H\otimes HS^\dagger \otimes S^\dagger$, to obtain the code with stabilizer group
    \begin{equation}
        \mathcal{S}
        \coloneqq
        \left\langle
        \texttt{IXZZ},
        \,\,
        \texttt{XZXI},
        \,\,
        \texttt{YIYX}
        \right\rangle.
    \end{equation}

    By definition, this code is LC-equivalent to \cref{cons:non_css_code_permutation_logical_s}. It has trivial transversal and permutation logical groups, but the full Clifford automorphism logical group. This does not contradict \cref{cons:non_css_code_permutation_logical_s}---which has trivial transversal, $C_2$ permutation, and full-Clifford automorphism logical groups---because while transversal and automorphism logical groups are invariant under LC-deformation of codes, the permutation logical group is not. Indeed, this code illustrates that the product structure of logical groups of CSS codes, either $\mathrm{Aut}_\mathcal{L}(\mathcal{C}) = \mathrm{Trans}_\mathcal{L}(\mathcal{C}) \rtimes \mathrm{Perm}_\mathcal{L}(\mathcal{C})$ or $\mathrm{Aut}_\mathcal{L}(\mathcal{C}) = \mathrm{Trans}_\mathcal{L}(\mathcal{C}) \rtimes \mathrm{Perm}_\mathcal{L}(\mathcal{C}).2$ (see \cref{thm:css_codes_auto_logical_groups}), does not hold for stabilizer codes in general.
    
    This code is unique in the sense that no $n < 4$ stabilizer code has trivial transversal logical group; and no other $n = 4$ code has trivial transversal but nontrivial automorphism logical group, up to $\mathrm{LC\Pi}$ code equivalence.
\end{example}

\begin{construction}
    [Unique minimum-$n$ codes hosting permutation logical $H$]
    \label{cons:non_css_code_permutation_logical_h}
    We present three $\db{8,1}$ non-CSS codes:
    \begin{itemize}
        
        \item First $\db{8,1,2}$ code. Take two copies of the $\db{4,1,2}$ non-CSS stabilizer code in \cref{cons:non_css_code_permutation_logical_s} and denote them codeblocks $a$ and $b$. Their stabilizer groups are $\mathcal{S}_a$ and $\mathcal{S}_b$, and their logical operators are $\overline{X}_a, \overline{Y}_a, \overline{Z}_a$ and $\overline{X}_b, \overline{Y}_b, \overline{Z}_b$. We add one extra stabilizer tying the two codeblocks together, $h \coloneqq -\overline{Z}_a \overline{Z}_b$. Thus the stabilizer group of the resulting $\db{8,1,2}$ code is
        \begin{equation}
            \left\langle
                \mathcal{S}_a,
                \,\,
                \mathcal{S}_b,
                \,\,
                h
            \right\rangle.
        \end{equation}
        
        Consider the logical basis
        \begin{equation}
            \overline{X}
            \coloneqq
            \overline{X}_a \overline{X}_b,
            \qquad
            \overline{Z}
            \coloneqq
            \overline{Y}_a\overline{X}_b .
        \end{equation}
    
        Let $\sigma_{ab}$ denote the physical permutation that swaps the two codeblocks $a$ and $b$. Then the qubit permutation $\pi \coloneqq \sigma_{ab} \circ \left(I_a \otimes \lambda_b\right)$, where $\lambda_b$ is the cyclic shift $\lambda=(4321)$ as in \cref{cons:non_css_code_permutation_logical_s} applied first to codeblock $b$ and then the two codeblocks swapped, implements a logical Hadamard.

        \item Second $\db{8,1,2}$ code. Let $\rho \coloneqq (12345678)$, and $g \coloneqq \texttt{IIXXIIXX}$ and $h \coloneqq \texttt{IIZIXXYX}$. We define our code by the stabilizer group by the cyclic shifts of $g$ and $h$,
        \begin{equation}
            \left\langle
                -\rho^m(g), \rho^m(h)
                :
                m \in \{0, \ldots, 7\}
            \right\rangle
            =
            \left\langle
            \begin{array}{c}
                -\texttt{IIXXIIXX}, \,\,
                -\texttt{XIIXXIIX}, \,\,
                -\texttt{XXIIXXII}, \\
                \texttt{IIZIXXYX}, \,\,
                \texttt{XIIZIXXY}, \\
                \texttt{YXIIZIXX}, \,\,
                \texttt{XYXIIZIX}
                \end{array}
            \right\rangle.
        \end{equation}

        Then in the logical basis
        \begin{equation}
            \overline X\coloneqq \texttt{ZYZYIIII},
            \qquad
            \overline Z\coloneqq \texttt{IZYZYIII} = \rho(\overline X),
        \end{equation}
        the qubit permutation $\rho$ implements a logical Hadamard.

        \item $\db{8,1,3}$ code. Let $\rho \coloneqq (12345678)$, and $g \coloneqq \texttt{IIXXIYXZ}$ and $h \coloneqq -\texttt{XXXXXXXX}$. We define our code by the stabilizer group by the cyclic shifts of $g$ and the unshifted $h$,
        \begin{equation}
            \left\langle
                \rho^m(g), h
                :
                m \in \{0, \ldots, 5\}
            \right\rangle.
        \end{equation}

        The remaining two cyclic shifts of $g$ are not independent. Then in the logical basis
        \begin{equation}
            \overline X\coloneqq \texttt{IIIXYIIZ},
            \qquad
            \overline Z\coloneqq \texttt{ZIIIXYII} = \rho(\overline X),
        \end{equation}
        the qubit permutation $\rho$ implements a logical Hadamard.
        
    \end{itemize}

    No $\db{<8,1}$ stabilizer code has the property that a qubit permutation implements exactly a logical $H$ gate up to global phase. Moreover, these are the only $\db{8,1}$ stabilizer codes  with such a property up to $\mathrm{LC\Pi}$ code equivalence.
\end{construction}

\subsubsection{Non-CSS codes achieving maximum permutation logical group}

\begin{construction}
    [Non-CSS codes supporting all logical $S$ and $\mathrm{CX}$ gates through permutations]
    \label{cons:non_css_code_permutation_all_logical_s_and_cx}
    Concatenate any stabilizer code that supports all logical $\mathrm{CX}$ gates through physical qubit permutations and logical $S$ gates through single-qubit $I, S, S^\dag$ gates (e.g.~the CSS codes in \cref{cons:concatenated_phantom_clifford}) at the outer level with the non-CSS code of \cref{cons:non_css_code_permutation_logical_s} at the inner level. The inner code implements the required $S, S^\dag$ gates through qubit permutations. Then the resulting code supports all logical $S$ and $\mathrm{CX}$ gates through physical qubit permutations.
\end{construction}

\subsubsection{Stabilizer \texorpdfstring{$k = 2$}{k=2} codes achieving maximum automorphism logical group}

\begin{example}
    [Trivial $\db{2,2,1}$ code]
    \label{cons:non_css_code_auto_22_general_linear}
    Two bare physical qubits trivially has the automorphism logical group $(\mathrm{PCl}_1 \times \mathrm{PCl}_1) \rtimes S_2$, corresponding to Cliffords on each physical qubit and a swap of the two qubits. This group is isomorphic to $S_3 \wr C_2 \cong O^+(4, \mathbb{F}_2)$. 
    In fact, the code can realize the logical gate group $\langle \mathbf{CX}_2 \rangle$ through automorphisms. For this to occur, we must take a nonseparable logical basis that involves Pauli products on both physical qubits. Consider:
    \begin{equation}
        \overline X_1=\texttt{ZZ},\qquad
        \overline X_2=\texttt{XX},\qquad
        \overline Z_1=\texttt{ZY},\qquad
        \overline Z_2=\texttt{YX}.
    \end{equation}

    Then we have the following logical gate implementations:
    \begin{equation}\begin{split}
        \mathrm{CX}_{12}
        &: \qquad
        \mathrm{SWAP} (S \otimes S),
        \\
        \mathrm{CX}_{21}
        &: \qquad
        \mathrm{SWAP} (HSH \otimes HSH)
        \\
        \mathrm{H}_1
        &: \qquad
        S \otimes I.
    \end{split}\end{equation}
\end{example}

\begin{example}
    [Unique minimum-$n$ stabilizer codes hosting $O^+(4, \mathbb{F}_2)$ automorphism logical group]
    \label{cons:non_css_code_auto_k_eq_2_h1_h2_cz}
    We present in \cref{tab:non_css_code_auto_k_eq_2_h1_h2_cz} a $\db{6,2,2}$ and $\db{9,2,3}$ non-CSS code, which are the unique smallest $k=2$ stabilizer codes that achieve the $O^+(4, \mathbb{F}_2)$ automorphism logical group at their distances, up to $\mathrm{LC\Pi}$ code equivalence. The unique smallest $d=1$ code is the $\db{2,2,1}$ trivial code in \cref{cons:non_css_code_auto_22_general_linear}.
        
    \begin{table}
        \begin{tabular}{p{1.7cm} p{2.5cm} p{3.2cm} p{2cm} p{3cm} p{4.7cm}}
            \toprule
            Code
            & Stab.~gens.
            & Logical basis
            & Gate
            & $\lambda$
            & Local Clifford layer
            \\
            \midrule
            \multirow{6}{*}[-8pt]{$\db{6,2,2}$}
            & \multirow{6}{*}[-8pt]{%
                $\begin{aligned}
                    & \texttt{ZIZIZZ} \\[-3pt]
                    & \texttt{XZZIIX} \\[-3pt]
                    & \texttt{IIXZXZ} \\[-3pt]
                    & \texttt{IXXXIX}
                \end{aligned}$
            }
            & \multirow{3}{*}{%
                $\begin{aligned}
                    \overline{X}_1 &= \texttt{ZXIIII} \\[-5pt]
                    \overline{X}_2 &= \texttt{XYXZII} \\[-5pt]
                    \overline{Z}_1 &= \texttt{YIYZZI} \\[-5pt]
                    \overline{Z}_2 &= \texttt{YZYIZI}
                \end{aligned}$
            }
            & $H_1$ 
            & $(25)$ 
            & $(YH)_1 X_3 H_4 H_6$
            \\
            \cmidrule{4-6}
            &
            &
            & $H_2$ 
            & $(46)$ 
            & $(HS^\dagger H)_2 (ZHS^\dagger H)_3 Y_4 (HS^\dagger H)_5$
            \\
            \cmidrule{4-6}
            &
            &
            & $\mathrm{CZ}$
            & $(15)(24)(36)$ 
            & $Y_3Y_6$
            \\
            \cmidrule{3-6}
            &
            &
            \multirow{3}{*}{%
                $\begin{aligned}
                    \overline{X}_1 &= \texttt{IIZZZI} \\[-5pt]
                    \overline{X}_2 &= \texttt{IZZIZI} \\[-5pt]
                    \overline{Z}_1 &= \texttt{XXXIII} \\[-5pt]
                    \overline{Z}_2 &= \texttt{ZXZZZI}
                \end{aligned}$
            }
            & $H_1$
            & $(25)$ 
            & $H_1H_4H_6$
            \\
            \cmidrule{4-6}
            &
            &
            & $S_1$
            & $(23)$ 
            & $(HSH)_1(HSH)_4(HS^\dagger H)_6$
            \\
            \cmidrule{4-6}
            &
            &
            & $\mathrm{SWAP}$
            & $(15)(24)(36)$ 
            & $I$
            \\
            \midrule
            \multirow{6}{*}[-8pt]{$\db{9,2,3}$}
            & \multirow{6}{*}[-8pt]{%
                $\begin{aligned}
                    & \texttt{XZIIYXIII} \\[-3pt]
                    & \texttt{IIXZIIYXI} \\[-3pt]
                    & \texttt{YXIIIXIIZ} \\[-3pt]
                    & \texttt{IIYXIIIXZ} \\[-3pt]
                    & \texttt{YYIIZYIII} \\[-3pt]
                    & \texttt{IIYYIIZYI} \\[-3pt]
                    & \texttt{XXXXZIZIX}
                \end{aligned}$
            }
            & \multirow{3}{*}{%
                $\begin{aligned}
                    \overline X_1 &= \texttt{IIZIYXIZI} \\[-5pt]
                    \overline X_2 &= \texttt{IYIIIYYXI} \\[-5pt]
                    \overline Z_1 &= \texttt{IXIXXIXII} \\[-5pt]
                    \overline Z_2 &= \texttt{IYIYIYIYI} 
                \end{aligned}$
            }
            & $H_1$
            & $(26)$
            & $H_1 Z_2 X_4 S_5 X_7 S_9$
            \\
            \cmidrule{4-6}
            &
            &
            & $H_2$
            & $(47)$
            & $S_3 X_4 X_5 Z_6 S_8^\dagger (XS)_9$
            \\
            \cmidrule{4-6}
            &
            &
            & $\mathrm{CZ}$
            & $(13)(24)(57)(68)$
            & $I$
            \\
            \cmidrule{3-6}
            &
            &
            \multirow{3}{*}{%
                $\begin{aligned}
                    \overline X_1 &= \texttt{YXIIYIIII} \\[-5pt]
                    \overline X_2 &= \texttt{IIYXIIYII} \\[-5pt]
                    \overline Z_1 &= \texttt{XXIIIYIII} \\[-5pt]
                    \overline Z_2 &= \texttt{IIXXIIIYI} 
                \end{aligned}$
            }
            & $H_1$
            & $(26)$
            & $H_1 Z_2 S_5 S_9^\dagger$
            \\
            \cmidrule{4-6}
            &
            &
            & $S_1$
            & $(25)$
            & $S_1^\dagger S_6^\dagger S_9^\dagger$
            \\
            \cmidrule{4-6}
            &
            &
            & $\mathrm{SWAP}$
            & $(13)(24)(57)(68)$
            & $I$
            \\
            \bottomrule
        \end{tabular}
        \caption{
            Minimum-$n$ non-CSS stabilizer codes hosting $O^+(4, \mathbb{F}_2)$ automorphism logical group. 
            All logical gate implementations are of the form $\smash{\overline{U} = U_\lambda (U_1 \otimes \cdots \otimes U_n)}$; we use the convention for writing permutations as defined in \cref{sec:constructions/preliminaries/permutations}. 
            The $O^+(4, \mathbb{F}_2)$ logical group has two natural gate set realizations: $O^+(4, \mathbb{F}_2) \cong \langle H_1, H_2, \mathrm{CZ}_{12}\rangle \cong \langle H_1, S_1, \mathrm{SWAP}_{12} \rangle$. 
            They are related by Clifford conjugation, and a code supporting one can be made to support the other by a change of logical basis. 
            We show logical bases and logical gate implementations for both.
        }
        \label{tab:non_css_code_auto_k_eq_2_h1_h2_cz}
    \end{table}
\end{example}

\subsection{Block length bounds at higher Clifford hierarchy levels}

\begin{theorem}
    [Block lengths of stabilizer codes with \(\smash{\mathbf D_k^{(t)}}\) logical group via transversal diagonal gates]
    \label{thm:stab_codes_trans_any_clifford_hierarchy_level_n_scaling}
    Let $\mathcal{C}$ be an $\db{n,k}$ stabilizer code and fix a Clifford hierarchy level $t \leq k$.
    Suppose that every element of \(\mathbf D_k^{(t)}\) is realized, modulo logical \(Z\)-type Paulis, by the logical action of a code-preserving tensor product of arbitrary physical diagonal single-qubit gates in some logical basis of $\mathcal{C}$.
    Then
    \begin{equation}
        n
        \geq
        \sum_{s=1}^{t}\binom{k}{s}.
        \label{eq:stab_codes_trans_any_clifford_hierarchy_level_n_scaling_1}
    \end{equation}
    Equality can be achieved at \(d=1\).
\end{theorem}

\begin{proof}
    For \(t=1\), the claimed bound is \(n\geq k\), which holds for every \(\db{n,k}\) stabilizer code. 
    For \(t\geq2\), by \cref{cor:diagonal_clifford_hierarchy_group_structure}, we have
    \begin{equation}
        \Delta\left(\mathbf D_k^{(t)}\right)
        =
        \sum_{s=1}^{t}\binom{k}{s}.
        \label{eq:stab_codes_trans_any_clifford_hierarchy_level_n_scaling_2}
    \end{equation}

    Now consider the group of code-preserving physical diagonal transversal gates modulo global phase, which is a closed subgroup of the \(n\)-torus
    \begin{equation}
        \mathbb T^n
        =
        \left\{
            \bigotimes_{j=1}^n \diag(1,e^{i\theta_j})
            :
            \theta_j\in\mathbb R/2\pi\mathbb Z
        \right\}.
    \end{equation}

    Fix the logical basis in which the hypothesis is stated.
    Let \(K\) be the closed subgroup consisting of those gates whose induced logical action lies in \(\widetilde{\mathbf D}_k^{(t)}\).
    By hypothesis, after taking these logical actions modulo logical \(Z\)-type Paulis, \(K\) maps onto \(\mathbf D_k^{(t)}\).
    Let \(K^\circ\) be the identity component of \(K\), namely the connected component containing the identity gate, and let \(K/K^\circ\) be its component group, whose elements correspond to the connected components of \(K\).
    Since \(K^\circ\) is connected while \(\widetilde{\mathbf D}_k^{(t)}\) is finite, \(K^\circ\) has trivial induced logical action, and hence \(\mathbf D_k^{(t)}\) is a quotient of the finite component group \(K/K^\circ\).

    A closed subgroup of an \(n\)-torus has component group generated by at most \(n\) elements~\cite[Thm.~9.2.2 and Cor.~9.4.3]{aussenhofer2022topological}.
    Therefore every finite quotient of its component group is also generated by at most \(n\) elements, and hence $\Delta\left(\mathbf D_k^{(t)}\right) \leq n$. 
    Combining this bound with \cref{eq:stab_codes_trans_any_clifford_hierarchy_level_n_scaling_2} gives \cref{eq:stab_codes_trans_any_clifford_hierarchy_level_n_scaling_1} for $t \geq 2$.
    Equality at \(d=1\) is achieved, for example, by \cref{cons:complete_hypergraph_css_generalized}.
\end{proof}

\begin{remark}
    [\cref{thm:stab_codes_trans_saturating_n_scaling} vs.~\cref{thm:stab_codes_trans_any_clifford_hierarchy_level_n_scaling}]
    \cref{thm:stab_codes_trans_any_clifford_hierarchy_level_n_scaling} yields $n \geq k(k+1)/2$ at the $t = 2$ Clifford level, where \(\smash{\mathbf D_k^{(2)} \cong \mathcal{U}(2k, \mathbb{F}_2)}\), exactly matching the bound in \cref{thm:stab_codes_trans_saturating_n_scaling}. But \cref{thm:stab_codes_trans_saturating_n_scaling} is a stronger result: while \cref{thm:stab_codes_trans_any_clifford_hierarchy_level_n_scaling} assumes that the logical gates are implemented by physical diagonal single-qubit gates, \cref{thm:stab_codes_trans_saturating_n_scaling} does not assume diagonality. On the other hand, \cref{thm:stab_codes_trans_any_clifford_hierarchy_level_n_scaling} applies to all $t \geq 2$ but \cref{thm:stab_codes_trans_saturating_n_scaling} treats only the Clifford level.
\end{remark}

\clearpage

\section{Universality}

To discuss the universality of gate sets, it is necessary to introduce additional notation. First, we account explicitly for Paulis rather than quotienting them out, as had been the case for much of the Clifford discussion in this manuscript. We define the $k$-qubit Clifford group modulo global phases, $\mathrm{Cl}_k^\circ  \coloneqq \mathrm{Cl}_k / U(1)$, not to be confused with $\mathrm{PCl}_k$ which is modulo Paulis and global phases. Following convention, we define $\PU(2^k)$ to be the unitary group on $k$ qubits modulo global phases. Second, for a group $G$, we denote its topological closure by $\overline{G}$.

\begin{lemma}
    [Single-qubit $D_4$ extension]
    \label{lem:universality_single_qubit_d4_extension}
    Let $V\in\PU(2)$ and $T\coloneqq\operatorname{diag}(1,e^{i\pi/4})$. 
    If $V \notin \mathrm{Cl}_1^\circ $ and $VZV^\dagger\neq\pm Z$, then $\overline{\langle X,S,V\rangle}$ is either $\PU(2)$ or $T \mathrm{Cl}_1^\circ  T^\dagger$.
\end{lemma}

\begin{proof}
    This follows from the classification of closed subgroups of $\PU(2)\cong \SO(3)$~\cite{olive2019effective}. Up to conjugacy, every nontrivial proper closed subgroup is either an axis-preserving subgroup, namely the rotations about a fixed axis or their normalizer preserving that axis as an unoriented line, a finite cyclic or rotational dihedral group, or one of three polyhedral rotation groups: the tetrahedral group of order $12$, octahedral group of order $24$, or icosahedral group of order $60$.
    
    Now $\langle X,S\rangle\cong D_4$ contains the quarter-turn $S$ about the $z$-axis. Any proper infinite closed subgroup of $\PU(2)$ containing $\langle X,S\rangle$ must therefore preserve the $z$-axis as an unoriented line. The same holds for any cyclic or rotational dihedral overgroup of $\langle X,S\rangle$, since in such a group an element of order four must rotate about the distinguished axis. Hence every element $U$ of any such overgroup satisfies $UZU^\dagger=\pm Z$, but these possibilities are excluded by $VZV^\dagger\neq\pm Z$. The tetrahedral and icosahedral rotation groups contain no element of order four, so they cannot contain $\langle X,S\rangle$. Hence the only remaining proper possibility is an octahedral rotation group.

    There are precisely two octahedral rotation groups containing the fixed subgroup $\langle X,S\rangle$. Indeed, let $O_0=\mathrm{Cl}_1^\circ $ be the standard octahedral rotation group. Since $S$ is a quarter-turn about the $z$-axis, the $z$-axis must be a fourfold axis of any octahedral rotation group containing $S$, while its other two fourfold axes form an orthogonal pair in the $xy$-plane. Hence every such group is of the form $O_\theta=R_z(\theta)O_0R_z(-\theta)$ for some $\theta$. The axes in the $xy$-plane about which $O_\theta$ contains half-turns occur at angles $\theta+j\pi/4$, with $j\in\{0,1,2,3\}$. Requiring the group also to contain $X$, the half-turn about the $x$-axis, therefore forces $\theta\in(\pi/4)\mathbb Z$. Since $O_{\theta+\pi/2}=O_\theta$, there are exactly two possibilities, represented by $\theta=0$ and $\theta=\pi/4$. These are respectively the Clifford group modulo global phases $\mathrm{Cl}_1^\circ $ and the rotated group $T\mathrm{Cl}_1^\circ T^\dagger$. The first is excluded by $V\notin\mathrm{Cl}_1^\circ$, proving the claim.
\end{proof}

\begin{theorem}
    [Universal single-qubit extensions of the $\langle \mathbf{X}_k, \mathbf{S}_k ,\mathbf{CX}_k \rangle$ gate set]
    \label{thm:universality_s_cx}
    Let $k\geq1$, and we work modulo global phases. For a single-qubit gate $V\in\PU(2)$ available on one arbitrary fixed qubit $i \in [k]$,
    \begin{equation}
        \overline{\langle \mathbf{X}_k, \mathbf{S}_k, \mathbf{CX}_k, V_i \rangle}
        =
        \PU(2^k)
        \quad\Longleftrightarrow\quad
        \begin{cases}
            VZV^\dagger\neq\pm Z, \;
            V\notin\mathrm{Cl}_1^\circ, \;
            V\notin T\mathrm{Cl}_1^\circ T^\dagger, \;
            & k=1,
            \\
            VZV^\dagger\neq\pm Z, \;
            V\notin\mathrm{Cl}_1^\circ, \;
            & k\geq2.
        \end{cases}
    \end{equation}
\end{theorem}

\begin{proof}
    We first prove necessity. If $V\in\mathrm{Cl}_1^\circ $, the generated gates remain Clifford, while if $VZV^\dagger=\pm Z$, every generator maps computational-basis states to computational-basis states up to phases, so the generated group cannot approximate a Hadamard gate; neither case is universal. For $k=1$, $V\in T\mathrm{Cl}_1^\circ T^\dagger$ is a further obstruction: since $\langle X,S\rangle\leq T\mathrm{Cl}_1^\circ T^\dagger$, adjoining such a $V$ leaves the generated group inside this finite group.

    Now we prove sufficiency. Suppose $V\notin\mathrm{Cl}_1^\circ $ and $VZV^\dagger\neq\pm Z$. By \cref{lem:universality_single_qubit_d4_extension}, $G_V\coloneqq\overline{\langle X,S,V\rangle}$ is either $\PU(2)$ or $T\mathrm{Cl}_1^\circ T^\dagger$. For $k=1$, the additional hypothesis excludes the latter, proving the claim. 

    Suppose now $k\geq2$. Since $\langle \mathbf{CX}_k \rangle$ furnishes all $\mathrm{SWAP}$ gates, $V$ is available on every qubit. If $G_V=\PU(2)$, arbitrary local unitaries together with $\langle \mathbf{CX}_k \rangle$ are universal~\cite{barenco1995elementary}. It remains to treat $G_V=T\mathrm{Cl}_1^\circ T^\dagger$. For analysis only, conjugate the generated group by $\Theta=T^{\otimes k}$---this is merely a change of basis and does not require $\Theta$ to be available. In the conjugated group we have all local Clifford gates, while $\mathrm{CX}_{ij}$ becomes $\mathrm{CA}_{ij}$, where the gate being controlled is $A \coloneqq T^\dagger X T$. Since the local gate $X_j$ is available and $AXA=-Y$, the group contains $\mathrm{CA}_{ij} X_j \mathrm{CA}_{ij} X_j=S_i\mathrm{CZ}_{ij}$. Hence it contains $\mathrm{CZ}_{ij}$, and therefore also $\mathrm{CX}_{ij}=H_j\mathrm{CZ}_{ij}H_j$, so it contains the full Clifford group on the $k$ qubits. Finally, $\mathrm{CA}_{ij}$ is non-Clifford, since $\smash{\mathrm{CA}_{ij} X_i \mathrm{CA}_{ij}^\dagger} = X_i A_j$ is not a Pauli operator. The Clifford group together with any non-Clifford unitary is universal~\cite{jozsa2013classical}, so the original group is universal as well.
\end{proof}

\begin{theorem}
    [Universal single-qubit extensions of the $\langle \mathbf{X}_k, \mathbf{S}_k ,\mathbf{CZ}_k \rangle$ gate set]
    \label{thm:universality_s_cz}
    Let $k\geq1$, and we work modulo global phases. For a single-qubit gate $V\in\PU(2)$ available at every qubit $i \in [k]$, denoted $\mathbf{V}_k$,
    \begin{equation}
        \overline{\langle \mathbf{X}_k, \mathbf{S}_k, \mathbf{CZ}_k, \mathbf{V}_k \rangle}
        =
        \PU(2^k)
        \quad\Longleftrightarrow\quad
        VZV^\dagger\neq\pm Z, \;
        V\notin\mathrm{Cl}_1^\circ, \;
        V\notin T\mathrm{Cl}_1^\circ T^\dagger.
    \end{equation}
\end{theorem}

\begin{proof}
    We first prove necessity. If $V\in\mathrm{Cl}_1^\circ $, the generated gates remain Clifford, while if $VZV^\dagger=\pm Z$, every generator maps computational-basis states to computational-basis states up to phases, so the generated group cannot approximate a Hadamard gate; neither case is universal.

    Lastly, suppose that $V\in T\mathrm{Cl}_1^\circ T^\dagger$. To analyze the generated group, make the fixed change of basis $\Theta\coloneqq T^{\otimes k}$---this does not require $\Theta$ to be an available gate. In this new basis, each $V_i$ becomes the Clifford gate $T^\dagger_i V_i T_i$ on qubit $i$, while $S_i$ and $\mathrm{CZ}_{ij}$ are unchanged because they commute with $\Theta$. Moreover, $X_i$ becomes $(T^\dagger X T)_i = e^{-i\pi/4} (XS)_i$ on qubit $i$, which is also Clifford. Hence $
        \Theta^\dagger
        \langle \mathbf{X}_k,\mathbf{S}_k,\mathbf{CZ}_k,\mathbf{V}_k\rangle
        \Theta
        \leq
        \mathrm{Cl}_k^\circ
    $.
    Thus the original generated group is contained in the conjugate finite group $\Theta\mathrm{Cl}_k^\circ\Theta^\dagger$ and cannot be universal. This proves necessity.

    Conversely, suppose $V\notin\mathrm{Cl}_1^\circ $, $VZV^\dagger\neq\pm Z$, and $V\notin T\mathrm{Cl}_1^\circ T^\dagger$. By \cref{lem:universality_single_qubit_d4_extension}, these conditions imply $\overline{\langle X,S,V\rangle}=\PU(2)$. Since the same gate $V$ is available independently on every qubit, $\overline{\langle \mathbf{X}_k, \mathbf{S}_k, \mathbf{CZ}_k, \mathbf{V}_k \rangle}$ therefore contains arbitrary single-qubit unitaries independently on every qubit. For $k=1$ this already proves the claim, while for $k\geq2$ arbitrary single-qubit unitaries together with all entangling $\mathrm{CZ}$ gates, which can be Hadamard-conjugated into  $\mathrm{CX}$ gates, are universal~\cite{barenco1995elementary}.
\end{proof}

\end{document}